\documentclass[11pt,a4paper]{article}

\usepackage[left=3cm,right=3cm,top=3cm,bottom=3cm]{geometry}
\usepackage[T1]{fontenc}
\usepackage[utf8]{inputenc}
\usepackage{microtype}

\usepackage{amsmath}
\usepackage{amssymb}
\usepackage{bbm}

\usepackage{graphicx}
\usepackage[table]{xcolor}
\usepackage{caption}
\usepackage{wrapfig}
\usepackage{pdflscape}
\usepackage{longtable}
\usepackage{multirow}
\usepackage{array}
\usepackage{ragged2e}
\usepackage{makecell}
\usepackage{booktabs}
\usepackage{siunitx}

\usepackage{tikz}
\usetikzlibrary{arrows.meta, positioning, fit, calc}

\usepackage{cite}
\usepackage{tocloft}
\usepackage{xurl}
\usepackage[hidelinks]{hyperref}

\usepackage[acronym,toc,nonumberlist,nogroupskip]{glossaries-extra}
\makeglossaries
\setabbreviationstyle[acronym]{long-short}
\newacronym{gpat}{GP-AT}{General Purpose Audio Tagging}
\newacronym{ev}{EV}{Emergency Vehicle}
\newacronym{at}{AT}{Audio Tagging}
\newacronym{seld}{SELD}{Sound Event Localization and Detection}
\newacronym{sed}{SED}{Sound Event Detection}
\newacronym{doa}{DOA}{Direction of Arrival}
\newacronym{dcase}{DCASE}{Detection and Classification of Acoustic Scenes and Events}
\newacronym{crnn}{CRNN}{Convolutional Recurrent Neural Network}
\newacronym{nsrp}{NSRP}{Neural Steered Response Power}
\newacronym{at2seld}{AT2SELD}{Audio-Tagging-to-Sound-Event-Localization-and-Detection}
\newacronym{ssl}{SSL}{Sound Source Localization}
\newacronym{sst}{SST}{Sound Source Tracking}
\newacronym{foa}{FOA}{First-Order Ambisonics}
\newacronym{srp}{SRP}{Steered Response Power}
\newacronym{accdoa}{ACCDOA}{Activity-Coupled Cartesian Direction of Arrival}
\newacronym{adpit}{ADPIT}{Auxiliary Duplicating Permutation Invariant Training}
\newacronym{pit}{PIT}{Permutation Invariant Training}
\newacronym{ein}{EIN}{Event-Independent Network}
\newacronym{mhsa}{MHSA}{Multi-Head Self-Attention}
\newacronym{bce}{BCE}{Binary Cross-Entropy}
\newacronym{mse}{MSE}{Mean Squared Error}

\newacronym{rnn}{RNN}{Recurrent Neural Network}
\newacronym{cnn}{CNN}{Convolutional Neural Network}
\newacronym{gru}{GRU}{Gated Recurrent Unit}
\newacronym{bigru}{BiGRU}{Bi-directional Gated Recurrent Unit}
\newacronym{dft}{DFT}{Discrete Fourier Transform}
\newacronym{relu}{ReLU}{Rectified Linear Unit}
\newacronym{itd}{ITD}{Interaural Time Difference}
\newacronym{ild}{ILD}{Interaural Level Difference}
\newacronym{gcc}{GCC}{Generalized Cross-Correlation}
\newacronym{gccphat}{GCC-PHAT}{Generalized Cross-Correlation with Phase Transform}
\newacronym{tdoa}{TDOA}{Time Difference of Arrival}
\newacronym{ansyn}{ANSYN}{Ambisonic, Anechoic and Synthetic Impulse Response}
\newacronym{mansyn}{MANSYN}{Moving Ambisonic, Anechoic and Synthetic Impulse Response}

\newacronym{lstm}{LSTM}{Long Short-Term Memory}
\newacronym{pirnn}{PI-RNN}{Permutation-Invariant Recurrent Neural Network}
\newacronym{mha}{MHA}{Multi-Head Attention}
\newacronym{mgu}{MGU}{Minimal Gated Unit}
\newacronym{ead}{EAD}{Event Activity Detection}
\newacronym{iv}{IV}{Intensity Vector}
\newacronym{stft}{STFT}{Short-Time Fourier Transform}

\newacronym{msca}{MSCA}{Multi-Scale Channel Attention}
\newacronym{asp}{ASP}{Attentive Statistics Pooling}
\newacronym{gap}{GAP}{Global Average Pooling}
\newacronym{ffn}{FFN}{Feed-Forward Network}
\newacronym{layernorm}{LayerNorm}{Layer Normalization}
\newacronym{acs}{ACS}{Audio Channel Swapping}
\newacronym{srir}{SRIR}{Spatial Room Impulse Response}
\newacronym{starss}{STARSS}{Sony-Tau Realistic Spatial Soundscapes}

\newacronym{maxcorr}{MaxCorr}{Maximum Cross Correlation}
\newacronym{tiou}{tIoU}{temporal Intersection over Union}
\newacronym{iir}{IIR}{Infinite Impulse Response}
\newacronym{mic}{MIC}{Microphone-array}
\newacronym{nas}{NAS}{Neural Architecture Search}

\newacronym{srpphat}{SRP-PHAT}{Steered Response Power with PHAse Transform}
\newacronym{motp}{MOTP}{Multiple Object Tracking Precision}
\newacronym{mota}{MOTA}{Multiple Object Tracking Accuracy}
\newacronym{snr}{SNR}{Signal-to-Noise Ratio}
\newacronym{rir}{RIR}{Room Impulse Response}
\newacronym{mlp}{MLP}{Multi-Layer Perceptron}
\newacronym{idft}{IDFT}{Inverse Discrete Fourier Transform}

\newacronym{epanns}{E-PANNs}{Efficient Pre-trained Audio Neural Networks}
\newacronym{tpit}{tPIT}{track-wise Permutation Invariant Training}
\newacronym{auroc}{AuROC}{Area under the Receiver Operating Characteristic curve}
\newacronym{yaml}{YAML}{YAML Ain't Markup Language}
\newacronym{cli}{CLI}{Command-Line Interface}
\newacronym{rms}{RMS}{Root Mean Square}
\newacronym{sma}{SMA}{Spherical Microphone Array}

\newacronym{mps}{MPS}{Metal Performance Shaders}
\newacronym{gflops}{GFLOPs}{Giga Floating-Point Operations}

\hypersetup{
    pdftitle={From General-Purpose Audio Tagging to Spatially Grounded Sound Event Localization and Detection},
    pdfauthor={Stefano Giacomelli, Stefano Damiano},
    pdfsubject={Sound Event Localization and Detection},
    pdfkeywords={Sound Event Localization and Detection (SELD), Audio Tagging (AT), Spatial Audio, Ambisonics, Computational Auditory Scene Analysis (CASA), Deep Learning (DL), Neural Networks (NN)}
}

\newcommand{\reporttitle}{From General-Purpose Audio Tagging to\\Spatially Grounded\\Sound Event Localization and Detection}
\newcommand{\reportsubtitle}{Technical Report}
\newcommand{\reportperiod}{March--May 2026}

\begin{document}

\pagenumbering{roman}

\begin{titlepage}
\centering

\begin{minipage}[c]{0.44\textwidth}
    \centering
    \includegraphics[height=2.1cm]{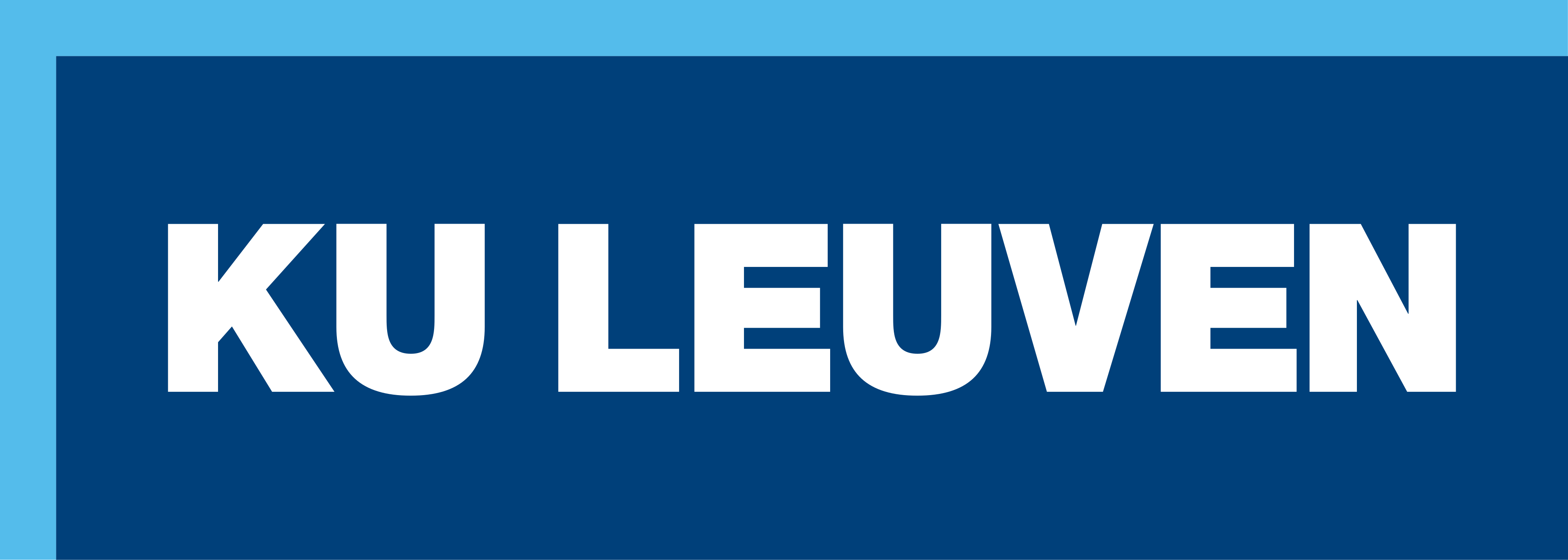}
\end{minipage}
\hfill
\begin{minipage}[c]{0.44\textwidth}
    \centering
    \includegraphics[height=2.9cm]{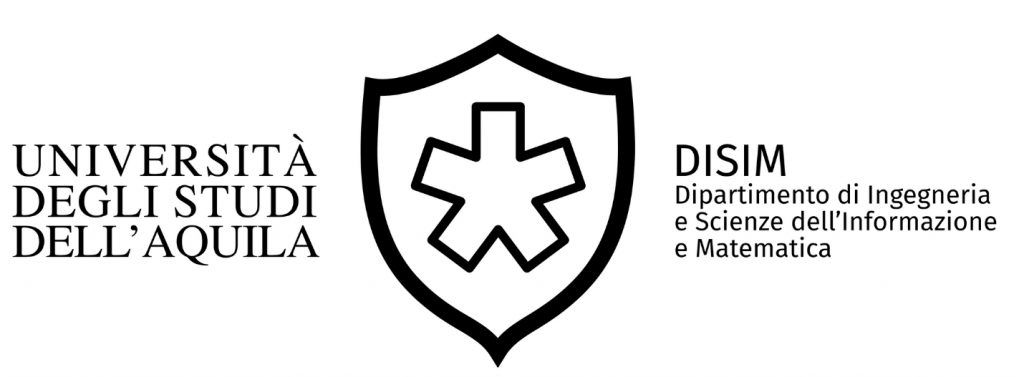}
\end{minipage}

\vspace{2.1cm}

{\LARGE\bfseries
\reporttitle
\par}

\vspace{0.75cm}

{\large\bfseries
\reportsubtitle
\par}

\vspace{1.55cm}

{\Large\bfseries
Stefano Giacomelli
\par}

\vspace{0.15cm}

{\small
DISIM Department, University of L'Aquila, Italy
\par}

\vspace{0.45cm}

{\Large\bfseries
Stefano Damiano
\par}

\vspace{0.15cm}

{\small
ESAT Department, KU Leuven, Belgium
\par}

\vspace{1.2cm}

{\large\bfseries
Scientific supervision
\par}

\vspace{0.35cm}

{\normalsize
\textbf{Claudia Rinaldi}\\
{\small CNIT, Research Unit of L'Aquila, Italy}
\par}

\vspace{0.25cm}

{\normalsize
\textbf{Fabio Graziosi}\\
{\small DISIM Department, University of L'Aquila, Italy}
\par}

\vspace{0.25cm}

{\normalsize
\textbf{Toon van Waterschoot}\\
{\small ESAT Department, KU Leuven, Belgium}
\par}

\vfill

{\normalsize
\textbf{ESAT - STADIUS Department, KU Leuven}\\
Audio Engineering Laboratory\\
\reportperiod
\par}

\vspace{0.65cm}

\begin{minipage}{0.88\textwidth}
\centering
{\footnotesize
This work was funded by the Italian Ministry of University and Research under Ministerial Decree No.~118/2023, within the Italian National Recovery and Resilience Plan (NRRP/PNRR), Mission~4, Component~1, Investment~4.1, ``PNRR Research'', CUP~E11I2300010000.
\par}
\end{minipage}

\end{titlepage}

\clearpage

\tableofcontents

\vfill
\noindent\textbf{GitHub Code Repository:} \url{https://github.com/StefanoGiacomelli/spatial_gpat}
\clearpage
\pagenumbering{arabic}

\section{Research Motivations and Scopes}
\label{sec:scope_logic}

\Gls{gpat} models pretrained on large-scale corpora such as AudioSet~\cite{audioset} have become effective semantic front-ends for downstream acoustic scene analysis. Their large vocabulary, broad acoustic coverage, and transferable representations make them suitable initialization points when target-domain annotations are limited or acoustically heterogeneous. In previous works on \gls{ev} detection, AudioSet-pretrained models were shown to provide robust semantic priors under realistic variability, and clip-level classifiers could be reinterpreted as practical temporal detectors by applying inference over successive analysis windows and concatenating the resulting predictions into segment-wise activity streams~\cite{giacomelli2026e2panns}. This strategy improves temporal usability without requiring a fully framewise architecture, but it remains intrinsically non-spatial: it does not estimate source direction, disambiguate simultaneous sources, or preserve source identity across time.

The present work investigates the next step in this progression: the extension of strong semantic \gls{at} backbones toward spatially grounded \gls{seld}. In this setting, acoustic scene understanding is no longer limited to detecting whether an event class is present, but requires estimating where each active source is located and how event activity evolves over time. The problem therefore requires an explicit interface between semantic evidence, multi-channel spatial cues, temporal context, and assignment mechanisms. This interface is central to \gls{seld}, where \gls{sed} and \gls{doa} estimation are jointly modeled within a single learning framework~\cite{adavanne2019seldnet}.

This transition is not a direct extension of \gls{at}. A window-based \gls{gpat} model provides local semantic evidence, but it does not impose temporal continuity beyond the analysis window and does not represent the spatial state of the scene. In dynamic or overlapping scenarios, the model must also distinguish whether multiple detections correspond to distinct sources, repeated observations of the same source, or competing assignments among available output tracks. Consequently, spatially grounded event analysis requires three additional capabilities: \textit{localization}, to associate event activity with source direction; \textit{temporal modeling}, to stabilize event and \gls{doa} estimates across consecutive frames; and \textit{assignment handling}, to manage concurrent sources without relying on arbitrary output ordering.

The \gls{dcase} community literature provides several complementary mechanisms for addressing these requirements. The original SELDnet formulation introduced a \gls{crnn} architecture with synchronized \gls{sed} and Cartesian \gls{doa} regression branches, thereby establishing a direct learnable association between class activity and source direction~\cite{adavanne2019seldnet}. Extensions to moving-source scenarios further showed that recurrent sequence modeling can support tracking-like behavior by smoothing event activity and localization trajectories over time~\cite{adavanne2019seldt_tracking}. More recent architectures refine this paradigm through permutation-aware recurrence, attention-based temporal modeling, Conformer blocks, activity-coupled Cartesian representations, and track-wise output spaces, improving the handling of temporal context, overlapping sources, and same-class spatial multiplicity~\cite{diazguerra2023pirnn,wang2022resnetconformer}.

At the same time, spatial learning remains constrained by the recording format and microphone geometry. Many neural localization systems operate on multi-channel time--frequency representations whose spatial interpretation is implicitly tied to a fixed array configuration. This limits portability across acquisition setups and motivates geometry-aware formulations in which spatial reasoning is conditioned on microphone placement or pairwise channel structure. Neural re-interpretations of classical localization operators, such as \gls{nsrp}, show that trainable systems can incorporate array metadata and multi-channel spatial cues while remaining compatible with arbitrary microphone geometries~\cite{grinstein2024neuralsrp}. These developments suggest that semantic transfer from \gls{gpat} should not be treated as an isolated initialization strategy, but as one component of a broader semantic--spatial architecture.

The scope of this report is therefore to study how pretrained semantic representations can be integrated into \gls{seld} systems and how their contribution interacts with spatial front-ends, temporal sequence models, output representations, and loss design. The objective is not to propose a single monolithic architecture in isolation, nor to reproduce a leaderboard-oriented comparison across all contemporary \gls{seld} systems. Instead, the work develops and evaluates a modular \gls{at2seld} framework in which semantic audio priors are combined with spatial processing and track-wise event localization, and complements this evaluation with a diagnostic analysis of the main failure modes that emerge during training and evaluation, including mitigation strategies and deployment-oriented metrics for interpreting their impact under practical operating conditions.

The main contributions are summarized as follows:
\begin{enumerate}
    \item a semantic-to-spatial \gls{seld} framework that reuses pretrained \gls{gpat} representations as high-level acoustic priors for localization-aware event analysis;

    \item a modular architecture search space combining semantic embeddings, spatial front-ends, temporal sequence modeling, and track-wise \gls{sed}/\gls{doa} output heads;

    \item an empirical comparison of candidate spatial and temporal modules under controlled training conditions, aimed at identifying robust architectural configurations rather than isolated benchmark scores;

    \item a diagnostic analysis of loss design and supervision strategies, with particular attention to inactive-target dominance in \gls{doa} regression and to activity-conditioned spatial supervision;

    \item an evaluation of calibration, threshold sensitivity, class imbalance, and dataset-dependent transfer effects, used to characterize the limits of semantic transfer in spatially grounded acoustic scene analysis.
\end{enumerate}

This organization positions \gls{seld} as a bridge between semantic audio understanding and spatial acoustic scene analysis. The following section formalizes the task taxonomy and output representations required for this bridge, distinguishing \gls{sed}, \gls{doa} estimation, \gls{seld}, \gls{ssl}, and \gls{sst} before introducing the representation choices adopted in the proposed framework.

\section{Task Taxonomy and Output Representations}
\label{sec:problem_taxonomy}

The extension from \gls{gpat} to spatially grounded acoustic scene analysis requires a precise separation between semantic detection, spatial estimation, and temporal association. A pretrained or fine-tuned \gls{gpat} model provides semantic evidence over an input segment or analysis window, but its output does not explicitly encode source direction, source identity, or trajectory continuity. These missing dimensions motivate a taxonomy in which \gls{sed}, \gls{doa} estimation, \gls{seld}, \gls{ssl}, and \gls{sst} are treated as related but non-equivalent formulations.

\Gls{sed} estimates the temporal activity of sound classes~\cite{sed_tutorial}. Given a set of \(C\) target classes and a frame index (\(t\)), the output is commonly represented as a multi-label activity vector:
\begin{equation}
\mathbf{y}^{\mathrm{SED}}_t =
[a_{1,t},\ldots,a_{C,t}]^\top
\in [0,1]^C
\end{equation}
where \(a_{c,t}\) denotes the confidence or posterior activity estimate for class \(c\) at frame \(t\), and \(C\) is the number of sound-event classes. This representation is naturally aligned with \gls{at} models operated in window-based mode, but it contains no spatial information and does not preserve source identity across adjacent frames or overlapping events.

\Gls{doa} estimation addresses the spatial component by estimating one or more source directions relative to a recording coordinate system. A direction can be represented in angular coordinates, using azimuth \(\phi\) and elevation \(\theta\), or in normalized Cartesian form:
\begin{equation}
\mathbf{r}_{i,t} =
[x_{i,t},y_{i,t},z_{i,t}]^\top,
\qquad
\|\mathbf{r}_{i,t}\|_2 = 1
\end{equation}
where \(\mathbf{r}_{i,t}\in\mathbb{R}^3\) is the unit Cartesian direction vector of source \(i\) at frame \(t\), and \(x_{i,t}\), \(y_{i,t}\), and \(z_{i,t}\) denote its 3D Cartesian components. Cartesian \gls{doa} regression is widely used in \gls{seld} because it avoids angular discontinuities and supports continuous optimization in Euclidean space. In contrast, localization-oriented formulations such as \gls{srp} may express the active sources directly as an unordered set of source positions:
\begin{equation}
U(t)=
\{\mathbf{u}_1(t),\ldots,\mathbf{u}_{N_t}(t)\}
\end{equation}
where \(U(t)\) is the set of active source positions at frame \(t\), \(\mathbf{u}_i(t)\) is the position or direction of source \(i\), and \(N_t\) is the number of active sources at that frame~\cite{grinstein2024neuralsrp}. This formulation clarifies the boundary between \gls{ssl}/\gls{sst} and \gls{seld}: \gls{ssl} estimates source positions, \gls{sst} preserves source trajectories over time, whereas \gls{seld} additionally associates these spatial estimates with semantic membership.

\begin{figure}[ht]
    \centering
    \includegraphics[width=0.52\linewidth]{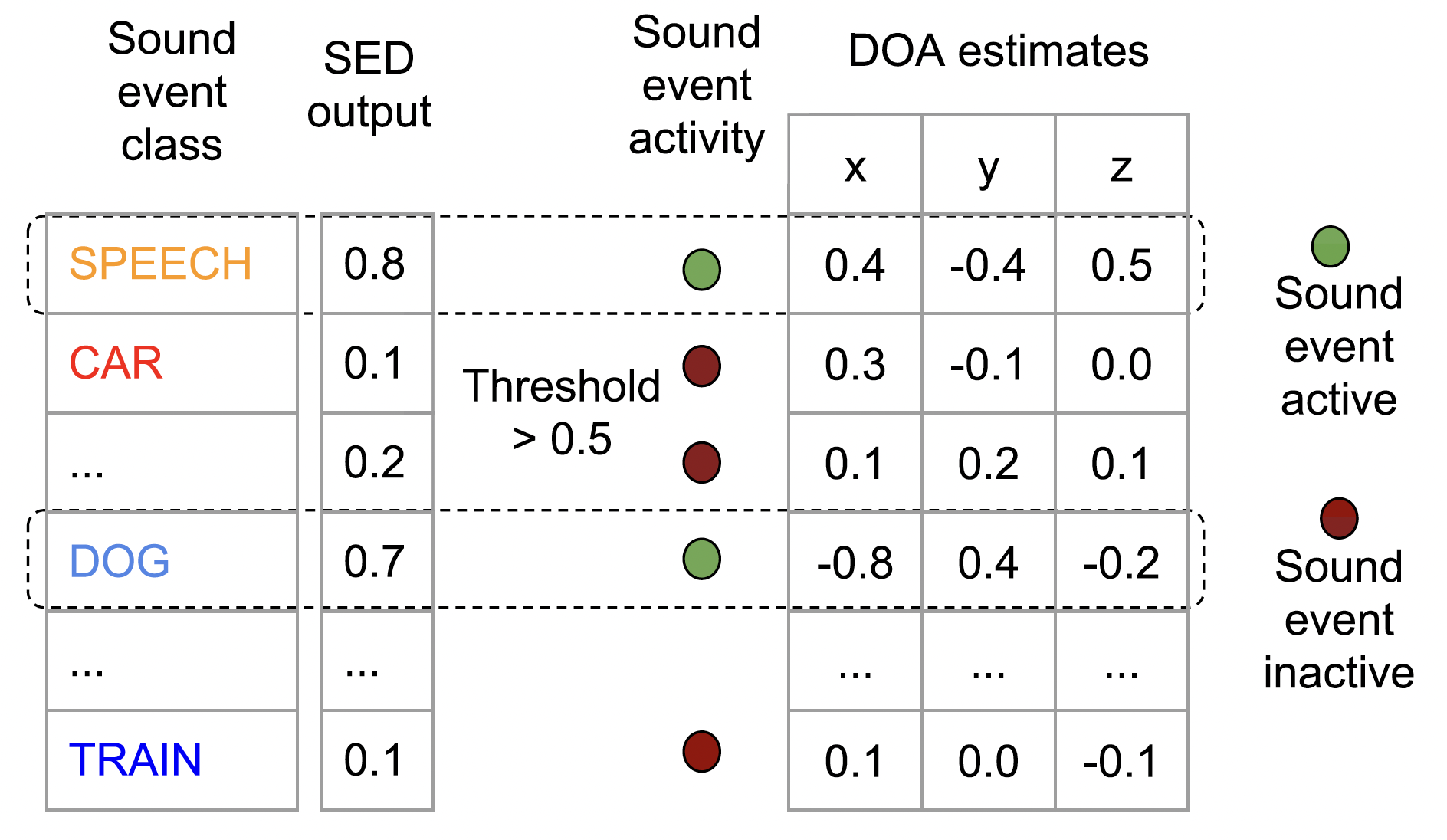}
    \caption{Class-wise two-branch \gls{seld} output format, with separate \gls{sed} activity estimates and Cartesian \gls{doa} vectors, adapted from S.~Adavanne et al., ``\textit{Sound Event Localization and Detection of Overlapping Sources Using Convolutional Recurrent Neural Networks}''~\cite{adavanne2019seldnet}.}
    \label{fig:seldnet_output}
\end{figure}

The original SELDnet formulation addresses joint semantic and spatial estimation through a two-branch output space composed of class-wise event activity and class-wise Cartesian \gls{doa} regression~\cite{adavanne2019seldnet}. For each frame \(t\), the output space can be written as:
\begin{equation}
\mathcal{Y}_{\mathrm{SELDnet}}
=
\left\{
\left(
\mathbf{y}^{\mathrm{SED}}_t,
\mathbf{Y}^{\mathrm{DOA}}_t
\right)
\,\middle|\,
\mathbf{y}^{\mathrm{SED}}_t \in [0,1]^C,\;
\mathbf{Y}^{\mathrm{DOA}}_t \in \mathbb{R}^{C\times 3}
\right\}
\end{equation}
where \(\mathbf{y}^{\mathrm{SED}}_t\) contains the framewise class activities, and \(\mathbf{Y}^{\mathrm{DOA}}_t\) contains one Cartesian \gls{doa} vector for each class. Equivalently, the \gls{doa} branch can be flattened as:
\begin{equation}
\mathbf{y}^{\mathrm{DOA}}_t =
[
(x_{1,t},y_{1,t},z_{1,t}),\ldots,
(x_{C,t},y_{C,t},z_{C,t})
]^\top
\in [-1,1]^{3C}
\end{equation}
where \((x_{c,t},y_{c,t},z_{c,t})\) denotes the Cartesian \gls{doa} estimate associated with class \(c\) at frame \(t\). This representation preserves a clean decomposition between activity estimation and spatial regression, typically optimized using \gls{bce} for the \gls{sed} branch and \gls{mse} for the \gls{doa} branch (Figure~\ref{fig:seldnet_output}). Its structural limitation is that only one \gls{doa} vector can be assigned to each class at a given frame, which makes the representation inadequate when multiple simultaneous sources of the same class are active at different locations.

\begin{figure}[ht]
    \centering
    \includegraphics[width=\linewidth]{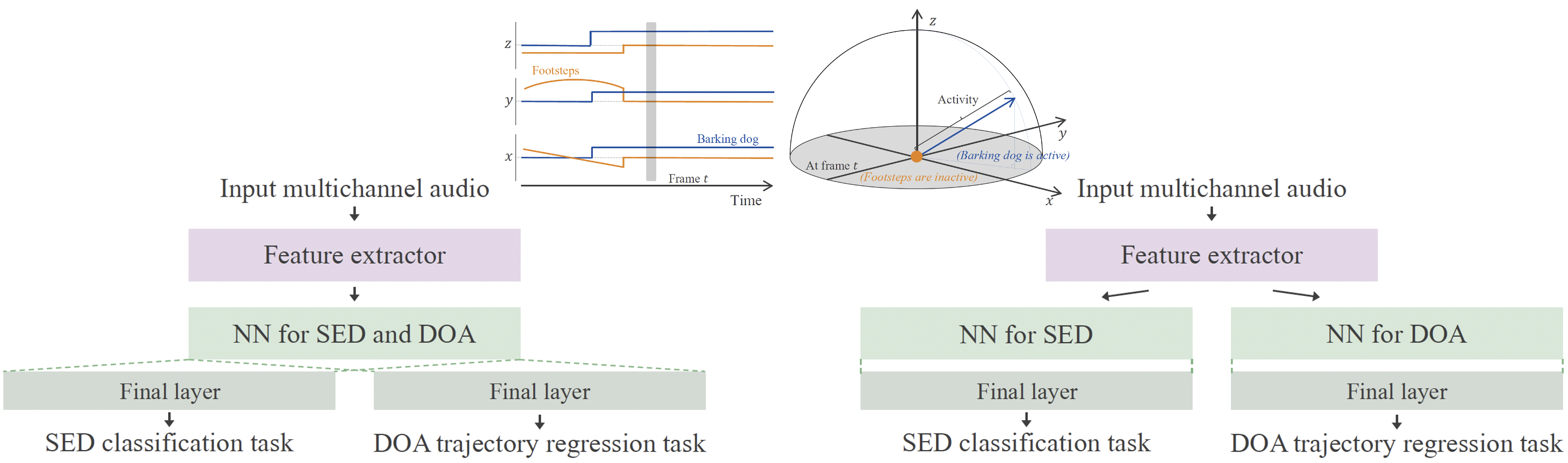}
    \caption{\Gls{accdoa} representation, where event activity is encoded by the norm of a Cartesian direction vector, adapted from K.~Shimada et al., ``\textit{ACCDOA: Activity-Coupled Cartesian Direction of Arrival Representation for Sound Event Localization and Detection}''~\cite{shimada2021accdoa}.}
    \label{fig:accdoa}
\end{figure}

This limitation motivated activity-coupled output representations in several \gls{dcase} \gls{seld} systems. In the \gls{accdoa} formulation, event activity and localization are merged into a single Cartesian vector whose norm inherently encodes activity~\cite{shimada2021accdoa}. For class \(c\) and frame \(t\), the target vector is defined as:
\begin{equation}
\mathbf{P}_{c,t}
=
a_{c,t}\mathbf{r}_{c,t}
\end{equation}
where \(\mathbf{P}_{c,t}\in\mathbb{R}^3\) is the \gls{accdoa} vector, \(a_{c,t}\in\{0,1\}\) is the binary activity target, and \(\mathbf{r}_{c,t}\in\mathbb{R}^3\) is the unit Cartesian \gls{doa} vector (Figure~\ref{fig:accdoa}). For active events, activity and direction can be recovered as:
\begin{equation}
a_{c,t}
=
\|\mathbf{P}_{c,t}\|_2,
\qquad
\mathbf{r}_{c,t}
=
\frac{\mathbf{P}_{c,t}}{\|\mathbf{P}_{c,t}\|_2}
\end{equation}
where the norm \(\|\mathbf{P}_{c,t}\|_2\) represents the activity magnitude and the normalized vector gives the estimated direction for active classes. When the event is inactive, the target collapses to the null vector; when it is active, the target is a unit vector pointing toward the source. This reformulation casts \gls{seld} as a single regression problem and removes the explicit need to balance separate \gls{sed} and \gls{doa} losses. However, in its original class-wise form, \gls{accdoa} still assumes at most one active source per class and frame.

Multi-\gls{accdoa} extends this idea to same-class spatial overlap by introducing a track dimension while preserving class-wise activity-coupled semantics~\cite{shimada2022multiaccdoa}. For \(N\) tracks, \(C\) classes, and \(T\) frames, the output tensor can be written as:
\begin{equation}
\mathbf{P}
\in
\mathbb{R}^{N\times C\times T\times 3},
\qquad
\mathbf{P}_{n,c,t}
=
a_{n,c,t}\mathbf{r}_{n,c,t}
\end{equation}
where \(\mathbf{P}_{n,c,t}\in\mathbb{R}^3\) is the activity-coupled Cartesian vector for track \(n\), class \(c\), and frame \(t\), \(a_{n,c,t}\) is the corresponding activity target, and \(\mathbf{r}_{n,c,t}\) is the unit Cartesian \gls{doa} vector. The track dimension enables multiple simultaneous instances of the same class, but it also introduces a permutation problem because several assignments of reference events to output tracks are equivalent.

\begin{figure}[ht]
    \centering
    \includegraphics[width=0.78\linewidth]{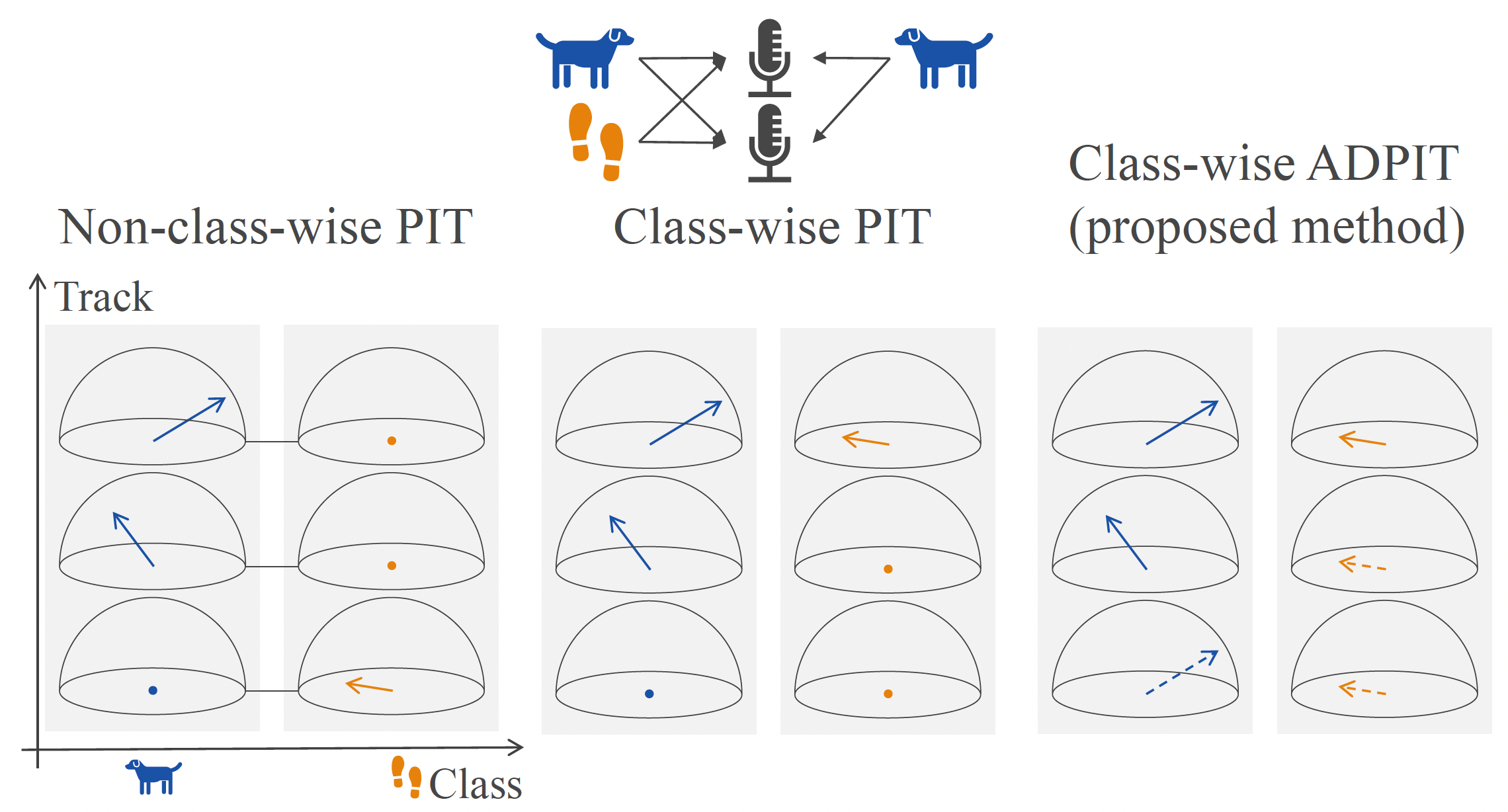}
    \caption{Multi-\gls{accdoa} and \gls{adpit} representation for same-class spatial overlap and permutation-aware supervision, adapted from K.~Shimada et al., ``\textit{Multi-ACCDOA: Localizing and Detecting Overlapping Sounds From the Same Class With Auxiliary Duplicating Permutation Invariant Training}''~\cite{shimada2022multiaccdoa}.}
    \label{fig:maccdoa}
\end{figure}

A class-wise \gls{pit} objective addresses this ambiguity by selecting, for each class and frame, the assignment that minimizes the regression error. A simplified formulation is:
\begin{equation}
\mathcal{L}_{\mathrm{PIT}}
=
\frac{1}{CT}
\sum_{c=1}^{C}
\sum_{t=1}^{T}
\min_{\alpha\in\mathcal{A}_{c,t}}
\ell^{\mathrm{ACCDOA}}_{\alpha,c,t}
\end{equation}
where \(\mathcal{L}_{\mathrm{PIT}}\) is the permutation-invariant loss, \(\mathcal{A}_{c,t}\) is the set of admissible track assignments for class \(c\) and frame \(t\), and \(\ell^{\mathrm{ACCDOA}}_{\alpha,c,t}\) is the assignment-specific \gls{accdoa} error. The assignment-specific error can be written as:
\begin{equation}
\ell^{\mathrm{ACCDOA}}_{\alpha,c,t}
=
\frac{1}{N}
\sum_{n=1}^{N}
\mathrm{MSE}
\left(
\hat{\mathbf{P}}_{\alpha(n),c,t},
\mathbf{P}^{*}_{n,c,t}
\right)
\end{equation}
where \(\hat{\mathbf{P}}_{\alpha(n),c,t}\) denotes the prediction assigned to reference track \(n\) under assignment \(\alpha\), and \(\mathbf{P}^{*}_{n,c,t}\) denotes the corresponding target. Standard class-wise \gls{pit} still assigns null vectors to inactive tracks, which can bias learning when the number of active same-class sources is smaller than the number of available tracks. \Gls{adpit} mitigates this issue by duplicating active targets across otherwise inactive tracks instead of using only zero-vector targets (Figure~\ref{fig:maccdoa}).

For \(M_{c,t}\) active sources of class \(c\) at frame \(t\), the number of candidate \gls{adpit} assignments can be expressed as:
\begin{equation}
K_{c,t}
=
\begin{cases}
{}^{N}P_{M_{c,t}}\,M_{c,t}^{\,N-M_{c,t}}, & M_{c,t}>0\\[4pt]
1, & M_{c,t}=0
\end{cases}
\end{equation}
where \(K_{c,t}\) is the number of candidate assignments, \({}^{N}P_{M_{c,t}}\) denotes the number of ordered selections of \(M_{c,t}\) tracks among \(N\), and \(M_{c,t}\) is the number of active same-class targets. During inference, duplicated outputs are typically consolidated by thresholding activity, computing angular similarity between active same-class outputs, and averaging sufficiently similar predictions.

\begin{figure}[ht]
    \centering
    \includegraphics[width=0.82\linewidth]{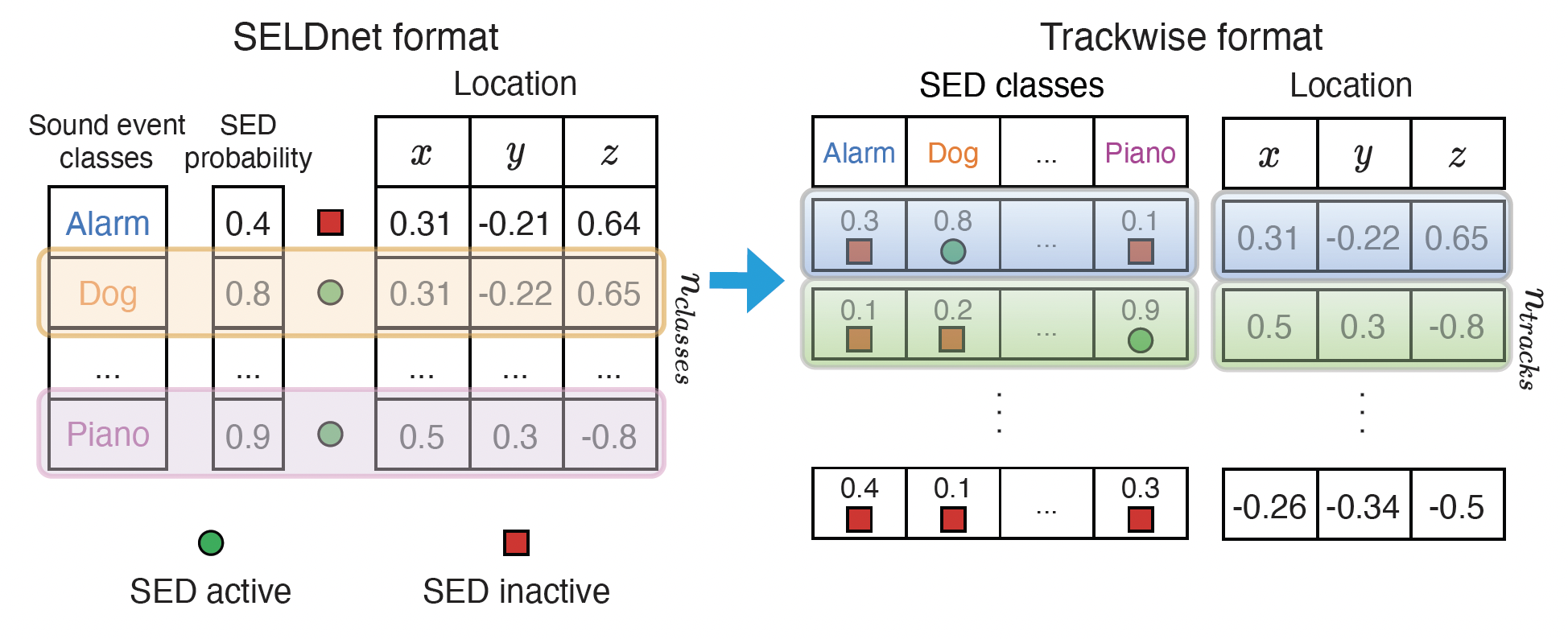}
    \caption{Comparison between class-wise \gls{seld} output formatting and track-wise event-independent formatting, adapted from N.~Cao et al., ``\textit{Event-Independent Network for Polyphonic Sound Event Localization and Detection}''~\cite{cao2020eventindependent}.}
    \label{fig:trackwise_out}
\end{figure}

A different strategy is adopted by track-wise event-independent formulations, such as \gls{ein}~\cite{cao2020eventindependent}. Instead of assigning one localization vector to each class, the model predicts a fixed number \(M\) of event-independent tracks, where each track contains at most one active event label and one corresponding location. The class-wise SELDnet-style output can be summarized as:
\begin{equation}
\mathcal{Y}_{\mathrm{class}}
=
\left\{
\left(
\mathbf{y}^{\mathrm{SED}}_t,
\mathbf{Y}^{\mathrm{DOA}}_t
\right)
\,\middle|\,
\mathbf{y}^{\mathrm{SED}}_t\in[0,1]^C,\;
\mathbf{Y}^{\mathrm{DOA}}_t\in\mathbb{R}^{C\times3}
\right\}
\end{equation}
where the model predicts one activity value and one \gls{doa} vector per class. The corresponding track-wise formulation is:
\begin{equation}
\mathcal{Y}_{\mathrm{track}}
=
\left\{
\left(
\mathbf{A}_t,
\mathbf{R}_t
\right)
\,\middle|\,
\mathbf{A}_t\in[0,1]^{M\times C},\;
\mathbf{R}_t\in\mathbb{R}^{M\times3}
\right\}
\end{equation}
where \(\mathbf{A}_t\) contains the class-activity probabilities for \(M\) event-independent tracks, and \(\mathbf{R}_t\) contains one Cartesian \gls{doa} vector per track (Figure~\ref{fig:trackwise_out}). This representation reduces redundant localization dimensions because the model estimates \(M\) locations rather than one location for every class. It also supports same-class overlap, since different tracks may contain the same event class with different source directions.

The cost of track-wise modeling is again the assignment ambiguity between reference events and predicted tracks. In \gls{ein}-V2, this ambiguity is addressed through \gls{pit} at frame or chunk level~\cite{cao2021einv2}. Let \(o\) denote either a frame index or a temporal chunk, and let \(\mathcal{P}(o)\) denote the set of admissible label-track assignments at index \(o\). The track-wise \gls{pit} loss can be written as:
\begin{equation}
\mathcal{L}^{\mathrm{PIT}}(o)
=
\min_{\alpha\in\mathcal{P}(o)}
\sum_{m=1}^{M}
\left[
\ell^{\mathrm{SED}}_{\alpha,m}(o)
+
\ell^{\mathrm{DOA}}_{\alpha,m}(o)
\right]
\end{equation}
where \(\ell^{\mathrm{SED}}_{\alpha,m}(o)\) is the detection loss for track \(m\) under assignment \(\alpha\), and \(\ell^{\mathrm{DOA}}_{\alpha,m}(o)\) is the corresponding localization loss. Frame-level \gls{pit} selects assignments independently for each frame, whereas chunk-level \gls{pit} constrains the assignment to remain fixed over a temporal region. The latter is more consistent with temporally extended events, because it discourages arbitrary track switching within a continuous source trajectory. \Gls{ein}-V2 further uses \gls{mhsa} to improve track separation and soft parameter sharing between \gls{sed} and \gls{doa} branches, allowing semantic and spatial features to interact through learnable cross-branch mixing~\cite{attention,cross-stitch}.

The distinction between output representation and temporal association is therefore essential. Two-branch SELDnet relies on recurrent layers to regularize sequential predictions, but its class-wise output cannot resolve same-class spatial overlap. \Gls{accdoa} couples event activity and localization into a single vector, but its basic form retains the one-location-per-class constraint. Multi-\gls{accdoa} introduces track multiplicity and \gls{adpit} to handle same-class overlap within an activity-coupled representation. Track-wise event-independent formulations instead reorganize the output space around source slots, resolving assignment ambiguity through \gls{pit} and track-separating temporal models.

Temporal continuity must also be distinguished from output formatting. Recurrent layers in SELDnet provide implicit smoothing and context aggregation, whereas explicit tracking formulations treat continuity as a stateful association problem. \Gls{pit} resolves assignment ambiguity at the loss level, but does not by itself guarantee persistent identity across time. Permutation-invariant recurrent models make this distinction explicit by replacing vector-valued recurrent states with unordered sets of embeddings, so that tracking becomes invariant to permutations of the input set and equivariant to permutations of the state set~\cite{diazguerra2023pirnn}.

The framework developed in the following sections adopts this taxonomy as the basis for \gls{at2seld}. Semantic \gls{at} representations provide class-level evidence, spatial front-ends encode multi-channel localization cues, temporal models aggregate framewise context, and track-wise \gls{sed}/\gls{doa} heads provide an output space capable of representing overlapping events without forcing localization to be replicated for every class independently.
\clearpage


\section{Related Works and Architectural Design Lineage}
\label{sec:related_work}
The methodological development of neural \gls{seld} systems can be read as a sequence of design responses to the constraints formalized in Section~\ref{sec:problem_taxonomy}: spatial feature extraction, temporal sequence modeling, semantic--spatial output representation, and assignment-aware supervision. This section follows the architectural lineage that connects these constraints to concrete model families, with emphasis on the mechanisms that inform the proposed \gls{at2seld} framework.

\subsection{\glsentryshort{crnn} Foundations for Joint Detection and Localization}
\label{subsec:related_crnn}

\begin{wrapfigure}{r}{0.65\textwidth}
    \centering
    \includegraphics[width=0.85\linewidth]{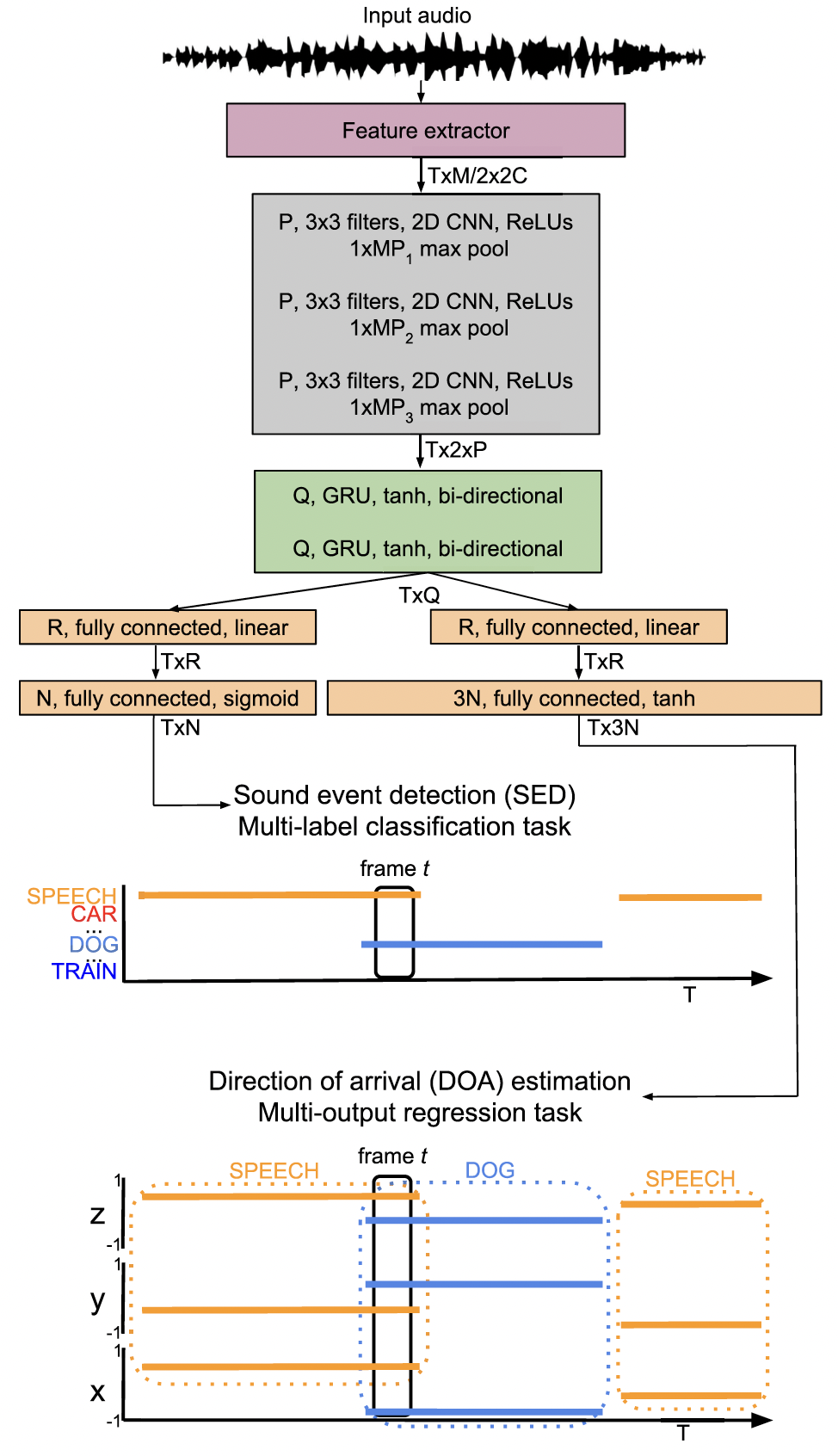}
    \caption{SELDnet model structure, with convolutional feature extraction, recurrent temporal modeling, and two synchronized \gls{sed}/\gls{doa} output branches, adapted from S.~Adavanne et al., ``\textit{Sound Event Localization and Detection of Overlapping Sources Using Convolutional Recurrent Neural Networks}''~\cite{adavanne2019seldnet}.}
    \label{fig:seldnet_model}
\end{wrapfigure}The \gls{crnn} formulation introduced by SELDnet established a reference neural formulation for joint \gls{seld} within a single trainable architecture~\cite{adavanne2019seldnet}. Its relevance is not only historical. SELDnet defines a reusable decomposition of the problem into multi-channel spectro-temporal encoding, recurrent context aggregation, and synchronized semantic--spatial prediction heads. This decomposition remains a reference point for later output representations, attention-based temporal models, and track-wise extensions, including the \gls{at2seld} architecture developed in this work.

SELDnet maps a sequence of multi-channel spectrogram frames to framewise \gls{sed} predictions and class-wise Cartesian \gls{doa} regression outputs. Given a multi-channel recording represented over \(T\) frames, its input tensor can be written as \(
\mathbf{X}
\in
\mathbb{R}^{T\times F\times 2C_{\mathrm{ch}}}
\), where \(\mathbf{X}\) is the spectrographic feature tensor, \(T\) is the number of temporal frames, \(F\) is the number of retained positive-frequency bins, and \(C_{\mathrm{ch}}\) is the number of microphone or Ambisonic channels. The factor \(2C_{\mathrm{ch}}\) arises from concatenating magnitude and phase spectrograms for each channel. In the original formulation, \(F=M/2\), where \(M\) denotes the \gls{dft} size after excluding the zeroth frequency bin~\cite{adavanne2019seldnet}.

The choice of magnitude and phase spectrograms is central to the original SELDnet design. Instead of using hand-crafted localization descriptors such as \gls{itd}, \gls{ild}, \gls{gcc}, or subspace-based pseudo-spectra, the network is expected to learn spatial cues directly from multi-channel spectro-temporal structure. Magnitude components carry strong semantic evidence related to source identity, timbre, and spectral energy distribution, whereas phase components preserve inter-channel structures that are informative for spatial localization. This design supports a degree of array generality because the input does not hard-code a specific localization algorithm, although adaptation to a different microphone geometry still requires appropriate training data or retraining.

Architecturally, SELDnet is organized as a cascade of convolutional, recurrent, and dense stages. The convolutional front-end applies local 2D convolutions over time and frequency while spanning the full channel-feature dimension (Figure~\ref{fig:seldnet_model}). Each convolutional block combines convolution, \gls{relu} activation, batch normalization, and pooling along the frequency axis. Temporal pooling is avoided in the early stages, so the temporal resolution required by the recurrent layers is preserved while the spectral dimension is progressively compressed. The resulting representation is reshaped into a frame-level embedding sequence and processed by \gls{bigru} units. The recurrent output is then shared by two fully connected branches: a sigmoid-activated \gls{sed} branch for multi-label event activity and a hyperbolic-tangent \gls{doa} branch whose Cartesian components are bounded in \([-1,1]\).

This architecture establishes three principles that remain relevant for later systems. \emph{(I)} event activity and localization are jointly optimized but remain explicitly separable at the output level; \emph{(II)} localization is formulated as continuous Cartesian regression rather than angular classification, avoiding discontinuities associated with azimuth wrap-around; and \emph{(III)} temporal continuity is delegated to a dedicated recurrent module rather than being absorbed entirely into convolutional context. The resulting model can be interpreted as a detect-and-localize system in which semantic confidence gates the interpretation of spatial estimates at inference time.

The training objective reflects this dual-branch organization. Let \(\mathbf{y}^{\mathrm{SED}}_t\) and \(\hat{\mathbf{y}}^{\mathrm{SED}}_t\) denote the \gls{sed} target and prediction at frame \(t\), and let \(\mathbf{y}^{\mathrm{DOA}}_t\) and \(\hat{\mathbf{y}}^{\mathrm{DOA}}_t\) denote the corresponding Cartesian \gls{doa} target and prediction. The SELDnet loss can be expressed as:
\begin{equation}
\mathcal{L}_{\mathrm{SELDnet}}
=
\lambda_{\mathrm{SED}}\mathcal{L}_{\mathrm{BCE}}
+
\lambda_{\mathrm{DOA}}\mathcal{L}_{\mathrm{MSE}}
\end{equation}
where \(\lambda_{\mathrm{SED}}\) and \(\lambda_{\mathrm{DOA}}\) are task-balancing weights, \(\mathcal{L}_{\mathrm{BCE}}\) is the multi-label detection loss, and \(\mathcal{L}_{\mathrm{MSE}}\) is the Cartesian localization loss. For \(C\) event classes and \(T\) frames, the detection term is:
\begin{equation}
\mathcal{L}_{\mathrm{BCE}}
=
-\frac{1}{TC}
\sum_{t=1}^{T}
\sum_{c=1}^{C}
\left[
y^{\mathrm{SED}}_{c,t}
\log
\hat{y}^{\mathrm{SED}}_{c,t}
+
\left(
1-y^{\mathrm{SED}}_{c,t}
\right)
\log
\left(
1-\hat{y}^{\mathrm{SED}}_{c,t}
\right)
\right]
\end{equation}
where \(y^{\mathrm{SED}}_{c,t}\in\{0,1\}\) is the binary activity target for class \(c\) at frame \(t\), and \(\hat{y}^{\mathrm{SED}}_{c,t}\in[0,1]\) is the corresponding predicted activity. The localization term is:
\begin{equation}
\mathcal{L}_{\mathrm{MSE}}
=
\frac{1}{3TC}
\sum_{t=1}^{T}
\sum_{c=1}^{C}
\left\|
\mathbf{r}_{c,t}
-
\hat{\mathbf{r}}_{c,t}
\right\|_2^2
\end{equation}
where \(\mathbf{r}_{c,t}\in\mathbb{R}^3\) is the target Cartesian \gls{doa} vector for class \(c\) at frame \(t\), and \(\hat{\mathbf{r}}_{c,t}\in\mathbb{R}^3\) is the predicted vector. Inactive classes are assigned zero-valued \gls{doa} targets, so the target representation is activity-dependent even though the localization loss may still be dominated by inactive entries if no explicit masking is applied. This design anticipates later activity-coupled representations, but the coupling remains mediated by the target convention and inference logic rather than being encoded directly in a single vector, as in \gls{accdoa}.

The original SELDnet experiments were trained with Adam~\cite{kingma2015adam} and early stopping based on the validation \gls{seld} score~\cite{prechelt2012earlystopping}. The reported best configuration on the \gls{ansyn} dataset used three convolutional layers with 64 filters, two recurrent layers with 128 units, one dense layer with 128 units in each branch, and frequency-axis pooling factors selected to preserve temporal resolution while compressing the spectral dimension~\cite{adavanne2019seldnet,sharath_adavanne_2018_1237703}.

\begin{wrapfigure}{l}{0.55\textwidth}
    \centering
    \includegraphics[width=0.80\linewidth]{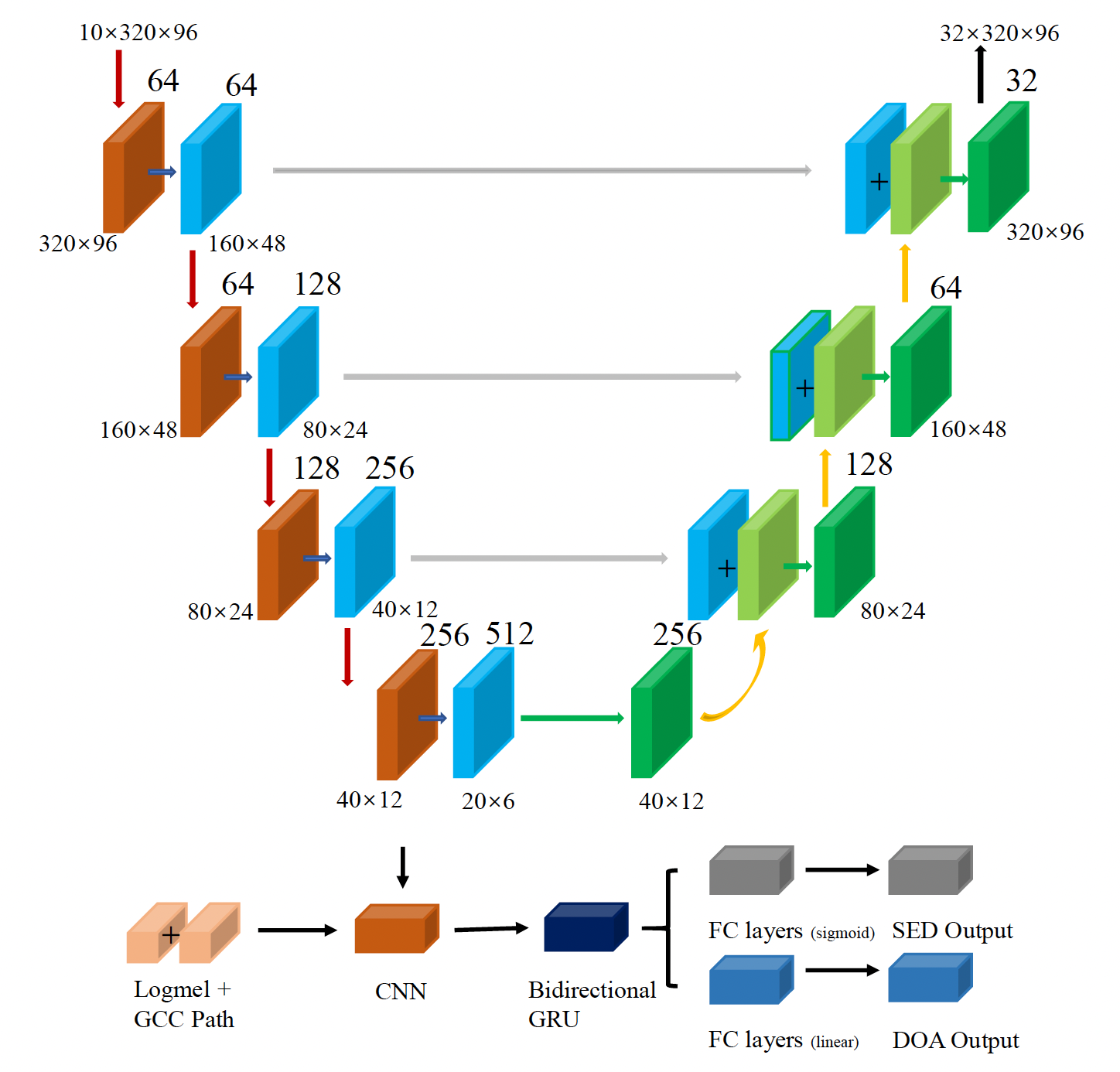}
    \caption{U-style recurrent neural network for \gls{seld}, combining multi-scale convolutional processing, skip connections, recurrent temporal modeling, and parallel \gls{sed}/\gls{doa} prediction branches, adapted from L.~Pi et al., ``\textit{U Recurrent Neural Network for Polyphonic Sound Event Detection and Localization}''~\cite{pi2020urnn}.}
    \label{fig:urnn_model}
\end{wrapfigure}
\begin{figure}[ht]
    \centering
    \includegraphics[width=0.82\linewidth]{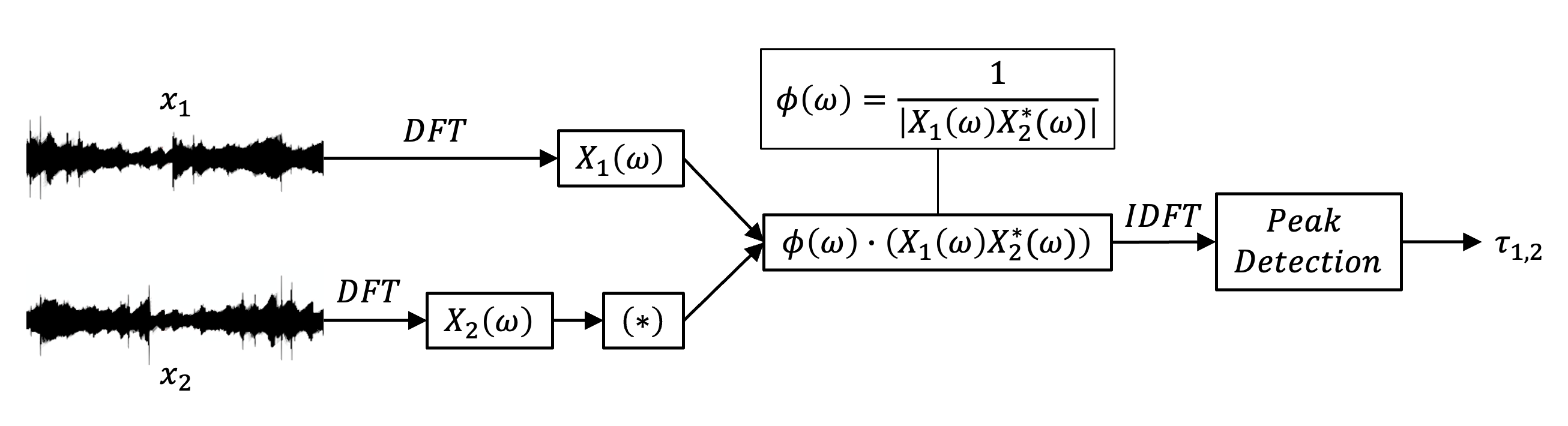}
    \caption{Schematic representation of \gls{gccphat}-based time-delay estimation, based on the generalized correlation framework of Knapp and Carter~\cite{gcc-phat}.}
    \label{fig:gcc_phat}
\end{figure}

An early refinement of this \gls{crnn} paradigm is the U-style recurrent neural network, which preserves the same joint \gls{sed}/\gls{doa} logic but strengthens feature extraction through U-Net-inspired skip connections~\cite{u-net_original} and multi-scale convolutional blocks~\cite{pi2020urnn}. This model uses log-mel spectral features together with \gls{gccphat} descriptors extracted from tetrahedral microphone-array recordings, thereby combining semantic spectral evidence with localization-sensitive inter-channel time-delay information. Its feature pipeline can be summarized as:
\begin{equation}
\mathbf{X}_{\mathrm{URNN}}
=
\left[
\mathbf{X}_{\mathrm{mel}},
\mathbf{X}_{\mathrm{GCC\mbox{-}PHAT}}
\right]
\end{equation}
where \(\mathbf{X}_{\mathrm{URNN}}\) denotes the input feature tensor, \(\mathbf{X}_{\mathrm{mel}}\) contains log-mel spectral features, and \(\mathbf{X}_{\mathrm{GCC\mbox{-}PHAT}}\) contains inter-channel \gls{gccphat} features (Figure~\ref{fig:gcc_phat}). In the reported configuration, the audio is processed at \(32~\mathrm{kHz}\), with \(10~\mathrm{ms}\) framing and 96 mel bands, while \gls{gccphat} features encode \gls{tdoa}-related information between microphone channels~\cite{pi2020urnn}. Unlike SELDnet, which emphasizes feature generality by using magnitude and phase spectrograms, this design injects stronger localization priors at the input level.

The U-style architecture modifies the convolutional hierarchy rather than the \gls{seld} task formulation. Its convolutional modules combine multiple kernel sizes, including \(3\times3\), \(1\times1\), and \(5\times5\), to capture local spectro-temporal patterns at different scales. Identity connections and decoder-style upsampling preserve lower-level feature information that may be relevant for localization, while the recurrent stage remains responsible for temporal aggregation. The final prediction still uses parallel \gls{sed} and \gls{doa} branches. This refinement is technically important because it shows that localization accuracy may depend on preserving low-level inter-channel and time-delay structures, whereas semantic discrimination often benefits from higher-level abstraction.

\begin{figure}[ht]
    \centering
    \includegraphics[width=0.78\linewidth]{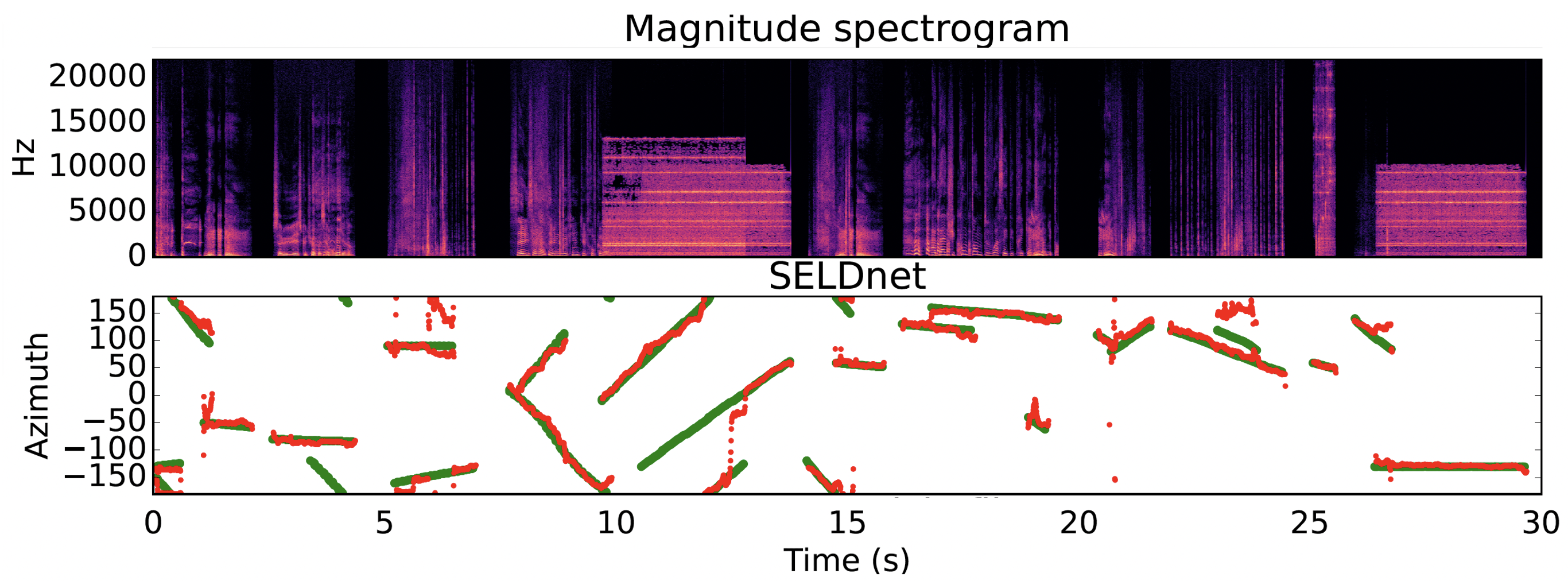}
    \caption{Example of SELDnet tracking behavior on a moving-source scene, with reference trajectories and predicted trajectories, adapted from S.~Adavanne et al., ``\textit{Localization, Detection and Tracking of Multiple Moving Sound Sources with a Convolutional Recurrent Neural Network}''~\cite{adavanne2019seldt_tracking}.}
    \label{fig:seldnet_tracking}
\end{figure}

The extension of SELDnet to moving-source scenarios makes explicit a property that is already latent in the \gls{crnn} formulation: recurrent layers do not only aggregate context, but can also regularize spatial trajectories over time~\cite{adavanne2019seldt_tracking}. In the moving-source setting, the architectural template remains substantially unchanged, with convolutional feature extraction, recurrent sequence modeling, and two output branches. The change lies in the training data and in the interpretation of the recurrent state. When trained on dynamic scenes, the recurrent module learns to map evolving spectro-spatial evidence into temporally coherent event activity and \gls{doa} trajectories (Figure~\ref{fig:seldnet_tracking}).

This dynamic formulation also exposes the structural limitation of class-wise outputs more clearly. Because SELDnet assigns a single Cartesian trajectory to each class, it cannot independently represent two simultaneous sources of the same class at different positions. In such conditions, the model may select one source, average incompatible spatial evidence, or switch between sources across time. The \gls{crnn} family therefore establishes both the first effective temporal modeling paradigm for joint \gls{seld} and the first clear motivation for richer output spaces, track-wise formulations, and permutation-aware supervision.

This lineage provides the architectural substrate for later \gls{seld} systems: spatial front-end processing, temporal aggregation, and coordinated semantic--spatial prediction. The specific way in which \gls{at2seld} inherits and modifies this substrate is synthesized in Section~\ref{subsec:related_at2seld_positioning}.

\subsection{Permutation-Aware and Track-Wise \glsentryshort{seld}}
\label{subsec:related_permutation_tracking}

The \gls{crnn} family reviewed in Section~\ref{subsec:related_crnn} shows that recurrent layers can regularize framewise \gls{sed} and \gls{doa} predictions by accumulating temporal context. This is sufficient when the output space assigns at most one trajectory to each class, but it does not fully address the structure of multi-source acoustic scenes. Once several concurrent sources are active, temporal coherence becomes an assignment problem: source estimates at the current frame must be associated with previous trajectories without relying on an arbitrary ordering of detections or output slots. This limitation motivates permutation-aware recurrent models and track-wise \gls{seld} architectures, which treat source identity, temporal association, and branch interaction as explicit modeling problems~\cite{diazguerra2023pirnn,cao2020eventindependent,cao2021einv2}.

Conventional recurrent layers, such as \gls{lstm} and \gls{gru} units, operate on ordered vectors. At frame \(t\), a recurrent model receives an input vector \(\mathbf{x}(t)\in\mathbb{R}^{d_x}\) and updates a hidden state \(\mathbf{h}(t)\in\mathbb{R}^{d_h}\). In a multi-source tracking scenario, this representation requires the set of detections to be serialized into a fixed order and the set of active trajectories to be compressed into a monolithic latent state. This is poorly aligned with \gls{seld}, where simultaneous sources are naturally unordered and where any permutation of equivalent source hypotheses should be treated as valid. The resulting mismatch appears both at the input level, because detections have no intrinsic ordering, and at the state level, because a single vector entangles all trajectories into a shared memory.

\Gls{pirnn} addresses this mismatch by replacing vector-valued inputs and states with unordered sets of source-wise embeddings~\cite{diazguerra2023pirnn}. Instead of processing one ordered input vector, the model operates on:
\begin{equation}
X(t)
=
\{\mathbf{x}_1(t),\mathbf{x}_2(t),\ldots,\mathbf{x}_{M_D}(t)\},
\qquad
H(t)
=
\{\mathbf{h}_1(t),\mathbf{h}_2(t),\ldots,\mathbf{h}_{M_H}(t)\}
\end{equation}
where \(X(t)\) is the set of detection embeddings at frame \(t\), \(H(t)\) is the set of recurrent track states, \(\mathbf{x}_i(t)\in\mathbb{R}^{d_x}\) denotes the embedding of detection \(i\), \(\mathbf{h}_j(t)\in\mathbb{R}^{d_h}\) denotes the latent state of track \(j\), \(M_D\) is the number of detections, and \(M_H\) is the number of maintained track states. This representation makes each recurrent state directly associated with one trajectory, avoids imposing an arbitrary detection order, and preserves the permutation structure of the tracking problem.

The central \gls{pirnn} operation is an attention-based \textit{soft} assignment between the previous track states and the union of current detections and previous states. Given \(X(t)\) and \(H(t-1)\), the context set is computed as:
\begin{equation}
C(t)
=
\mathrm{MHA}
\left(
H(t-1),
X(t)\cup H(t-1),
X(t)\cup H(t-1)
\right)
\end{equation}
where \(C(t)\) is the set of context vectors assigned to the tracked trajectories, and \(\mathrm{MHA}(\cdot)\) denotes \gls{mha}. With query matrix \(Q\), key matrix \(K\), and value matrix \(V\), the attention operation is:
\begin{equation}
\mathrm{Attention}(Q,K,V)
=
\mathrm{softmax}
\left(
\frac{QK^\top}{\sqrt{d_k}}
\right)
V
\end{equation}
where \(d_k\) is the key dimensionality and the softmax is applied row-wise. The multi-head version concatenates several such attention heads:
\begin{equation}
\mathrm{MHA}(Q,K,V)
=
\mathrm{Concat}
\left(
\mathrm{head}_1,\ldots,\mathrm{head}_{N_h}
\right)
W^{O}
\end{equation}
where \(N_h\) is the number of attention heads, \(W^{O}\) is the output projection, and each head is computed from learned query, key, and value projections. In this setting, attention 
\begin{wrapfigure}{r}{0.60\textwidth}
    \centering
    \includegraphics[width=0.5\textwidth]{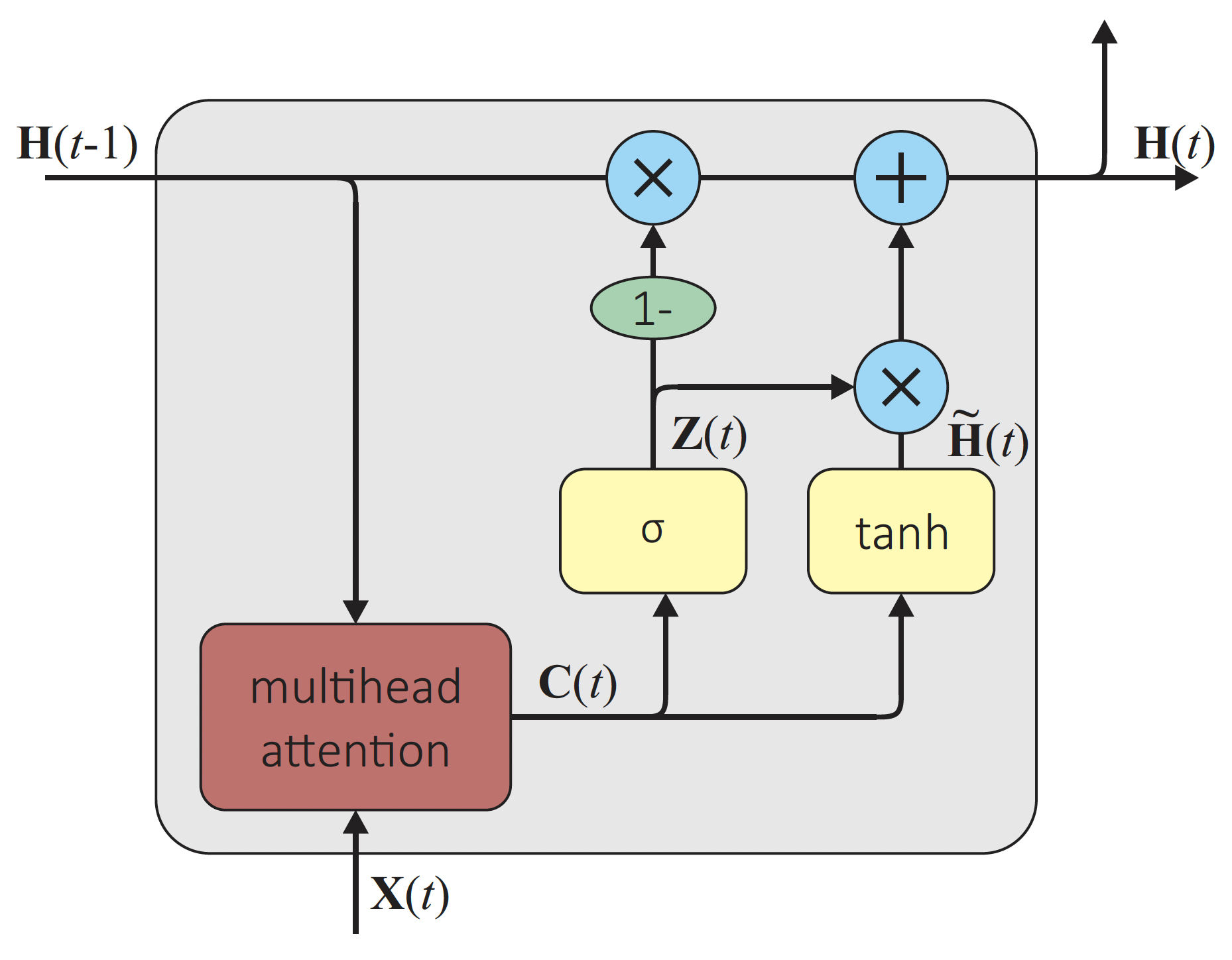}
    \caption{\Gls{pirnn} attention module for soft association between current detections and recurrent track states, adapted from D.~Díaz-Guerra et al., \textit{``Permutation Invariant Recurrent Neural Networks for Sound Source Tracking Applications''}~\cite{diazguerra2023pirnn}.}
    \label{fig:pirnn_mha}
\end{wrapfigure}
is not only a temporal context operator; it acts as a differentiable assignment mechanism that determines how current source evidence is attached to existing track states (Figure~\ref{fig:pirnn_mha}). Each previous track state queries the current detections and previous states jointly, producing one context vector per maintained trajectory. The recurrent update is then applied track-wise. In the \gls{pirnn} formulation, this update is based on a gated mechanism related to the \gls{mgu}~\cite{zhou2016mgu}. For track \(i\), the update can be written as:
\begin{equation}
\begin{split}
\mathbf{h}_i(t)
=
&\left[
1-\mathbf{z}_i(t)
\right]
\odot
\mathbf{h}_i(t-1)
+\\
&+\mathbf{z}_i(t)
\odot
\tilde{\mathbf{h}}_i(t)
\end{split}
\end{equation}
with:
\begin{equation}
\mathbf{z}_i(t)
=
\sigma
\left(
\mathbf{c}_i(t)W^{z}
\right),
\qquad
\tilde{\mathbf{h}}_i(t)
=
\tanh
\left(
\mathbf{c}_i(t)W^{h}
\right)
\end{equation}
where \(\mathbf{h}_i(t)\) is the updated state of track \(i\), \(\mathbf{h}_i(t-1)\) is the previous state, \(\mathbf{c}_i(t)\) is the corresponding attention-derived context vector, \(\mathbf{z}_i(t)\) is the update gate, \(\tilde{\mathbf{h}}_i(t)\) is the candidate state, \(W^{z}\) and \(W^{h}\) are learned projection matrices, \(\sigma(\cdot)\) denotes the sigmoid function, and \(\odot\) denotes element-wise multiplication. This preserves the logic of gated recurrence while aligning the hidden-state structure with the set structure of multi-source tracking.

\begin{figure}[ht]
    \centering
    \includegraphics[width=0.95\linewidth]{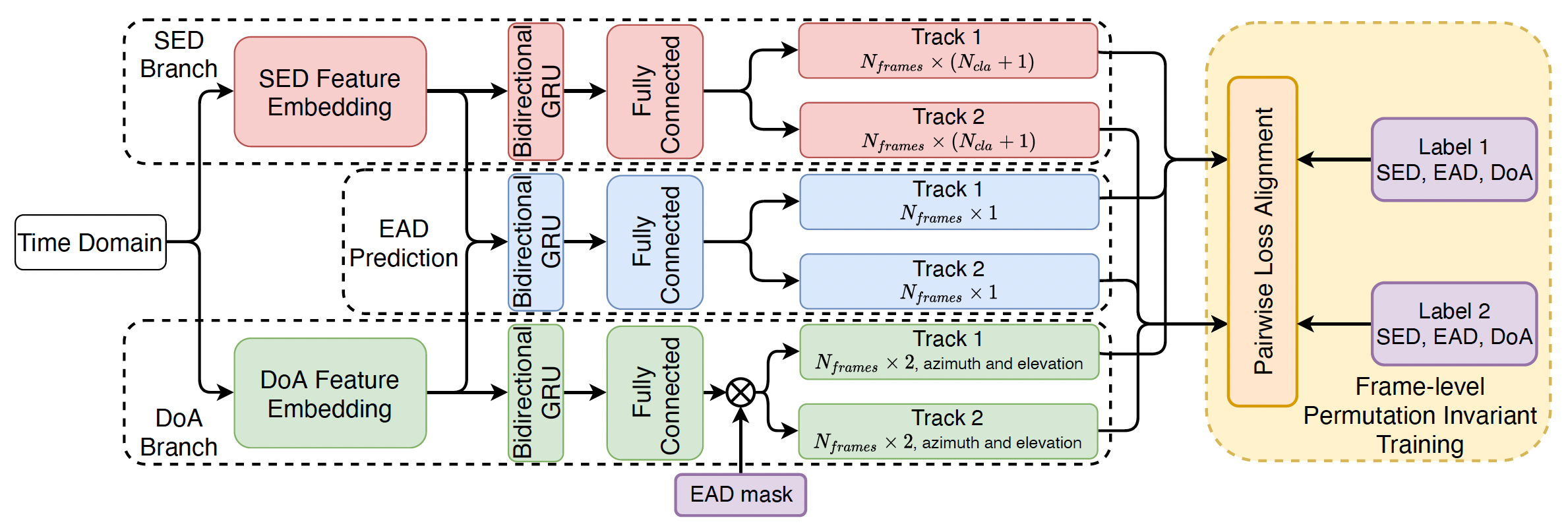}
    \caption{\Gls{ein}-V1 model structure, with event-independent tracks and an auxiliary \gls{ead} branch coupling \gls{sed} and \gls{doa} representations, adapted from Y.~Cao et al., ``\textit{Event-Independent Network for Polyphonic Sound Event Localization and Detection}''~\cite{cao2020eventindependent}.}
    \label{fig:ein_v1}
\end{figure}

A related but distinct strategy is represented by \gls{ein}, which reformulates \gls{seld} around event-independent tracks rather than class-wise localization vectors~\cite{cao2020eventindependent}. As discussed in Section~\ref{sec:problem_taxonomy}, the track-wise output space reduces redundant localization dimensions and enables same-class overlap by allowing different tracks to host the same event class with different \gls{doa} estimates. The first \gls{ein} formulation (Figure~\ref{fig:ein_v1}) adds a third branch called \gls{ead} in addition to the \gls{sed} and \gls{doa} branches. The purpose of \gls{ead} is not merely to detect whether a track is active, but to couple semantic and spatial evidence in a branch-aware activity estimate that can regulate \gls{doa} regression.

In \gls{ein}-V1, each track predicts one class among \(C+1\) alternatives, where the additional class corresponds to silence. For track \(m\) at frame \(t\), the classification loss can be written as:
\begin{equation}
\ell^{\mathrm{SED}}_{m,t}
=
-
\log
\left[
\frac{
e^
{\left(
s^{\mathrm{SED}}_{m,t,c_t}
\right)}
}{
\sum_{j\in\mathcal{J}}
e^
{\left(
s^{\mathrm{SED}}_{m,t,j}
\right)}
}
\right]
\end{equation}
where \(s^{\mathrm{SED}}_{m,t,j}\) is the class logit for class \(j\), \(c_t\) is the target class assigned to track \(m\) at frame \(t\), and \(\mathcal{J}\) is the set of event classes plus the silence class. The total \gls{sed} loss is accumulated across tracks and frames:
\begin{equation}
\mathcal{L}^{\mathrm{SED}}
=
\sum_{m=1}^{M}
\sum_{t=1}^{T}
\ell^{\mathrm{SED}}_{m,t}
\end{equation}
where \(M\) is the number of event-independent tracks and \(T\) is the number of frames. The \gls{ead} branch receives a representation derived from both \gls{sed} and \gls{doa} embeddings and predicts a binary activity variable for each track. During training, the activity mask used for localization is given by the reference \gls{ead} target:
\begin{equation}
M^{\mathrm{EAD}}_{m,t}
=
y^{\mathrm{EAD}}_{m,t}
\end{equation}
where \(M^{\mathrm{EAD}}_{m,t}\in\{0,1\}\) is the activity mask for track \(m\) at frame \(t\), and \(y^{\mathrm{EAD}}_{m,t}\) is the corresponding ground-truth activity. During inference, the mask is obtained by combining the predicted \gls{ead} activity with the predicted non-silence class assignment:
\begin{equation}
M^{\mathrm{EAD}}_{m,t}
=
\mathbbm{1}
\left[
\hat{y}^{\mathrm{EAD}}_{m,t}>\tau_{\mathrm{EAD}}
\right]
\cap
\mathbbm{1}
\left[
\arg\max_{j\in\mathcal{J}}
\hat{p}^{\mathrm{SED}}_{m,t,j}
\neq
\mathrm{sil}
\right]
\end{equation}
where \(\hat{y}^{\mathrm{EAD}}_{m,t}\) is the predicted event-activity probability, \(\tau_{\mathrm{EAD}}\) is the \gls{ead} threshold, \(\hat{p}^{\mathrm{SED}}_{m,t,j}\) is the predicted class posterior, and \(\mathrm{sil}\) denotes the silence class. The \gls{doa} loss is then evaluated only over active tracks:
\begin{equation}
\mathcal{L}^{\mathrm{DOA}}
=
\frac{
\sum_{m=1}^{M}
\sum_{t=1}^{T}
M^{\mathrm{EAD}}_{m,t}
\left\|
\mathbf{r}_{m,t}
-
\hat{\mathbf{r}}_{m,t}
\right\|_{p}
}{
\sum_{m=1}^{M}
\sum_{t=1}^{T}
M^{\mathrm{EAD}}_{m,t}
}
\end{equation}
where \(\mathbf{r}_{m,t}\) and \(\hat{\mathbf{r}}_{m,t}\) are the target and predicted \gls{doa} coordinates for track \(m\) at frame \(t\), and \(\|\cdot\|_{p}\) denotes the norm adopted for the localization error. This mechanism anticipates a central diagnostic issue in this work: spatial regression should be conditioned on event activity, otherwise inactive tracks can dominate the localization objective and bias the model toward degenerate low-norm predictions.

\Gls{ein}-V2 removes the auxiliary \gls{ead} branch and addresses branch consistency through a more direct multi-task design~\cite{cao2021einv2}. Its architecture combines a VGG-style convolutional front-end, separate \gls{sed} and \gls{doa} branches, frequency-wise global average pooling, \gls{mhsa} for track separation, and final fully connected prediction heads. The track-wise assignment ambiguity is handled through \gls{pit}, applied either independently at each frame or over temporal chunks (Section~\ref{sec:problem_taxonomy}). This change moves the model from auxiliary activity-mediated coupling toward representation-level and loss-level mechanisms for track consistency.

\begin{figure}[ht]
    \centering
    \includegraphics[width=0.78\linewidth]{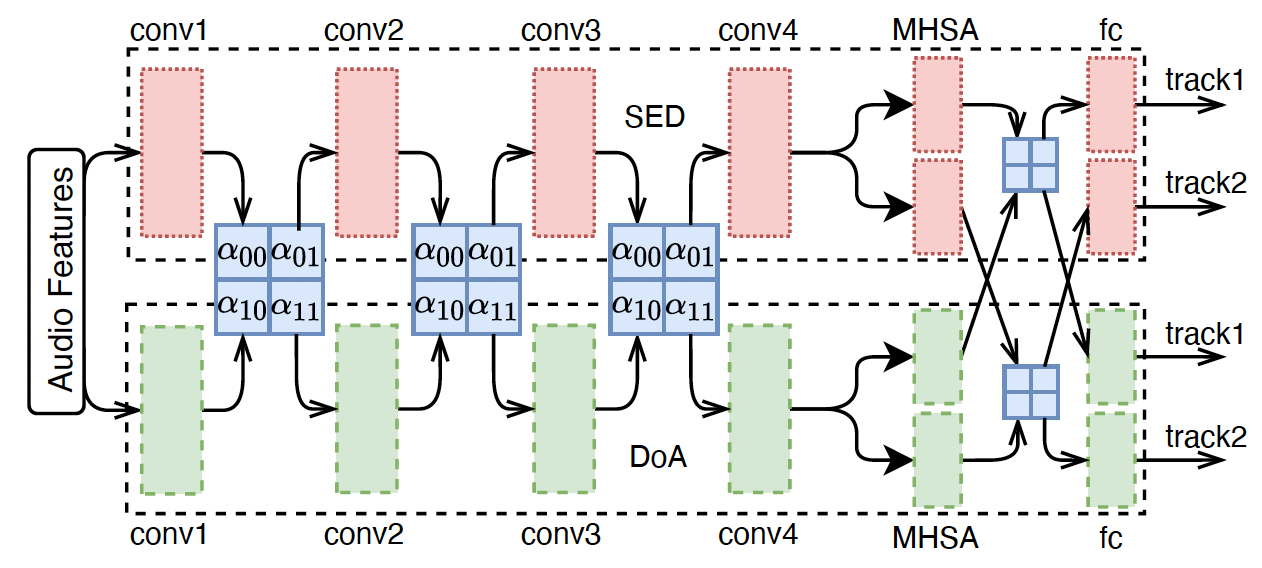}
    \caption{\Gls{ein}-V2 model structure, with track-wise \gls{seld} prediction, \gls{mhsa}-based track separation, and soft parameter sharing between \gls{sed} and \gls{doa} branches, adapted from Y.~Cao et al., \textit{``An Improved Event-Independent Network for Polyphonic Sound Event Localization and Detection''}~\cite{cao2021einv2}.}
    \label{fig:ein_v2}
\end{figure}

The use of \gls{mhsa} in \gls{ein}-V2 differs from its role in \gls{pirnn}. In \gls{pirnn}, attention performs soft temporal association between detections and existing tracks. In \gls{ein}-V2, attention separates latent track representations within a frame-structured \gls{seld} model. For an input sequence \(X\in\mathbb{R}^{T\times D}\), self-attention can be written as:
\begin{equation}
\mathrm{SA}(X)
=
\mathrm{softmax}
\left(
\frac{
(X+P)W^{Q}
\left[
(X+P)W^{K}
\right]^\top
}{
\sqrt{d_k}
}
\right)
XW^{V}
\end{equation}
where \(X\) is the input feature sequence, \(P\) is a positional encoding, \(W^{Q}\), \(W^{K}\), and \(W^{V}\) are learned query, key, and value projections, and \(d_k\) is the key dimensionality. The multi-head output is:
\begin{equation}
\mathrm{MHSA}(X)
=
\mathrm{Concat}
\left(
\mathrm{SA}_1(X),\ldots,\mathrm{SA}_{N_h}(X)
\right)
W^{O}
+
\mathbf{b}^{O}
\end{equation}
where \(N_h\) is the number of attention heads, \(W^{O}\) is the output projection, and \(\mathbf{b}^{O}\) is the output bias. Attention block is placed after convolutional processing and before the final track-wise heads, where it supports latent separation among event-independent tracks (Figure~\ref{fig:ein_v2}).

A further contribution of \gls{ein}-V2 is \textit{soft parameter sharing} between \gls{sed} and \gls{doa} branches. Instead of forcing the two subtasks to use either completely shared or completely independent representations, cross-stitch-style mixing allows the branches to exchange information through learned coefficients~\cite{cross-stitch}. If \(\mathbf{x}^{\mathrm{SED}}\) and \(\mathbf{x}^{\mathrm{DOA}}\) denote feature maps from the two branches, the mixed representations are:
\begin{equation}
\label{eq:cross_stitch_soft_sharing}
\begin{bmatrix}
\hat{\mathbf{x}}^{\mathrm{SED}}\\
\hat{\mathbf{x}}^{\mathrm{DOA}}
\end{bmatrix}
=
\boldsymbol{\alpha}
\begin{bmatrix}
\mathbf{x}^{\mathrm{SED}}\\
\mathbf{x}^{\mathrm{DOA}}
\end{bmatrix}
\end{equation}
where \(\boldsymbol{\alpha}\in\mathbb{R}^{2\times2}\) is a learned mixing matrix, \(\mathbf{x}^{\mathrm{SED}}\) and \(\mathbf{x}^{\mathrm{DOA}}\) are the input feature maps, and \(\hat{\mathbf{x}}^{\mathrm{SED}}\) and \(\hat{\mathbf{x}}^{\mathrm{DOA}}\) are the branch-mixed feature maps. This mechanism is particularly relevant for \gls{at2seld}, because it provides a principled precedent for selective interaction between semantic and spatial representations rather than unconditional feature fusion.

The branch specialization in \gls{ein}-V2 is also reflected in the input features. The \gls{sed} branch uses Log-Mel spectrograms, while the \gls{doa} branch uses Log-Mel features together with \glspl{iv} computed from \gls{foa} channels. Let \(W(f,t)\), \(X(f,t)\), \(Y(f,t)\), and \(Z(f,t)\) denote the \gls{stft} coefficients of the \gls{foa} channels at frequency bin \(f\) and frame \(t\). A frequency-domain \gls{iv} estimate can be written as:
\begin{equation}
\mathbf{I}(f,t)
=
\frac{1}{\rho_0 c}
\Re
\left\{
W^{*}(f,t)
\begin{bmatrix}
X(f,t)\\
Y(f,t)\\
Z(f,t)
\end{bmatrix}
\right\}
\end{equation}
where \(\mathbf{I}(f,t)\in\mathbb{R}^{3}\) is the acoustic intensity vector, \(\rho_0\) is the air density, \(c\) is the speed of sound, \(W^{*}(f,t)\) denotes the complex conjugate of the omnidirectional \gls{foa} component, and \(\Re\{\cdot\}\) extracts the real part. The vector is then normalized and projected onto the Mel scale:
\begin{equation}
\mathbf{I}^{\mathrm{mel}}(k,t)
=
-
\sum_{f}
H_{\mathrm{mel}}(k,f)
\frac{
\mathbf{I}(f,t)
}{
\|\mathbf{I}(f,t)\|_2
}
\end{equation}
where \(\mathbf{I}^{\mathrm{mel}}(k,t)\) is the Mel-scaled normalized intensity vector at Mel bin \(k\), \(H_{\mathrm{mel}}(k,f)\) is the Mel filterbank coefficient, and \(\|\cdot\|_2\) is the Euclidean norm. This construction produces spatial features aligned with the temporal and spectral resolution of the Log-Mel representation while retaining direction-sensitive information from the \gls{foa} channels.

Taken together, \gls{pirnn}, \gls{ein}, and \gls{ein}-V2 show that multi-source \gls{seld} requires more than recurrent smoothing: it requires track-structured outputs, assignment-aware supervision, and controlled interaction between semantic and spatial branches. These principles are reused in \gls{at2seld} through track-wise \gls{sed}/\gls{doa} heads and semantic--spatial interaction mechanisms, as discussed in Section~\ref{subsec:related_at2seld_positioning}.

\subsection{Attention and Conformer-based \glsentryshort{seld} Systems}
\label{subsec:related_conformer}

The transition from \gls{crnn}-based \gls{seld} to attention- and Conformer-based systems reflects a broader methodological shift in the \gls{dcase} literature. Early models primarily established the feasibility of joint event detection and localization on controlled or synthetic spatial scenes. More recent systems are designed for real-world recordings, limited annotation budgets, stronger reverberation, overlapping sources, background interference, and tighter evaluation protocols. Under these conditions, performance depends on the coordinated design of spatial input features, hierarchical spectro-temporal encoders, temporal sequence models, output representations, augmentation policies, and training schedules.

A stable design choice in several recent high-performing systems is the use of \gls{foa}-derived features combining Log-Mel spectrograms and \glspl{iv}. In the DCASE 2022 ResNet-Conformer system and in later related variants, four-channel Log-Mel features are extracted from the \gls{foa} channels and concatenated with three-channel \gls{iv} features, yielding a seven-channel input tensor~\cite{wang2022resnetconformer,xue2023mscarcnet,vo2024resnetconformer_einv2}. The resulting feature tensor can be written as:
\begin{equation}
\mathbf{X}_{\mathrm{FOA}}
=
\left[
\mathbf{X}_{\mathrm{mel}}^{W},
\mathbf{X}_{\mathrm{mel}}^{X},
\mathbf{X}_{\mathrm{mel}}^{Y},
\mathbf{X}_{\mathrm{mel}}^{Z},
\mathbf{I}^{X}_{\mathrm{mel}},
\mathbf{I}^{Y}_{\mathrm{mel}},
\mathbf{I}^{Z}_{\mathrm{mel}}
\right]
\in
\mathbb{R}^{7\times T\times K}
\end{equation}
where \(\mathbf{X}_{\mathrm{FOA}}\) is the input feature tensor, \(W,X,Y,Z\) denote the \gls{foa} channels, \(\mathbf{X}_{\mathrm{mel}}^{(\cdot)}\) are channel-wise Log-Mel spectrograms, \(\mathbf{I}^{X}_{\mathrm{mel}}\), \(\mathbf{I}^{Y}_{\mathrm{mel}}\), and \(\mathbf{I}^{Z}_{\mathrm{mel}}\) are Mel-scaled \gls{iv} components, \(T\) is the number of temporal frames, and \(K\) is the number of Mel bands. This representation preserves semantic spectral evidence through the Log-Mel channels while providing explicit direction-sensitive cues through the \gls{iv} channels. In the reviewed \gls{dcase} systems, such features are typically extracted from \gls{foa} audio resampled at \(24~\mathrm{kHz}\), using a \(1024\)-point transform with a \(40~\mathrm{ms}\) Hann window and \(20~\mathrm{ms}\) hop length~\cite{wang2022resnetconformer,vo2024resnetconformer_einv2}.

The move from shallow convolutional stacks to residual front-ends is a second recurring element of this family. Residual connections allow deeper encoders to be optimized while preserving low-level spectro-spatial information through identity paths~\cite{resnet}. In the DCASE 2022 system, a modified ResNet encoder extracts high-level features before a linear projection maps the representation to the dimensionality required by the Conformer module~\cite{wang2022resnetconformer}. In MSCA-RCnet, the ResNet backbone is organized as a stem block followed by four residual stages with increasing channel dimensions, frequency-axis pooling, and residual shortcuts~\cite{xue2023mscarcnet}. The 2024 ResNet-Conformer system adopts a related residual hierarchy inside an \gls{ein}-V2-inspired dual-branch structure, using successive residual blocks before Conformer-based temporal modeling and final output prediction~\cite{vo2024resnetconformer_einv2}.

\begin{wrapfigure}{l}{0.55\textwidth}
    \centering
    \includegraphics[width=0.53\textwidth]{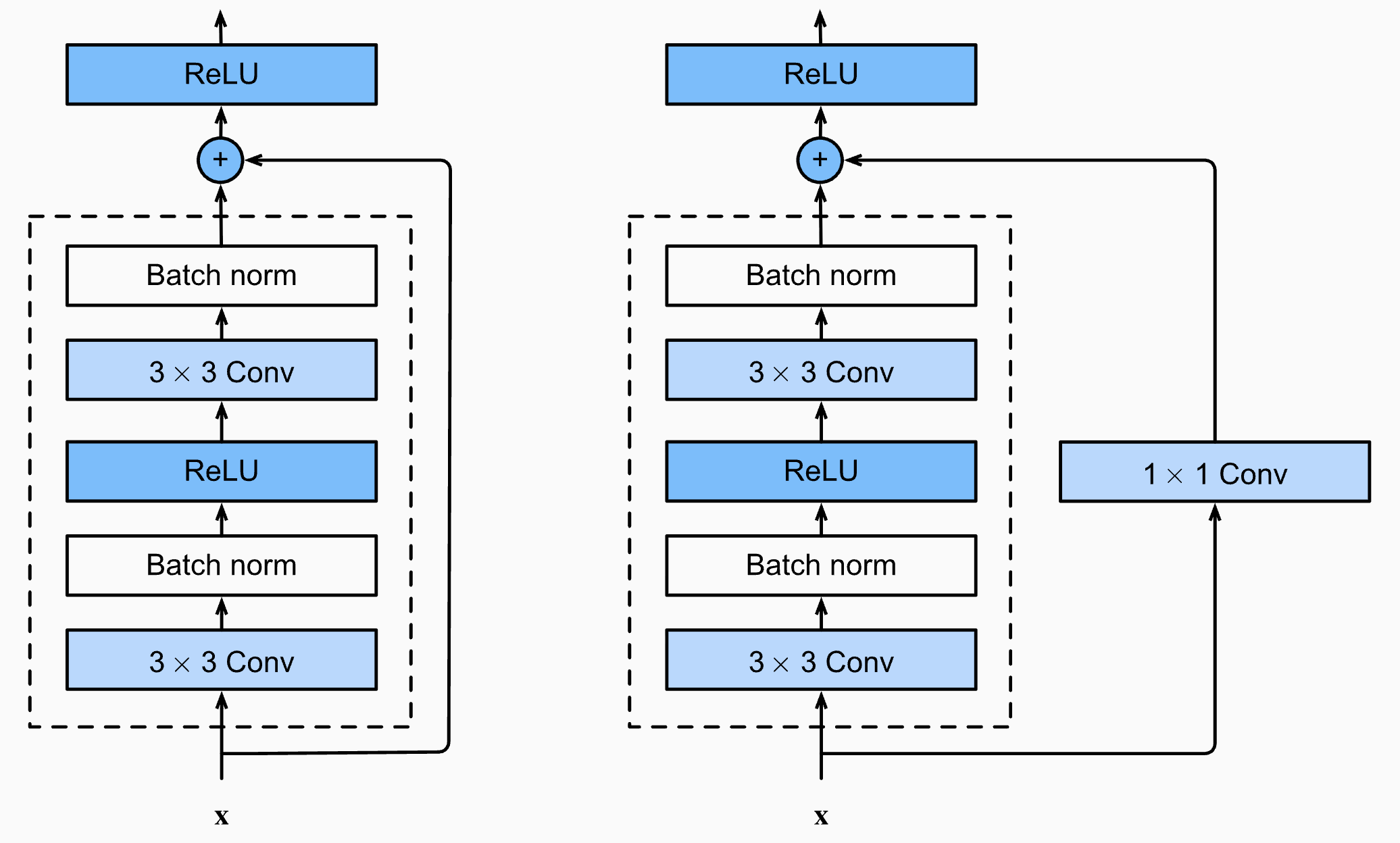}
    \caption{Residual block structure with identity shortcut, adapted from A.~Zhang et al., \textit{``Dive into Deep Learning''}~\cite{zhang2023dive}.}
    \label{fig:resblock}
\end{wrapfigure}

Given an input feature map \(\mathbf{X}\), a residual block computes:
\begin{equation}
\mathbf{Y}
=
\mathcal{F}(\mathbf{X};\boldsymbol{\theta})
+
\mathcal{S}(\mathbf{X})
\end{equation}
where \(\mathbf{Y}\) is the block output, \(\mathcal{F}(\cdot;\boldsymbol{\theta})\) denotes the learned residual transformation parameterized by \(\boldsymbol{\theta}\), and \(\mathcal{S}(\cdot)\) is the shortcut path, implemented either as an identity mapping or as a projection when the number of channels or the resolution changes (Figure~\ref{fig:resblock}). In \gls{seld}, this structure is useful because low-level spatial cues, such as inter-channel phase or intensity patterns, may be degraded by repeated pooling if they are not preserved through skip pathways. Residual encoders therefore act as hierarchical spectro-spatial processors: they transform \gls{foa} features into compact latent representations while retaining information needed by the localization branch.

\begin{figure}[ht]
    \centering
    \includegraphics[width=0.86\linewidth]{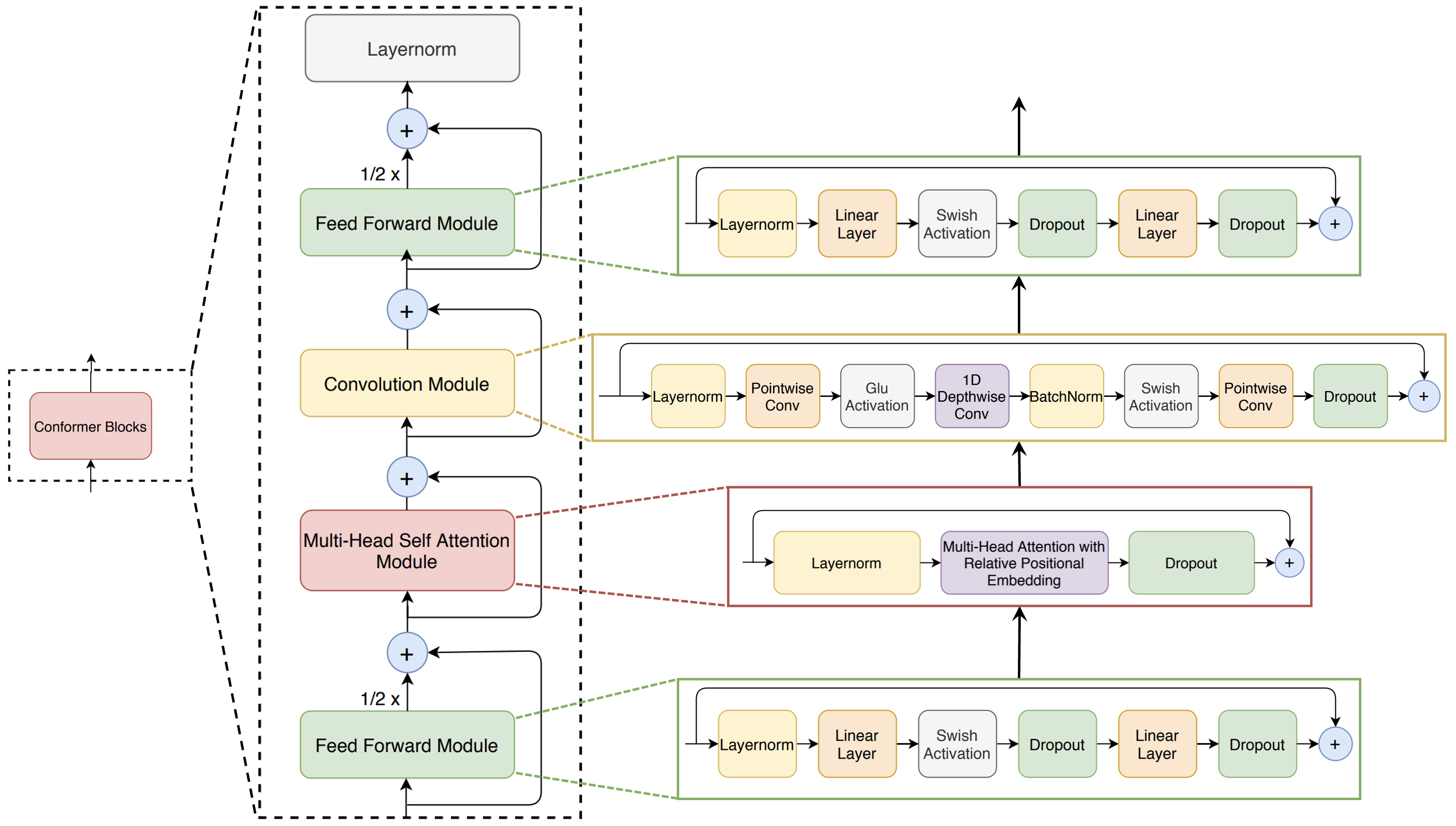}
    \caption{Conformer module structure, combining feed-forward, self-attention, convolutional, and normalization stages, adapted from A.~Gulati et al., \textit{``Conformer: Convolution-Augmented Transformer for Speech Recognition''}~\cite{gulati2020conformer}.}
    \label{fig:conformer}
\end{figure}

The defining sequence operator in this architectural family is the Conformer~\cite{gulati2020conformer}. Unlike a purely recurrent module, which propagates state sequentially, or a purely attention-based module, which emphasizes global token interactions, the Conformer combines self-attention with local convolutional processing inside a residual block. Its canonical structure applies a half-step \gls{ffn}, \gls{mhsa}, a convolutional module, a second half-step \gls{ffn}, and \gls{layernorm}. A compact formulation is:
\begin{align}
\mathbf{Z}_1
&=
\mathbf{X}
+
\frac{1}{2}
\mathrm{FFN}(\mathbf{X})
\\
\mathbf{Z}_2
&=
\mathbf{Z}_1
+
\mathrm{MHSA}(\mathbf{Z}_1)
\\
\mathbf{Z}_3
&=
\mathbf{Z}_2
+
\mathrm{Conv}(\mathbf{Z}_2)
\\
\mathbf{Y}
&=
\mathrm{LayerNorm}
\left(
\mathbf{Z}_3
+
\frac{1}{2}
\mathrm{FFN}(\mathbf{Z}_3)
\right)
\end{align}
where \(\mathbf{X}\) is the input sequence, \(\mathbf{Y}\) is the Conformer output, \(\mathrm{FFN}(\cdot)\) denotes a position-wise feed-forward network, \(\mathrm{MHSA}(\cdot)\) is multi-head self-attention, \(\mathrm{Conv}(\cdot)\) is the convolutional sub-block, and \(\mathrm{LayerNorm}(\cdot)\) denotes layer normalization. In \gls{seld}, this combination is well matched to the temporal structure of acoustic scenes: \gls{mhsa} captures longer-range context and event co-occurrence patterns, while the convolutional sub-block preserves local continuity, onset structure, and short-term dynamics (Figure~\ref{fig:conformer}).

In the DCASE 2022 system, the Conformer is inserted after the ResNet encoder, and temporal pooling is postponed until after the Conformer so that sequence modeling operates at a sufficiently high temporal resolution~\cite{wang2022resnetconformer}. MSCA-RCnet follows the same general logic by combining residual encoding, Conformer layers, and attentive temporal compression~\cite{xue2023mscarcnet}. The 2024 ResNet-Conformer architecture also replaces earlier recurrent or \gls{mhsa}-only temporal blocks with Conformer layers inside a two-branch \gls{ein}-V2-inspired design, reinforcing the role of Conformer blocks as the dominant modern alternative to recurrent temporal smoothing~\cite{vo2024resnetconformer_einv2}.

Attention in these systems is not limited to temporal self-attention. MSCA-RCnet introduces \gls{msca} to recalibrate feature channels inside the residual encoder, with the goal of preserving salient spatial and spectral information while suppressing background interference~\cite{xue2023mscarcnet}. Given an input feature map \(\mathbf{X}\in\mathbb{R}^{C\times T\times F}\), \gls{msca} combines a global channel descriptor and a local channel descriptor. The global descriptor is obtained by \gls{gap}:
\begin{equation}
g(\mathbf{X})
=
\frac{1}{TF}
\sum_{t=1}^{T}
\sum_{f=1}^{F}
\mathbf{X}_{:,t,f}
\end{equation}
where \(g(\mathbf{X})\in\mathbb{R}^{C}\) is the global channel descriptor, \(C\) is the number of channels, \(T\) is the number of temporal frames, \(F\) is the number of frequency bins, and \(\mathbf{X}_{:,t,f}\) denotes the channel vector at time--frequency position \((t,f)\). The global and local channel contexts are then computed as:
\begin{align}
\mathbf{G}(\mathbf{X})
&=
\beta
\left(
\mathrm{Conv}_2
\left(
\mathrm{ReLU}
\left(
\beta
\left(
\mathrm{Conv}_1
\left(
g(\mathbf{X})
\right)
\right)
\right)
\right)
\right)
\\
\mathbf{L}(\mathbf{X})
&=
\beta
\left(
\mathrm{Conv}_2
\left(
\mathrm{ReLU}
\left(
\beta
\left(
\mathrm{Conv}_1(\mathbf{X})
\right)
\right)
\right)
\right)
\end{align}
where \(\mathbf{G}(\mathbf{X})\) is the global channel context, \(\mathbf{L}(\mathbf{X})\) is the local channel context, \(\mathrm{Conv}_1(\cdot)\) and \(\mathrm{Conv}_2(\cdot)\) are \(1\times1\) convolutional projections implementing a bottleneck and channel restoration, \(\beta(\cdot)\) denotes batch normalization, and \(\mathrm{ReLU}(\cdot)\) is the rectified linear activation. The refined output is:
\begin{equation}
\mathbf{Y}
=
\mathbf{X}
\otimes
\sigma
\left(
\mathbf{G}(\mathbf{X})
\oplus
\mathbf{L}(\mathbf{X})
\right)
\end{equation}
where \(\mathbf{Y}\) is the recalibrated feature map, \(\sigma(\cdot)\) is the sigmoid activation, \(\oplus\) denotes broadcast addition, and \(\otimes\) denotes element-wise multiplication. This mechanism uses attention as feature-map calibration rather than as sequence association: it determines which channels should be emphasized after combining local and global context.

\begin{wrapfigure}{r}{0.55\textwidth}
    \centering
    \includegraphics[width=0.4\textwidth]{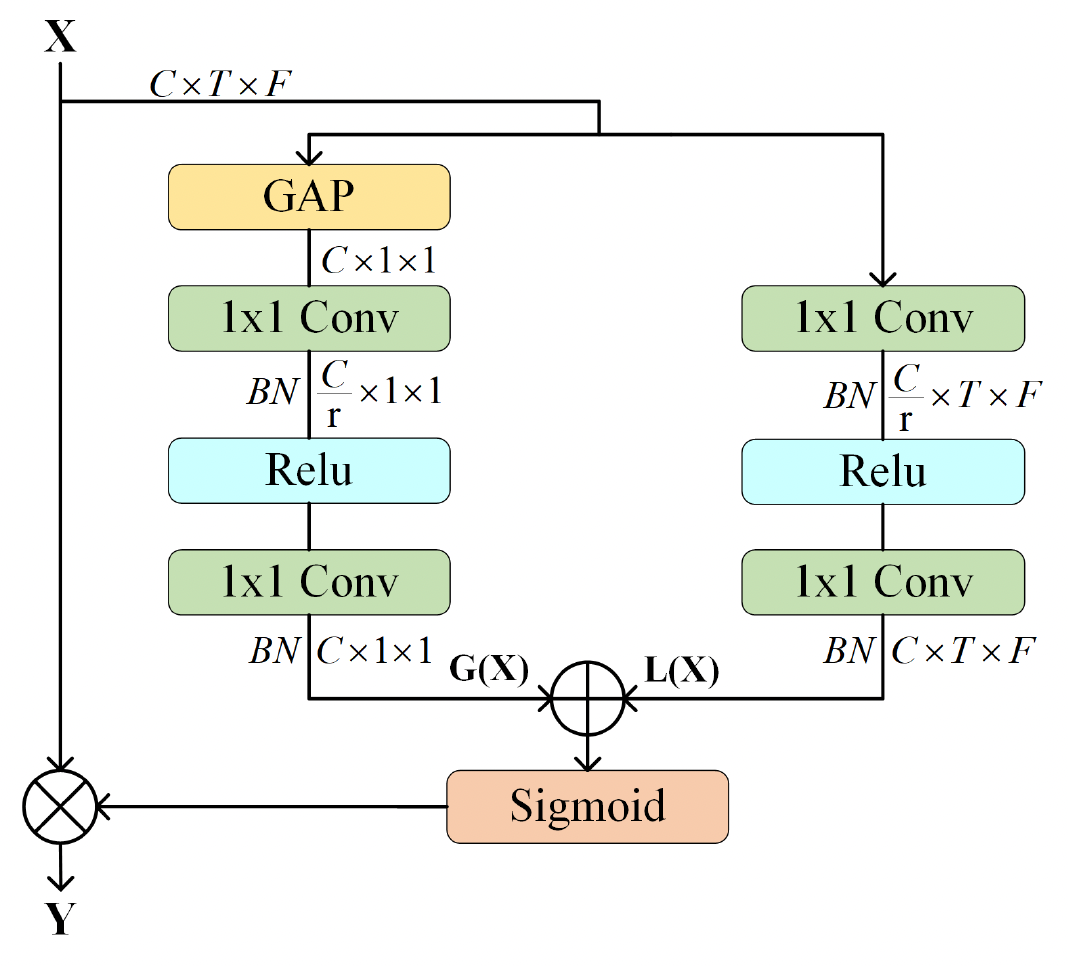}
    \caption{\Gls{msca} module for joint local and global channel recalibration, adapted from L.~Xue et al., \textit{``Resnet-Conformer Network Using Multi-Scale Channel Attention for Sound Event Localization and Detection in Real Scenes''}~\cite{xue2023mscarcnet}.}
    \label{fig:msca}
\end{wrapfigure}

The \gls{msca} block (Figure~\ref{fig:msca}) is relevant to \gls{at2seld} because it provides a compact example of spatial-stream recalibration before higher-level sequence modeling. In a semantic-to-spatial transfer setting, this distinction matters: semantic embeddings may provide strong class evidence, but localization still depends on preserving and selectively amplifying channel-dependent structures in the spatial branch.

MSCA-RCnet further introduces \gls{asp} to address temporal-resolution mismatch between high-rate feature sequences and lower-rate \gls{seld} labels~\cite{xue2023mscarcnet}. In datasets such as STARSS22 and STARSS23, frame-level acoustic features may be computed at a finer temporal resolution than the annotation grid~\cite{politis2022starss22,politis_2023_7880637}. Direct temporal pooling enforces dimensional compatibility, but may discard informative transient structure. \Gls{asp} instead computes attention weights over short temporal regions and derives weighted first- and second-order statistics. For frame-level embeddings \(\mathbf{h}_1,\ldots,\mathbf{h}_N\), the attention score and normalized weight are:
\begin{align}
e_n
&=
\mathbf{v}^{\top}
\tanh
\left(
W\mathbf{h}_n
+
\mathbf{b}
\right)
\\
\alpha_n
&=
\frac{
e^{(e_n)}
}{
\sum_{i=1}^{N}e^{(e_i)}
}
\end{align}
where \(e_n\) is the scalar attention score for frame \(n\), \(\alpha_n\) is the normalized attention weight, \(\mathbf{h}_n\) is the frame embedding, \(W\) is a learned projection matrix, and \(\mathbf{v}\) and \(\mathbf{b}\) are learned parameters. The pooled statistics are:
\begin{align}
\boldsymbol{\mu}
&=
\sum_{n=1}^{N}
\alpha_n
\mathbf{h}_n
\\
\boldsymbol{\sigma}
&=
\sqrt{
\sum_{n=1}^{N}
\alpha_n
\left(
\mathbf{h}_n
\odot
\mathbf{h}_n
\right)
-
\boldsymbol{\mu}
\odot
\boldsymbol{\mu}
}
\end{align}
where \(\boldsymbol{\mu}\) is the attention-weighted mean, \(\boldsymbol{\sigma}\) is the attention-weighted standard deviation, and \(\odot\) denotes element-wise multiplication. In this context, \gls{asp} is not merely a pooling layer; it is a temporal compression mechanism that preserves information about which frames are most informative before aligning the representation with the label resolution.

The output formats used in these systems also reflect the evolution described in Section~\ref{sec:problem_taxonomy}. The DCASE 2022 ResNet-Conformer system trains both \gls{accdoa} and Multi-\gls{accdoa} variants and fuses their predictions, using the former for sharper event boundaries and the latter for improved same-class overlap handling~\cite{wang2022resnetconformer}. MSCA-RCnet combines a Multi-\gls{accdoa} model with an \gls{ein}-V2-based Multi-\gls{accdoa} variant and averages their outputs~\cite{xue2023mscarcnet}. The 2024 system extends the Multi-\gls{accdoa} representation to include source distance, thereby moving from \gls{seld} toward joint localization, detection, and distance estimation~\cite{vo2024resnetconformer_einv2}. For track \(n\), class \(c\), and frame \(t\), the target can be written as:
\begin{equation}
\mathbf{y}_{n,c,t}
=
\left[
a_{n,c,t}\mathbf{r}_{n,c,t},
d_{n,c,t}
\right]
\end{equation}
where \(\mathbf{y}_{n,c,t}\) is the extended output target, \(a_{n,c,t}\in\{0,1\}\) is the activity target, \(\mathbf{r}_{n,c,t}\in\mathbb{R}^{3}\) is the unit Cartesian \gls{doa} vector, and \(d_{n,c,t}\in[0,\infty)\) is the source distance in meters. The model is trained with an \gls{mse}-\gls{adpit} objective, preserving permutation-aware supervision while extending the regression target beyond direction alone~\cite{vo2024resnetconformer_einv2}.

The optimization protocols of these systems are equally informative. Because manually annotated real \gls{seld} data are scarce, high-performing \gls{dcase} systems rely on aggressive but physically meaningful data expansion. The DCASE 2022 system augments the small official real dataset through \gls{foa}-based \gls{acs}, generating directional variants that preserve Ambisonic consistency, and then adds large-scale simulated multi-channel data constructed from AudioSet~\cite{audioset}, ESC-50~\cite{piczak2015esc50}, FSD50K~\cite{fonseca2022fsd50k}, and TAU-NIGENS \gls{srir}~\cite{politis2021tau_nigens,wang2022resnetconformer}. The system is trained on \(20~\mathrm{s}\) segments with Adam~\cite{kingma2015adam}, a three-step learning-rate schedule, mini-batches of 16 samples, and separate \gls{mse} objectives for \gls{accdoa} and \gls{mse}-\gls{adpit} supervision for Multi-\gls{accdoa}~\cite{wang2022resnetconformer}. MSCA-RCnet combines real STARSS23 data with approximately \(200~\mathrm{h}\) of synthetic content, maintains a mixture of real and simulated scenes during training, and applies AugMix~\cite{hendrycks2020augmix}, \gls{acs}, and test-time \gls{acs} augmentation~\cite{xue2023mscarcnet}. The 2024 architecture adopts a multi-phase strategy in which the model is first trained on synthetic data and then fine-tuned on the official development set with \gls{acs}, random cutout, noise injection, time--frequency masking, frequency shifting, SpecAugment~\cite{park2019specaugment}, random mixing, Adam optimization, learning-rate scheduling, large mini-batches, and early stopping~\cite{vo2024resnetconformer_einv2}.

These systems reveal a coherent design pattern rather than a collection of challenge-specific heuristics. \Gls{acs} increases directional diversity while preserving the physical structure of \gls{foa} signals. Synthetic multi-channel generation compensates for the scarcity of annotated real spatial corpora and provides controllable overlap, direction, and reverberation conditions. Multi-phase optimization separates the acquisition of generic spatial structure from adaptation to real acoustic scenes. Attention and Conformer blocks strengthen temporal and feature-selective modeling, but they do not remove the need for explicit spatial features, activity-coupled outputs, or permutation-aware supervision.

For \gls{at2seld}, this lineage provides two complementary lessons. First, strong temporal models and residual encoders are valuable only when the input representation and target structure preserve the spatial information required by the localization task. Second, high-performing \gls{dcase} systems obtain robustness not only from architecture depth, but also from augmentation, calibration, and training design. The proposed framework therefore uses Conformer- and attention-inspired modules as components of a controlled semantic-to-spatial design space, rather than treating them as isolated high-capacity replacements for the \gls{crnn} template. In parallel, it draws from recent \gls{dcase}-oriented systems the methodological emphasis on spatially consistent preprocessing, Ambisonics-aware augmentation, and staged training protocols, using these elements to maximize the utility of \gls{seld}-only datasets without relying exclusively on large external and weakly labeled corpora.

\subsection{Raw-Waveform and Learnable Front-Ends}
\label{subsec:related_raw_waveform}

The systems reviewed in Sections~\ref{subsec:related_crnn}--\ref{subsec:related_conformer} mostly rely on explicit time--frequency front-ends, such as magnitude--phase spectrograms, Log-Mel features, \glspl{iv}, or \gls{gccphat}-like descriptors. A different design direction is represented by raw-waveform \gls{seld} systems, which attempt to learn semantic and spatial representations directly from multi-channel sound-pressure signals. This family does not simply replace the \gls{stft} by a deeper convolutional front-end. Rather, it asks whether the feature extraction stage itself can be parameterized so that spectral selectivity, inter-channel phase structure, temporal extent, and localization evidence are learned jointly with the downstream task.

Two representative examples are SoundDet~\cite{he2021sounddet} and SoundDoA~\cite{he2022sounddoa}. Both systems abandon the conventional \gls{stft}-to-Mel pipeline, but they do so through different assumptions. SoundDet introduces a compact phase-sensitive filter bank and reformulates moving-source \gls{seld} as dense event proposal scoring. SoundDoA instead constructs a learnable complex-valued time--frequency representation using parameterized Gabor filters and processes the resulting sound-object representation through a multi-task, track-wise architecture. The comparison between the two systems is useful because it separates two questions that are often conflated: whether the front-end should be learned from waveform, and whether the output should remain framewise or become event-proposal-based.

\begin{figure}[ht]
    \centering
    \includegraphics[width=\linewidth]{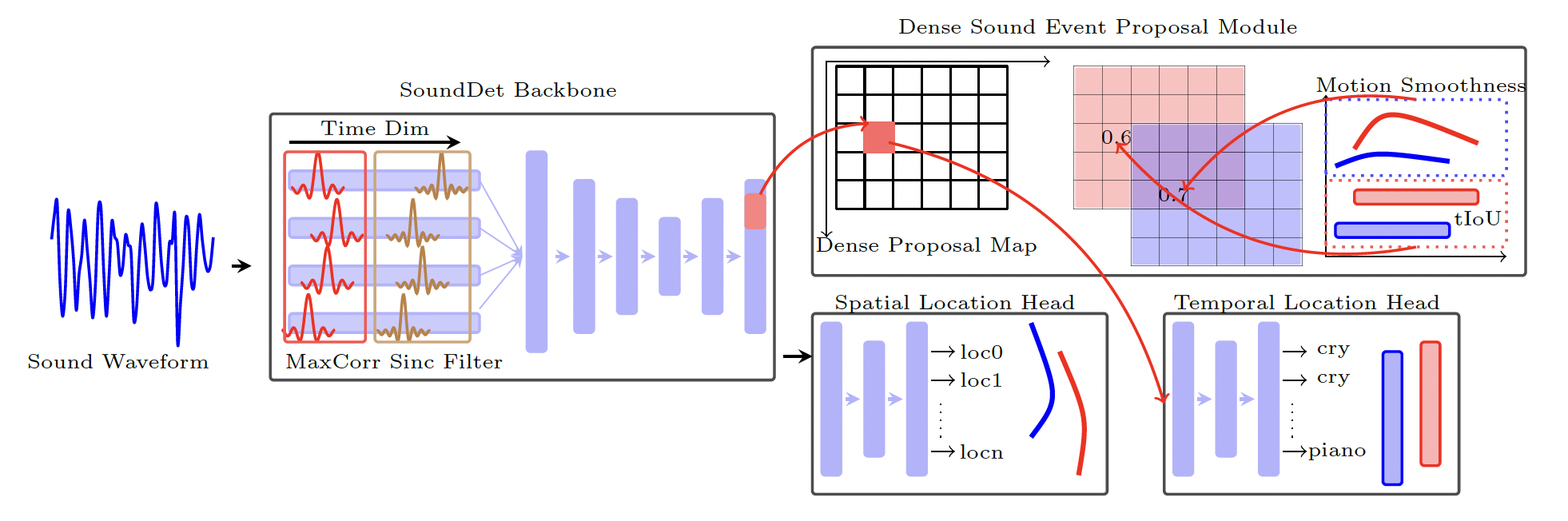}
    \caption{SoundDet model structure, with raw-waveform feature extraction, framewise localization, and dense event proposal scoring, adapted from Y.~He et al., \textit{``SoundDet: Polyphonic Moving Sound Event Detection and Localization from Raw Waveform''}~\cite{he2021sounddet}.}
    \label{fig:sounddet}
\end{figure}

SoundDet formulates polyphonic moving \gls{seld} directly from raw multi-channel waveforms by treating each event as a complete spatio-temporal object characterized by onset, offset, class label, and spatial trajectory~\cite{he2021sounddet}. The architecture contains a raw-waveform backbone, a framewise spatial localization head, and a dense event proposal module (Figure~\ref{fig:sounddet}). This is a substantial departure from standard framewise \gls{seld}: temporal extent and spatial smoothness are not reconstructed only by thresholding local predictions, but are explicitly scored at the level of candidate events.

The SoundDet front-end begins with the \gls{maxcorr} filter bank, a learnable multi-channel extension of a sinc-based band-pass filtering. A rectangular band-pass filter is parameterized as:
\begin{equation}
k[n;f_1,f_2]
=
2f_2\,\mathrm{sinc}(2\pi f_2 n)
-
2f_1\,\mathrm{sinc}(2\pi f_1 n)
\end{equation}
where \(k[n;f_1,f_2]\) is the discrete-time band-pass filter, \(n\) is the sample index, and \(f_1\) and \(f_2\) are learnable lower and upper cutoff frequencies. The multi-channel extension introduces channel-specific learnable time shifts:
\begin{equation}
\mathbf{g}
\left[
n;f_1,f_2,\tau_1,\ldots,\tau_{C_{\mathrm{ch}}}
\right]
=
\left[
k[n+\tau_1;f_1,f_2],
\ldots,
k[n+\tau_{C_{\mathrm{ch}}};f_1,f_2]
\right]
\end{equation}
where \(\mathbf{g}[\cdot]\) is the multi-channel \gls{maxcorr} filter, \(C_{\mathrm{ch}}\) is the number of input channels, and \(\tau_i\) is the learnable time shift applied to channel \(i\). For a four-channels input, each filter is therefore controlled by two spectral cutoff parameters and four temporal-shift parameters. The resulting filters are frequency-selective and phase-sensitive, so they can encode between-channel delay structure with fewer parameters than unconstrained 1D or 2D convolutional kernels.

\begin{wrapfigure}{l}{0.50\textwidth}
    \centering
    \includegraphics[width=0.45\textwidth]{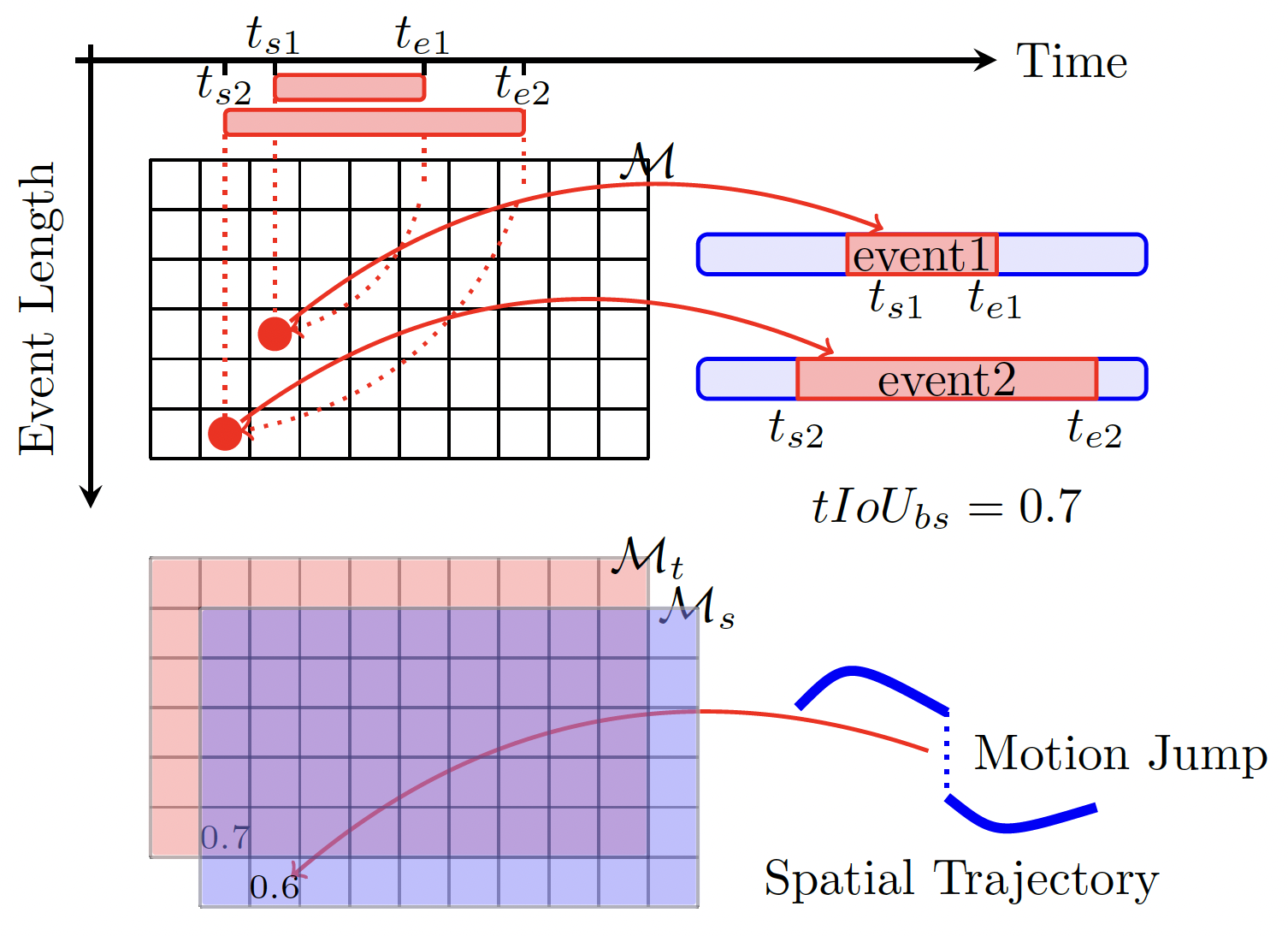}
    \caption{SoundDet dense event proposal module, adapted from Y.~He et al., \textit{``SoundDet: Polyphonic Moving Sound Event Detection and Localization from Raw Waveform''}~\cite{he2021sounddet}.}
    \label{fig:proposal_module}
\end{wrapfigure}
The learned waveform representation is processed by a 1D convolutional encoder--decoder backbone with skip connections. The encoder progressively reduces temporal resolution while increasing channel dimensionality, whereas the decoder partially restores temporal resolution to produce frame-level features aligned with the annotation grid. These features feed two prediction components: a framewise localization head and a dense event proposal module. The latter organizes candidate events into a proposal matrix \(M\), where each cell \(C_{i,j}\) represents a temporal hypothesis with start index \(j\) and duration \(i\). The proposal score is derived from temporal overlap and motion consistency (Figure~\ref{fig:proposal_module}).

The temporal component is based on \gls{tiou}. Given a predicted interval \([t_{s1},t_{e1}]\) and a reference interval \([t_{s2},t_{e2}]\), the temporal overlap is:
\begin{equation}
\mathrm{tIoU}
=
\frac{
\max\{0,\min(t_{e1},t_{e2})-\max(t_{s1},t_{s2})\}
}{
\max(t_{e1},t_{e2})-\min(t_{s1},t_{s2})
}
\end{equation}
where \(t_{s1}\) and \(t_{e1}\) are the predicted start and end times, and \(t_{s2}\) and \(t_{e2}\) are the corresponding reference times. Motion smoothness is measured along the spatial trajectory by:
\begin{equation}
M_s(i,j)
=
\max_{k=0,\ldots,i-1}
d
\left(
\mathbf{l}_{k}
-
\mathbf{l}_{k+1}
\right)
\end{equation}
where \(M_s(i,j)\) is the motion-smoothness score for the proposal starting at \(j\) with duration \(i\), \(d(\cdot)\) is a spatial displacement measure, and \(\mathbf{l}_k\) denotes the predicted source location at temporal index \(k\). A boundary-sensitive variant of \gls{tiou} is further defined as:
\begin{equation}
\mathrm{tIoU}_{\mathrm{bs}}
=
\mathrm{tIoU}
e^
{-w(1-\mathrm{tIoU})}
\end{equation}
where \(\mathrm{tIoU}_{\mathrm{bs}}\) is the boundary-sensitive temporal overlap score and \(w\) is a weighting factor controlling the penalty for boundary mismatch. This event-proposal mechanism is the distinctive contribution of SoundDet: it converts frame-level waveform features and framewise localization into event-level hypotheses evaluated by temporal completeness and spatial consistency.

\begin{figure}[ht]
    \centering
    \includegraphics[width=\linewidth]{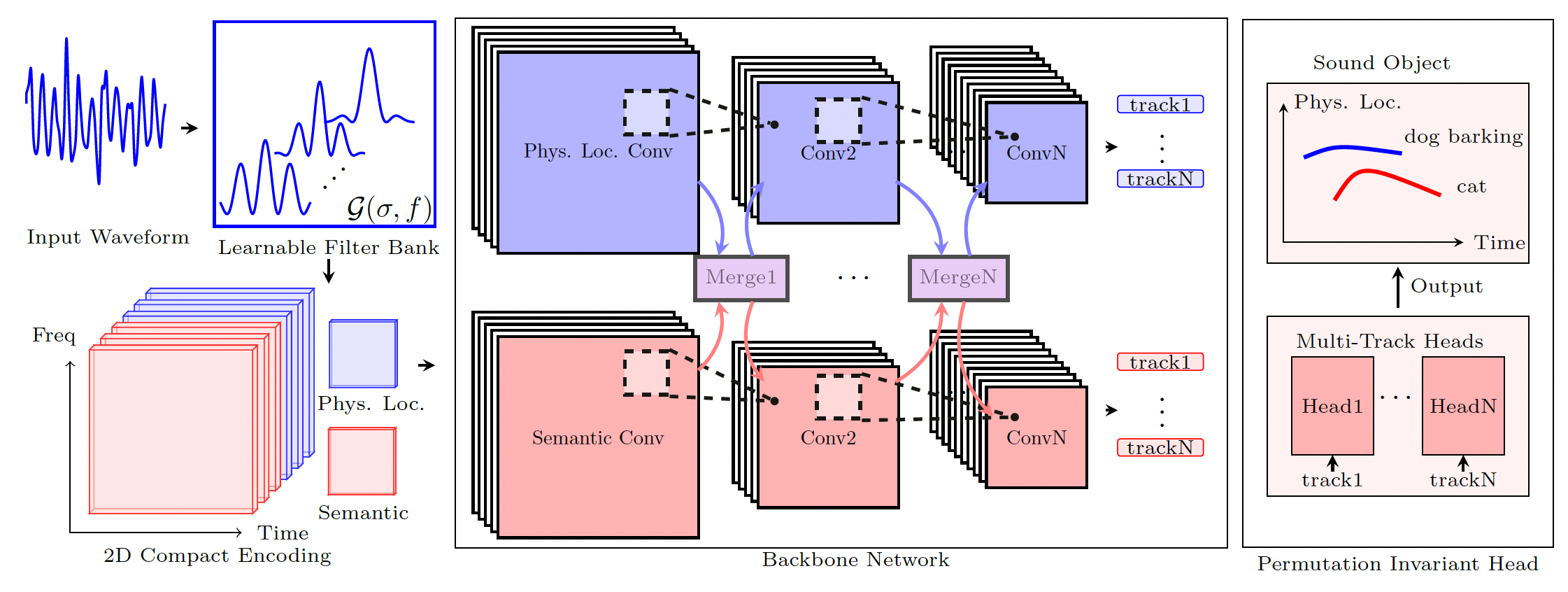}
    \caption{SoundDoA model structure, with learnable Gabor-domain front-end, semantic--spatial enhancement, and track-wise multi-task prediction, adapted from Y.~He and A.~Markham, \textit{``SoundDoA: Learn Sound Source Direction of Arrival and Semantics from Sound Raw Waveforms''}~\cite{he2022sounddoa}.}
    \label{fig:sound_doa}
\end{figure}

SoundDoA follows a different route: rather than introducing event proposals, it learns a complex-valued time--frequency representation from raw waveforms and keeps a track-wise \gls{seld} output structure closer to the \gls{ein} family~\cite{he2022sounddoa}. The front-end is based on parameterized Gabor filters (Figure~\ref{fig:sound_doa}). Each filter is defined as:
\begin{equation}
g_n(t)
=
e^{j2\pi\eta_n t}
\frac{1}{\sqrt{2\pi}\sigma_n}
e^
{-\frac{t^2}{2\sigma_n^2}}
\end{equation}
where \(g_n(t)\) is the \(n\)-th complex Gabor filter, \(t\) is continuous time or the corresponding discrete-time sample variable, \(\eta_n\) is the learnable center frequency, and \(\sigma_n\) is the learnable Gaussian bandwidth. A bank of \(N\) filters with finite temporal support is convolved with the raw waveform to obtain a compact complex time--frequency representation. The use of Gabor filters preserves an interpretable band-pass structure, supports a principled time--frequency localization trade-off, and avoids fixing the analysis window independently of the task. In practice, the center frequencies and bandwidths are initialized according to a Log-Mel-like distribution, with narrower filters at low frequencies and wider filters at high frequencies~\cite{he2022sounddoa}.

\begin{wrapfigure}{r}{0.58\textwidth}
    \centering
    \includegraphics[width=0.56\textwidth]{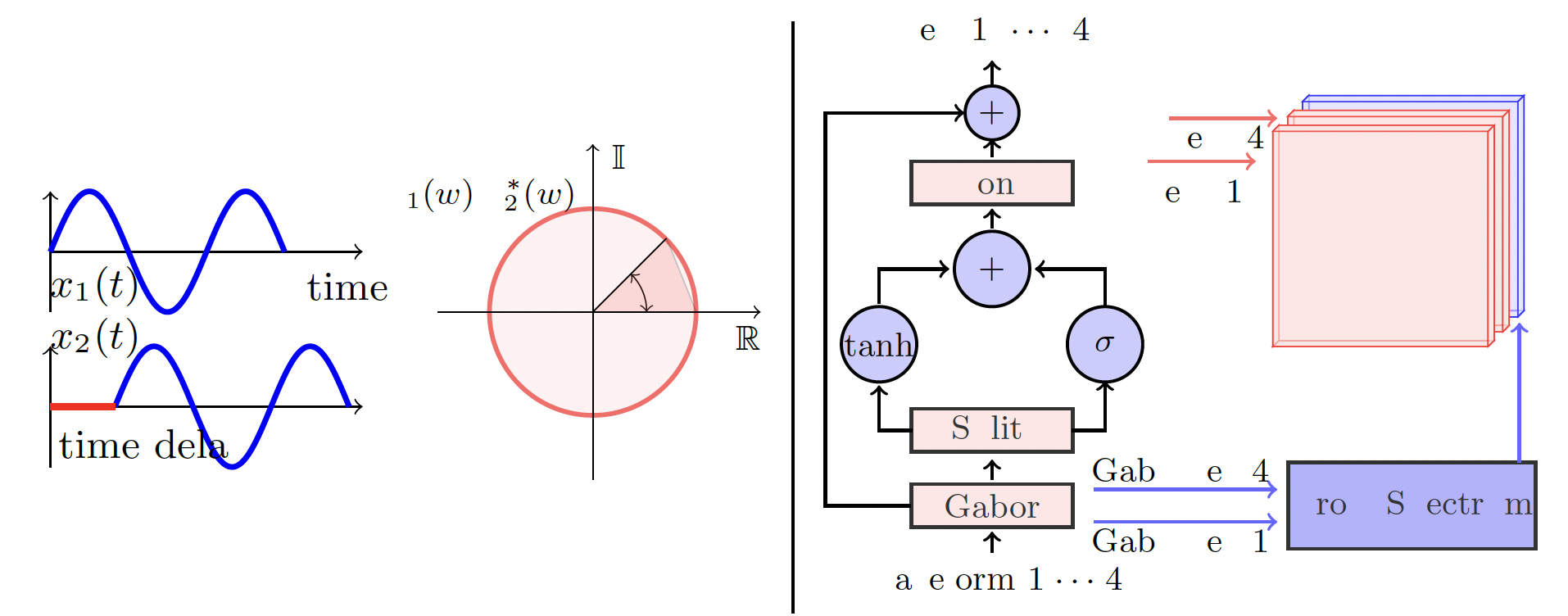}
    \caption{SoundDoA learnable front-end and enhancement module, adapted from Y.~He and A.~Markham, \textit{``SoundDoA: Learn Sound Source Direction of Arrival and Semantics from Sound Raw Waveforms''}~\cite{he2022sounddoa}.}
    \label{fig:sounddoa_frontend}
\end{wrapfigure}

SoundDoA further refines the learned representation through a formant-oriented enhancement stage (Figure~\ref{fig:sounddoa_frontend}). Let \(F\) denote the learned time--frequency representation and let \(f_n\) denote its \(n\)-th frequency bin. A smoothed formant trace is computed through a first-order \gls{iir} recursion:
\begin{equation}
\mathrm{fr}(f_n)
=
(1-s)f_n
+
s\,\mathrm{fr}(f_{n-1})
\end{equation}
where \(\mathrm{fr}(f_n)\) is the smoothed trace at bin \(n\), \(s\) is a learnable smoothing coefficient, and \(f_n\) is the input frequency-bin representation. The detail component is obtained by subtraction:
\begin{equation}
\mathrm{dt}(F)
=
F
-
\mathrm{fr}(F)
\end{equation}
where \(\mathrm{dt}(F)\) denotes the residual detail representation. The enhanced feature is then computed as:
\begin{equation}
\tilde{F}
=
\sigma
\left(
\mathrm{fr}(F)
\right)
+
\tanh
\left(
\mathrm{dt}(F)
\right)
\end{equation}
where \(\tilde{F}\) is the enhanced representation, \(\sigma(\cdot)\) is the sigmoid activation, and \(\tanh(\cdot)\) is the hyperbolic tangent activation. The intended effect is to preserve semantic traces such as slowly varying formant-like structures while retaining fine local detail.

Spatial information is encoded in the learned Gabor domain through cross-spectrum operations. Given two channels with learned frequency-domain representations \(F_1(\omega)\) and \(F_2(\omega)\), the inter-channel phase relation is represented by:
\begin{equation}
C_{12}(\omega)
=
F_2(\omega)F_1^{*}(\omega)
\end{equation}
where \(C_{12}(\omega)\) is the cross-spectrum between the two channels, \(F_1(\omega)\) and \(F_2(\omega)\) are the complex spectra, and \(F_1^{*}(\omega)\) denotes the complex conjugate of \(F_1(\omega)\). The exact spatial encoding depends on the recording format. For \gls{foa}, SoundDoA uses cross-spectra between the omnidirectional channel and the directional channels to obtain an \gls{iv}-like representation. For generalized \gls{mic} array recordings, it uses phase relations between real and imaginary components to obtain a \gls{gccphat}-like cue. These spatial channels are stacked with the semantic time--frequency representation to form a preliminary sound-object representation.

The downstream SoundDoA architecture differs from SoundDet in that it does not score dense temporal proposals. Instead, it uses two parallel sub-networks with layer-wise communication, one oriented toward semantic label learning and the other toward spatial localization. The resulting representations are organized into event-independent tracks, and the final prediction is produced by a permutation-invariant multi-track head. In this sense, SoundDoA is closer to the track-wise \gls{seld} formulations discussed in Section~\ref{subsec:related_permutation_tracking}: the front-end is learned from waveform, but the output remains structured around semantic--spatial tracks rather than around proposal-level event objects.

Raw-waveform systems clarify an important boundary for \gls{at2seld}. They demonstrate that spectral analysis, phase-sensitive spatial encoding, and temporal event organization can be partially learned instead of being fixed entirely by a handcrafted \gls{stft} or Log-Mel pipeline. Although these systems do not directly address the transfer of pretrained \gls{gpat} representations into spatially grounded \gls{seld}, they motivate the inclusion of learnable signal-analysis modules in the spatial branch of the proposed search space. In the first \gls{nas} stage, this possibility is explicitly tested by allowing the experimental grid to compare conventional feature-based spatial front-ends with learned waveform-inspired or signal-analysis-inspired modules derived from this lineage. The raw-waveform literature is therefore not treated as an alternative to semantic transfer, but as evidence that the spatial branch should not be fixed \textit{a priori}.

\subsection{Geometry-Aware Neural Localization}
\label{subsec:related_geometry}

The previous subsections reviewed \gls{seld} systems in which localization is learned together with semantic event detection, either through class-wise \gls{sed}/\gls{doa} branches, track-wise output spaces, activity-coupled representations, or learnable raw-waveform front-ends. A complementary line of work isolates the spatial subproblem more explicitly. In this setting, the objective is not to infer event classes, but to estimate one or more source directions or positions from multi-channel recordings while accounting for reverberation, noise, microphone geometry, and source motion. This distinction is important for \gls{at2seld}, because the spatial branch can benefit from localization-specific inductive biases even when semantic evidence is supplied by a pretrained \gls{gpat} model.

A first reference point is DOANet, which addresses multi-source \gls{ssl} and tracking through a differentiable tracking-based training procedure~\cite{adavanne2021differentiabletracking}. The model processes multi-channel acoustic features with a \gls{crnn} and predicts a fixed maximum number \(N_{\max}\) of Cartesian \gls{doa} trajectories together with track-activity estimates. The key contribution is not the feature extractor alone, but the training objective: rather than minimizing a pointwise regression loss independently for each output, DOANet uses differentiable surrogates of tracking metrics so that localization precision, detection quality, and identity consistency are optimized jointly.

Let \(\tilde{X}_t=[\tilde{\mathbf{x}}_1(t),\ldots,\tilde{\mathbf{x}}_{M_t}(t)]\) denote the set of predicted \gls{doa} vectors at frame \(t\), and let \(X_t=[\mathbf{x}_1(t),\ldots,\mathbf{x}_{N_t}(t)]\) denote the corresponding set of reference source directions. The pairwise distance matrix is:
\begin{equation}
D_t(i,j)
=
d
\left(
\tilde{\mathbf{x}}_i(t),
\mathbf{x}_j(t)
\right)
\end{equation}
where \(D_t(i,j)\) is the distance between predicted source \(i\) and reference source \(j\) at frame \(t\), and \(d(\cdot,\cdot)\) is a spatial distance measure, typically Euclidean distance between Cartesian unit vectors or angular distance. Given an assignment matrix \(A_t\), the framewise localization error is:
\begin{equation}
LE_t
=
\frac{
\left\|
A_t\odot D_t
\right\|_1
}{
\left\|
A_t
\right\|_1
}
\end{equation}
where \(LE_t\) is the localization error at frame \(t\), \(A_t\) is the assignment matrix, \(D_t\) is the pairwise distance matrix, \(\odot\) denotes element-wise multiplication, and \(\|\cdot\|_1\) denotes the entrywise \(L_1\) norm. Across a sequence, the \gls{motp} and \gls{mota} metrics can be written as:
\begin{align}
\mathrm{MOTP}
&=
\frac{
\sum_t
\left\|
A_t\odot D_t
\right\|_1
}{
\sum_t K_t
}
\\
\mathrm{MOTA}
&=
1
-
\frac{
\sum_t
\left(
\mathrm{FP}_t
+
\mathrm{FN}_t
+
\mathrm{IDS}_t
\right)
}{
\sum_t N_t
}
\end{align}
\begin{wrapfigure}{l}{0.58\textwidth}
    \centering
    \includegraphics[width=0.47\textwidth]{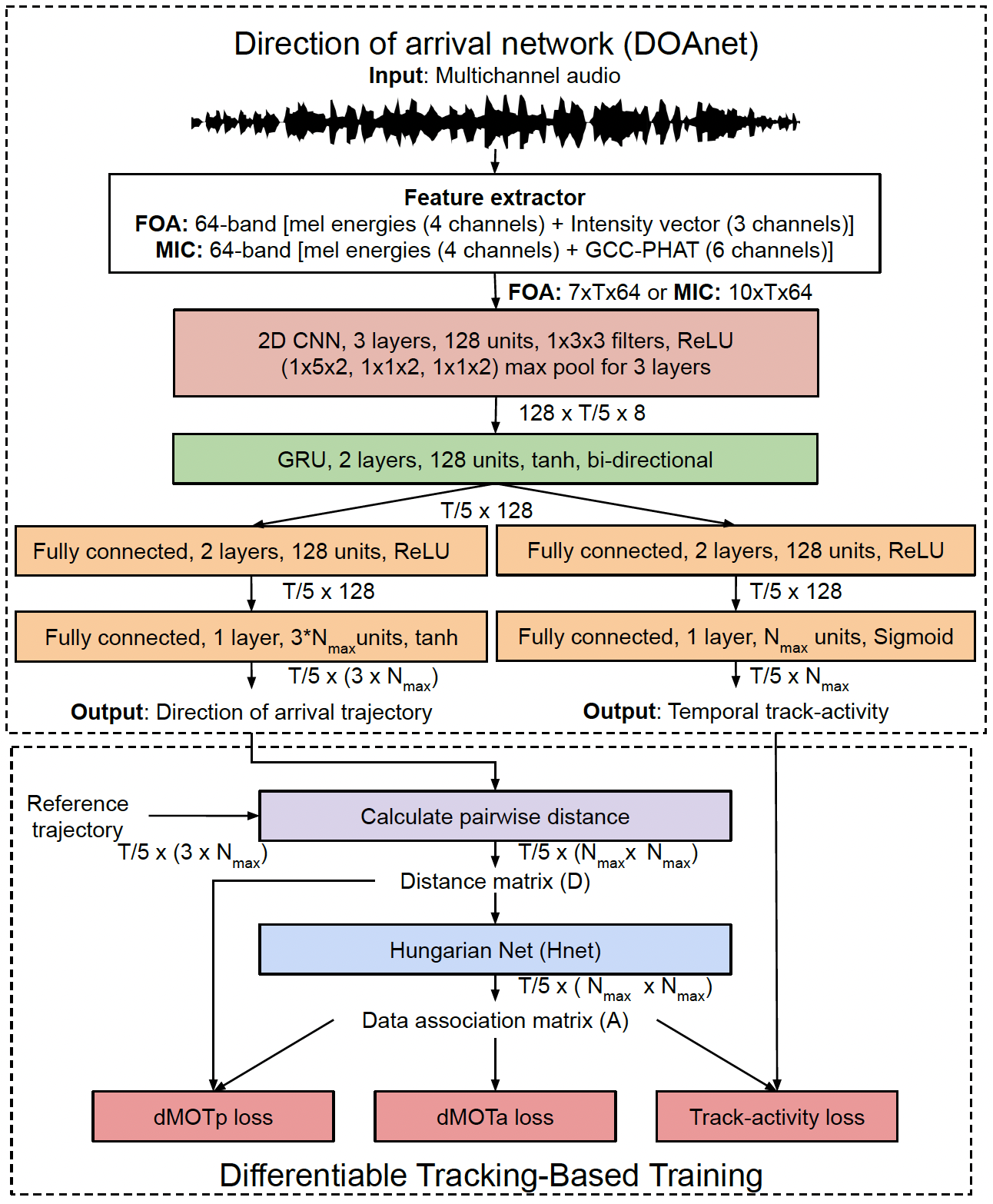}
    \caption{DOANet model structure for differentiable tracking-based \gls{ssl}, adapted from S.~Adavanne et al., \textit{``Differentiable Tracking-Based Training of Deep Learning Sound Source Localizers''}~\cite{adavanne2021differentiabletracking}.}
    \label{fig:doanet}
\end{wrapfigure}
where \(K_t\) is the number of matched source pairs, \(\mathrm{FP}_t\) is the number of false positives, \(\mathrm{FN}_t\) is the number of false negatives, \(\mathrm{IDS}_t\) is the number of identity switches, and \(N_t\) is the number of reference sources at frame \(t\). In standard tracking evaluation, \(A_t\) is obtained through a Hungarian assignment. DOANet introduces a differentiable approximation of this assignment step so that the model can be trained with objectives related to \(\mathrm{MOTP}\), \(\mathrm{MOTA}\), and track activity.

For \gls{foa} input, the feature tensor concatenates four Mel-band energy channels with three \gls{iv} components. For tetrahedral \gls{mic} input, it combines four mel-band energy channels with six \gls{gccphat} channels computed from all microphone pairs. The feature extractor is a \gls{crnn}: convolutional layers encode local spectro-spatial patterns, recurrent layers model temporal structure, and the final representation is split into a Cartesian trajectory branch and a track-activity branch (Figure~\ref{fig:doanet}). This design is fully trainable and naturally supports moving sources, but the learned spatial representation remains tied to the input format and array configuration used during training.

A second strategy is represented by Cross3D, which explicitly computes \gls{srpphat} spatial maps before applying a neural network~\cite{srp-phat_3d}. Instead of learning localization directly from spectrogram-like features, Cross3D treats the \gls{srpphat} map as a spatial likelihood image and uses a causal 3D \gls{cnn} to interpret its temporal evolution. This design preserves a strong beamforming prior while allowing the network to exploit the structure of the full map rather than only its maximum.

\begin{figure}[ht]
    \centering
    \includegraphics[width=\linewidth]{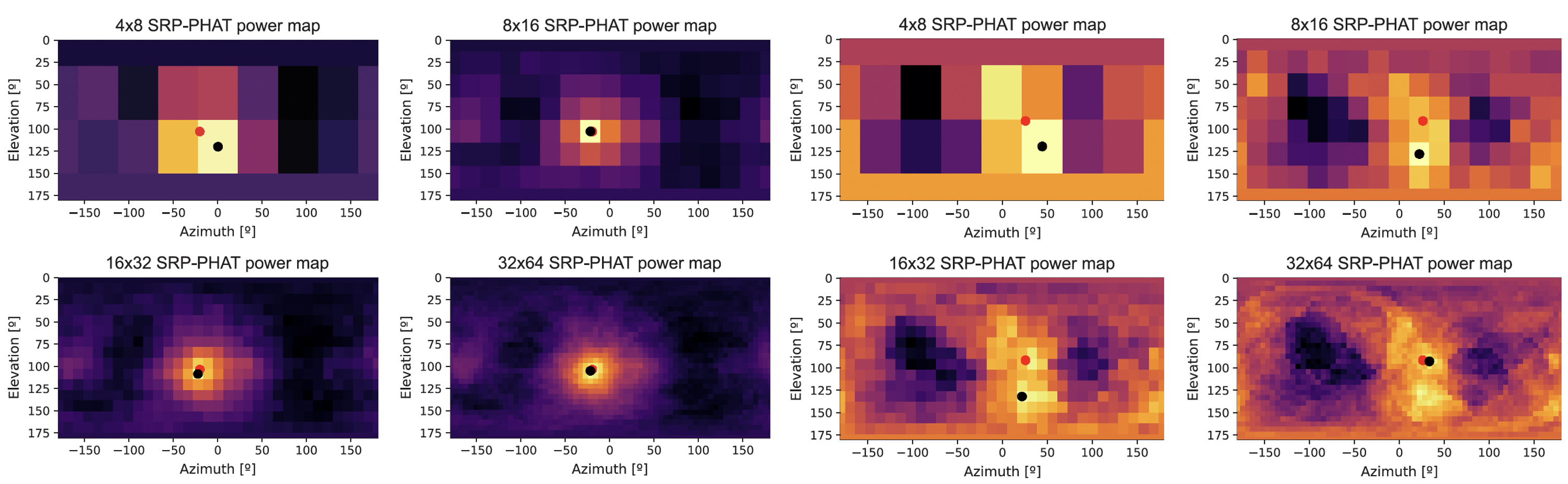}
    \caption{\Gls{srpphat} maps under favorable and challenging acoustic conditions, adapted from D.~Díaz-Guerra et al., \textit{``Robust Sound Source Tracking Using SRP-PHAT and 3D Convolutional Neural Networks''}~\cite{srp-phat_3d}.}
    \label{fig:srp_phat_comparison}
\end{figure}
The classical signal model assumes that the \(n\)-th microphone signal is:
\begin{equation}
x_n(t)
=
a_s(t)
\ast
h_n(\boldsymbol{\theta}_s,t)
+
v_n(t)
\end{equation}
where \(x_n(t)\) is the microphone signal, \(a_s(t)\) is the source signal, \(h_n(\boldsymbol{\theta}_s,t)\) is the \gls{rir} from source position \(\boldsymbol{\theta}_s\) to microphone \(n\), \(v_n(t)\) is additive noise, and \(\ast\) denotes convolution. The source direction is estimated by maximizing the \gls{srp}:
\begin{equation}
\hat{\boldsymbol{\theta}}_s
=
\arg\max_{\boldsymbol{\theta}}
P(\boldsymbol{\theta})
\end{equation}
where \(\hat{\boldsymbol{\theta}}_s\) is the estimated source direction and \(P(\boldsymbol{\theta})\) is the spatial power response for candidate direction \(\boldsymbol{\theta}\). In frequency-domain form:
\begin{equation}
P(\boldsymbol{\theta})
=
\int_{-\infty}^{+\infty}
\left|
\sum_{n=0}^{N-1}
G_n(\omega)
X_n(\omega)
e^{-j\omega\tau_n(\boldsymbol{\theta})}
\right|^2
d\omega
\end{equation}
where \(X_n(\omega)\) is the Fourier transform of the \(n\)-th microphone signal, \(G_n(\omega)\) is a beamforming filter, \(\tau_n(\boldsymbol{\theta})\) is the propagation delay associated with candidate direction \(\boldsymbol{\theta}\), and \(N\) is the number of microphones. The practical \gls{srp} computation can be expressed in terms of pairwise cross-correlations:
\begin{equation}
P(\boldsymbol{\theta})
=
2\pi
\sum_{n=0}^{N-1}
\sum_{m=0}^{N-1}
R_{n,m}
\left(
\Delta\tau_{n,m}(\boldsymbol{\theta})
\right)
\end{equation}
where \(R_{n,m}(\tau)\) is the \gls{gcc} between microphones \(n\) and \(m\), and \(\Delta\tau_{n,m}(\boldsymbol{\theta})=\tau_n(\boldsymbol{\theta})-\tau_m(\boldsymbol{\theta})\) is the relative delay predicted by the candidate direction. The \gls{gcc} is:
\begin{equation}
R_{n,m}(\tau)
=
\frac{1}{2\pi}
\int_{-\infty}^{+\infty}
\Psi_{n,m}(\omega)
X_n(\omega)
X_m^{*}(\omega)
e^{j\omega\tau}
d\omega
\end{equation}
where \(\Psi_{n,m}(\omega)\) is a frequency-domain weighting function, and \(X_m^{*}(\omega)\) is the complex conjugate of \(X_m(\omega)\). When \(\Psi_{n,m}(\omega)\) is chosen as the phase-transform weighting, the method becomes \gls{srpphat}. As shown in Figure~\ref{fig:srp_phat_comparison}, clean conditions tend to produce a dominant spatial peak, whereas reverberation and low \gls{snr} produce multiple local maxima and structured artifacts. Cross3D exploits this observation by learning from the full map rather than relying only on its argmax.

\begin{figure}[ht]
    \centering
    \includegraphics[width=0.92\linewidth]{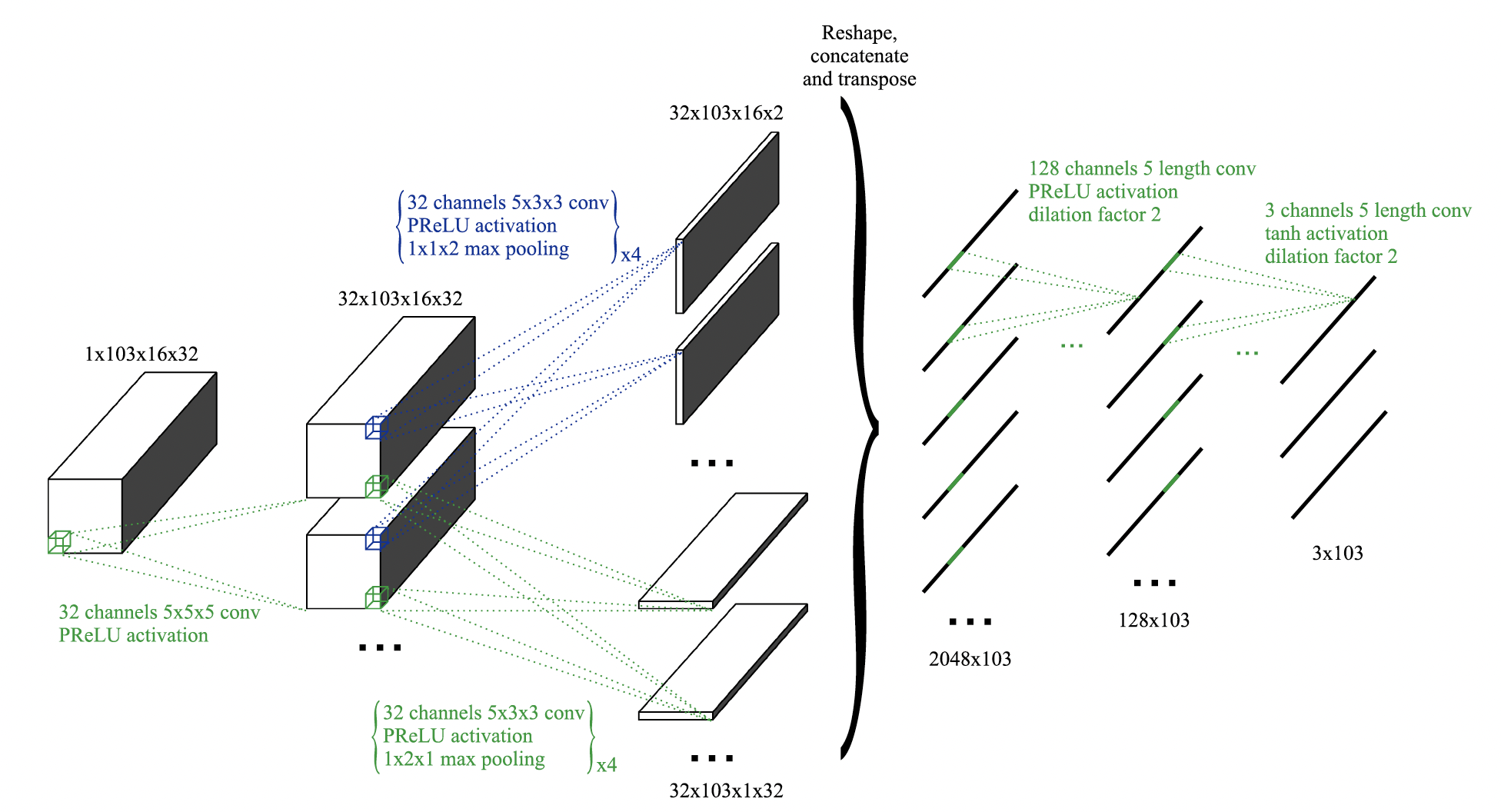}
    \caption{Cross3D architecture for causal \gls{srpphat}-map tracking, adapted from D.~Díaz-Guerra et al., \textit{``Robust Sound Source Tracking Using SRP-PHAT and 3D Convolutional Neural Networks''}~\cite{srp-phat_3d}.}
    \label{fig:cross3d}
\end{figure}
The Cross3D input is a temporally ordered tensor of \gls{srpphat} maps \(
\mathbf{M}
\in
\mathbb{R}^{C_{\mathrm{map}}\times T\times N_{\theta}\times N_{\phi}}
\) where \(\mathbf{M}\) is the input map tensor, \(C_{\mathrm{map}}\) is the number of map channels, \(T\) is the temporal context length, and \(N_{\theta}\) and \(N_{\phi}\) are the elevation and azimuth resolutions. The first channel contains the \gls{srpphat} map, while the remaining channels encode the coordinates of the map maximum:
\begin{equation}
M_{2,t,i,j}
=
\hat{\theta}^{\mathrm{SRP}}_t,
\qquad
M_{3,t,i,j}
=
\hat{\phi}^{\mathrm{SRP}}_t
\end{equation}
where \(M_{2,t,i,j}\) and \(M_{3,t,i,j}\) are the replicated elevation and azimuth estimates at map position \((i,j)\) and time \(t\), and \(\hat{\theta}^{\mathrm{SRP}}_t\) and \(\hat{\phi}^{\mathrm{SRP}}_t\) are the coordinates of the \gls{srp} maximum. This gives the network direct access to the coarse non-differentiable estimate while still allowing it to learn from the distributed spatial pattern of the map. Architecturally, Cross3D applies causal 3D convolutions over time, elevation, and azimuth, then uses complementary pooling branches to avoid destroying directional structure along both spatial axes simultaneously. The final output is a sequence of Cartesian unit-vector components trained with Euclidean localization loss.

Neural-\gls{srp} occupies an intermediate position between generic neural localizers and explicit \gls{srpphat}-map networks~\cite{grinstein2024neuralsrp}. It preserves the pairwise decomposition of \gls{srp}, but replaces the analytical global map with learned pairwise encoders and a trainable global decoder. For a candidate position \(p=[p_x,p_y,p_z]^\top\), the pairwise \gls{srp} contribution of microphones \(i\) and \(j\) can be written as:
\begin{equation}
\mathrm{SRP}_{i,j}
\left(
p;x_i,x_j
\right)
=
\left(
x_i\star x_j
\right)
\left(
\tau_{i,j}(p)
\right)
\end{equation}
where \(\mathrm{SRP}_{i,j}(p;x_i,x_j)\) is the pairwise spatial response, \(x_i\) and \(x_j\) are the microphone signals, \(\star\) denotes cross-correlation, and \(\tau_{i,j}(p)\) is the theoretical \gls{tdoa} for candidate position \(p\). The delay is:
\begin{equation}
\tau_{i,j}(p)
=
\frac{f_s}{c}
\left(
\left\|
v_i-p
\right\|_2
-
\left\|
v_j-p
\right\|_2
\right)
\end{equation}
where \(f_s\) is the sampling frequency, \(c\) is the speed of sound, and \(v_i\) and \(v_j\) are the 3D coordinates of microphones \(i\) and \(j\). The global \gls{srp} response is obtained by summing over all microphone pairs:
\begin{equation}
\mathrm{SRP}
\left(
p;\{x_1,\ldots,x_M\}
\right)
=
\sum_{i=1}^{M}
\sum_{j=i+1}^{M}
\mathrm{SRP}_{i,j}
\left(
p;x_i,x_j
\right)
\end{equation}
where \(M\) is the number of microphones. The source position is then estimated by:
\begin{equation}
\hat{p}
=
\arg\max_{p}
\mathrm{SRP}
\left(
p;\{x_1,\ldots,x_M\}
\right)
\end{equation}
where \(\hat{p}\) is the estimated source position. This decomposition is the central inductive bias of Neural-\gls{srp}: localization is factorized into local pairwise evidence and global aggregation.

The input to Neural-\gls{srp} is the \gls{gccphat} feature computed for every microphone pair. For microphones \(i\) and \(j\), the pairwise feature is:
\begin{equation}
g_{i,j}
=
\mathrm{IDFT}
\left(
\frac{X_i}{|X_i|}
\odot
\frac{X_j^{*}}{|X_j|}
\right)
\end{equation}
where \(g_{i,j}\) is the pairwise \gls{gccphat} feature, \(X_i\) and \(X_j\) are the \gls{dft} spectra of microphone signals \(x_i\) and \(x_j\), \(|\cdot|\) denotes element-wise magnitude, \(X_j^{*}\) is the complex conjugate of \(X_j\), \(\mathrm{IDFT}(\cdot)\) is the inverse \gls{dft}, and \(\odot\) denotes element-wise multiplication. The full input tensor stacks all microphone-pair features in \(
\mathbf{G}
\in
\mathbb{R}^{\frac{M(M-1)}{2}\times T\times G}
\) where \(\mathbf{G}\) is the stacked pairwise feature tensor, \(M(M-1)/2\) is the number of microphone pairs, \(T\) is the number of frames, and \(G\) is the number of retained central correlation delays. The delay dimension is selected according to the maximum theoretical \gls{tdoa} of the array:
\begin{equation}
G
=
2
\max_{1\leq i<j\leq M}
\left\{
\frac{
\|v_i-v_j\|_2 f_s
}{
c
}
\right\}
+
2G_0
\end{equation}
where \(G_0\geq0\) is an extension margin beyond the strict theoretical delay range. This representation is geometry-aware without fixing the global array layout: each pair contributes a local delay cue, and the corresponding microphone coordinates are reintroduced explicitly as metadata.

\begin{figure}[ht]
    \centering
    \includegraphics[width=0.90\linewidth]{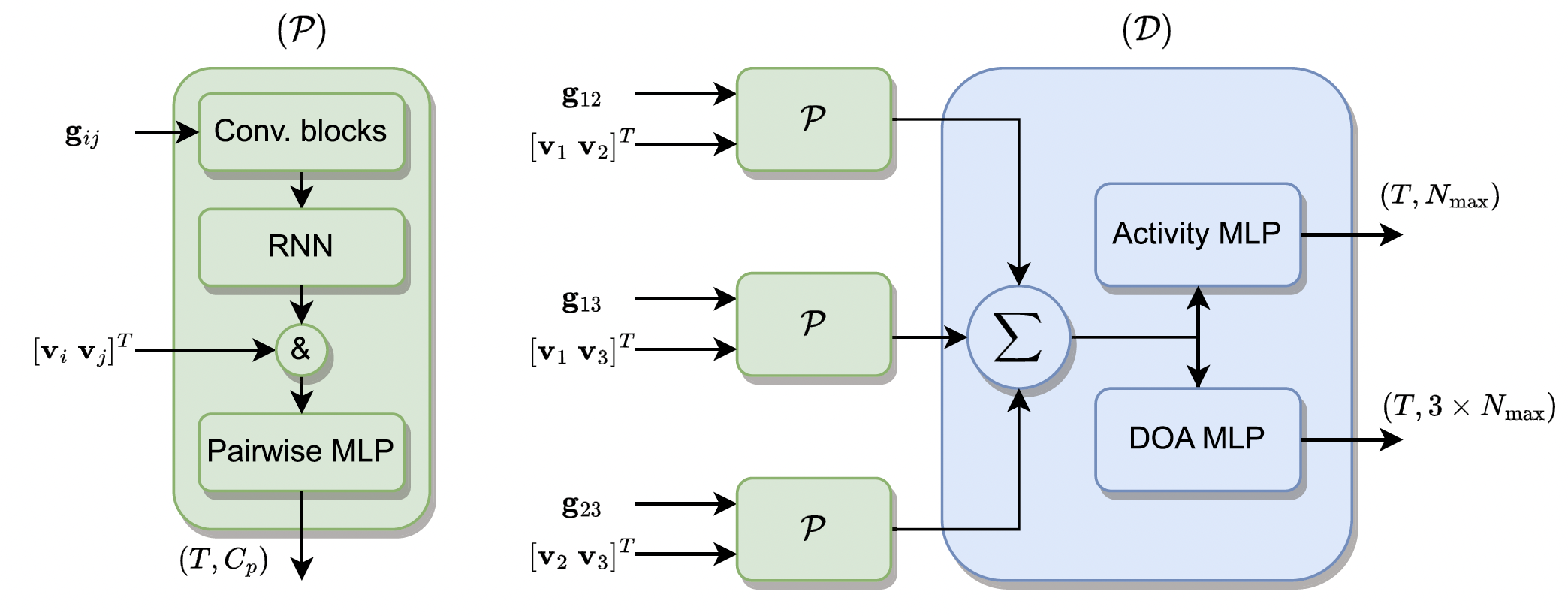}
    \caption{Neural-\gls{srp} architecture with shared pairwise encoding, microphone-coordinate metadata fusion, and summation-based global decoding, adapted from E.~Grinstein et al., \textit{``The Neural-SRP Method for Positional Sound Source Localization''}~\cite{grinstein2024neuralsrp}.}
    \label{fig:neural_srp}
\end{figure}
Neural-\gls{srp} is divided into a pairwise network \(P\) and a global decoder \(D\) (Figure~\ref{fig:neural_srp}). The pairwise network processes each \(g_{i,j}\) with shared parameters, using convolutional blocks over the correlation-delay axis and causal temporal processing. After the pairwise acoustic feature has been encoded, the microphone coordinates \((v_i,v_j)\) are concatenated to the latent representation and passed through a pairwise \gls{mlp}. The resulting pairwise likelihood features are summed over microphone pairs, preserving the additive structure of classical \gls{srp}. The global decoder then predicts source activity and source position from the aggregated representation.

For the single-source case, the Neural-\gls{srp} loss at one time instant is:
\begin{equation}
\mathcal{L}
\left(
U,\hat{U},z,\hat{z}
\right)
=
\alpha z_1
\left\|
\mathbf{u}_1-\hat{\mathbf{u}}_1
\right\|_2
+
\beta
\mathrm{BCE}
\left(
z_1,\hat{z}_1
\right)
\end{equation}
where \(\mathbf{u}_1\) and \(\hat{\mathbf{u}}_1\) are the target and predicted source positions or directions, \(z_1\) and \(\hat{z}_1\) are the target and predicted activity values, and \(\alpha\) and \(\beta\) are task-balancing weights. The localization term is multiplied by the reference activity \(z_1\), so inactive frames do not contribute to the spatial error. This activity-conditioned localization loss is directly relevant to the diagnostic analysis in this report, where inactive-target dominance is shown to affect \gls{doa} learning.

For multiple simultaneous sources, Neural-\gls{srp} must also solve an assignment problem. Let \(U=[\mathbf{u}_1,\ldots,\mathbf{u}_{N}]\) be the target source matrix, let \(\hat{U}=[\hat{\mathbf{u}}_1,\ldots,\hat{\mathbf{u}}_{\hat{N}}]\) be the predicted source matrix, and let \(A\) be an assignment matrix. The multi-source localization term can be written as:
\begin{equation}
\mathcal{L}_{\mathrm{DOA}}
\left(
U,\hat{U},z
\right)
=
\frac{
\left\|
D\odot A
\right\|_1
}{
|z|
}
\end{equation}
where \(D\) is the pairwise distance matrix with entries \(D_{i,j}=\|\mathbf{u}_i-\hat{\mathbf{u}}_j\|_2\), \(A\) is the assignment matrix, \(\odot\) denotes element-wise multiplication, and \(|z|\) is the number of active reference sources. The optimal assignment minimizes this distance, but exact Hungarian matching is not differentiable; Neural-\gls{srp} therefore adopts a neural approximation of the assignment step during backpropagation. This places it in conceptual continuity with the permutation-aware \gls{seld} models discussed in Section~\ref{subsec:related_permutation_tracking}, although the problem is solved in a purely spatial and class-agnostic setting.

The geometry-aware localization lineage shows that spatial reasoning benefits from assignment-aware objectives, full-map spatial evidence, pairwise processing, and microphone-coordinate metadata. These observations complement the class-aware \gls{seld} literature by clarifying which localization-specific inductive biases should be preserved when a spatial branch is coupled with pretrained semantic representations.

\subsection{Positioning of \glsentryshort{at2seld}}
\label{subsec:related_at2seld_positioning}

The reviewed literature identifies the main architectural and methodological axes along which contemporary \gls{seld} systems have evolved. \Gls{crnn} systems establish the joint semantic--spatial processing template, permutation-aware and track-wise models expose assignment as a structural property of multi-source scenes, Conformer-based systems demonstrate the relevance of residual spectro-spatial encoding and high-capacity temporal modeling, raw-waveform approaches show that signal-analysis stages can be learned rather than fixed, and geometry-aware localization models preserve spatial inductive biases through pairwise processing, \gls{srp}-like aggregation, and microphone-coordinate metadata. \Gls{at2seld} is positioned at the intersection of these directions, but with a different central hypothesis: pretrained semantic representations from \gls{gpat} can guide spatially grounded \gls{seld} only if they are integrated with explicit spatial processing, controlled temporal modeling, track-wise supervision, and deployment-aware diagnostics.

Formally, the design space explored in \gls{at2seld} can be interpreted as a family of candidate architectures:
\begin{equation}
\mathcal{A}
=
\left(
\mathcal{E}_{\mathrm{sem}},
\mathcal{E}_{\mathrm{spat}},
\mathcal{F}_{\mathrm{int}},
\mathcal{T},
\mathcal{H},
\mathcal{L}
\right)
\end{equation}

where \(\mathcal{A}\) denotes a candidate \gls{seld} model, \(\mathcal{E}_{\mathrm{sem}}\) is the pretrained semantic encoder inherited from the \gls{gpat} branch, \(\mathcal{E}_{\mathrm{spat}}\) is the spatial front-end or spatial encoder, \(\mathcal{F}_{\mathrm{int}}\) is the semantic--spatial interaction mechanism, \(\mathcal{T}\) is the temporal organization module, \(\mathcal{H}\) is the track-wise prediction head, and \(\mathcal{L}\) is the multi-task supervision objective. This formulation makes explicit that the work is not an unconstrained search for the largest model, but a controlled study of how semantic priors, spatial signal processing, temporal sequence modeling, and loss design interact.

The semantic axis differentiates \gls{at2seld} from most \gls{dcase}-oriented systems reviewed above. In standard \gls{seld} training, semantic discrimination is learned primarily from \gls{seld} labels, possibly supported by synthetic data and augmentation. In \gls{at2seld}, the semantic branch starts from a pretrained \gls{gpat} model trained on large-scale weakly labeled audio and is therefore treated as a high-level acoustic prior rather than as a randomly initialized detector. The central question is whether such a prior can adapt and improve spatially grounded event analysis when it is fine-tuned, coupled, or regularized under \gls{seld} supervision, without collapsing the spatial branch into a purely semantic representation.

The spatial axis inherits complementary elements from the reviewed systems. From \gls{crnn} and Conformer-based \gls{seld}, it preserves explicit multi-channel feature processing and Ambisonics-aware input representations. From raw-waveform and learnable-front-end systems, it inherits the idea that the first stage of spatial analysis should not be fixed a priori: the initial \gls{nas} stage explicitly tests whether learned signal-analysis-inspired modules can improve the spatial branch relative to conventional feature-based processing. From Neural-\gls{srp} and related geometry-aware localization systems, it inherits the methodological principle that spatial cues should remain physically interpretable whenever possible, especially when phase, delay, channel-pair, or geometry-dependent information is used.

The temporal and output axes are shaped by the limitations of class-wise \gls{seld} representations. As discussed in Section~\ref{sec:problem_taxonomy}, class-wise two-branch outputs and basic \gls{accdoa} formulations cannot independently represent same-class spatial overlap without additional tracks or assignment mechanisms. \Gls{at2seld} therefore adopts track-wise \gls{sed}/\gls{doa} prediction heads, preserving a separation between activity estimation and Cartesian localization while allowing multiple concurrent events to be represented through separate and finite tracks. This places the framework closer to \gls{ein}-style and permutation-aware formulations than to purely class-wise \gls{seld} outputs, while retaining a dual-head structure that supports diagnostic analysis of \gls{sed} and \gls{doa} errors separately.

The training and preprocessing strategy is also inherited from recent \gls{dcase} systems, but adapted to the constraints of semantic-to-spatial transfer. Ambisonics-aware augmentation, spatially consistent preprocessing, staged training, class balancing, and threshold calibration are treated as methodological components of the framework rather than as post-hoc performance heuristics. This is particularly important because the work focuses on maximizing the utility of \gls{seld}-focused datasets and controlled augmentation, rather than relying unconditionally on large external corpora at the \gls{seld} stage. Reference results from the reviewed systems are therefore reported only as contextual anchors in Appendix~\ref{app:reference_baselines}, Table~\ref{tab:app_reference_baselines}; they are not interpreted as a direct controlled comparison, since datasets, splits, input formats, augmentation policies, ensembles, and evaluation protocols differ substantially across works.

The methodological role of \gls{at2seld} is consequently diagnostic as much as architectural. The framework evaluates whether semantic transfer improves localization-aware event analysis, but also identifies when it fails: inactive-target dominance in \gls{doa} regression, threshold sensitivity, class imbalance, cross-dataset transfer, and dependence on spatial preprocessing. In this sense, \gls{at2seld} does not simply combine a pretrained \gls{at} model with a \gls{seld} one. It leveraged the reviewed approaches to construct a controlled semantic-to-spatial experimental framework in which the contribution of each component can be isolated, evaluated, and interpreted under deployment-operating conditions.

\clearpage


\section{Framework Design \& Methodology}
\label{sec:experimental_design}
The previous sections defined the conceptual and architectural basis for coupling pretrained semantic audio representations with spatially grounded \gls{seld}. Section~\ref{subsec:related_at2seld_positioning} positioned \gls{at2seld} as a controlled semantic-to-spatial transfer framework, rather than as an unconstrained search for a larger \gls{seld} model. The purpose of the experimental design is therefore to translate this positioning into a reproducible evaluation protocol in which semantic priors, spatial front-ends, temporal modules, track-wise supervision, and optimization strategies are compared under shared target representations and consistent training conditions.

The experimental campaign is methodological rather than leaderboard-oriented. Its objective is not to claim direct superiority over all reference systems reviewed in Section~\ref{sec:related_work}, whose datasets, augmentation policies, output formats, and evaluation protocols differ substantially. Instead, the study investigates how a pretrained \gls{gpat} backbone can be integrated into a \gls{seld} pipeline and which architectural or optimization choices make this transfer effective. The resulting analysis supports two complementary goals: \emph{(I)} identifying robust design choices for semantic-to-spatial \gls{seld}; and \emph{(II)} diagnosing the failure modes that limit deployment-oriented reliability and evaluating mitigation strategies for class imbalance, activity-threshold sensitivity, inactive-target dominance in \gls{doa} regression, and dataset-dependent transfer.


\subsection{Objectives and Experimental Protocol}
\label{subsec:exp_objectives}

The central objective is to determine whether pretrained \gls{gpat} representations can improve spatially grounded event analysis when they are coupled with explicit multi-channel processing and track-wise \gls{seld} supervision. The reference semantic branch is based on \gls{epanns}, selected for its favorable accuracy--efficiency trade-off and for its demonstrated suitability in deployable \gls{at} scenarios~\cite{singh2023epanns, epanns_filtering, giacomelli2026e2panns}. In the present framework, this branch is not used as an external classifier or post-processing stage. It acts as a high-level semantic prior whose representations interact with spatial features, temporal modules, and multi-task losses during \gls{seld} training.

For this reason, the experimental procedure is organized as an informed and staged \gls{nas} process~\cite{elsken2019nas_survey}. The term \emph{informed} is used here in a restricted methodological sense: the candidate modules are not sampled from an unrestricted architectural space, but are selected from the design lineages analyzed in Section~\ref{sec:related_work}. \Gls{crnn} systems motivate the decomposition into spatial encoding, temporal aggregation, and synchronized prediction heads; track-wise and permutation-aware systems motivate source-slot representations and assignment-aware supervision; Conformer-based systems motivate residual spectro-spatial encoding, high-capacity temporal modeling, and Ambisonics-aware augmentation; raw-waveform systems motivate the evaluation of learnable signal-analysis-inspired spatial modules; and geometry-aware localization systems motivate activity-conditioned spatial losses and physically meaningful localization cues (Table~\ref{tab:spatial_modules_inventory}).

The resulting experimental logic is ruled by four research questions:
\begin{itemize}
    \item \textbf{RQ1}: \textit{Which spatial input representation most effectively exposes localization-relevant information to the model?} (among explicit time--frequency descriptors, learned signal-analysis modules, and spatial encoders based on phase-sensitive or channel-dependent processing)

    \item \textbf{RQ2}: \textit{How should spatial information be processed before interacting with the semantic branch?}

    \item \textbf{RQ3}: \textit{Should semantic and spatial streams interact only through shared \gls{seld} supervision, or does explicit feature-level coupling improve semantic-to-spatial transfer?}

    \item \textbf{RQ4}: \textit{How should temporal modules and track-wise prediction heads represent sequential consistency and overlapping events under permutation-aware supervision?}
\end{itemize}

These questions define the staged structure of the experimental campaign and establish the methodological criteria used to interpret the subsequent results. Stage~1 addresses \textbf{RQ1} and \textbf{RQ2} through a shallow screening of spatial front-ends, spatial-processing modules, and temporal smoothing operators under a common training setup. Stage~2 further investigates \textbf{RQ2} by deepening the strongest configurations, in order to determine where additional capacity is beneficial and where it introduces unnecessary computational or optimization cost. Stage~3 addresses \textbf{RQ3} by studying regularization and semantic--spatial interaction, including whether explicit feature bridges improve or destabilize \gls{seld} performance. The final diagnostic stage complements \textbf{RQ1}--\textbf{RQ4} by moving beyond architecture selection and analyzing the selected configurations under loss calibration, class balancing, active-only spatial supervision, temporal-context variation, and deployment-oriented thresholding.

This structure deliberately separates architectural selection from diagnostic interpretation. Early stages identify promising configurations under controlled assumptions; later stages test whether the observed behavior is robust to supervision choices, dataset composition, and operating-point calibration. This separation is essential because a semantic prior may improve class evidence while still leaving localization fragile, or conversely stabilize event detection without improving track-wise spatial consistency. The experimental design therefore treats \gls{at2seld} as a structured investigation of semantic-to-spatial transfer, in which performance tables are interpreted together with error modes, calibration behavior, and dataset-dependent generalization.


\subsection{Informed \glsentryshort{nas} Staging Rationale}
\label{subsec:exp_staging}

The \gls{at2seld} experimental campaign is organized as an informed multi-stage \gls{nas} procedure. The term \emph{staging} denotes that the architecture space is not evaluated as a single exhaustive grid, but is progressively restricted and refined according to the evidence produced by each phase. This design is required because the complete search space combines heterogeneous variables: spatial input representations, early spatial encoders, late track-wise abstraction modules, temporal smoothing operators, semantic--spatial bridges and their hyper-parameters, regularization policies, and loss formulations. Evaluating all combinations simultaneously would be computationally demanding and would obscure the contribution of each processing level, making the results difficult to analyze and interpret.

The staging logic is also motivated by the state-of-the-art analysis in Section~\ref{sec:related_work}. Several design choices are well established as effective in recent \gls{seld} systems, however, the literature does not always provide a controlled characterization of where these mechanisms should be inserted, how they interact, or whether their benefits persist when the spatial branch is coupled with a \gls{gpat} prior. Each stage of the proposed search therefore focuses on a specific under-characterized axis: spatial representation, depth allocation, semantic--spatial interaction, and final-system diagnostic characterization. The staged design follows directly from the research questions introduced in Section~\ref{subsec:exp_objectives}. The resulting protocol is therefore sequential, reducing uncertainty about a specific design axis before the next axis is explored.

\begin{figure}[ht]
    \centering
    \includegraphics[width=\linewidth]{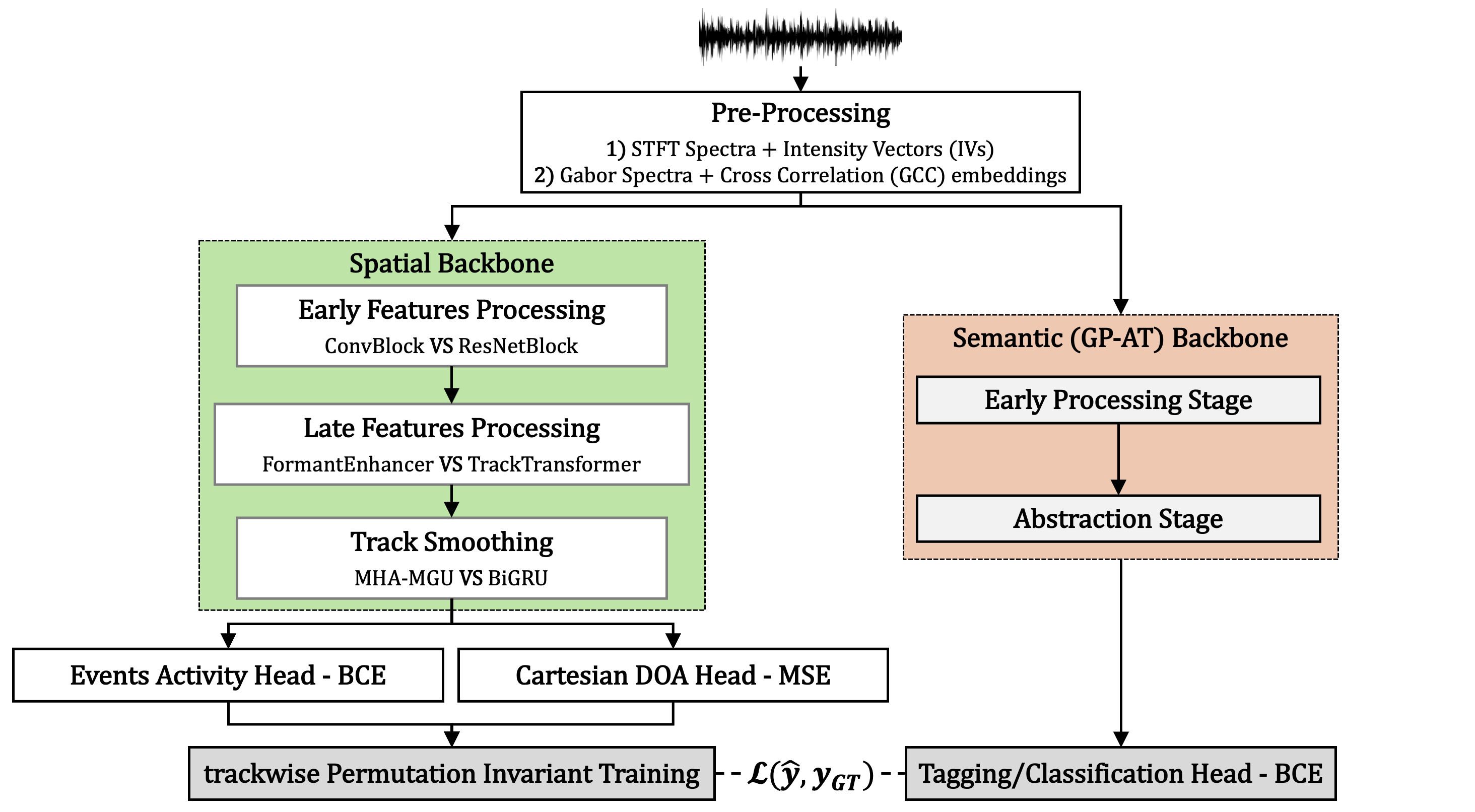}
    \caption{\Gls{nas} Stage~1: shallow grid search over spatial front-end families, early spatial encoders, late track-wise abstraction modules, and temporal smoothing operators.}
    \label{fig:nas_stage_1}
\end{figure}

\textbf{Stage~1} addresses the compatibility between spatial input representations and shallow processing modules, performing a shallow architectural screening under a shared training protocol. Two spatial front-end families are compared: \emph{(I)} a spectral \gls{foa} branch based on \gls{stft} magnitude, phase, and \gls{iv} features; and \emph{(II)} a Gabor-based branch inspired by learnable raw-waveform front-ends (Section~\ref{subsec:related_raw_waveform}), in which learned filterbank spectra are combined with cross-spectrum spatial embeddings. For each family, the search strategy instantiates one module in the early spatial stage, one module in the late track-wise abstraction stage, and one module in the temporal smoothing stage (Figure~\ref{fig:nas_stage_1}). The objective is to determine which module families are compatible with the semantic-to-spatial setting before increasing depth or introducing explicit feature bridges.

At this stage, the semantic and spatial branches interact only through the joint optimization objective. The \gls{gpat} branch provides pretrained semantic representations and an auxiliary presence signal, whereas the spatial branch is responsible for producing track-wise \gls{sed}/\gls{doa} predictions. No explicit feature-level fusion is introduced in Stage~1. This makes the first stage a controlled test of two questions left open by the reviewed design lineages: whether the spatial branch can benefit of semantic prior through loss-level coupling alone, and whether a learnable signal-analysis front-end can compete with explicit \gls{foa} time--frequency descriptors in the spatial branch.

\begin{figure}[ht]
    \centering
    \includegraphics[width=\linewidth]{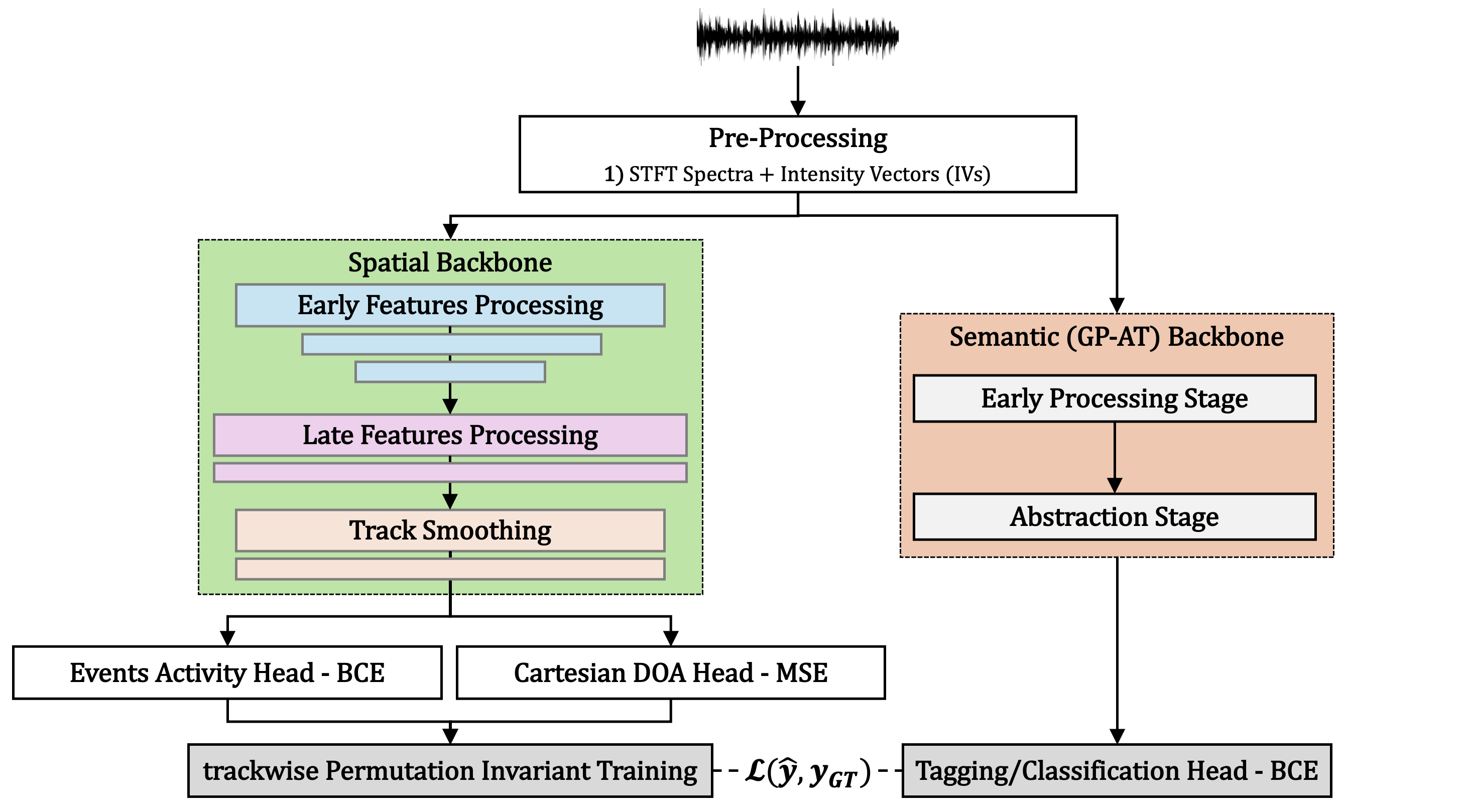}
    \caption{\Gls{nas} Stage~2: controlled depth allocation over the early spatial stage, late track-wise abstraction stage, and temporal smoothing stage.}
    \label{fig:nas_stage_2}
\end{figure}

\textbf{Stage~2} studies the effect of controlled depth allocation once the strongest shallow topology has been identified in Stage~1. The purpose is not to enlarge the architecture uniformly, but to determine where additional capacity is beneficial. Three depth axes are considered: the early spatial encoder, the late track-wise abstraction module, and the temporal smoothing operator. This distinction is important because these stages process different types of information. The early stage still operates on dense spatial time--frequency features; the late stage organizes the representation closer to track-wise latent states; the smoothing stage refines temporal consistency after track-wise abstraction has already occurred.

To preserve comparability across configurations, Stage~2 uses shape-preserving depth policies. In the early spatial stage, stacked encoders follow a bottleneck channel schedule: internal channel capacity may increase across intermediate layers, but the final layer projects to the reference interface expected by the subsequent stage. In the late abstraction stage, depth is introduced either by stacking shape-preserving modules or by increasing the number of internal layers of the selected Transformer-like operators, while keeping the external tensor interface unchanged. In the smoothing stage, recurrent modules are deepened by increasing the number of recurrent layers, whereas attention-based smoothers are deepened by cascading homologous blocks (Figure~\ref{fig:nas_stage_2}). This strategy makes the comparison more interpretable: performance changes can be related primarily to depth allocation rather than to uncontrolled changes in layer-wise interfacing.

\begin{figure}[ht]
    \centering
    \includegraphics[width=\linewidth]{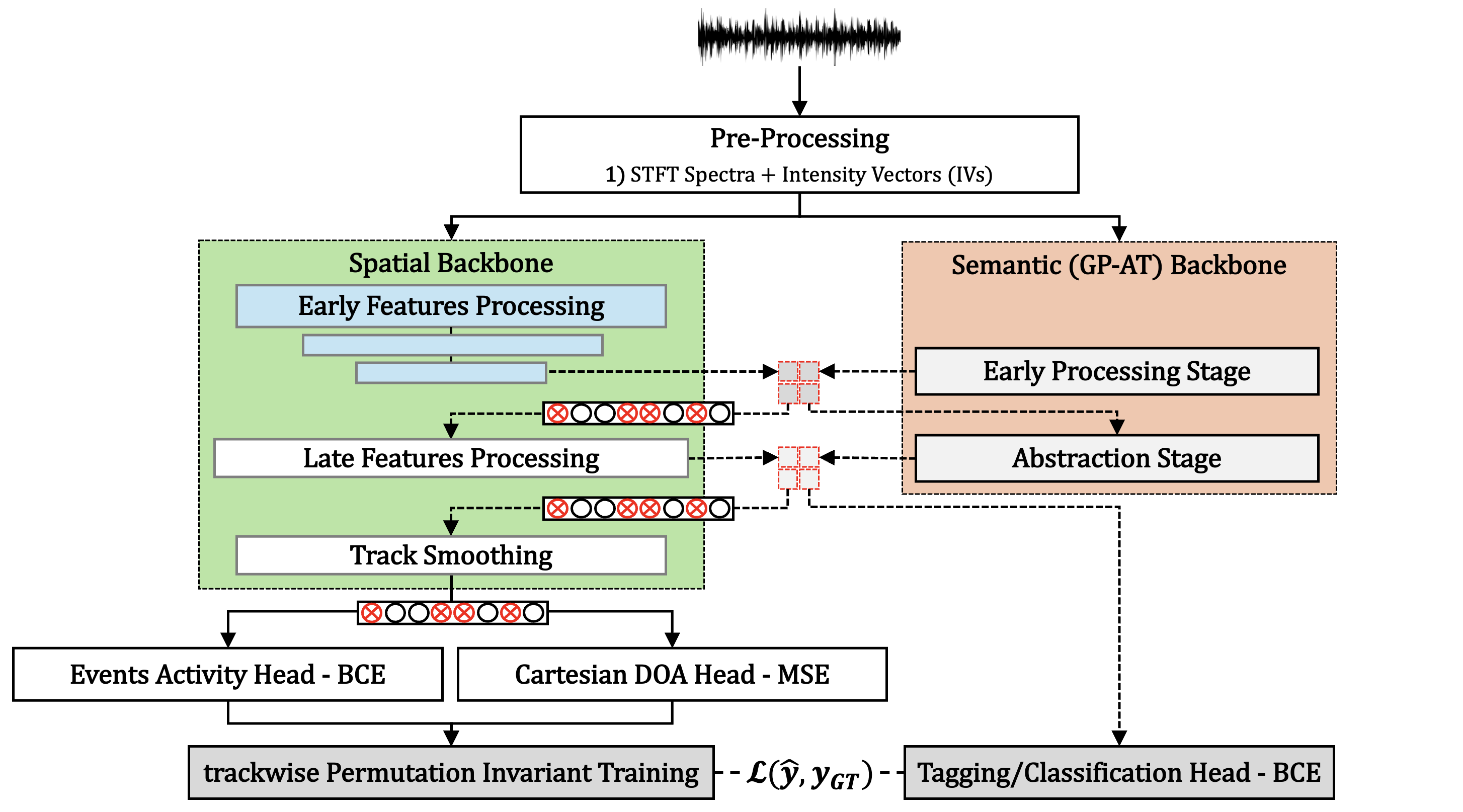}
    \caption{\Gls{nas} Stage~3: regularization and semantic--spatial interaction search through controlled cross-branch bridge insertion.}
    \label{fig:nas_stage_3}
\end{figure}

\textbf{Stage~3} evaluates whether the best deep topology benefits from stronger regularization and explicit semantic--spatial interaction. The motivation comes from the track-wise and multi-task systems reviewed in Section~\ref{subsec:related_permutation_tracking}, where \gls{sed} and \gls{doa} branches are allowed to exchange information through controlled sharing mechanisms rather than being fully merged. However, the effect of such interaction is not fully characterized when one branch is initialized from a pretrained semantic model and the other is responsible for spatial inference. In \gls{at2seld}, the idea is therefore adapted to a different setting: the interacting streams are not two task-specific \gls{seld} branches, but an aggregated semantic branch and a multi-channel spatial branch. The bridge is therefore custom-designed to preserve the physical structure of the spatial representation while injecting temporally aligned semantic evidence.

Let:
\begin{equation}
\mathbf{S}\in\mathbb{R}^{B\times C_s\times T_s\times F_s},
\qquad
\mathbf{E}\in\mathbb{R}^{B\times C_e\times T_e\times F_e}
\end{equation}
where \(\mathbf{S}\) is the spatial feature map, \(\mathbf{E}\) is the semantic feature map, \(B\) is the batch size, \(C_s\) and \(C_e\) are the spatial and semantic channel counts, \(T_s\) and \(T_e\) are the temporal resolutions, and \(F_s\) and \(F_e\) are the corresponding frequency or latent-frequency dimensions. The asymmetry between the two tensors is part of the bridge design. The spatial stream retains a time--frequency lattice tied to multi-channel localization cues, whereas the semantic stream encodes class-discriminative representations whose latent-frequency axis is not physically tied to microphone-array geometry. The bridge therefore aligns and conditions the streams without treating their frequency axes as equivalent.

The bridge begins with spatial preconditioning. The spatial feature map is processed by an inverted depth-wise separable multi-scale attention block:
\begin{equation}
\tilde{\mathbf{S}}
=
\Phi_s(\mathbf{S})
\end{equation}
where \(\Phi_s(\cdot)\) denotes the spatial preconditioning operator. Operationally, \(\Phi_s\) first expands the channel dimension with a point-wise \(1\times1\) convolution:
\begin{equation}
\mathbf{U}
=
\textrm{ReLU}
\left(
\beta
\left(
W_{\mathrm{exp}} * \mathbf{S}
\right)
\right)
\end{equation}
where \(\mathbf{U}\in\mathbb{R}^{B\times C'\times T_s\times F_s}\) is the expanded spatial tensor, \(W_{\mathrm{exp}}\) is the learnable \(1\times1\) expansion kernel, \(C'\) is the expanded channel count, \(*\) denotes convolution, and \(\beta(\cdot)\) denotes batch normalization. The expanded tensor is then processed by parallel depth-wise convolutions~\cite{howard2017mobilenets} with different receptive fields:
\begin{equation}
\mathbf{U}^{(m)}_{b,c,t,f}
=
\sum_{\Delta t=-r_t^{(m)}}^{r_t^{(m)}}
\sum_{\Delta f=-r_f^{(m)}}^{r_f^{(m)}}
K^{(m)}_{\mathrm{dw},c,\Delta t,\Delta f}
\mathbf{U}_{b,c,t+\Delta t,f+\Delta f}
\end{equation}
where \(\mathbf{U}^{(m)}\) is the output of the \(m\)-th depth-wise branch, \(K^{(m)}_{\mathrm{dw}}\) is the channel-specific depth-wise kernel, \(r_t^{(m)}\) and \(r_f^{(m)}\) define the temporal and frequency radii of the kernel, and \((b,c,t,f)\) index batch, channel, time, and frequency. The multi-scale responses are aggregated and projected back to the bridge interface:
\begin{equation}
\tilde{\mathbf{S}}
=
\mathbf{S}
+
W_{\mathrm{proj}}
*
\left(
\sum_{m=1}^{M_{\mathrm{ms}}}
\mathbf{U}^{(m)}
\right)
\end{equation}
where \(W_{\mathrm{proj}}\) is the point-wise projection kernel and \(M_{\mathrm{ms}}\) is the number of multi-scale depth-wise branches. This spatial preconditioning step follows the same local/global recalibration principle discussed for \gls{msca} in Section~\ref{subsec:related_conformer}, but is used here as a bridge-specific preprocessing operator. Its purpose is to refine spatial-channel salience before semantic information is injected.

The semantic stream is then aligned to the spatial temporal grid. Since \(T_e\) and \(T_s\) may differ, the semantic tensor is resized along time using linear interpolation:
\begin{equation}
\widetilde{E}_{b,c,t',f}
=
(1-\alpha_{t'})
E_{b,c,t_0,f}
+
\alpha_{t'}
E_{b,c,t_1,f}
\end{equation}
where \(\widetilde{E}_{b,c,t',f}\) is the temporally resized semantic tensor at target frame \(t'\), \(E_{b,c,t_0,f}\) and \(E_{b,c,t_1,f}\) are the neighboring semantic frames, and \(\alpha_{t'}\) is the interpolation coefficient. The continuous source-frame coordinate is:
\begin{equation}
\tau(t')
=
\left(
\frac{t'-1/2}{T_s}
\right)
T_e
-
\frac{1}{2},
\qquad
t_0=\lfloor\tau(t')\rfloor,
\qquad
t_1=t_0+1,
\qquad
\alpha_{t'}=\tau(t')-t_0
\end{equation}
where boundary indices are clipped to the valid temporal range. After temporal alignment, the semantic latent-frequency axis is collapsed by arithmetic averaging:
\begin{equation}
\bar{E}_{b,c,t,1}
=
\frac{1}{F_e}
\sum_{f=1}^{F_e}
\widetilde{E}_{b,c,t,f}
\end{equation}
where \(\bar{E}\in\mathbb{R}^{B\times C_e\times T_s\times 1}\) is the frequency-collapsed semantic tensor. This operation removes the semantic branch frequency axis before fusion, because this axis is not interpreted as a localization-bearing spatial frequency axis. The collapsed representation is then broadcast over the spatial frequency dimension:
\begin{equation}
\tilde{E}_{b,c,t,f_s}
=
\bar{E}_{b,c,t,1},
\qquad
f_s=1,\ldots,F_s
\end{equation}
where \(\tilde{E}\in\mathbb{R}^{B\times C_e\times T_s\times F_s}\) is the broadcast semantic conditioning tensor. This operation injects semantic evidence at the same temporal resolution as the spatial branch while preserving the spatial branch frequency lattice.

The two tensors are projected to a common bridge dimensionality:
\begin{equation}
\mathbf{S}_{b}
=
P_s(\tilde{\mathbf{S}}),
\qquad
\mathbf{E}_{b}
=
P_e(\tilde{\mathbf{E}})
\end{equation}
where \(P_s(\cdot)\) and \(P_e(\cdot)\) are learned \(1\times1\) channel projections. The bridge dimensionality \(C_b\) is selected according to the interaction point: \(C_b=64\) for the post-early bridge and \(C_b=128\) for the post-late bridge. The projected tensors satisfy \(
\mathbf{S}_{b},\mathbf{E}_{b}
\in
\mathbb{R}^{B\times C_b\times T_s\times F_s}
\) where \(C_b\) is the shared bridge-channel dimensionality.

The projected tensors are concatenated channel-wise and passed through a joint refinement block:
\begin{equation}
\mathbf{H}
=
\Psi
\left(
\left[
\mathbf{S}_{b},
\mathbf{E}_{b}
\right]
\right)
\in
\mathbb{R}^{B\times 2C_b\times T_s\times F_s}
\end{equation}
where \([\cdot,\cdot]\) denotes channel-wise concatenation and \(\Psi(\cdot)\) is a second inverted depth-wise separable multi-scale refinement block. Unlike the first preconditioning block, which processes only the spatial stream, \(\Psi\) operates on the concatenated semantic--spatial bridge tensor. Its role is to allow local time--frequency context and channel-wise interactions to be computed before explicit cross-stitch mixing. The refined tensor is split into two groups:
\begin{equation}
\mathbf{H}
=
\left[
\mathbf{S}_{\mathrm{mid}},
\mathbf{E}_{\mathrm{mid}}
\right],
\qquad
\mathbf{S}_{\mathrm{mid}},\mathbf{E}_{\mathrm{mid}}
\in
\mathbb{R}^{B\times C_b\times T_s\times F_s}
\end{equation}
where \(\mathbf{S}_{\mathrm{mid}}\) and \(\mathbf{E}_{\mathrm{mid}}\) are context-refined spatial and semantic bridge tensors.

The cross-branch exchange is then performed by a channel-wise cascade cross-stitch operator. For each bridge channel \(c\), a learnable matrix:
\begin{equation}
\mathbf{A}^{(c)}
=
\begin{bmatrix}
\alpha^{(c)}_{11} & \alpha^{(c)}_{12}\\
\alpha^{(c)}_{21} & \alpha^{(c)}_{22}
\end{bmatrix}
\end{equation}
controls the amount of within-branch preservation and cross-branch exchange. The spatial update is computed first:
\begin{equation}
\mathbf{S}^{(c)}_{\mathrm{cs}}
=
\alpha^{(c)}_{11}
\mathbf{S}^{(c)}_{\mathrm{mid}}
+
\alpha^{(c)}_{12}
\mathbf{E}^{(c)}_{\mathrm{mid}}
\end{equation}
where \(\mathbf{S}^{(c)}_{\mathrm{cs}}\) is the cross-stitched spatial channel. The semantic update is then computed from the updated spatial channel:
\begin{equation}
\mathbf{E}^{(c)}_{\mathrm{cs}}
=
\alpha^{(c)}_{21}
\mathbf{S}^{(c)}_{\mathrm{cs}}
+
\alpha^{(c)}_{22}
\mathbf{E}^{(c)}_{\mathrm{mid}}
\end{equation}
where \(\mathbf{E}^{(c)}_{\mathrm{cs}}\) is the cross-stitched semantic channel. This cascade differs from a fully symmetric cross-stitch layer: the spatial stream first receives semantic conditioning, and the semantic stream is subsequently corrected using the localization-aware spatial update. This ordering reflects the intended asymmetry of the bridge, where semantic evidence conditions spatial processing but is also refined before being returned to the semantic pathway.

The bridge outputs are finally mapped back to the original branch interfaces through residual reinjection. The spatial correction is projected to the original spatial-channel count and added to the input spatial tensor:
\begin{equation}
\mathbf{S}_{\mathrm{out}}
=
\mathbf{S}
+
Q_s
\left(
\mathbf{S}_{\mathrm{cs}}
\right)
\end{equation}
where \(Q_s(\cdot)\) is a learned \(1\times1\) projection from \(C_b\) bridge channels to \(C_s\) spatial channels. The semantic correction is first averaged over the spatial frequency axis:
\begin{equation}
\bar{\mathbf{E}}_{\mathrm{cs}}
=
\mathcal{C}_{F_s}
\left(
\mathbf{E}_{\mathrm{cs}}
\right)
\end{equation}
where \(\mathcal{C}_{F_s}(\cdot)\) denotes averaging over the \(F_s\) spatial-frequency bins. It is then projected back to the semantic channel space and temporally resized to the original semantic resolution:
\begin{equation}
\mathbf{E}_{\mathrm{out}}
=
\mathbf{E}
+
\mathcal{R}_{T_e}
\left(
Q_e
\left(
\bar{\mathbf{E}}_{\mathrm{cs}}
\right)
\right)
\end{equation}
where \(Q_e(\cdot)\) is a learned projection from \(C_b\) bridge channels to \(C_e\) semantic channels, and \(\mathcal{R}_{T_e}(\cdot)\) restores the temporal resolution to \(T_e\). The residual form keeps bridge insertion shape-preserving and experimentally controllable: disabling the bridge recovers the original branch interfaces, whereas enabling it introduces a localized, temporally aligned semantic--spatial correction.

Stage~3 tests the bridge at different insertion points. An \textit{early bridge} is applied after the early spatial encoder, where the spatial representation still preserves dense time--frequency cues. A \textit{late bridge} is applied after the late track-wise abstraction stage, where the spatial representation is closer to the prediction heads. A combined configuration uses both interaction points. This design makes it possible to distinguish low-level semantic conditioning from higher-level semantic--spatial latent fusion.

\begin{figure}[ht]
    \centering
    \includegraphics[width=\linewidth]{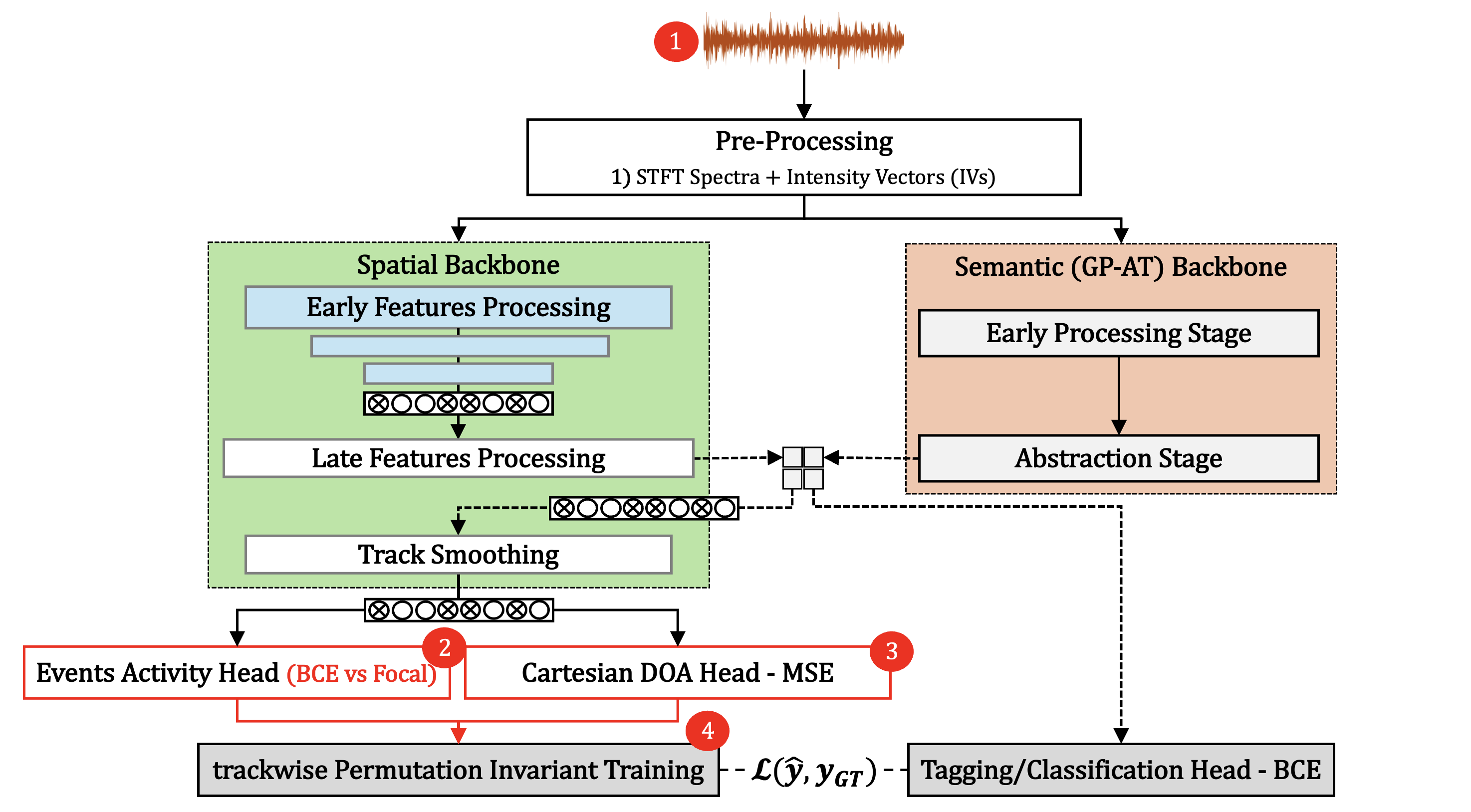}
    \caption{\gls{nas} Stage~4: diagnostic characterization over data balancing \textcolor{red}{(1)}, activity-loss calibration \textcolor{red}{(2)}, activity-conditioned \gls{doa} supervision \textcolor{red}{(3)}, operating-point calibration \textcolor{red}{(4)}, and cross-dataset evaluation.}
    \label{fig:nas_stage_4}
\end{figure}

\textbf{Stage~4} is a diagnostic characterization stage rather than a pure architecture-search phase. Once the most promising configurations from the previous stages have been identified, the analysis shifts from module selection to characterizing how the selected systems behave under alternative supervision, data-balancing, temporal-context, and calibration conditions. This stage is motivated by a limitation that recurs across the reviewed literature: architectural improvements are often reported through aggregate \gls{seld} scores, whereas the interaction between loss design, class imbalance, task de-coupling, inactive targets, and operating-point selection is less extensively characterized.

The first diagnostic axis concerns the activity loss, comparing standard \gls{bce} against focal re-weighting to evaluate whether rare or difficult events receive sufficient gradient support. The second axis concerns class distribution, where balanced training regimes are used to test whether improved exposure to underrepresented classes alleviates the long-tailed behavior (typical of real-scene data). The third axis concerns \gls{doa} supervision: active-only spatial losses are evaluated to avoid regression dominance by inactive targets, a failure mode anticipated in Sections~\ref{subsec:related_permutation_tracking} and~\ref{subsec:related_geometry}.

The remaining characterization axes evaluate the selected systems beyond the original \gls{nas} grid. Standalone \gls{seld} ablations test whether a conventional \gls{sed}/\gls{doa}-only configuration provides stronger task specialization than the proposed \gls{at2seld} design, where semantic representations are injected into the spatial pathway. Temporal-awareness experiments examine whether the fixed \(10\)-s context, aligned with the \gls{at} pretraining regime, is sufficient, or whether targeted increases in temporal capacity and input duration improve generalization. Finally, deployment-oriented threshold analysis evaluates the sensitivity of the selected models to operating-point calibration. These diagnostics characterize the final selections not only by their aggregated \gls{seld} scores, but also by their supervision sensitivity, temporal-context dependence, calibration behavior, and deployment-relevant operating limits.


\subsection{Track-Wise \glsentryshort{seld} Supervision and Multi-Task Objective}
\label{subsec:exp_supervision_objective}

The staged search described in Section~\ref{subsec:exp_staging} requires a common supervision interface across all candidate architectures. For this reason, all models are trained with a track-wise \gls{seld} target representation derived from Multi-\gls{accdoa}-style tensors, but the prediction space is deliberately decoupled into separate activity and localization heads. This choice preserves the ability to represent same-class overlap through independent tracks, while allowing \gls{sed} and \gls{doa} errors to be analyzed separately during the diagnostic stages. The use and construction of the target tensors from dataset metadata is described in the following Section.

For each mini-batch, the spatial supervision tensor is \(
\mathbf{T}_{s}
\in
\mathbb{R}^{B\times T\times N\times C\times 3}
\) where \(B\) is the batch size, \(T\) is the number of supervision frames, \(N\) is the maximum number of output tracks, \(C\) is the number of event classes in the dataset vocabulary, and the last dimension stores 3D Cartesian \gls{doa} coordinates. The tensor follows an activity-coupled convention: active events have non-zero Cartesian vectors, whereas inactive class-track entries are represented by null vectors. However, the model does not directly optimize a Multi-\gls{accdoa} regression output. Instead, it predicts two independent tensors:
\begin{equation}
\hat{\mathbf{Z}}
\in
\mathbb{R}^{B\times T\times N\times C},
\qquad
\hat{\mathbf{R}}
\in
\mathbb{R}^{B\times T\times N\times C\times 3}
\end{equation}
where \(\hat{\mathbf{Z}}\) contains raw activity logits and \(\hat{\mathbf{R}}\) contains Cartesian \gls{doa} estimates. The localization head uses a final hyperbolic-tangent activation so that each coordinate is bounded in \([-1,1]\). The binary activity target is recovered from the target-vector norm:
\begin{equation}
A_{b,t,n,c}
=
\mathbbm{1}
\left[
\left\|
\mathbf{T}_{s,b,t,n,c,:}
\right\|_2
>
\tau_{\mathrm{act}}
\right],
\qquad
\tau_{\mathrm{act}}=0.5
\end{equation}
where \(A_{b,t,n,c}\in\{0,1\}\) is the activity target for batch element \(b\), frame \(t\), track \(n\), and class \(c\), while \(\tau_{\mathrm{act}}\) is the activity threshold used to recover binary targets from the activity-coupled representation. The \gls{doa} target is \(
\mathbf{R}_{b,t,n,c}
=
\mathbf{T}_{s,b,t,n,c,:}
\) where \(\mathbf{R}_{b,t,n,c}\in\mathbb{R}^{3}\) denotes the Cartesian target associated with the same class-track entry. The decoupled formulation was adopted because it makes the contribution of activity classification and spatial regression explicit, and because preliminary Multi-\gls{accdoa} optimization led to unstable minimization in the considered transfer setting.

The user-defined track dimension introduces a permutation ambiguity: for a given class and frame, several assignments between predicted tracks and reference tracks are equivalent (Section \ref{subsec:related_permutation_tracking}). The base objective therefore uses \gls{tpit}. For each batch element \(b\), frame \(t\), and class \(c\), the set of admissible track permutations is denoted by \(\Pi_N\). Given a permutation \(\pi\in\Pi_N\), the unweighted assignment cost is:
\begin{equation}
\mathcal{C}_{b,t,c}(\pi)
=
\lambda_{\mathrm{pit}}
\mathcal{C}^{\mathrm{SED}}_{b,t,c}(\pi)
+
\left(
1-\lambda_{\mathrm{pit}}
\right)
\mathcal{C}^{\mathrm{DOA}}_{b,t,c}(\pi)
\end{equation}
where \(\lambda_{\mathrm{pit}}\in[0,1]\) balances the activity and localization terms during permutation selection. The activity assignment cost is:
\begin{equation}
\mathcal{C}^{\mathrm{SED}}_{b,t,c}(\pi)
=
\frac{1}{N}
\sum_{n=1}^{N}
\mathrm{BCE}
\left(
\hat{Z}_{b,t,\pi(n),c},
A_{b,t,n,c}
\right)
\end{equation}
where \(\hat{Z}_{b,t,\pi(n),c}\) is the activity logit assigned to reference track \(n\), and \(\mathrm{BCE}(\cdot,\cdot)\) is the unweighted binary cross-entropy. The localization assignment cost is:
\begin{equation}
\mathcal{C}^{\mathrm{DOA}}_{b,t,c}(\pi)
=
\frac{1}{N}
\sum_{n=1}^{N}
\left\|
\hat{\mathbf{R}}_{b,t,\pi(n),c}
-
\mathbf{R}_{b,t,n,c}
\right\|_2^2
\end{equation}
where \(\hat{\mathbf{R}}_{b,t,\pi(n),c}\) is the predicted Cartesian \gls{doa} vector assigned to reference track \(n\). The selected permutation is:
\begin{equation}
\pi^{\star}_{b,t,c}
=
\arg\min_{\pi\in\Pi_N}
\mathcal{C}_{b,t,c}(\pi)
\end{equation}
where \(\pi^{\star}_{b,t,c}\) is the minimum-cost assignment for batch element \(b\), frame \(t\), and class \(c\). This assignment is used only to align predictions and targets before the final gradient loss is computed.

The optimization loss is evaluated after applying the selected permutation. For the activity branch, class imbalance can be handled through weighted \gls{bce}:
\begin{equation}
\mathcal{L}^{\mathrm{SED}}_{\mathrm{wBCE}}
=
-\frac{1}{BTNC}
\sum_{b,t,n,c}
\left[
w_c^{+}
A_{b,t,n,c}
\log
\sigma
\left(
\hat{Z}_{b,t,\pi^{\star}(n),c}
\right)
+
w_c^{-}
\left(
1-A_{b,t,n,c}
\right)
\log
\left(
1-
\sigma
\left(
\hat{Z}_{b,t,\pi^{\star}(n),c}
\right)
\right)
\right]
\end{equation}
where \(w_c^{+}\) and \(w_c^{-}\) are class-dependent positive and negative weights, and \(\sigma(\cdot)\) is the sigmoid activation. Importantly, these weights are applied after permutation selection. The assignment cost remains unweighted so that large class-imbalance factors do not dominate the permutation search or suppress the contribution of the \gls{doa} term.

The base localization loss is computed on the same selected assignment:
\begin{equation}
\mathcal{L}^{\mathrm{DOA}}_{\mathrm{all}}
=
\frac{1}{BTNC}
\sum_{b,t,n,c}
\left\|
\hat{\mathbf{R}}_{b,t,\pi^{\star}(n),c}
-
\mathbf{R}_{b,t,n,c}
\right\|_2^2
\end{equation}
where all class-track entries contribute to the regression loss, including inactive entries whose target vector is zero. Since this formulation can make the localization objective dominated by inactive targets, the diagnostic stage also evaluates an activity-conditioned variant:
\begin{equation}
\mathcal{L}^{\mathrm{DOA}}_{\mathrm{act}}
=
\frac{
\sum_{b,t,n,c}
A_{b,t,n,c}
\left\|
\hat{\mathbf{R}}_{b,t,\pi^{\star}(n),c}
-
\mathbf{R}_{b,t,n,c}
\right\|_2^2
}{
3\sum_{b,t,n,c}A_{b,t,n,c}
+
\epsilon
}
\end{equation}
where \(\epsilon>0\) avoids division by zero in segments without active targets. This active-only formulation is consistent with the localization losses discussed in Section~\ref{subsec:related_geometry}: spatial regression is evaluated only when a source is active, preventing inactive class-track entries from dominating the \gls{doa} gradient.

The activity loss can also be replaced by a \textit{focal objective} when rare or difficult events remain under-emphasized~\cite{focal_loss}. Let \(p=\sigma(\hat{z})\) be the predicted activity probability for a generic logit \(\hat{z}\), and let \(a\in\{0,1\}\) be the corresponding binary target. The target-aligned probability is:
\begin{equation}
p_t
=
a p
+
(1-a)(1-p)
\end{equation}
where \(p_t\) is high when the prediction is correct for the target class. The focal loss is:
\begin{equation}
\mathcal{L}_{\mathrm{Focal}}
\left(
\hat{z},a
\right)
=
-\alpha_t
\left(
1-p_t
\right)^{\gamma}
\log
\left(
p_t
\right)
\end{equation}
where \(\gamma\geq0\) is the focusing parameter and \(\alpha_t\) is the optional class-balancing coefficient:
\begin{equation}
\alpha_t
=
\alpha a
+
(1-\alpha)(1-a)
\end{equation}
where \(\alpha\in[0,1]\) controls the balance between positive and negative targets. In the diagnostic experiments, \(\gamma=2\) is used when focal re-weighting is enabled. This objective reduces the relative contribution of already confident activity decisions and increases the influence of difficult or underrepresented examples.

The track-wise \gls{seld} loss is therefore:
\begin{equation}
\mathcal{L}_{\mathrm{tPIT}}
=
\lambda_{\mathrm{pit}}
\mathcal{L}^{\mathrm{SED}}
+
\left(
1-\lambda_{\mathrm{pit}}
\right)
\mathcal{L}^{\mathrm{DOA}}
\end{equation}
where \(\mathcal{L}^{\mathrm{SED}}\) denotes either weighted \gls{bce} or focal activity loss, and \(\mathcal{L}^{\mathrm{DOA}}\) denotes either the all-entry or active-only localization loss, depending on the experimental stage. This formulation separates the mechanism used to choose the track assignment from the loss used for gradient optimization, which is necessary when class weights or activity masks are introduced after the assignment step.

In addition to the track-wise \gls{seld} heads, the architecture includes an auxiliary semantic-presence branch associated with the pretrained \gls{gpat} prior. Let \(
\hat{\mathbf{P}}_{e}
\in
\mathbb{R}^{B\times T_{\mathrm{sem}}\times C}
\) where \(\hat{\mathbf{P}}_{e}\) contains semantic class-presence logits at the temporal resolution of the semantic branch. The target is obtained by temporally aligning the spatial activity tensor and collapsing over tracks:
\begin{equation}
Y^{\mathrm{AT}}_{b,u,c}
=
\max_{n}
A_{b,\nu(u),n,c}
\end{equation}
where \(Y^{\mathrm{AT}}_{b,u,c}\in\{0,1\}\) is the semantic-presence target for semantic frame \(u\), and \(\nu(u)\) maps semantic frame \(u\) to the nearest spatial supervision frame. The auxiliary semantic loss is:
\begin{equation}
\mathcal{L}_{\mathrm{sem}}
=
\mathcal{L}^{\mathrm{wBCE}}
\left(
\hat{\mathbf{P}}_{e},
\mathbf{Y}^{\mathrm{AT}}
\right)
\end{equation}
where \(\mathcal{L}^{\mathrm{wBCE}}\) is weighted multi-label binary cross-entropy. The role of this branch is to preserve class-discriminative semantic information inherited from the pretrained \gls{at} backbone while supporting the spatial heads in learning localization-aware event prediction. The complete objective is:
\begin{equation}
\mathcal{L}_{\mathrm{tot}}
=
\lambda_s
\mathcal{L}_{\mathrm{tPIT}}
+
\lambda_e
\mathcal{L}_{\mathrm{sem}}
\end{equation}
where \(\lambda_s\) and \(\lambda_e\) control the relative contribution of track-wise \gls{seld} supervision and auxiliary semantic-presence supervision. In the adopted setting, the spatial term remains the primary optimization driver, while the semantic term acts as a regularizer that constrains the transferred representation to retain high-level class evidence.

Training is performed with variable mini-batch sizes determined by the memory footprint of each architecture, while gradient accumulation is adjusted to preserve an effective batch size of \(16\) samples. Depending on the experimental stage, the optimizer is Adam~\cite{kingma2015adam} or AdamW~\cite{loshchilov2019decoupled}. The initial learning rate is set to \(
\eta_{\mathrm{start}}
=
5\times10^{-4}
\) where \(\eta_{\mathrm{start}}\) denotes the initial learning rate. Stage-dependent annealing schedules are then applied during training, including cosine-style schedules inspired by warm-restart optimization~\cite{loshchilov2017sgdr}. Model selection relies on checkpointing and early-stopping criteria adapted to the corresponding search phase, using the validation \gls{seld} score as the principal criterion.

During validation and test, activity logits are converted into probabilities through a sigmoid activation. A track-class prediction is considered active when \(
\sigma
\left(
\hat{Z}_{b,t,n,c}
\right)
\geq
\delta_{\mathrm{act}}
\) where \(\delta_{\mathrm{act}}\) is the decision threshold used at inference. Active predicted \gls{doa} vectors are matched to active reference \gls{doa} vectors of the same class using Hungarian assignment with angular distance as the matching cost. For predicted vector \(\hat{\mathbf{r}}\) and reference vector \(\mathbf{r}\), the angular distance in degrees is:
\begin{equation}
\theta
\left(
\hat{\mathbf{r}},
\mathbf{r}
\right)
=
\frac{180}{\pi}
\arccos
\left(
\mathrm{clip}
\left(
\frac{
\hat{\mathbf{r}}^{\top}\mathbf{r}
}{
\|\hat{\mathbf{r}}\|_2
\|\mathbf{r}\|_2
},
-1,
1
\right)
\right)
\end{equation}
where \(\theta(\hat{\mathbf{r}},\mathbf{r})\) is the angular error, and the clipping operation prevents numerical errors outside the valid arccosine domain. A matched pair is counted as a true positive only when its angular deviation is below \(20^\circ\). Otherwise, unmatched predictions and references contribute to the corresponding false-positive and false-negative counts.

The class-dependent localization error is:
\begin{equation}
\mathrm{LE}_{\mathrm{CD}}
=
\frac{1}{|\mathcal{C}_{\mathrm{TP}}|}
\sum_{c\in\mathcal{C}_{\mathrm{TP}}}
\frac{1}{TP_c}
\sum_{k=1}^{TP_c}
\theta_{c,k}
\end{equation}
where \(\theta_{c,k}\) is the angular error of the \(k\)-th true-positive match for class \(c\), \(TP_c\) is the number of true-positive localizations for class \(c\), and \(\mathcal{C}_{\mathrm{TP}}\) is the set of classes with at least one true-positive match. The class-dependent localization recall is:
\begin{equation}
\mathrm{LR}_{\mathrm{CD}}
=
\frac{1}{|\mathcal{C}_{\mathrm{R}}|}
\sum_{c\in\mathcal{C}_{\mathrm{R}}}
\frac{TP_c}{N^{\mathrm{ref}}_c}
\end{equation}
where \(N^{\mathrm{ref}}_c\) is the number of active reference instances of class \(c\), and \(\mathcal{C}_{\mathrm{R}}\) is the set of classes with at least one reference instance. Detection performance under the same \(20^\circ\) localization tolerance is summarized by the location-dependent error rate:
\begin{equation}
\mathrm{ER}_{20^\circ}
=
\frac{
S+D+I
}{
N_{\mathrm{ref}}
}
\end{equation}
where \(S\), \(D\), and \(I\) denote substitutions, deletions, and insertions accumulated under the angular tolerance, and \(N_{\mathrm{ref}}\) is the total number of reference event instances. The corresponding F-score is:
\begin{equation}
\mathrm{F}_{20^\circ}
=
\frac{
2TP
}{
2TP+FP+FN
}
\end{equation}
where \(TP\), \(FP\), and \(FN\) are the global true-positive, false-positive, and false-negative counts under the same matching criterion. These quantities are combined into the standard \gls{seld} error score:
\begin{equation}
\mathrm{SELD}_{\mathrm{err}}
=
\frac{
\mathrm{ER}_{20^\circ}
+
\left(
1-\mathrm{F}_{20^\circ}
\right)
+
\mathrm{LE}_{\mathrm{CD}}/180
+
\left(
1-\mathrm{LR}_{\mathrm{CD}}
\right)
}{4}
\end{equation}
where \(\mathrm{SELD}_{\mathrm{err}}\in[0,1]\) is minimized during model selection. In parallel, the auxiliary semantic-presence branch is monitored through multi-label \gls{at} metrics, including accuracy, F-score, \gls{auroc}, and per-class confusion matrices. This dual evaluation protocol is necessary because the proposed framework must be assessed both as a spatially grounded \gls{seld} system and as a semantic-transfer architecture that should not discard the class-discriminative information inherited from the pretrained backbone.


\subsection{Datasets and Data Pipeline}
\label{subsec:exp_datasets_pipeline}

The \gls{at2seld} campaign relies on multiple \gls{foa} \gls{seld} corpora that expose the proposed framework to different spatial, acoustic, and annotation regimes. The primary datasets are STARSS23, TAU-NIGENS Spatial Sound Events 2021, and TAU Spatial Sound Events 2019. STARSS23 provides the real-world scene reference conditions~\cite{shimada2023starss23,politis_2023_7880637}; TAU-NIGENS2021 provides controlled synthetic moving-source scenes with strong spatial and polyphonic variability~\cite{politis2021tau_nigens}; and TAU2019 provides a fixed-source reverberant condition useful for separating basic localization consistency from the additional complexity of moving trajectories~\cite{adavanne2019multiroom}. TAU-NIGENS Spatial Sound Events 2020 is used as a complementary diagnostic corpus in the last experimental and diagnostic stage. It occupies an intermediate regime between TAU2019 and TAU-NIGENS2021: it introduces reverberant dynamic scenes and moving sources, but remains closer to the earlier DCASE synthetic-generation paradigm than to the later TAU-NIGENS2021 setting~\cite{politis2020tau_nigens}.

\subsubsection{Roles and Annotation Regimes}
\label{subsubsec:dataset_roles}

Table~\ref{tab:dataset_summary} summarizes the methodological role of each corpus used in the study. Detailed native and windowed-regime statistics are reported in Appendix~\ref{app:dataset_statistics}.

\begin{table}[ht]
\centering
\caption{Summary of the datasets used in the \gls{at2seld} study.}
\label{tab:dataset_summary}
\scriptsize
\renewcommand{\arraystretch}{1.45}
\setlength{\tabcolsep}{3.0pt}
\resizebox{\textwidth}{!}{%
\begin{tabular}{|
>{\RaggedRight\arraybackslash}p{2.85cm}|
>{\RaggedRight\arraybackslash}p{3.45cm}|
>{\centering\arraybackslash}p{1.75cm}|
>{\RaggedRight\arraybackslash}p{2.55cm}|
>{\centering\arraybackslash}p{1.00cm}|
>{\centering\arraybackslash}p{1.00cm}|
>{\RaggedRight\arraybackslash}p{2.45cm}|
>{\centering\arraybackslash}p{1.65cm}|}
\hline
\textbf{Dataset} &
\textbf{Characteristics} &
\textbf{Format} &
\textbf{Total Duration} &
\textbf{Classes} &
\textbf{SR} &
\textbf{Annotations} &
\textbf{Reference} \\
\hline

\mbox{STARSS23} &
Real spatial sound scenes &
\makecell[c]{4 channels\\FOA, MIC} &
\makecell[tl]{168 dev clips\\$\sim 7$h $22$min} &
13 &
24 kHz &
100 ms framewise &
\cite{shimada2023starss23,politis_2023_7880637} \\
\hline

\mbox{TAU-NIGENS2021} &
Synthetic, spatially rendered, moving sources and interferers &
\makecell[c]{4 channels\\FOA, MIC} &
\makecell[tl]{800 clips\\1 min per clip} &
12 &
24 kHz &
100 ms framewise &
\cite{politis2021tau_nigens} \\
\hline

\mbox{TAU-NIGENS2020} &
Synthetic, reverberant, static and moving sources &
\makecell[c]{4 channels\\FOA, MIC} &
\makecell[tl]{800 clips\\1 min per clip} &
14 &
24 kHz &
100 ms framewise &
\cite{politis2020tau_nigens} \\
\hline

\mbox{TAU2019} &
Synthetic, RIR-based, stationary sources &
\makecell[c]{4 channels\\FOA, MIC} &
\makecell[tl]{400 dev clips\\100 eval clips\\1 min per clip} &
11 &
48 kHz &
\makecell[tl]{event intervals\\+ $\phi,\theta$} &
\cite{adavanne2019multiroom} \\
\hline

\end{tabular}%
}
\end{table}

STARSS23 is selected as the in-domain reference because it consists of real-world spatial sound scenes recorded in 4-channels \gls{foa} and tetrahedral \gls{mic} formats. Its annotations provide framewise multi-label event activity and source directions over \(13\) target classes. The dataset exposes the models to realistic scene variability, room-dependent acoustics, human behavioral diversity, long-tailed class distributions, mixed static and moving events, and class-dependent spatial structure~\cite{shimada2023starss23,politis_2023_7880637}. These properties make STARSS23 the most challenging and relevant corpus for assessing whether semantic-to-spatial transfer remains useful under realistic recording conditions.

TAU-NIGENS2021 is used as a controlled dynamic out-domain condition. It contains synthetic spatial scenes generated from isolated events out of the Neural Information Processing group GENeral sounds (NIGENS) dataset, composed with measured \gls{srir}s, moving-source trajectories, reverberation, ambient noise, and additional localized interfering events outside the target taxonomy~\cite{politis2021tau_nigens}. Compared with STARSS23, it provides stronger control over spatial rendering and event composition, while still stressing the model through source motion, overlap, and non-target directional interference. It is therefore useful for evaluating whether the selected \gls{at2seld} configurations learn transferable spatial mechanisms rather than only real-scene priors.

TAU-NIGENS2020 is retained as a complementary diagnostic corpus. It was introduced for the \gls{dcase} 2020 \gls{seld} task as a more challenging successor to TAU2019, with a wider range of acoustical conditions and moving sources in approximately half of the active events~\cite{politis2020tau_nigens}. Its scenes are synthesized from measured multichannel \glspl{rir} captured with a slowly moving excitation source, and both static and moving events are rendered before ambient noise recorded on location is added. The corpus uses \(14\) classes from NIGENS, provides both \gls{foa} and \gls{mic} formats, and includes one-minute recordings sampled at \(24~\mathrm{kHz}\). Within this study, TAU-NIGENS2020 is not the principal development condition, but supports Stage~4 diagnostics by providing an intermediate dynamic regime between the fixed-source TAU2019 corpus and the moving trajectories of TAU-NIGENS2021 corpus.

TAU2019 is used as a fixed-source out-domain benchmark. In contrast to STARSS23 and the TAU-NIGENS datasets, TAU2019 contains stationary sources spatialized through measured \glspl{rir}. Its development set contains four cross-validation folds, and its evaluation set provides additional held-out one-minute recordings~\cite{adavanne2019multiroom}. Scenes are synthesized from measured spatial responses in five indoor environments, using fixed azimuth--elevation--distance combinations and controlled temporal overlap. This makes TAU2019 valuable for testing whether a model can isolate basic event-localization consistency when trajectory complexity is removed.

The four corpora therefore probe distinct failure modes: STARSS23 stresses ecological variability and class imbalance; TAU-NIGENS2021 stresses dynamic complexity and non-target interference; TAU-NIGENS2020 provides an intermediate reverberant moving-source condition; and TAU2019 isolates fixed-source localization under synthetic but measured reverberation. Their joint use prevents the analysis from being interpreted only as adaptation to one dataset family.

\begin{figure}[ht]
    \centering
    \includegraphics[width=\linewidth]{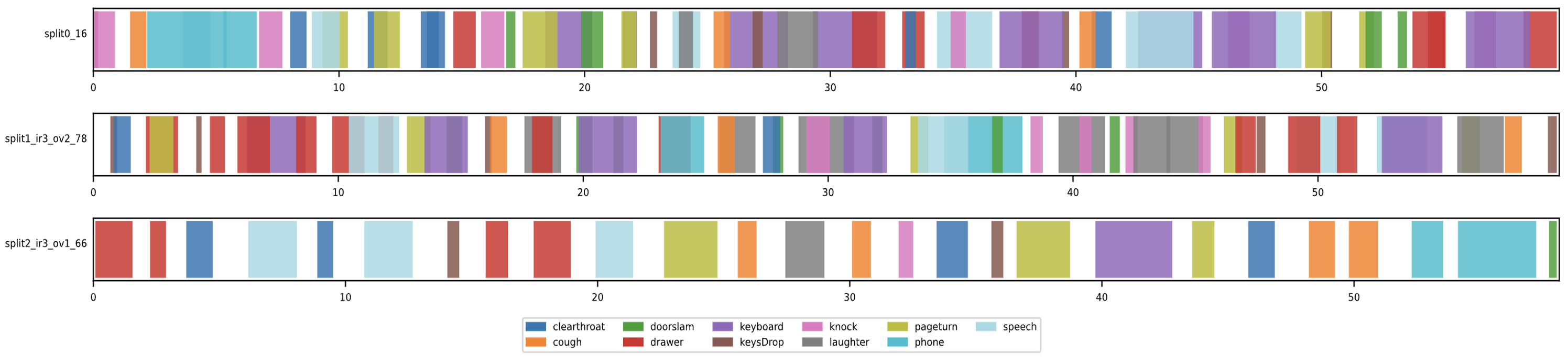}
    \caption{TAU2019 temporal polyphony examples: \texttt{split0} is used as test set, \texttt{split1} as validation set, and \texttt{split2-3} as training set.}
    \label{fig:tau2019_polyphony}
\end{figure}

A first layer of unification concerns the target interface. All datasets are converted to a track-wise Cartesian target tensor \(
\mathbf{Y}
\in
\mathbb{R}^{T\times N_{\mathrm{tracks}}\times C\times 3}
\) where \(T\) is the target temporal resolution, \(N_{\mathrm{tracks}}\) is the maximum number of track slots, \(C\) is the class vocabulary of the corresponding corpus, and the last dimension stores Cartesian \gls{doa} coordinates. This tensor interface is the common representation exposed to the models; the actual optimization objective remains the track-wise multi-task loss defined in Section~\ref{subsec:exp_supervision_objective}. Within this representation, TAU2019 differs from the other corpora mainly in annotation structure: its sources are stationary within each active event interval, so the Cartesian \gls{doa} remains constant over the event duration (Figure~\ref{fig:tau2019_polyphony}).

A second layer of unification concerns temporal context: the \gls{at2seld} experiments are conducted in a fixed \(10\)-s windowed regime rather than on full audio clips. This choice is partly practical, because it enables shared batching conditions, stable memory requirements, and controlled model-to-model comparison. It is also methodological: \(10\)-s excerpts match the native operating scale of many pretrained \gls{at} backbones, while still providing sufficient temporal support to test whether spatial localization and track continuity degrade when the longer \(20\)--\(40\)s excerpts often used in \gls{dcase} \gls{seld} challenges training are not available.

This segmentation introduces a distinction between native dataset statistics and the statistics effectively seen by the models after windowing. The dataset analysis is therefore carried out at two levels: native statistics computed directly from the released annotations, and window-based statistics computed after conversion to \(10\)-s segments. This distinction is important because segmentation can change the effective class balance, overlap structure, and temporal distribution of active events without changing the underlying label semantics. In trajectory-oriented corpora, windowing regularizes the event density seen by the models across time and direction, whereas in TAU2019 the effect is more limited because scenes are already built from stationary sources with lower maximum overlap and more stable spatio-temporal occupancy. The corresponding statistics are extensively reported in Appendix~\ref{app:dataset_statistics}.

Beyond class coverage and overlap, the spatial label geometry differs substantially across the datasets. In STARSS23, \gls{doa} distributions are strongly class-dependent because sources are tied to real objects, human activity, and room layout. TAU-NIGENS2021 and TAU-NIGENS2020 derive their spatial diversity from controlled rendering techniques and over measured trajectories and \glspl{rir} (Figure~\ref{fig:starss23_vs_tau_nigens2021}), rather than from real-world scene semantics. TAU2019 is more discretized, with stationary source positions selected from a fixed set of measured locations (Figure~\ref{fig:tau2019_doa}).

\clearpage

\vspace*{\fill}

\begin{center}
\captionsetup{type=figure}
\includegraphics[width=\linewidth]{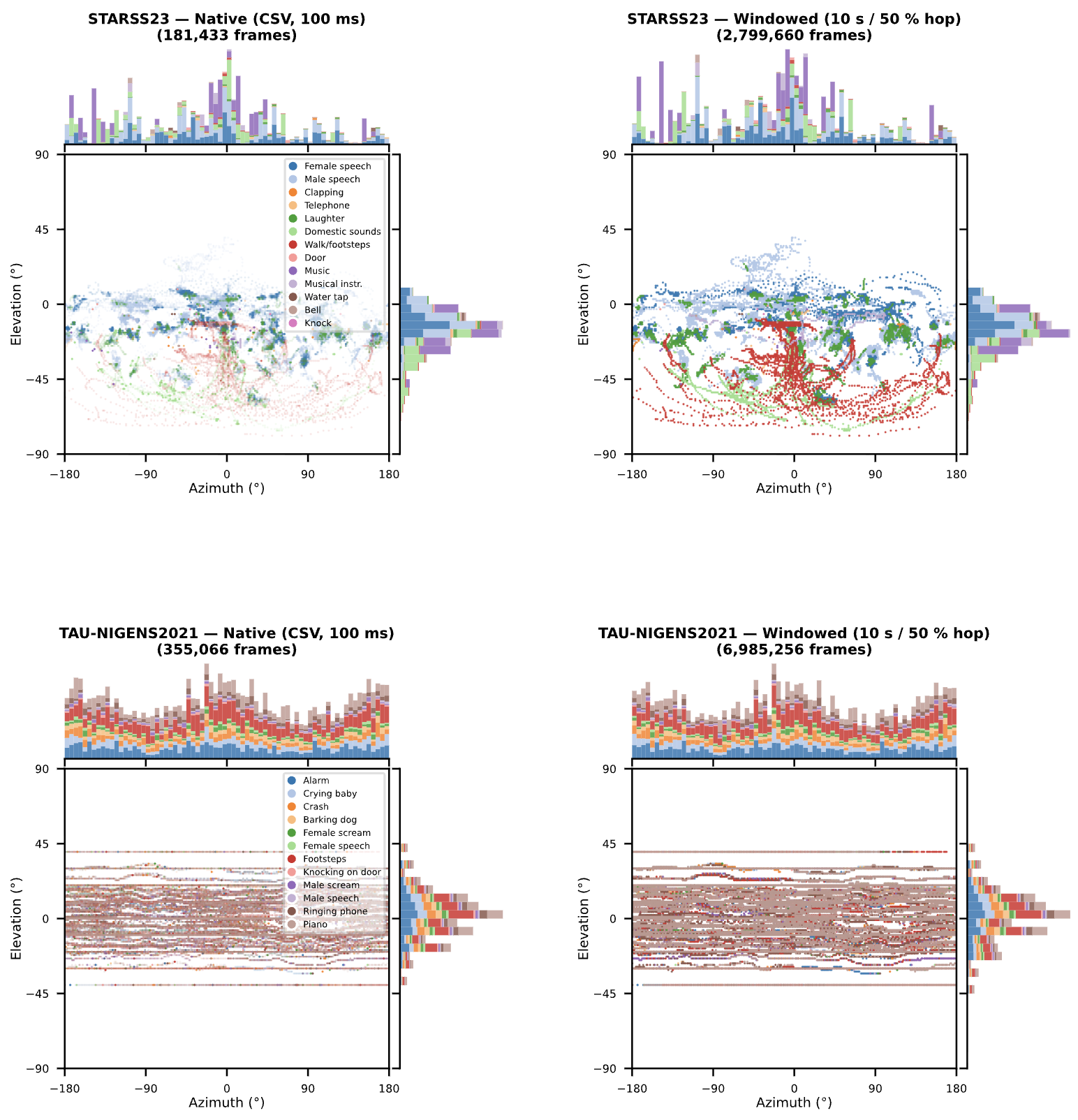}
\captionof{figure}{STARSS23 and TAU-NIGENS2021 \gls{doa} marginal distributions. Event density is represented by scatter-point color saturation.}
\label{fig:starss23_vs_tau_nigens2021}
\end{center}

\vspace*{\fill}
\clearpage

\vspace*{\fill}

\begin{center}
\captionsetup{type=figure}
\includegraphics[width=\linewidth]{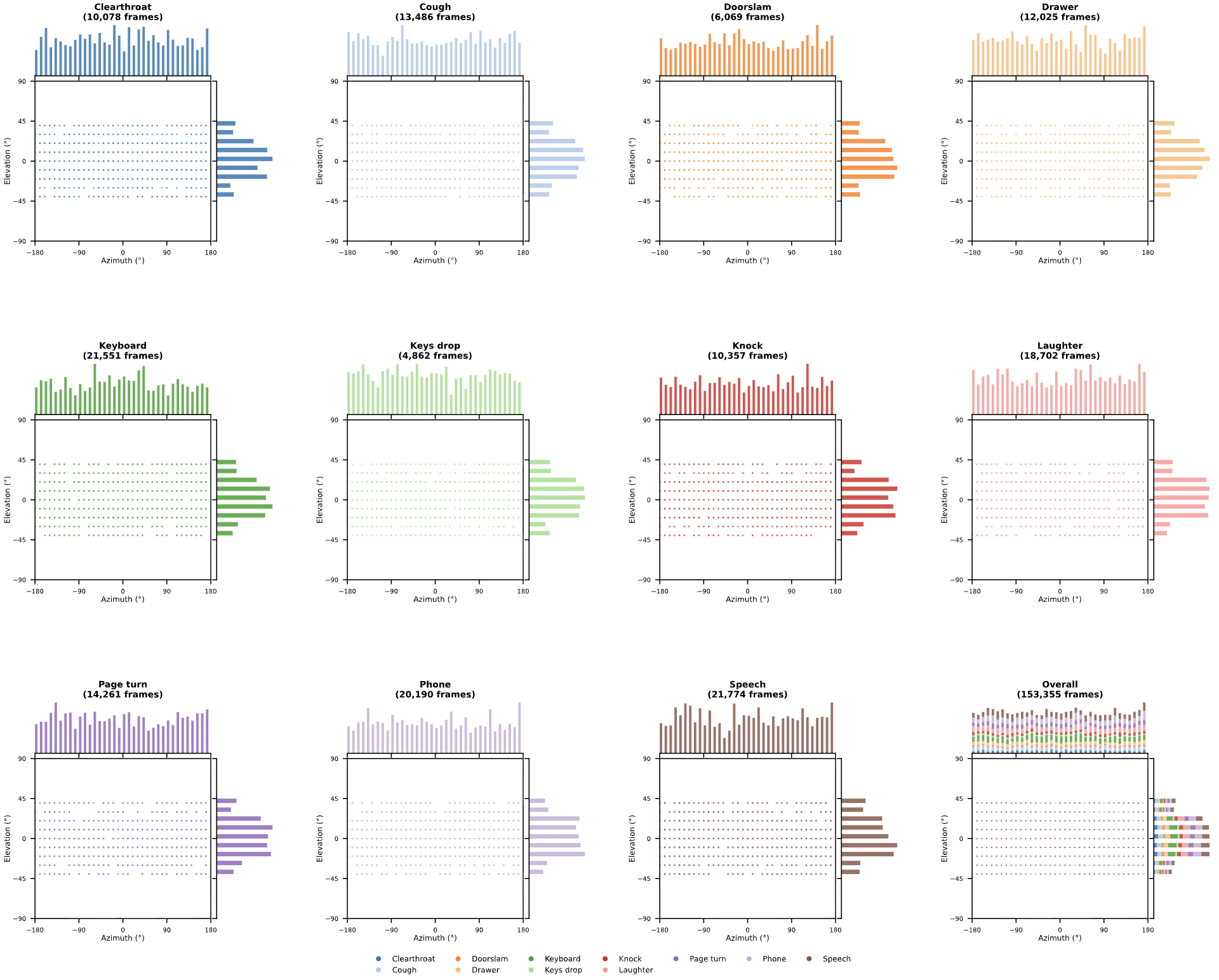}
\captionof{figure}{TAU2019 \gls{doa} marginal distributions by class. Event density is represented by scatter-point color saturation.}
\label{fig:tau2019_doa}
\end{center}

\vspace*{\fill}
\clearpage

\subsubsection{Splits, Windowing, and Metadata Parsing}
\label{subsubsec:splits_windowing_metadata}

\begin{figure*}[h]
\centering
\resizebox{\textwidth}{!}{%
\begin{tikzpicture}[
    >=Latex,
    flow/.style={-Latex, thick}
]


\node[
    draw, rounded corners, thick, align=center, fill=gray!10,
    text width=2.9cm, minimum height=1.2cm
] (yaml) at (0,0)
{Dataset \texttt{.yaml}\\Experiment \texttt{.yaml}\\CLI overrides};

\node[
    draw, rounded corners, thick, align=center, fill=white,
    text width=4.2cm, minimum height=1.2cm,
    right=0.5cm of yaml
] (cfg)
{Experiment namespace\\\((f_{\mathrm{model}},h,T_w,H_w,N_{\mathrm{tracks}})\)};

\node[
    draw, rounded corners, thick, align=center, fill=white,
    text width=3.2cm, minimum height=1.2cm,
    right=1.0cm of cfg
] (clips)
{RAM clip collection\\train/val/test split\\\(\mathcal{C}=\{(a_i,m_i)\}\)};

\node[
    draw, rounded corners, thick, align=center, fill=white,
    text width=3.5cm, minimum height=1.2cm,
    right=0.5cm of clips
] (items)
{Flat indexing\\windowing / full-clip\\\(\mathcal{I}=\{(i,n_i^{\mathrm{start}},b_i)\}\)};

\node[
    draw, rounded corners, thick, align=center, fill=green!5,
    text width=3.4cm, minimum height=1.2cm,
    right=0.5cm of items
] (target)
{Track-wise\\Multi-ACCDOA\\\(\mathbf{T}\in\mathbb{R}^{T_{\mathrm{mel}}\times N\times C\times 3}\)};

\node[
    draw, rounded corners, thick, align=center, fill=white,
    text width=4.4cm, minimum height=1.2cm,
    right=0.5cm of target
] (aug)
{Audio augmentation\\FOA: 16patterns / gain scaling / noise injection};

\node[
    draw, rounded corners, thick, align=center, fill=white,
    text width=2.9cm, minimum height=1.2cm,
    right=0.5cm of aug
] (batch)
{Batching\\causal 0-padding};

\node[
    draw, rounded corners, thick, align=center, fill=white,
    text width=3.3cm, minimum height=1.2cm,
    right=1.0cm of batch
] (modelin)
{Resampling \((f_{\mathrm{model}})\)\\mono reduction};

\node[
    draw, rounded corners, thick, align=center, fill=red!5,
    text width=4.9cm, minimum height=1.2cm,
    right=0.5cm of modelin
] (feat)
{$\mathbf{X}\in\mathbb{R}^{B\times 4\times S_{\mathrm{model}}}$\\
$\bar{\mathbf{X}}\in\mathbb{R}^{B\times S_{\mathrm{model}}}$\\
$\mathbf{S}_{\mathrm{FOA}}\in\mathbb{R}^{B\times (4\cdot2+3)\times T_{\mathrm{mel}}\times 256}$\\
$\mathbf{S}_{\mathrm{Gabor}}\in\mathbb{R}^{B\times 7\times T_{\mathrm{out}}\times N_{\mathrm{filters}}}$};


\draw[flow] (yaml) -- (cfg);
\draw[flow, draw=red] (cfg) -- (clips);
\draw[flow] (clips) -- (items);
\draw[flow] (items) -- (target);
\draw[flow] (target) -- (aug);
\draw[flow] (aug) -- (batch);
\draw[flow, draw=red] (batch) -- (modelin);
\draw[flow] (modelin) -- (feat);


\node[
    draw, rounded corners, thick, align=center, fill=blue!5,
    text width=4.2cm, minimum height=1.2cm,
    below=1.6cm of clips,
    xshift=-4.4cm
] (starss)
{STARSS23\\room-aware train/val split\\13 native classes};

\node[
    draw, rounded corners, thick, align=center, fill=blue!5,
    text width=3.6cm, minimum height=1.2cm,
    right=0.2cm of starss
] (tau21)
{TAU-NIGENS2021\\official DCASE splits\\12 native classes};

\node[
    draw, rounded corners, thick, align=center, fill=blue!5,
    text width=3.6cm, minimum height=1.2cm,
    right=0.2cm of tau21
] (tau19)
{TAU2019\\4-fold CV\\11 native classes};

\draw[flow] (starss.north) -- ($(clips.south)+(-0.65,0)$);
\draw[flow] (tau21.north) -- ($(clips.south)+(-0.10,0)$);
\draw[flow] (tau19.north) -- ($(clips.south)+(0.45,0)$);


\node[
    draw, dashed, rounded corners,
    inner sep=0.1cm,
    fit=(yaml)(cfg),
    label=above:{\textbf{Experiment Set-up}}
] {};

\node[
    draw, dashed, rounded corners,
    inner sep=0.1cm,
    fit=(clips)(items)(target)(aug)(batch),
    label=above:{\textbf{Data Pipeline}}
] {};

\node[
    draw, dashed, rounded corners,
    inner sep=0.1cm,
    fit=(modelin)(feat),
    label=above:{\textbf{Model Post-Processing}}
] {};

\end{tikzpicture}%
}
\caption{\Gls{at2seld} data pipeline, from configuration assembly and split construction to metadata parsing, windowed item generation, track-wise target construction, waveform-level augmentation, batching, model-side resampling, and spatial/semantic feature extraction.}
\label{fig:nas_data_pipeline}
\end{figure*}
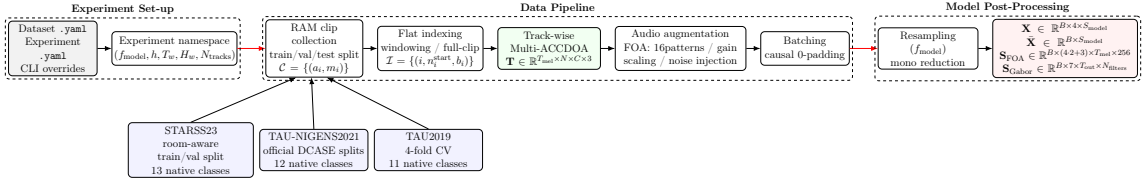

The unified data pipeline is implemented as a configuration-driven sequence of operations (Figure~\ref{fig:nas_data_pipeline}). The pipeline is organized around three levels: experiment configuration, dataset indexing and target generation, and model-side feature construction. This separation ensures that architecture-specific processing remains comparable across datasets while preserving corpus-specific split and metadata conventions.

Before any waveform is loaded, the pipeline assembles a scripted configuration object by merging experiment-level descriptors, dataset-specific descriptors, and runtime execution parameters. Dataset configuration files specify corpus-dependent entries such as file paths, split conventions, augmentation probabilities, class-imbalance options, and fold identifiers. \gls{cli} overrides are then applied to the subset of parameters intended to remain externally adjustable at launch time. The resulting \texttt{Python} configuration object controls the complete data flow, including dataset paths, model sampling rate, front-end hop length, number of track slots, window length and hop, augmentation parameters, STARSS23 room-aware validation options, and TAU2019 fold selection.

The first dataset operation is the construction of an ordered list of waveform/metadata pairs \(
\mathcal{C}
=
\{(a_i,m_i)\}_{i=1}^{N_{\mathrm{clips}}}
\) where \(\mathcal{C}\) is the clip list, \(a_i\) is the path to the \gls{foa} waveform of clip \(i\), \(m_i\) is the corresponding metadata file, and \(N_{\mathrm{clips}}\) is the number of collected clips. All subsequent indexing operations refer to clip indices in \(\mathcal{C}\), independently of the original folder hierarchy.

For STARSS23, the official development-test subset is used for testing, while the development-training folders are aggregated and partitioned with a room-aware strategy. Clips are grouped by site and room, and a reproducible subset is assigned to validation within each room group:
\begin{equation}
n_{\mathrm{val}}
=
\max
\left(
1,
\mathrm{round}
\left(
n_{\mathrm{room}}\rho_{\mathrm{val}}
\right)
\right)
\end{equation}
where \(n_{\mathrm{val}}\) is the number of validation clips selected from a room group, \(n_{\mathrm{room}}\) is the number of clips in that group, and \(\rho_{\mathrm{val}}\) is the validation ratio, set by default to \(20\%\). The remaining clips are assigned to training. This strategy avoids validation sets dominated by arbitrary number of rooms meanwhile preserving the intentional design of the corpus.

For TAU-NIGENS2021, the official \gls{dcase} split structure is used directly. TAU-NIGENS2020 follows the corresponding DCASE 2020 development setup when used for diagnostics, with development splits allocated to training, validation, and testing according to the official evaluation convention~\cite{politis2020tau_nigens}. TAU2019 follows a dedicated logic. In standalone mode, it is handled through the original four-fold cross-validation protocol over the development set. In mixed-data mode, a fixed split is used: development splits 2--4 for training, development split 1 for validation, and the evaluation subset as test. This avoids overlap between mixed training data and held-out evaluation material.

After clip collection, each metadata file is parsed and stored in memory as a list of events. For STARSS23, TAU-NIGENS2021, and TAU-NIGENS2020, valid rows are converted directly into tuples \(
e
=
(f,c,s,\phi,\theta)
\) where \(e\) is one annotated event entry, \(f\) is the native metadata frame index at \(100\) ms resolution, \(c\) is the class index, \(s\) is the source index, and \(\phi\) and \(\theta\) are azimuth and elevation in degrees. TAU2019 follows a different convention because metadata are stored as onset--offset intervals. Each interval is expanded into \(100\) ms framewise entries, and a source index is assigned per class according to order of appearance in the file, so that simultaneous same-class instances can be mapped to distinct tracks.

The dataset then constructs a flat item index \(
\mathcal{I}
=
\{(i,n_i^{\mathrm{start}})\}_{i=1}^{N_{\mathrm{items}}}
\) where \(\mathcal{I}\) is the item list, \(i\) is the clip index, \(n_i^{\mathrm{start}}\) is the starting sample at the native dataset sampling rate, and \(N_{\mathrm{items}}\) is the number of generated training or evaluation items. Two operating modes are supported: in \textit{full-clip} mode, each clip contributes one item; in \textit{windowed mode}, each clip is segmented into fixed-duration windows. If the waveform has \(S_{\mathrm{orig}}\) samples, the window length and hop size are:
\begin{equation}
W
=
\mathrm{round}
\left(
T_w f_{\mathrm{orig}}
\right),
\qquad
H
=
\mathrm{round}
\left(
H_w f_{\mathrm{orig}}
\right)
\end{equation}
where \(W\) and \(H\) are the window length and hop size in samples, \(T_w\) and \(H_w\) are the corresponding durations in seconds, and \(f_{\mathrm{orig}}\) is the original dataset sampling rate. Window start positions are generated recursively:
\begin{equation}
n_{k+1}^{\mathrm{start}}
=
n_k^{\mathrm{start}}
+
H,
\qquad
n_0^{\mathrm{start}}=0
\end{equation}
where \(n_k^{\mathrm{start}}\) is the start sample of the \(k\)-th window. The last window is retained even if it is shorter than \(W\), and missing samples are zero-padded at collation time.

At retrieval time, the waveform segment is loaded as \(
\mathbf{x}
\in
\mathbb{R}^{C_{\mathrm{ch}}\times S}
\) where \(\mathbf{x}\) is the waveform tensor, \(C_{\mathrm{ch}}=4\) for \gls{foa} channels \((W,X,Y,Z)\), and \(S\) is the number of samples in the selected item. STARSS23, TAU-NIGENS2021, and TAU-NIGENS2020 operate natively at \(24~\mathrm{kHz}\). TAU2019 is downsampled from \(48~\mathrm{kHz}\) to \(24~\mathrm{kHz}\) at dataset level when needed, so that subsequent model-side processing remains shared across corpora.

\subsubsection{Track-Wise Target Construction}
\label{subsubsec:target_construction}

For each item, the annotation list of the corresponding clip is converted into a track-wise Cartesian target tensor \(
\mathbf{T}
\in
\mathbb{R}^{T_{\mathrm{mel}}\times N_{\mathrm{tracks}}\times C_{\mathrm{cls}}\times 3}
\) where \(\mathbf{T}\) is the item-level target tensor, \(T_{\mathrm{mel}}\) is the number of \gls{at} model-aligned temporal frames, \(N_{\mathrm{tracks}}\) is the number of track slots, \(C_{\mathrm{cls}}\) is the class vocabulary of the active dataset, and the last dimension stores Cartesian \gls{doa} coordinates. Active sources are encoded by unit-norm Cartesian vectors, whereas inactive entries remain equal to \((0,0,0)\). This target is Multi-\gls{accdoa}-compatible at the tensor level, but the optimization uses the decoupled activity and localization heads defined in Section~\ref{subsec:exp_supervision_objective}.

If the waveform presented to the model front-end has effective sampling rate \(f_{\mathrm{eff}}\) and \(S_{\mathrm{eff}}\) samples, the number of target frames is:
\begin{equation}
T_{\mathrm{mel}}
=
1+
\left\lfloor
\frac{
\mathrm{round}
\left(
S_{\mathrm{eff}}
\frac{f_{\mathrm{model}}}{f_{\mathrm{eff}}}
\right)
}{
h
}
\right\rfloor
\end{equation}
where \(f_{\mathrm{model}}\) is the model sampling rate and \(h\) is the front-end hop length in samples at \(f_{\mathrm{model}}\). For fixed windows, this reduces to:
\begin{equation}
T_{\mathrm{mel}}
=
1+
\left\lfloor
\frac{
\mathrm{round}
\left(
W
\frac{f_{\mathrm{model}}}{f_{\mathrm{orig}}}
\right)
}{
h
}
\right\rfloor
\end{equation}
where \(W\) is the window length in samples at the original dataset rate.

Each annotation direction is converted from azimuth/elevation to Cartesian coordinates:
\begin{equation}
x
=
\sin(\phi)\cos(\theta),
\qquad
y
=
\cos(\phi)\cos(\theta),
\qquad
z
=
\sin(\theta)
\end{equation}
where \(\phi\) is azimuth, \(\theta\) is elevation, and angles are expressed in radians. To align model frames with metadata frames, the temporal coordinate of the \(t\)-th model frame is:
\begin{equation}
\tau_t
=
\frac{
n_{\mathrm{start}}^{(\mathrm{model})}
+
th
}{
f_{\mathrm{model}}
}
\end{equation}
where \(\tau_t\) is the time coordinate in seconds, \(h\) is the model-frame hop size, and \(n_{\mathrm{start}}^{(\mathrm{model})}\) is the window start converted to model-rate samples:
\begin{equation}
n_{\mathrm{start}}^{(\mathrm{model})}
=
\mathrm{round}
\left(
n_{\mathrm{start}}
\frac{
f_{\mathrm{model}}
}{
f_{\mathrm{orig}}
}
\right)
\end{equation}

The corresponding \(100\) ms metadata frame is:
\begin{equation}
k_t
=
\left\lfloor
\frac{\tau_t}{0.1}
\right\rfloor
\end{equation}
where \(k_t\) is the metadata frame index. Since annotations are defined at \(100\) ms resolution, each metadata frame typically spans approximately ten model frames (when \(f_{\mathrm{model}}=32~\mathrm{kHz}\) and \(h=320\) samples).

Within each metadata frame, active events are ordered by source index. Track slot \(0\) receives the smallest source index, track slot \(1\) the next source index, and so forth. If more sources are active than the available number of track slots, the excess events are discarded. This remains rare in the considered settings, where the maximum track capacity is chosen to cover the observed polyphony of the training and diagnostic corpora.

\subsubsection{Ambisonics-Aware Augmentation and Batching}
\label{subsubsec:augmentation_batching}

Waveform-level augmentation is applied only on the training split and before model-side resampling. Validation and test samples are never augmented. The augmentation module is designed specifically for \gls{foa} input and acts jointly on the raw 4-channel waveform and on the associated Cartesian target tensor. Three mutually exclusive augmentation families are supported. 

\begin{landscape}
\begin{figure}[p]
    \centering
    \includegraphics[width=0.97\linewidth]{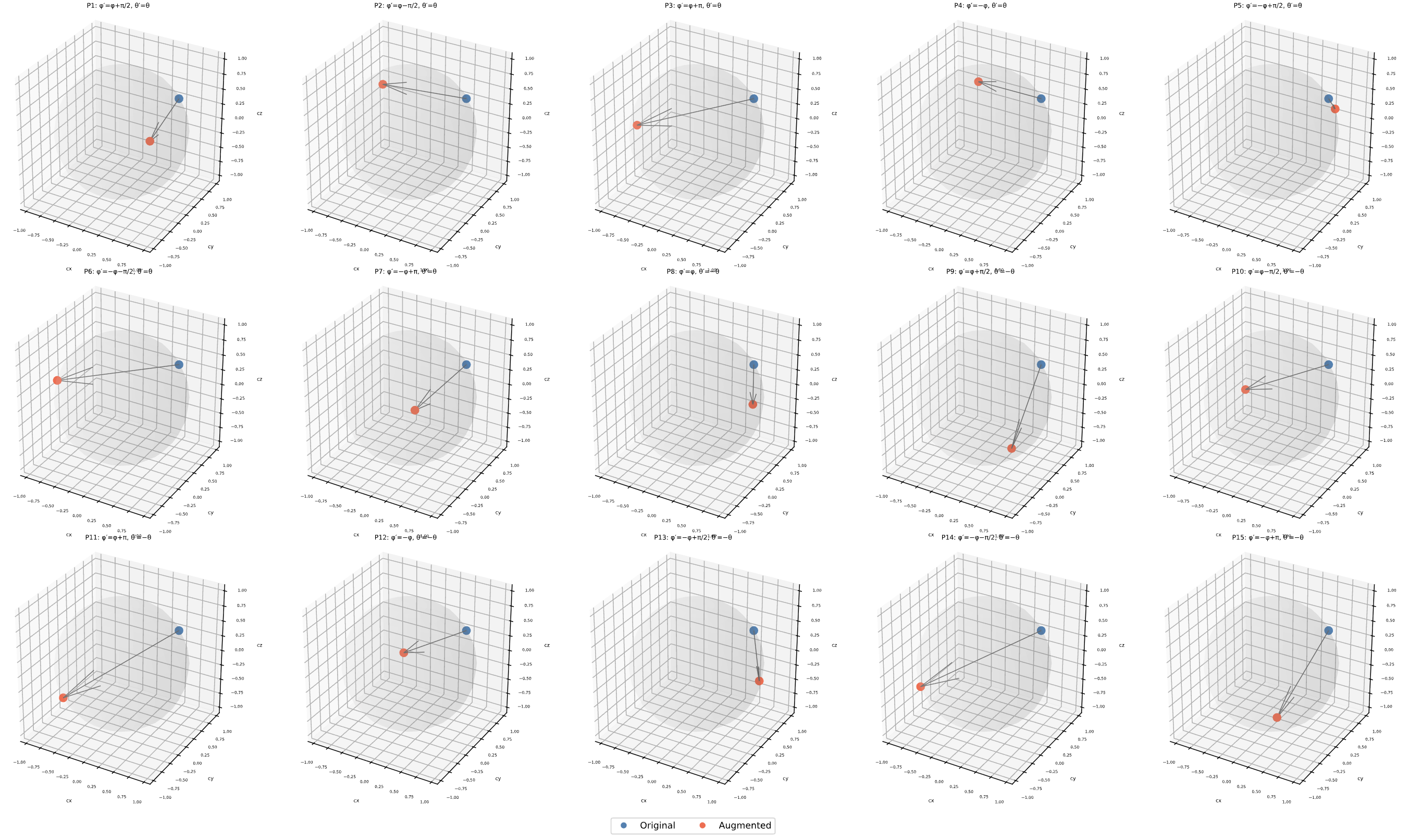}
    \caption{Effect of the \textit{16patterns} augmentation on a fixed Cartesian position with azimuth \(35^\circ\) and elevation \(20^\circ\). The \textcolor{blue}{blue point} denotes the original position, whereas the \textcolor{orange}{orange point} denotes the transformed position.}
    \label{fig:foa_aug_fixed}
\end{figure}
\end{landscape}

\begin{landscape}
\begin{figure}[p]
    \centering
    \includegraphics[width=0.97\linewidth]{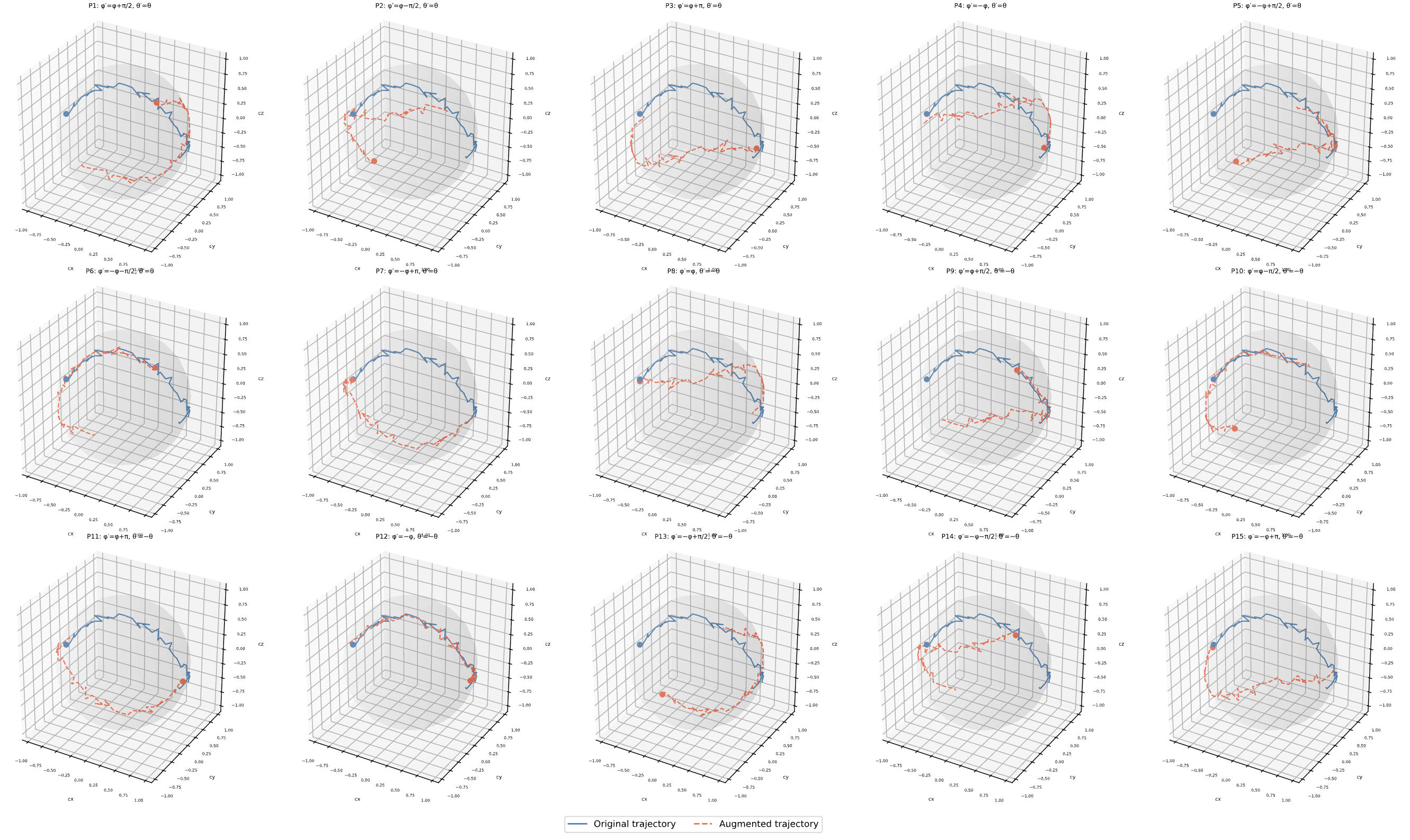}
    \caption{Effect of the \textit{16patterns} augmentation on a synthetic spatial trajectory. The \textcolor{blue}{blue curve} denotes the original trajectory, whereas the \textcolor{orange}{orange curve} denotes the transformed trajectory.}
    \label{fig:foa_aug_trajectories}
\end{figure}
\end{landscape}

The first is a \textit{symmetry-based \gls{foa} transformation} derived from Ambisonics channel symmetries~\cite{10068271}. One pattern is sampled from the set of non-identity transforms, and the Cartesian targets are updated coherently. The identity operation is represented by the null-augmentation branch. The full set of supported \textit{16patterns} is reported in Table~\ref{tab:foa_16patterns}, its effect on fixed source positions in Figure~\ref{fig:foa_aug_fixed} and on spatial trajectories in Figure~\ref{fig:foa_aug_trajectories}.

\newcolumntype{C}[1]{>{\centering\arraybackslash}m{#1}}
\newcommand{\thetashift}{-0.6cm}
\newcolumntype{T}[1]{>{\hspace*{\thetashift}\centering\arraybackslash}m{#1}}

\begin{table}[ht]
\centering
\caption{\Gls{foa} symmetry \textit{16patterns} supported by the augmentation module. The identity operation is represented by the null-augmentation branch. The omnidirectional channel \(W\) is left unchanged.}
\vspace{0.1cm}
\label{tab:foa_16patterns}
\footnotesize
\renewcommand{\arraystretch}{1.12}
\setlength{\tabcolsep}{3.2pt}
\begin{tabular}{|C{0.7cm}|C{1.45cm}|T{1.45cm}||C{1.15cm}|C{1.15cm}|C{1.15cm}||C{1.15cm}|C{1.15cm}|C{1.15cm}|}
\hline
\multirow{2}{*}{\textbf{ID}} &
\multicolumn{2}{c||}{\textbf{Geometric interpretation}} &
\multicolumn{3}{c||}{\textbf{FOA channel transform}} &
\multicolumn{3}{c|}{\textbf{Cartesian target transform}} \\
\cline{2-9}
& \(\phi'\) & \(\theta'\) & \(X'\) & \(Y'\) & \(Z'\) & \(x'\) & \(y'\) & \(z'\) \\
\hline
\hline
\rowcolor{gray!15}
0  & \(\phi\) & \(\theta\) & \(X\) & \(Y\) & \(Z\) & \(x\) & \(y\) & \(z\) \\
\hline
1  & \(\phi+\frac{\pi}{2}\) & \(\theta\) & \(-Y\) & \(X\) & \(Z\) & \(y\) & \(-x\) & \(z\) \\
\hline
2  & \(\phi-\frac{\pi}{2}\) & \(\theta\) & \(Y\) & \(-X\) & \(Z\) & \(-y\) & \(x\) & \(z\) \\
\hline
3  & \(\phi+\pi\) & \(\theta\) & \(-X\) & \(-Y\) & \(Z\) & \(-x\) & \(-y\) & \(z\) \\
\hline
4  & \(-\phi\) & \(\theta\) & \(X\) & \(-Y\) & \(Z\) & \(x\) & \(-y\) & \(z\) \\
\hline
5  & \(-\phi+\frac{\pi}{2}\) & \(\theta\) & \(Y\) & \(X\) & \(Z\) & \(y\) & \(x\) & \(z\) \\
\hline
6  & \(-\phi-\frac{\pi}{2}\) & \(\theta\) & \(-Y\) & \(-X\) & \(Z\) & \(-y\) & \(-x\) & \(z\) \\
\hline
7  & \(-\phi+\pi\) & \(\theta\) & \(-X\) & \(Y\) & \(Z\) & \(-x\) & \(y\) & \(z\) \\
\hline
8  & \(\phi\) & \(-\theta\) & \(X\) & \(Y\) & \(-Z\) & \(x\) & \(y\) & \(-z\) \\
\hline
9  & \(\phi+\frac{\pi}{2}\) & \(-\theta\) & \(-Y\) & \(X\) & \(-Z\) & \(y\) & \(-x\) & \(-z\) \\
\hline
10 & \(\phi-\frac{\pi}{2}\) & \(-\theta\) & \(Y\) & \(-X\) & \(-Z\) & \(-y\) & \(x\) & \(-z\) \\
\hline
11 & \(\phi+\pi\) & \(-\theta\) & \(-X\) & \(-Y\) & \(-Z\) & \(-x\) & \(-y\) & \(-z\) \\
\hline
12 & \(-\phi\) & \(-\theta\) & \(X\) & \(-Y\) & \(-Z\) & \(x\) & \(-y\) & \(-z\) \\
\hline
13 & \(-\phi+\frac{\pi}{2}\) & \(-\theta\) & \(Y\) & \(X\) & \(-Z\) & \(y\) & \(x\) & \(-z\) \\
\hline
14 & \(-\phi-\frac{\pi}{2}\) & \(-\theta\) & \(-Y\) & \(-X\) & \(-Z\) & \(-y\) & \(-x\) & \(-z\) \\
\hline
15 & \(-\phi+\pi\) & \(-\theta\) & \(-X\) & \(Y\) & \(-Z\) & \(-x\) & \(y\) & \(-z\) \\
\hline
\end{tabular}
\end{table}

The second augmentation family is global \textit{amplitude gain scaling}:
\begin{equation}
g
\sim
\mathcal{U}
\left(
g_{\min},
g_{\max}
\right),
\qquad
\mathbf{x}'=g\mathbf{x}
\end{equation}
where \(g\) is the sampled unitary gain, \(\mathbf{x}\) is the input waveform, and \(\mathbf{x}'\) is the gain-scaled waveform. 

The third augmentation family is \textit{noise injection}. Zero-mean Gaussian noise is added to all channels with standard deviation:
\begin{equation}
\sigma_{\varepsilon}
=
\frac{
\mathrm{RMS}
\left(
\mathbf{x}
\right)
}{
10^{\mathrm{SNR}_{\mathrm{dB}}/20}
}
\end{equation}
where \(\sigma_{\varepsilon}\) is the noise standard deviation, \(\mathrm{RMS}(\mathbf{x})\) is the \gls{rms} amplitude of the waveform, and \(\mathrm{SNR}_{\mathrm{dB}}\) is the target \gls{snr} in decibels. The augmented waveform is:
\begin{equation}
\mathbf{x}'
=
\mathbf{x}
+
\boldsymbol{\varepsilon},
\qquad
\varepsilon_{c,t}
\sim
\mathcal{N}
\left(
0,
\sigma_{\varepsilon}^{2}
\right)
\end{equation}
where \(\boldsymbol{\varepsilon}\) is the sampled noise tensor and \(\varepsilon_{c,t}\) denotes the noise sample at channel \(c\) and time index \(t\).

The augmentation families are sampled \textit{categorically} rather than independently. If the nominal probabilities are \(p_{\mathrm{pat}}\), \(p_{\mathrm{gain}}\), and \(p_{\mathrm{noise}}\), then the null operation has probability:
\begin{equation}
p_{\mathrm{none}}
=
1-
\left(
p_{\mathrm{pat}}
+
p_{\mathrm{gain}}
+
p_{\mathrm{noise}}
\right)
\end{equation}
where \(p_{\mathrm{none}}\) is the probability of leaving the sample unchanged. At most one augmentation is applied to each retrieved training item at runtime. In the dataset configuration used for the \gls{seld} corpora, the default probabilities are:
\begin{equation}
p_{\mathrm{pat}}=0.5,
\qquad
p_{\mathrm{gain}}=0.2,
\qquad
p_{\mathrm{noise}}=0.2,
\qquad
p_{\mathrm{none}}=0.1
\end{equation}

Balanced and mixed-data variants introduced in Stage~4 reuse the same augmentation module, but apply it after their own sampling or mixing logic.

In windowed mode, all items have the same waveform and target lengths, so standard tensor stacking is sufficient:
\begin{equation}
\mathbf{X}_{\mathrm{batch}}
\in
\mathbb{R}^{B\times 4\times W},
\qquad
\mathbf{T}_{\mathrm{batch}}
\in
\mathbb{R}^{B\times T_{\mathrm{mel}}\times N_{\mathrm{tracks}}\times C_{\mathrm{cls}}\times 3}
\end{equation}
where \(\mathbf{X}_{\mathrm{batch}}\) is the batched waveform tensor and \(\mathbf{T}_{\mathrm{batch}}\) is the batched target tensor. In full-clip mode, waveform and target lengths may vary across items (according to variable file lengths), so the custom collation routine pads both tensors to the maximum length in the batch:
\begin{equation}
\mathbf{X}_{\mathrm{pad}}
\in
\mathbb{R}^{B\times 4\times S_{\max}},
\qquad
\mathbf{T}_{\mathrm{pad}}
\in
\mathbb{R}^{B\times T_{\max}\times N_{\mathrm{tracks}}\times C_{\mathrm{cls}}\times 3}
\end{equation}
where \(S_{\max}\) and \(T_{\max}\) are the maximum waveform and target lengths inside the batch. No explicit length mask is passed downstream. Padded target frames remain zero and therefore do not support active localization targets.

\subsubsection{Model-Side Feature Interfaces}
\label{subsubsec:model_feature_interfaces}

The PyTorch~\cite{10.5555/3454287.3455008} \texttt{DataLoader} delivers \gls{foa} waveform tensors at the effective dataset sampling rate. The first operation inside the model is resampling to the model sampling rate \(f_{\mathrm{model}}\), typically \(32~\mathrm{kHz}\), according to the \gls{gpat} backbone pretraining requirements. The semantic branch receives a monophonic reduction obtained by channel averaging:
\begin{equation}
\bar{\mathbf{X}}
=
\frac{1}{4}
\sum_{c=1}^{4}
\mathbf{X}_{c}
\in
\mathbb{R}^{B\times S_{\mathrm{model}}}
\end{equation}
where \(\bar{\mathbf{X}}\) is the monophonic waveform forwarded to the pretrained \gls{at} backbone, \(\mathbf{X}_{c}\) is the \(c\)-th \gls{foa} channel, and \(S_{\mathrm{model}}\) is the number of samples after model-side resampling. The full 4-channel tensor is preserved for the spatial branch.

In the spectral setting, the \gls{foa} waveform is transformed through a channel-wise \gls{stft} front-end with \(n_{\mathrm{fft}}=512\), Hann analysis window, and hop length \(h\) expressed in samples at model rate. Magnitude and phase spectra are extracted from the one-sided complex spectrogram after DC-bin removal (Section~\ref{subsec:related_crnn} and stacked along the channel dimension, yielding \(2\times4=8\) channels. In addition, the directional channels are combined with the omnidirectional component to derive \glspl{iv} (Section~\ref{subsec:related_permutation_tracking}), producing three further spatial channels. The resulting spectral input tensor is:
\begin{equation}
\mathbf{S}_{\mathrm{FOA}}
\in
\mathbb{R}^{B\times (4\cdot2+3)\times T_{\mathrm{mel}}\times 256}
\end{equation}
where \(\mathbf{S}_{\mathrm{FOA}}\) is the spectral spatial input tensor. Magnitude, phase, and \gls{iv} groups are normalized independently through 2D instance normalization, preserving their distinct numerical ranges while stabilizing numerical optimization.

In the Gabor-based setting, the spatial branch does not receive the \gls{stft} tensor. Instead, the multi-channel waveform is processed by a learnable Gabor filterbank front-end that produces a \gls{foa}-oriented representation composed of one omnidirectional spectral channel and six cross-spectrum channels:
\begin{equation}
\mathbf{S}_{\mathrm{Gabor}}
\in
\mathbb{R}^{B\times 7\times T_{\mathrm{out}}\times N_{\mathrm{filters}}}
\end{equation}
where \(\mathbf{S}_{\mathrm{Gabor}}\) is the Gabor-domain spatial tensor, \(T_{\mathrm{out}}\) is the output temporal resolution of the learnable front-end, and \(N_{\mathrm{filters}}\) is the number of learned filters (Section~\ref{subsec:related_raw_waveform}). This interface allows Stage~1 to compare conventional explicit \gls{foa} descriptors with learnable representations under the same track-wise supervision.

\clearpage

\clearpage
\clearpage

\clearpage


\section{\glsentryshort{nas} Results}
\label{sec:nas_results}
The experimental design defined in Section~\ref{sec:experimental_design} organizes the proposed \gls{at2seld} framework as an informed, staged architecture search. The present section reports the results of Stages~1--3, following the same progression: shallow spatial-branch screening, controlled depth allocation, and semantic--spatial interaction analysis. The objective is not to enumerate isolated configurations, but to identify which architectural choices consistently improve semantic-to-spatial \gls{seld} under the shared target interface and multi-task objective defined in Sections~\ref{subsec:exp_supervision_objective} and~\ref{subsec:exp_datasets_pipeline}.

All results in this section are interpreted as architecture-selection evidence. Stage~1 identifies viable spatial front-end and shallow module families; Stage~2 tests where additional capacity should be allocated; Stage~3 evaluates whether explicit semantic--spatial bridges improve the selected topology. The final diagnostic characterization of the selected configurations, including loss design, balancing, threshold calibration, and cross-dataset transfer, is reported separately in Section~\ref{sec:diagnostic_characterization}.


\subsection{Stage 1: Shallow Spatial-Branch Screening}
\label{subsec:nas_stage1}

Stage~1 was designed to answer two preliminary questions. The first concerns the computational envelope of the candidate modules: before composing them into complete \gls{seld} systems, each operator must be characterized in terms of parameter count, memory footprint, arithmetic complexity, and latency scaling. The second concerns the shallow semantic-to-spatial regime: with all branches kept shallow and without explicit semantic--spatial feature bridges, the experiment tests whether the spatial branch can already exploit the pretrained \gls{gpat} prior through shared optimization, and which front-end and module families are compatible with the track-wise \gls{seld} objective.

Accordingly, Stage~1 is divided into an isolated module-profiling phase and a shallow \gls{nas} grid search. The complete module inventory is reported in Appendix~\ref{app:spatial_modules} (Table~\ref{tab:spatial_modules_inventory}). All shallow grid runs use the shared training setup in Table~\ref{tab:stage1_common_params} and the hardware platform summarized in Table~\ref{tab:hardware_platform}.

\begin{table}[ht]
\centering
\caption{Common training parameters adopted in the Stage~1 shallow \gls{nas} experiments.}
\label{tab:stage1_common_params}
\footnotesize
\renewcommand{\arraystretch}{1.08}
\setlength{\tabcolsep}{4.2pt}
\begin{tabular}{|p{3.4cm}|p{2.2cm}|p{8.7cm}|}
\hline
\textbf{Parameter} & \textbf{Value} & \textbf{Description} \\
\hline
Random seed & \(42\) & Global experiment, dataset, and model initialization seed \\
\hline
Validation fraction & \(0.2\) & Fraction of STARSS23 development clips assigned to validation split \\
\hline
Max epochs & \(20\) & Upper bound on training epochs \\
\hline
Batch size & \(16\) & Mini-batch size (in samples) used during training \\
\hline
Optimizer & Adam & Adaptive Momentum-estimation optimizer~\cite{kingma2015adam} \\
\hline
Learning rate \(\eta\) & \(5\times10^{-4}\) & Fixed learning rate used during the shallow screening stage \\
\hline
Early-stopping monitor & \gls{seld} score & Validation metric used to drive early stopping (lower$\rightarrow$better) \\
\hline
Early-stopping patience & \(10\) epochs & Number of epochs without validation improvement before interrupting training \\
\hline
Input regime & windowed & Window size: \(10\mathrm{s}\); window hop: \(2.5\mathrm{s}\) \\
\hline
\gls{tpit} loss weight & \(\lambda_s=0.8\) & Weight assigned to the track-wise \gls{seld} loss in the total objective \\
\hline
\gls{at} \gls{bce} loss weight & \(\lambda_e=0.2\) & Weight assigned to the auxiliary semantic-presence loss in the total objective \\
\hline
\gls{sed}/\gls{doa} mixing & \(\lambda_{\mathrm{SED}}=0.6\) & Relative contribution of the activity term inside the track-wise \gls{seld} loss; the \gls{doa} term contributes with \(1-\lambda_{\mathrm{SED}}=0.4\) \\
\hline
Angular threshold & \(20^\circ\) & Angular tolerance used in \gls{seld} matching and evaluation \\
\hline
\end{tabular}
\end{table}

\begin{table}[ht]
\centering
\caption{Hardware platform used to execute the \gls{at2seld} experiments.}
\label{tab:hardware_platform}
\footnotesize
\renewcommand{\arraystretch}{1.08}
\setlength{\tabcolsep}{4.2pt}
\begin{tabular}{|p{3.3cm}|p{8.5cm}|}
\hline
\textbf{Component} & \textbf{Specifications} \\
\hline
CPUs & \(2\times\) Intel Xeon Gold 6140 Skylake, 2.30 GHz \(18\times\) cores \\
\hline
Cluster memory & \(1\times\) NUMA domain and \(1\times\) L3 cache per CPU \\
\hline
System memory & 192 GB RAM \\
\hline
Memory per core & 5000 MB \\
\hline
Local storage & 200 GB SSD \\
\hline
\hline
GPU model & NVIDIA Tesla P100 SXM2, NVLink-optimized, 1.3 GHz \\
\hline
GPU memory & 16 GB CoWoS HBM2 \\
\hline
GPU PCIe bandwidth & 32 GB/s \\
\hline
Cluster partitions & \texttt{gpu\_p100} and \texttt{gpu\_p100\_long} on the \textit{Genius} cluster \\
\hline
\end{tabular}

\vspace{0.1cm}
\parbox{\linewidth}{\centering\footnotesize
The resources and services used in this work were provided by the Flemish Supercomputer Center (VSC), funded by the Research Foundation -- Flanders (FWO) and the Flemish Government.
}
\end{table}

\subsubsection{Module Profiling}
\label{subsubsec:stage1_module_profiling}

The preliminary profiling phase characterizes the computational behavior of the candidate spatial operators before they are embedded into full \gls{at2seld} pipelines. Each module is instantiated independently using the representative hyper-parameter configuration adopted in the original implementation and evaluated on synthetic inputs whose shapes match the expected internal interfaces. The analyzed operators include preprocessing modules, early spatial encoders, late track-wise abstraction modules, temporal smoothing blocks, and feature-interaction components.

For each instantiated module, the total and trainable parameter counts are computed directly from the corresponding \texttt{torch.nn.Module}. The static memory footprint associated with learnable parameters is estimated as:
\begin{equation}
M_{\mathrm{p}}
=
\frac{1}{1024^2}
\sum_{p\in\Theta}
N_p s_p
\end{equation}
where \(M_{\mathrm{p}}\) is the parameter-memory footprint in MegaBytes, \(\Theta\) is the set of module parameters, \(N_p\) is the number of scalar elements of parameter \(p\), and \(s_p\) is the storage size of each scalar element in bytes. This estimate measures only static parameter memory; it does not include activation memory, optimizer states, or peak training memory.

Arithmetic complexity is estimated with \texttt{torchinfo} using a single forward pass with batch size \(B=1\). Let \(\mathrm{MAC}\) denote the number of multiply--accumulate operations returned by the summary routine. The corresponding \gls{gflops} estimate is:
\begin{equation}
\mathrm{GFLOPs}
=
\frac{2\,\mathrm{MAC}}{10^9}
\end{equation}
where the factor \(2\) follows the standard convention of counting one multiplication and one addition per multiply--accumulate operation. Although this value is approximate and backend-dependent execution may differ, it provides a consistent indicator of arithmetic burden across the candidate modules.

Latency is profiled separately from arithmetic complexity. For each module and each batch size \(B\in\{1,2,4,8,16\}\), the profiling routine executes two warm-up forward passes followed by twenty measured forward passes with gradients disabled. On CUDA devices, elapsed time is measured through synchronized event timestamps, whereas on CPU and \gls{mps} backends it is measured through high-resolution wall-clock timers. For each backend, the stored timing samples are summarized by: mean latency, median latency, standard deviation, empirical 90th percentile, skewness, and kurtosis. This allows the comparison to reflect not only average runtime but also latency dispersion and distributional asymmetry.

\begin{figure}[ht]
    \centering
    \includegraphics[width=\linewidth]{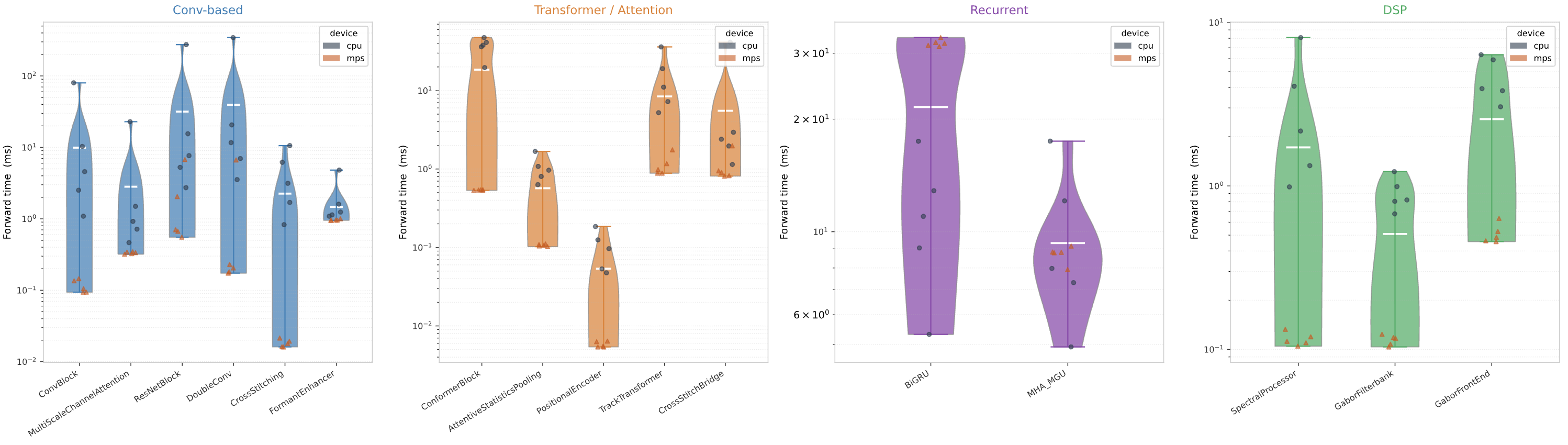}
    \caption{Technical profiling of the candidate spatial modules. \textbf{Black dots}: \(5\times\) random CPU profiling runs; \textbf{\textcolor{red}{red triangles}}: \(5\times\) random \gls{mps} profiling runs.}
    \label{fig:spatial_modules_profiling}
\end{figure}

The profiling results (Figure~\ref{fig:spatial_modules_profiling}) reveal a strongly non-uniform computational landscape. Parameter-free or nearly parameter-free operators, such as the \texttt{SpectralProcessor} and the \texttt{PositionalEncoder}, define the low-cost boundary of the space. Lightweight local modules, including \texttt{ConvBlock}, \texttt{MSCA}, and \texttt{FormantEnhancementModule}, remain compact in both parameter count and static memory. In contrast, high-capacity sequence and abstraction blocks dominate the parameter and memory envelope. The \texttt{TrackTransformer} is the heaviest individual operator, with approximately \(12.61\)M parameters and \(48.10\) MB of static parameter memory, whereas the \texttt{ConformerBlock} is the second most demanding module, with approximately \(1.52\)M parameters and \(5.81\) MB.

Temporal smoothing operators occupy an intermediate but practically important regime. The \texttt{BiGRU} has approximately \(396\,000\) parameters and \(1.51\) MB of parameter memory, whereas \texttt{MHA-MGU} is lighter in trainable weights, with approximately \(100\,000\) parameters, but remains non-negligible in latency. This discrepancy shows that trainable parameter count alone is not a sufficient proxy for deployability. Attention-style recurrent updates may remain slower than more conventional recurrent smoothing even when their static parameter footprint is smaller.

The profiling phase therefore constrains the interpretation of the subsequent shallow grid. Late track-wise abstraction and temporal smoothing dominate the computational envelope more than front-end operators. Any gain produced by richer late processing or more sophisticated smoothing must consequently be interpreted against a runtime and latency penalty. Conversely, a heavy standalone module such as \texttt{TrackTransformer} can still be viable at full-model level when paired with efficient early and smoothing components. This observation directly motivates the controlled Stage~1 grid, where module families are evaluated as complete semantic-to-spatial pipelines rather than only as isolated operators.

Backend behavior also provides a practical constraint. The \gls{mps} backend accelerates many convolutional, feed-forward, and Transformer-style components, but recurrent modules benefit less consistently and may even become slower in some configurations. Since the main training campaign is executed on CUDA hardware, the profiling results are not used as absolute runtime predictors for all experiments. They are instead used to identify relative complexity regimes and to avoid interpreting a validation improvement without considering the computational structure that produced it.

\subsubsection{Shallow Grid Results}
\label{subsubsec:stage1_shallow_grid}

The shallow grid evaluates two spatial front-end families under the same training protocol (Table~\ref{tab:stage1_common_params}). The spectral family uses explicit \gls{foa} time--frequency descriptors composed of \gls{stft} magnitude, phase, and \glspl{iv} (Section~\ref{subsubsec:model_feature_interfaces}). The Gabor family uses a learnable signal-analysis front-end that combines filterbank spectra and cross-spectrum spatial embeddings. Each front-end is paired with one \textit{early spatial encoder}, one \textit{late abstraction module}, and one \textit{temporal smoothing operator}. The early alternatives are \texttt{ConvBlock} and \texttt{ResNetBlock}; the late alternatives are \texttt{FormantEnhancement} and \texttt{TrackTransformer}; and the smoothing alternatives are \texttt{BiGRU} and \texttt{MHA-MGU}.

Tables~\ref{tab:stage1_shallow_spec} and~\ref{tab:stage1_shallow_gab} report shallow-stage results on STARSS23, ranked by best validation \gls{seld} score. Configuration names are normalized preserving their shorthand module identifiers: \texttt{spe} and \texttt{gab} denote spectral and Gabor front-ends, \texttt{cb} and \texttt{rn} denote convolutional and residual early encoders, \texttt{fe} and \texttt{tt} denote formant-enhancement and track-transformer late modules, and \texttt{bg} and \texttt{mg} denote \texttt{BiGRU} and \texttt{MHA-MGU} smoothing.

\begin{table}[ht]
\centering
\caption{Stage~1 \gls{at2seld} results on STARSS23 for the spectral front-end family.}
\label{tab:stage1_shallow_spec}
\footnotesize
\renewcommand{\arraystretch}{1.05}
\setlength{\tabcolsep}{3.5pt}
\resizebox{\textwidth}{!}{%
\begin{tabular}{|p{2.35cm}|c|c|c|c|c|c|c|c|}
\hline
\textbf{Configuration} & \textbf{Early} & \textbf{Late} & \textbf{Smooth} & \textbf{Val SELD \(\downarrow\)} & \textbf{Duration} & \textbf{Latency} & \textbf{Params} & \textbf{GFLOPs} \\
\hline
\rowcolor{gray!15}
\texttt{spe\_rn\_tt\_bg} & rn & tt & bg & 0.48 & 1.58 & 269.7 & 28.74 & 59.15 \\
\hline
\texttt{spe\_cb\_tt\_bg} & cb & tt & bg & 0.53 & 1.35 & 176.6 & 28.70 & 36.32 \\
\texttt{spe\_rn\_tt\_mg} & rn & tt & mg & 0.58 & 5.34 & 550.4 & 28.09 & 57.52 \\
\texttt{spe\_cb\_tt\_mg} & cb & tt & mg & 0.64 & 5.15 & 496.7 & 28.05 & 34.69 \\
\texttt{spe\_rn\_fe\_bg} & rn & fe & bg & 0.65 & 3.86 & 420.3 & 28.07 & 72.12 \\
\texttt{spe\_cb\_fe\_bg} & cb & fe & bg & 0.71 & 3.63 & 338.6 & 28.03 & 49.30 \\
\texttt{spe\_cb\_fe\_mg} & cb & fe & mg & 0.75 & 7.20 & 651.7 & 27.65 & 48.26 \\
\texttt{spe\_rn\_fe\_mg} & rn & fe & mg & 0.76 & 7.36 & 747.1 & 27.69 & 71.09 \\
\hline
\end{tabular}%
}

\vspace{0.1cm}
\parbox{0.95\linewidth}{\centering\footnotesize
Duration = training duration in hours; Latency = single-sample inference latency in ms; Params = trainable parameters in millions; GFLOPs = estimated single-forward arithmetic complexity.
}
\end{table}

\begin{table}[ht]
\centering
\caption{Stage~1 \gls{at2seld} results on STARSS23 for the Gabor front-end family.}
\label{tab:stage1_shallow_gab}
\footnotesize
\renewcommand{\arraystretch}{1.05}
\setlength{\tabcolsep}{3.5pt}
\resizebox{\textwidth}{!}{%
\begin{tabular}{|p{2.35cm}|c|c|c|c|c|c|c|c|}
\hline
\textbf{Configuration} & \textbf{Early} & \textbf{Late} & \textbf{Smooth} & \textbf{Val SELD \(\downarrow\)} & \textbf{Duration} & \textbf{Latency} & \textbf{Params} & \textbf{GFLOPs} \\
\hline
\rowcolor{gray!5}
\texttt{gab\_cb\_tt\_bg} & cb & tt & bg & 0.73 & 1.36 & 154.9 & 28.70 & 33.83 \\
\hline
\texttt{gab\_rn\_tt\_bg} & rn & tt & bg & 0.74 & 1.50 & 206.6 & 28.74 & 45.17 \\
\texttt{gab\_cb\_tt\_mg} & cb & tt & mg & 0.76 & 4.98 & 453.7 & 28.04 & 32.20 \\
\texttt{gab\_rn\_fe\_mg} & rn & fe & mg & 0.79 & 5.51 & 565.9 & 27.69 & 49.99 \\
\texttt{gab\_cb\_fe\_bg} & cb & fe & bg & 0.80 & 2.06 & 228.5 & 28.03 & 39.69 \\
\texttt{gab\_cb\_fe\_mg} & cb & fe & mg & 0.81 & 5.40 & 525.3 & 27.65 & 38.65 \\
\texttt{gab\_rn\_tt\_mg} & rn & tt & mg & 0.81 & 5.16 & 507.6 & 28.09 & 43.54 \\
\texttt{gab\_rn\_fe\_bg} & rn & fe & bg & 0.82 & 2.16 & 279.1 & 28.07 & 51.03 \\
\hline
\end{tabular}%
}

\vspace{0.1cm}
\parbox{0.95\linewidth}{\centering\footnotesize
Duration = training duration in hours; Latency = single-sample inference latency in ms; Params = trainable parameters in millions; GFLOPs = estimated single-forward arithmetic complexity.
}
\end{table}

The first result is that the spectral front-end clearly dominates the Gabor-based alternative in the shallow regime. The best spectral model, \texttt{spe\_rn\_tt\_bg}, reaches a validation \gls{seld} score of \(0.48\), whereas the best Gabor model, \texttt{gab\_cb\_tt\_bg}, reaches \(0.73\). This gap is not an isolated outlier: averaged over all shallow configurations, the spectral family remains substantially stronger than the Gabor family. The result indicates that explicit \gls{stft} magnitude and phase, combined with \gls{foa} \glspl{iv}, expose localization-relevant information more effectively than the learned Gabor and \gls{gcc} interface when the downstream architecture is still shallow. In this setting, the advantage of a learnable front-end does not compensate for the loss of the explicit spatial structure provided by the spectral \gls{foa} representation.

\begin{figure}[ht]
    \centering
    \includegraphics[width=0.82\linewidth]{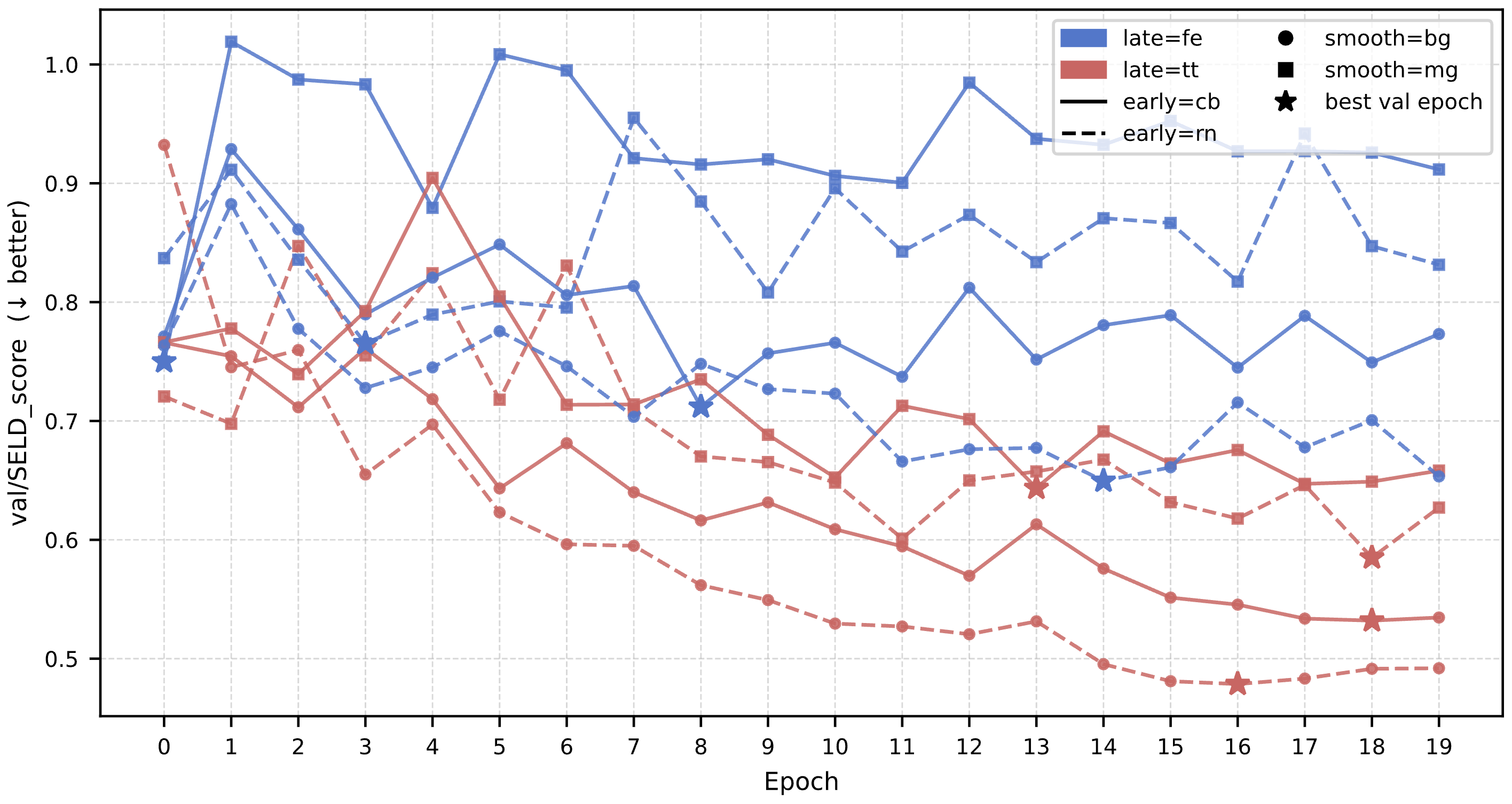}
    \caption{Stage~1 \gls{at2seld} results on STARSS23: validation \gls{seld} score trends for the spectral front-end family.}
    \label{fig:shallow_spe_score_val}
\end{figure}

The second result concerns the late track-wise abstraction module. Within both front-end families, \texttt{TrackTransformer} consistently outperforms \texttt{FormantEnhancement}. In the spectral family, the mean validation \gls{seld} score is approximately \(0.56\) for configurations using \texttt{tt}, compared with approximately \(0.72\) for configurations using \texttt{fe}. In the Gabor family, the difference is smaller but remains systematic. This supports the hypothesis that explicit track-wise attentive abstraction is better aligned with the \gls{tpit} supervision interface than formant-oriented local refinement, whose design is more suitable for enhancing spectral traces than for organizing class-track spatial hypotheses.

The third result concerns temporal smoothing. \texttt{BiGRU} is more stable and more efficient than \texttt{MHA-MGU}. In the spectral family, \texttt{bg} configurations reach a mean validation \gls{seld} score of approximately \(0.59\), compared with approximately \(0.69\) for \texttt{mg} configurations. The computational evidence is consistent with this performance trend: all \texttt{MHA-MGU} configurations require substantially longer training and higher single-sample latency than their \texttt{BiGRU} counterparts. Thus, the theoretically lighter parameterization of \texttt{MHA-MGU} does not translate into practical efficiency or better generalization in the shallow \gls{at2seld} setting.

The early spatial module has a more moderate effect. In the spectral family, \texttt{ResNetBlock} is preferable to \texttt{ConvBlock} in the best-performing configuration, especially when paired with \texttt{TrackTransformer} and \texttt{BiGRU}. However, the effect is weaker than the front-end and late-module choices. In the Gabor family, the average difference between \texttt{rn} and \texttt{cb} is small and sometimes favors the simpler convolutional block. This suggests that, before depth is introduced, the dominant determinants of shallow performance are the spatial input representation and the track-wise abstraction module, whereas the exact early operator becomes more decisive only after the representation is deepened in Stage~2.

\begin{figure}[ht]
    \centering
    \includegraphics[width=\linewidth]{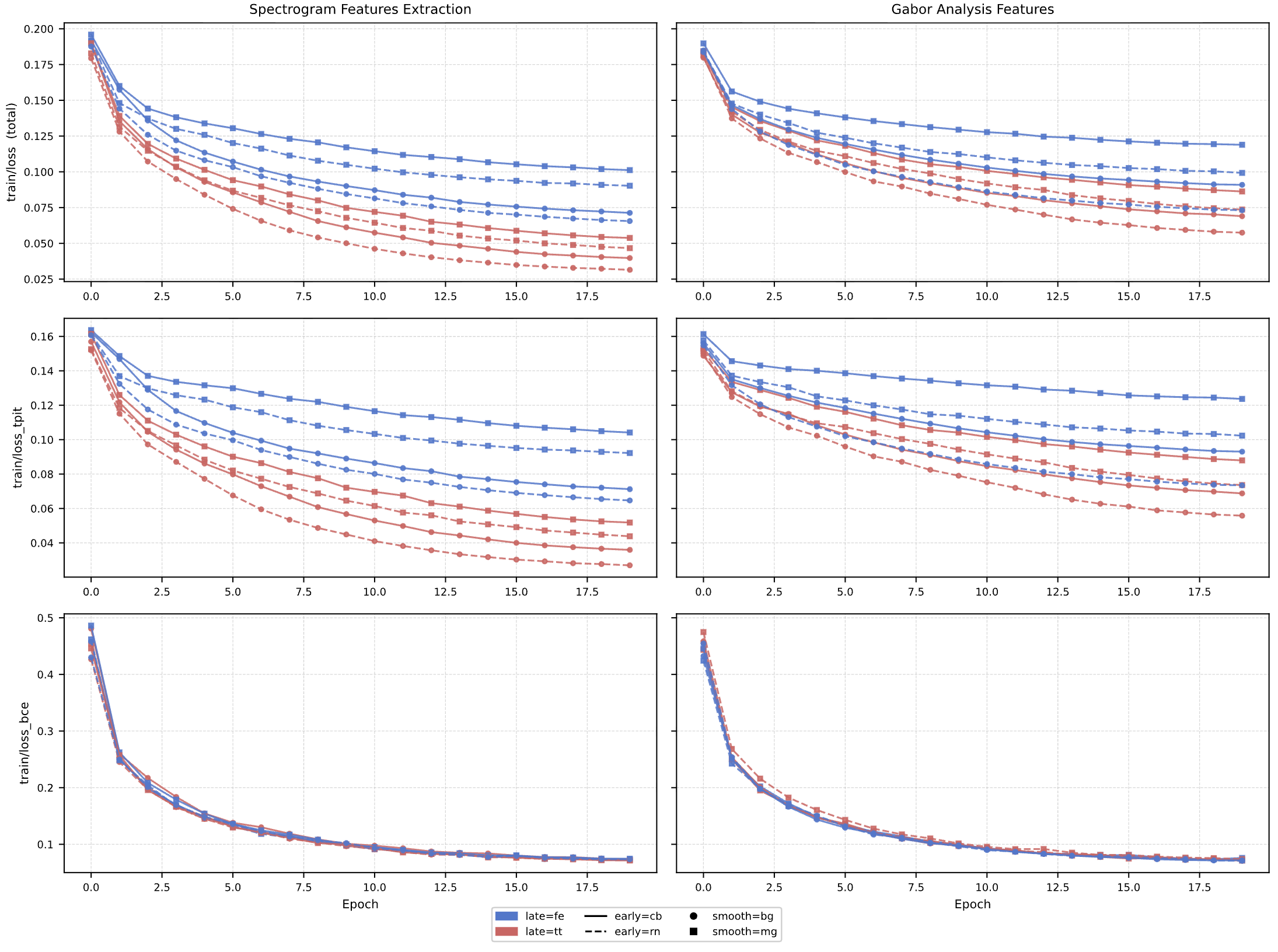}
    \caption{Stage~1 \gls{at2seld} shallow-grid results on STARSS23: training-loss trends. \texttt{loss\_bce} denotes the auxiliary semantic-presence loss, while \texttt{loss\_tpit} denotes the track-wise \gls{seld} loss.}
    \label{fig:shallow_train_loss}
\end{figure}

The optimization curves in Figures~\ref{fig:shallow_spe_score_val} and~\ref{fig:shallow_train_loss} reinforce the same interpretation. The two best spectral \texttt{tt+bg} configurations exhibit comparatively regular validation trajectories, with the best configuration reaching its optimum around epoch \(16\). Several \texttt{mg}-based configurations instead show slower optimization, weaker validation improvement, or early stagnation despite decreasing training loss. More generally, the shallow configurations that generalize best are those in which the track-wise \gls{seld} loss and the auxiliary \gls{at} presence loss decrease in a coupled manner. Configurations in which the two objectives become weakly aligned do not convert semantic evidence into stronger localization-aware predictions.

\begin{table}[ht]
\centering
\caption{Stage~1 \gls{at2seld} \gls{gpat} semantic summary.}
\label{tab:stage1_shallow_semantic_summary}
\scriptsize
\renewcommand{\arraystretch}{1.05}
\setlength{\tabcolsep}{3.5pt}
\resizebox{0.82\textwidth}{!}{%
\begin{tabular}{|p{2.35cm}|c||c|c|c|c|}
\hline
\textbf{Configuration} & \textbf{Val SELD \(\downarrow\)} & \textbf{Accuracy} & \textbf{F1} & \textbf{AUROC} & \textbf{Best epoch} \\
\hline
\texttt{spe\_rn\_tt\_bg} & \textbf{0.48} & \textbf{0.95} & \textbf{0.57} & \textbf{0.92} & 16 \\
\hline
\texttt{spe\_cb\_tt\_bg} & 0.53 & \textbf{0.95} & \textbf{0.57} & \textbf{0.92} & 18 \\
\texttt{spe\_rn\_tt\_mg} & 0.58 & \textbf{0.95} & 0.56 & 0.91 & 18 \\
\texttt{spe\_cb\_tt\_mg} & 0.64 & \textbf{0.95} & 0.55 & \textbf{0.92} & 13 \\
\texttt{spe\_rn\_fe\_bg} & 0.65 & \textbf{0.95} & 0.56 & \textbf{0.92} & 14 \\
\texttt{spe\_cb\_fe\_bg} & 0.71 & 0.94 & 0.50 & 0.87 & 8 \\
\texttt{gab\_cb\_tt\_bg} & 0.73 & \textbf{0.95} & 0.55 & \textbf{0.92} & 14 \\
\texttt{gab\_rn\_tt\_bg} & 0.74 & \textbf{0.95} & 0.52 & 0.89 & 9 \\
\texttt{spe\_cb\_fe\_mg} & 0.75 & 0.92 & 0.44 & 0.90 & 0 \\
\texttt{gab\_cb\_tt\_mg} & 0.76 & 0.94 & 0.47 & 0.87 & 1 \\
\texttt{spe\_rn\_fe\_mg} & 0.77 & 0.93 & 0.48 & \textbf{0.92} & 3 \\
\texttt{gab\_rn\_fe\_mg} & 0.79 & 0.94 & 0.49 & 0.86 & 5 \\
\texttt{gab\_cb\_fe\_bg} & 0.80 & \textbf{0.95} & \textbf{0.57} & \textbf{0.92} & 19 \\
\texttt{gab\_cb\_fe\_mg} & 0.81 & 0.92 & 0.43 & 0.86 & 2 \\
\texttt{gab\_rn\_tt\_mg} & 0.81 & 0.93 & 0.51 & 0.91 & 3 \\
\texttt{gab\_rn\_fe\_bg} & 0.82 & \textbf{0.95} & 0.56 & \textbf{0.92} & 10 \\
\hline
\end{tabular}%
}

\vspace{0.1cm}
\parbox{0.92\linewidth}{\centering\footnotesize
Accuracy, F1, and AUROC are computed on the auxiliary \gls{gpat} branch. Best epochs are 0-indexed. \textbf{Bold values} denote the best observed values for the specific metric.
}
\end{table}

The semantic summary in Table~\ref{tab:stage1_shallow_semantic_summary} shows that classification metrics are comparatively stable across many shallow configurations. Several models reach similar semantic accuracy and \gls{auroc}, but differ substantially in validation \gls{seld} scores. This is a central Stage~1 finding: preserving semantic class evidence is necessary but not sufficient. The quality of semantic-to-spatial transfer depends on how the spatial branch represents multi-channel cues, how track-wise abstraction organizes them, and how temporal smoothing stabilizes them.


\subsection{Stage 2: Controlled Depth Allocation}
\label{subsec:nas_stage2}

Stage~2 deepens the strongest shallow architecture identified in Stage~1. The objective is not to expand the full search space again, but to determine where additional capacity should be allocated once the front-end and module family have already been selected. Three depth axes are therefore isolated: the early spatial encoder, the late track-wise abstraction module, and the temporal smoothing stage. This design follows the staging rationale introduced in Section~\ref{subsec:exp_staging}: each phase should reduce uncertainty about one architectural axis before introducing additional degrees of freedom.

All Stage~2 configurations preserve the same semantic branch, target interface, evaluation procedure, and dataset split used in Stage~1. Table~\ref{tab:stage2_common_params} reports only the parameters changed with respect to the shallow search. The main modifications are the longer optimization horizon, gradient accumulation to preserve the effective batch size under larger models, and a stronger emphasis on the track-wise \gls{seld} objective relative to the auxiliary \gls{at} semantic-presence loss.

\begin{table}[ht]
\centering
\caption{Stage~2 training parameters (changed w.r.t. Stage~1). Unreported parameters remain identical to Table~\ref{tab:stage1_common_params}.}
\label{tab:stage2_common_params}
\footnotesize
\renewcommand{\arraystretch}{1.08}
\setlength{\tabcolsep}{4.2pt}
\begin{tabular}{|p{3.4cm}|p{1.7cm}|p{9cm}|}
\hline
\textbf{Parameter} & \textbf{Value} & \textbf{Rationale} \\
\hline
Max epochs & \(100\) & Extends the training horizon required by deeper spatial and temporal modules \\
\hline
Batch size & \(2\) -- \((16)\) & Physical mini-batch size of \(2\), with gradient accumulation preserving an effective batch size of \(16\) \\
\hline
Early-stopping patience & \(20\) epochs & Allows deeper configurations to converge without prematurely stopping slow but improving runs \\
\hline
\gls{tpit} \gls{seld} loss weight & \(\lambda_s=0.9\) & Increases the contribution of the spatially grounded track-wise \gls{seld} objective \\
\hline
\gls{at} \gls{bce} loss weight & \(\lambda_e=0.1\) & Reduces the auxiliary semantic-presence contribution while preserving semantic regularization \\
\hline
\gls{sed}/\gls{doa} mixing & \(\lambda_{\mathrm{SED}}=0.7\) & Assigns larger relative weight to activity estimation inside the track-wise \gls{seld} loss; the \gls{doa} term contributes with \(1-\lambda_{\mathrm{SED}}=0.3\) \\
\hline
\end{tabular}
\end{table}

\subsubsection{Depth-Performance Trade-Off}
\label{subsubsec:stage2_depth_tradeoff}

The Stage~2 search explores seven structured variants around the selected Stage~1 topology. The early spatial encoder is expanded from one to three stacked \texttt{ResNetBlock}-based layers, the late abstraction stage is expanded from one to two internal \texttt{TrackTransformer} layers, and the temporal smoothing stage is expanded from one to two internal \texttt{BiGRU} layers. Each configuration is denoted as \texttt{eX\_lY\_sZ}, where \(X\), \(Y\), and \(Z\) values indicate the depth of stages.

\begin{table}[ht]
\centering
\caption{Stage~2 \gls{at2seld} results on STARSS23.}
\label{tab:stage2_deep_spec}
\footnotesize
\renewcommand{\arraystretch}{1.05}
\setlength{\tabcolsep}{3.5pt}
\resizebox{\textwidth}{!}{%
\begin{tabular}{|p{3.15cm}|c|c|c|c|c|c|c|c|}
\hline
\textbf{Configuration} & \textbf{Early} & \textbf{Late} & \textbf{Smooth} & \textbf{Val SELD \(\downarrow\)} & \textbf{Duration} & \textbf{Latency} & \textbf{Params} & \textbf{GFLOPs} \\
\hline
\rowcolor{gray!15}
\texttt{e3\_l1\_s1} & 3 & 1 & 1 & 0.357 & 18.59 & 4006.1 & 30.00 & 755.82 \\
\hline
\rowcolor{gray!8}
\texttt{e2\_l1\_s1} & 2 & 1 & 1 & 0.360 & 11.08 & 571.3 & 29.00 & 200.10 \\
\texttt{e2\_l2\_s1} & 2 & 2 & 1 & 0.362 & 11.23 & 581.7 & 29.40 & 200.70 \\
\texttt{e3\_l2\_s1} & 3 & 2 & 1 & 0.366 & 17.71 & 3782.0 & 30.40 & 756.41 \\
\texttt{e1\_l2\_s1} & 1 & 2 & 1 & 0.374 & 8.16 & 279.1 & 29.14 & 59.74 \\
\texttt{e1\_l1\_s2} & 1 & 1 & 2 & 0.380 & 9.01 & 319.4 & 29.53 & 60.93 \\
\texttt{e1\_l2\_s2} & 1 & 2 & 2 & 0.391 & 8.68 & 335.0 & 29.93 & 61.52 \\
\hline
\end{tabular}%
}

\vspace{0.1cm}
\parbox{0.95\linewidth}{\centering\footnotesize
Duration = training duration in hours; Latency = single-sample inference latency in ms; Params = trainable parameters in millions; GFLOPs = estimated single-forward arithmetic complexity.
}
\end{table}

Table~\ref{tab:stage2_deep_spec} shows that controlled deepening substantially improves the validation \gls{seld} score with respect to the selected shallow configuration. The best Stage~2 model, \texttt{at2seld\_e3\_l1\_s1}, reaches \(0.357\), compared with \(0.48\) for the Stage~1 reference \texttt{spe\_rn\_tt\_bg}. The improvement corresponds to an absolute reduction of approximately \(0.12\) \gls{seld} points, indicating that the shallow architecture was not yet capacity-saturated.

The dominant effect is associated with the early spatial encoder. Increasing the number of early \texttt{ResNetBlock} layers from one to two produces most of the observed gain. The two-layer early configuration \texttt{e2\_l1\_s1} reaches \(0.360\), nearly matching the best three-layer configuration. A further increase to three early layers gives the best score, but the marginal validation improvement is small relative to the computational cost. This indicates that the early spatial encoder is the main representational bottleneck in this regime: refining low-level and mid-level \gls{foa} time--frequency cues before track-wise abstraction is more beneficial than increasing later processing stages.

By contrast, deepening the late \texttt{TrackTransformer} does not provide a systematic advantage. For the two-layer early encoder, \texttt{e2\_l1\_s1} and \texttt{e2\_l2\_s1} obtain nearly identical validation scores, with the single-layer late module slightly ahead. The same trend appears in the three-layer early setting, where \texttt{e3\_l1\_s1} outperforms \texttt{e3\_l2\_s1}. Once the early spatial representation is sufficiently structured, the single \texttt{TrackTransformer} layer appears adequate for organizing track-wise latent information under \gls{tpit} supervision. Additional late attentive depth mainly increases computational cost without a proportional gain.

The smoothing stage follows the same pattern. Increasing the \texttt{BiGRU} depth from one to two layers does not improve the validation \gls{seld} score. The comparison between \texttt{e1\_l2\_s1} and \texttt{at2seld\_e1\_l2\_s2} is particularly informative because the early and late stages are held fixed while only the recurrent smoother is deepened. The deeper smoother is consistently worse, suggesting that temporal refinement saturates quickly once the upstream spatial representation has already been organized.

\begin{figure}[ht]
    \centering
    \includegraphics[width=\linewidth]{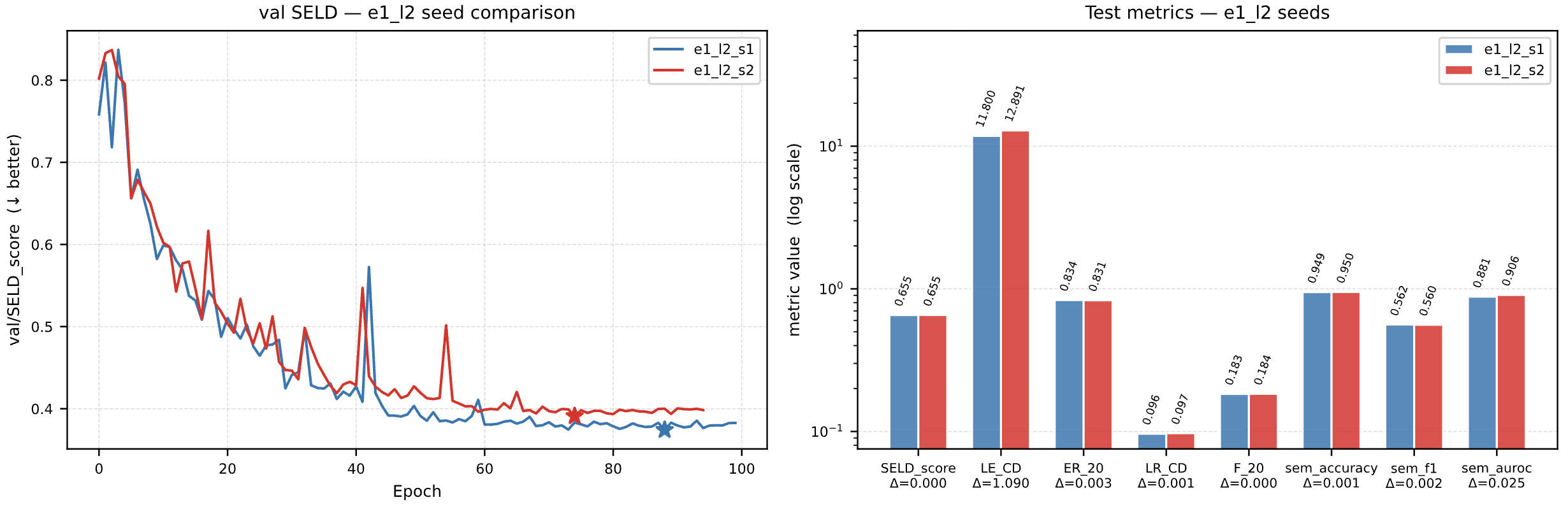}
    \caption{Stage~2 \gls{at2seld} validation \gls{seld} trends for the \texttt{BiGRU} deepening analysis.}
    \label{fig:deep_bigru_smoothing_seed}
\end{figure}

Figure~\ref{fig:deep_bigru_smoothing_seed} confirms this interpretation at the optimization-trajectory level. The two-layer \texttt{BiGRU} does not produce a delayed improvement after longer training; instead, it remains above the shallower smoother for most of the trajectory. This suggests that the limitation is not insufficient training time, but the limited utility of additional recurrent depth in the selected topology. In practical terms, temporal smoothing should be kept compact unless the upstream representation or the input-context regime is changed.

The resulting Stage~2 trade-off is therefore not between shallow and deep models in general, but between where depth is allocated. The best predictive configuration is \texttt{e3\_l1\_s1}, but the two-layer early alternative \texttt{e2\_l1\_s1} is nearly \textit{Pareto-equivalent} with a much lower computational burden. Its validation \gls{seld} score differs by only \(0.003\), while latency decreases from approximately \(4.0\)s to \(0.57\)s and arithmetic complexity decreases from approximately \(756\) to \(200\) GFLOPs. Stage~2 therefore identifies the three-layer early encoder as the strongest validation configuration and the two-layer early encoder as the most favorable accuracy--efficiency compromise.

\subsubsection{Optimization Dynamics}
\label{subsubsec:stage2_optimization_dynamics}

\begin{figure}[ht]
    \centering

    \begin{minipage}[t]{0.55\textwidth}
        \vspace{-11.4cm}
        {The loss-component trajectories (Figures~\ref{fig:deep_seld_val}--\ref{fig:deep_losses}) are more diagnostic than the aggregate score alone. The validation activity loss evolves in a comparatively narrow range but tends to worsen after an initial decrease, whereas the localization component follows a smoother and more regular minimization trend. This means that the improvement of the validation \gls{seld} score is not driven primarily by better semantic discrimination or more stable activity estimation. Instead, the spatial regression continues to improve and compensates for the fragility of the activity branch.}
        
        \vspace{0.64cm}
        
        \includegraphics[width=\linewidth]{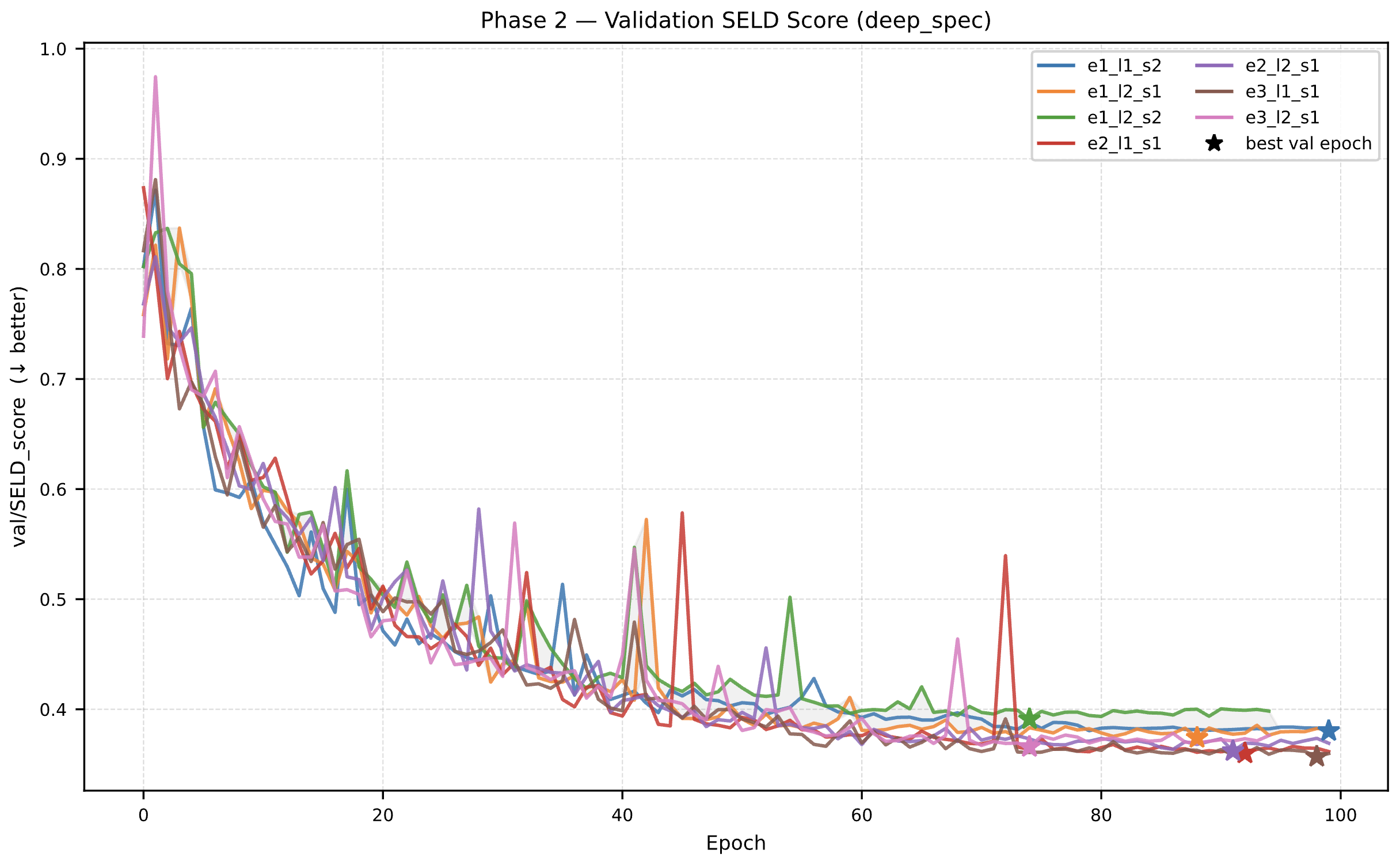}
        \caption{Stage~2 \gls{at2seld} validation \gls{seld} score trends on STARSS23.}
        \label{fig:deep_seld_val}
    \end{minipage}
    \begin{minipage}[t]{0.44\textwidth}
        \centering
        \includegraphics[width=\linewidth]{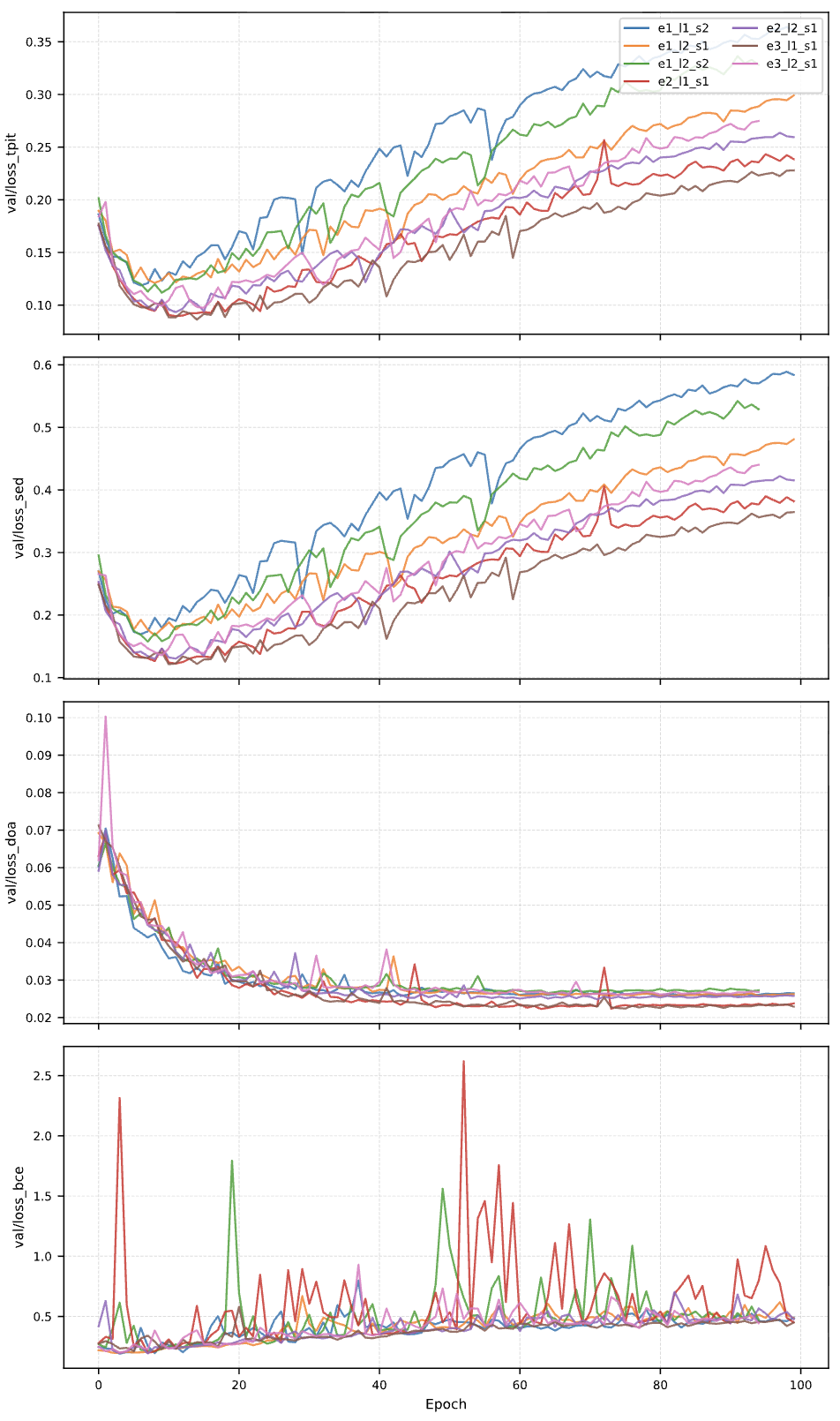}
        \caption{Stage~2 \gls{at2seld} validation loss-component trends on STARSS23.}
        \label{fig:deep_losses}
    \end{minipage}
\end{figure}

The validation trajectories provide additional evidence that the deep regime is more optimization-demanding than the shallow one. As shown in Figure~\ref{fig:deep_seld_val}, the strongest configurations do not reach their best validation \gls{seld} score in the early epochs. Instead, the best models continue improving over most of the \(100\)-epoch horizon, with the top configurations reaching their best checkpoints late in training. This justifies the Stage~2 increase in maximum epochs and patience reported in Table~\ref{tab:stage2_common_params}.

This asymmetry is important because Stage~2 models use a track-wise target interface derived from Multi-\gls{accdoa}-style tensors, but optimize decoupled activity and localization heads, as formalized in Section~\ref{subsec:exp_supervision_objective}. Under this formulation, activity and localization are expected to cooperate, but they can still exhibit different optimization dynamics. Stage~2 makes this separation visible: deeper spatial processing improves the \gls{doa}-related component more reliably than the activity component, indicating that the semantic-to-spatial coupling remains weak when the branches interact only through the joint loss.

\begin{table}[ht]
\centering
\caption{Stage~2 \gls{at2seld} semantic-presence summary.}
\label{tab:stage2_deep_semantic_summary}
\footnotesize
\renewcommand{\arraystretch}{1.05}
\setlength{\tabcolsep}{3.5pt}
\resizebox{0.80\textwidth}{!}{%
\begin{tabular}{|p{3.15cm}|c||c|c|c|c|}
\hline
\textbf{Configuration} & \textbf{Val SELD \(\downarrow\)} & \textbf{Accuracy} & \textbf{F1} & \textbf{AUROC} & \textbf{Best epoch} \\
\hline
\texttt{e3\_l1\_s1} & 0.357 & 0.95 & 0.56 & 0.91 & 98 \\
\texttt{e2\_l1\_s1} & 0.360 & 0.95 & 0.56 & 0.87 & 92 \\
\hline
\texttt{e2\_l2\_s1} & 0.362 & 0.95 & 0.53 & 0.84 & 91 \\
\texttt{e3\_l2\_s1} & 0.366 & 0.95 & 0.54 & 0.87 & 74 \\
\texttt{e1\_l2\_s1} & 0.374 & 0.95 & 0.56 & 0.88 & 88 \\
\texttt{e1\_l1\_s2} & 0.380 & 0.95 & 0.57 & 0.91 & 99 \\
\texttt{e1\_l2\_s2} & 0.391 & 0.95 & 0.56 & 0.91 & 74 \\
\hline
\end{tabular}%
}

\vspace{0.1cm}
\parbox{0.92\linewidth}{\centering\footnotesize
Accuracy, F1, and AUROC are computed on the auxiliary semantic-presence branch. Best epochs are 0-indexed.
}
\end{table}

Table~\ref{tab:stage2_deep_semantic_summary} confirms that deeper spatial processing does not substantially change the semantic-presence branch. Accuracy remains around \(0.95\), F1 remains in the \(0.53\)--\(0.57\) range, and \gls{auroc} does not exhibit a systematic improvement with depth. This stability is not a negative result by itself: it shows that the pretrained \gls{gpat} branch is not degraded by deeper spatial training. However, it also shows that Stage~2 improvements are not caused by stronger auxiliary semantic recognition. The gains arise mainly from improved spatial representation and localization behavior.

This observation provides the direct motivation for Stage~3. Up to this point, the semantic and spatial streams are optimized jointly but do not exchange information internally. Increasing the weight of the auxiliary semantic-presence loss is not a principled solution, because preliminary runs indicate that excessive semantic weighting degrades the \gls{seld} branch, which remains the primary objective. The next stage therefore introduces explicit feature interaction through controlled cross-branch bridges, testing whether feature-level exchange can improve the selected topology without collapsing the two streams into a single undifferentiated representation.


\subsection{Stage 3: Regularization and Semantic--Spatial Interaction}
\label{subsec:nas_stage3}

Stage~3 addresses two limitations that remain visible after controlled depth allocation: \emph{(I)} the strongest Stage~2 configurations improve \gls{seld} performance mainly through spatial representation, while the activity branch remains comparatively fragile; \emph{(II)} semantic and spatial streams are still coupled only through the shared training objective, despite the methodological role assigned to semantic-to-spatial transfer in Section~\ref{subsec:exp_objectives}. Stage~3 therefore evaluates whether stronger regularization and explicit cross-branch interaction can improve the selected architecture without changing the input representation, the target interface, or the track-wise supervision protocol defined in Sections~\ref{subsec:exp_supervision_objective} and~\ref{subsec:exp_datasets_pipeline}.

The search is centered on the Stage~2 topology with a three-layer residual early spatial encoder, one \texttt{TrackTransformer} late abstraction module, and one \texttt{BiGRU} temporal smoother. This architecture corresponds to the strongest internal validation configuration identified in Section~\ref{subsec:nas_stage2}. Stage~3 keeps this backbone fixed and varies only the regularized interaction pattern: \textit{no bridge}, \textit{late bridge only}, \textit{early bridge only}, and \textit{simultaneous early--late bridges}. This isolates the effect of semantic--spatial interaction placement while preserving the spatial architecture and training protocol.

\subsubsection{Interaction Configurations and Computational Cost}
\label{subsubsec:stage3_config_cost}

Regularization is introduced directly into the selected topology. In the early spatial encoder, each \texttt{ResNetBlock} preserves its residual structure but adds spatial dropout after the final residual addition and activation.

\vspace{0.5cm}

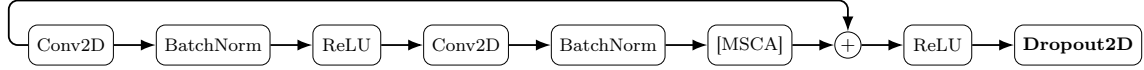
\begin{figure}[ht]
\centering
\resizebox{\textwidth}{!}{%
\begin{tikzpicture}[
    node distance=0.65cm,
    block/.style={
        draw,
        rounded corners,
        align=center,
        minimum height=0.62cm,
        minimum width=1.05cm,
        font=\scriptsize
    },
    plus/.style={
        draw,
        circle,
        inner sep=1.0pt,
        font=\scriptsize
    },
    arrow/.style={-Latex, thick},
    skip/.style={-Latex, thick, rounded corners}
]

\node[block] (conv1) {Conv2D};
\node[block, right=of conv1] (bn1) {BatchNorm};
\node[block, right=of bn1] (relu1) {ReLU};
\node[block, right=of relu1] (conv2) {Conv2D};
\node[block, right=of conv2] (bn2) {BatchNorm};
\node[block, right=of bn2] (msca) {[MSCA]};
\node[plus, right=of msca] (add) {$+$};
\node[block, right=of add] (relu2) {ReLU};
\node[block, right=of relu2] (drop) {\textbf{Dropout2D}};

\draw[arrow] (conv1) -- (bn1);
\draw[arrow] (bn1) -- (relu1);
\draw[arrow] (relu1) -- (conv2);
\draw[arrow] (conv2) -- (bn2);
\draw[arrow] (bn2) -- (msca);
\draw[arrow] (msca) -- (add);
\draw[arrow] (add) -- (relu2);
\draw[arrow] (relu2) -- (drop);

\draw[skip]
    (conv1.west) -- ++(-0.30,0)
    |- ([yshift=0.48cm]add.north)
    -- (add.north);

\end{tikzpicture}%
}
\caption{Regularized \texttt{ResNetBlock} used in Stage~3. The skip path represents the residual shortcut added before the final activation and spatial dropout.}
\label{fig:stage3_regularized_resblock}
\end{figure}
The dropout probability is set to \(0.4\), so that regularization acts on entire time--frequency feature maps rather than on independent scalar activations. In the late abstraction stage, dropout with probability \(0.3\) is applied inside the \texttt{TrackTransformer} encoder layer, affecting both the self-attention and feed-forward sub-blocks. The smoothing stage remains a single-layer \texttt{BiGRU}; since standard recurrent dropout is inactive for a single recurrent layer, dropout is applied explicitly to the smoother output after combining the forward and backward streams.

Explicit semantic--spatial interaction is implemented through the custom \texttt{CrossStitch} bridge described in Section~\ref{subsec:exp_staging}. Two insertion points are evaluated. The early bridge is placed after the early spatial encoder and after the early semantic group of the pretrained \gls{gpat} branch. The late bridge is placed after the spatial late abstraction stage and after the semantic late group, before temporal smoothing and before the auxiliary semantic-presence head. The early bridge therefore operates on high-resolution spatial feature maps with bridge dimensionality \(C_b=64\), whereas the late bridge operates after spatial abstraction with bridge dimensionality \(C_b=128\).

\begin{table}[ht]
\centering
\caption{Stage~3 \gls{at2seld} regularization and cross-branch interaction configurations.}
\label{tab:stage3_reg_configs}
\footnotesize
\renewcommand{\arraystretch}{1.05}
\setlength{\tabcolsep}{3.8pt}
\resizebox{\textwidth}{!}{%
\begin{tabular}{|p{2.4cm}|c|c|p{6.9cm}|c|c|c|}
\hline
\textbf{Configuration} & \textbf{Early CS} & \textbf{Late CS} & \textbf{Description} & \textbf{Params} & \textbf{GFLOPs} & \textbf{Latency} \\
\hline
\texttt{e3\_t1\_cs00} & No & No &
Regularized baseline: \texttt{ResNetBlock}$\times3$ with dropout \(0.4\), \texttt{TrackTransformer}$\times1$ with dropout \(0.3\), and \texttt{BiGRU}$\times1$ with output dropout \(0.3\). &
28.92 & 755.82 & 3720.9 \\
\hline
\texttt{e3\_t1\_cs01} & No & \textbf{Yes} &
Late cross-stitching only, inserted after the spatial \texttt{TrackTransformer} and after the semantic late group. &
29.38 & 756.43 & 3755.5 \\
\hline
\texttt{e3\_t1\_cs10} & \textbf{Yes} & No &
Early cross-stitching only, inserted after the \texttt{ResNetBlock}$\times3$ encoder and after the first semantic convolutional group. &
28.98 & 766.59 & 4069.4 \\
\hline
\texttt{e3\_t1\_cs11} & \textbf{Yes} & \textbf{Yes} &
Both early and late cross-stitch bridges enabled. &
29.44 & 767.20 & 4889.0 \\
\hline
\end{tabular}%
}

\vspace{0.1cm}
\parbox{0.95\linewidth}{\centering\footnotesize
Params = trainable parameters in millions; GFLOPs = estimated single-forward arithmetic complexity; Latency = single-sample inference latency in ms.
}
\end{table}

Table~\ref{tab:stage3_reg_configs} shows that the computational cost of interaction depends mainly on the insertion point. The late bridge is comparatively inexpensive: enabling late-only cross-stitching increases the regularized baseline by approximately \(0.46\)M trainable parameters, \(0.61\) GFLOPs, and \(35\) ms of single-sample latency. This limited overhead is consistent with its placement after spatial abstraction, where the feature representation has already been compressed into a track-wise latent structure. By contrast, the early bridge adds only \(0.06\)M parameters but increases the arithmetic cost by more than \(10\) GFLOPs and raises latency by approximately \(350\) ms, because it operates before frequency-axis compression on high-resolution spatial maps. The early--late configuration combines both overheads and reaches the highest latency. Early semantic--spatial interaction is therefore both computationally less favorable and representationally more invasive.

\subsubsection{\glsentryshort{seld} Effects of Cross-Branch Interaction}
\label{subsubsec:stage3_seld_effects}

Table~\ref{tab:stage3_results} reports Stage~3 validation and test results. Model selection remains validation-driven: test metrics are computed only for the checkpoint selected by the best validation \gls{seld} score.

\begin{table}[ht]
\centering
\caption{Stage~3 \gls{at2seld} results on STARSS23.}
\label{tab:stage3_results}
\footnotesize
\renewcommand{\arraystretch}{1.05}
\setlength{\tabcolsep}{3.5pt}
\resizebox{\textwidth}{!}{%
\begin{tabular}{|p{2.85cm}|c|c|c|c|c|c|c|c|}
\hline
\textbf{Configuration} &
\textbf{Val SELD \(\downarrow\)} &
\textbf{Test SELD \(\downarrow\)} &
\textbf{LE\(_{\mathrm{CD}}\) \(\downarrow\)} &
\textbf{ER\(_{20^\circ}\) \(\downarrow\)} &
\textbf{LR\(_{\mathrm{CD}}\) \(\uparrow\)} &
\textbf{F\(_{20^\circ}\) \(\uparrow\)} &
\textbf{Duration} &
\textbf{Best epoch} \\
\hline
\rowcolor{gray!15}
\texttt{e3\_t1\_cs01} &
\cellcolor{green!35}0.385 &
\cellcolor{green!30}0.624 &
\cellcolor{green!30}12.49 &
\cellcolor{green!35}0.796 &
\cellcolor{green!35}0.142 &
\cellcolor{green!40}0.226 &
18.82 & 86 \\
\hline
\texttt{e3\_t1\_cs00} &
0.592 &
0.708 &
14.55 &
0.901 &
0.050 &
0.102 &
18.96 & 81 \\
\texttt{e3\_t1\_cs10} &
\cellcolor{red!12}0.608 &
\cellcolor{red!8}0.715 &
\cellcolor{green!22}13.15 &
\cellcolor{red!25}0.935 &
\cellcolor{gray!6}0.050 &
\cellcolor{red!8}0.098 &
23.63 & 90 \\
\texttt{e3\_t1\_cs11} &
\cellcolor{red!16}0.614 &
\cellcolor{green!5}0.705 &
\cellcolor{green!22}13.18 &
\cellcolor{red!8}0.905 &
\cellcolor{green!10}0.058 &
\cellcolor{red!6}0.099 &
23.65 & 94 \\
\hline
\end{tabular}%
}

\vspace{0.1cm}
\parbox{0.95\linewidth}{\centering\footnotesize
Cell colors are assigned relative to the regularized no-stitch baseline \texttt{e3\_t1\_cs00}: \textcolor{green}{green} denotes improvement, \textcolor{red}{red} denotes degradation, and \textcolor{gray}{gray} denotes negligible variation. \textbf{Val~SELD} is the best validation score achieved during training. Test metrics are computed on the test split using the checkpoint selected by validation \gls{seld}. Duration is reported in hours. Best epochs are 0-indexed.
}
\end{table}

The regularized no-stitch baseline, \texttt{e3\_t1\_cs00}, degrades substantially with respect to the unregularized Stage~2 endpoint. Its validation \gls{seld} score increases to \(0.592\), and its test \gls{seld} score is \(0.708\). This behavior is consistent with the purpose of Stage~3: the adopted dropout regime is intentionally strong, so that the experiment tests whether explicit feature exchange can compensate for regularization-induced loss of spatial discriminative capacity. Without such compensation, the regularized architecture remains optimizable but loses part of the representational strength acquired in Stage~2.

\begin{figure}[ht]
    \centering
    \includegraphics[width=\linewidth]{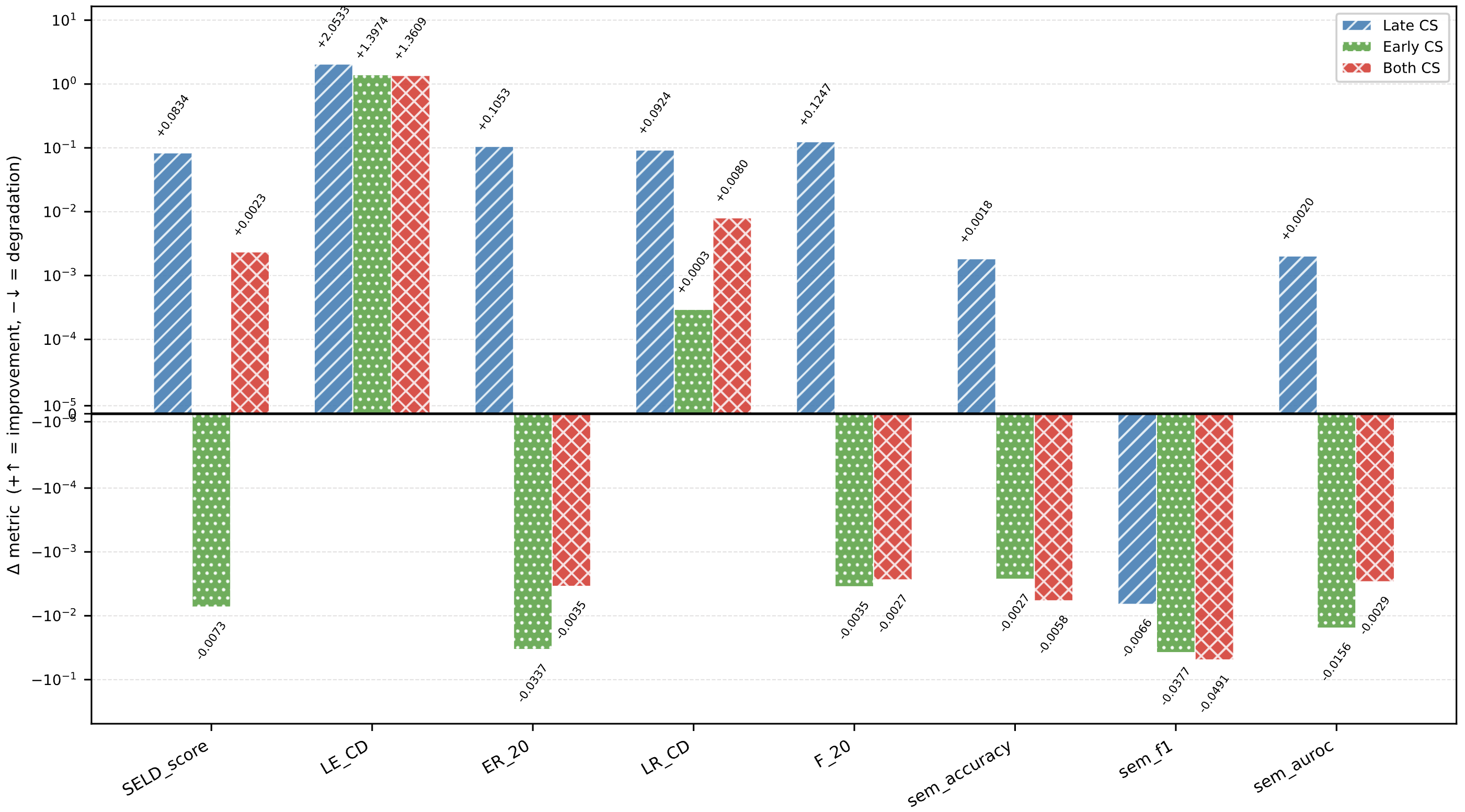}
    \caption{Stage~3 metric deltas with respect to the regularized no-stitch baseline. Values are reported as absolute score differences.}
    \label{fig:stage3_delta_bar}
\end{figure}

The main result is that late cross-stitching compensates for this degradation. The configuration \texttt{e3\_t1\_cs01} obtains the best Stage~3 validation score (\(0.385\)), and the best test score (\(0.624\)). Relative to the regularized baseline, the test \gls{seld} score improves by \(0.084\) absolute points. The improvement is also visible in the component metrics: \(\mathrm{LE}_{\mathrm{CD}}\) decreases from \(14.55^\circ\) to \(12.49^\circ\), \(\mathrm{ER}_{20^\circ}\) decreases from \(0.901\) to \(0.796\), \(\mathrm{LR}_{\mathrm{CD}}\) increases from \(0.050\) to \(0.142\), and \(\mathrm{F}_{20^\circ}\) increases from \(0.102\) to \(0.226\). These changes indicate that late interaction improves both localization quality and location-aware detection, 
\begin{wrapfigure}{l}{0.6\textwidth}
    \centering
    \includegraphics[width=\linewidth]{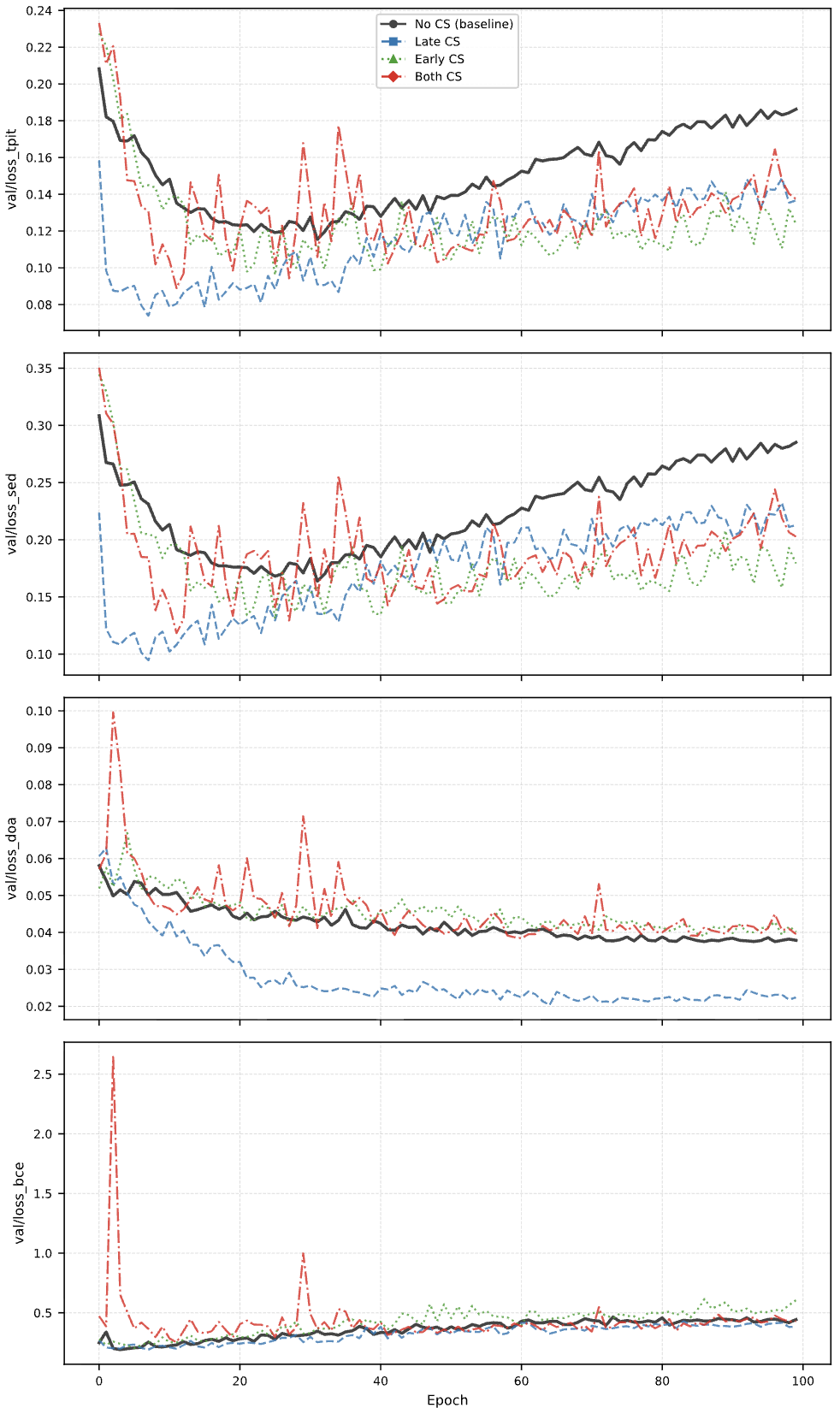}
    \caption{Stage~3 validation loss-component curves.}
    \label{fig:stage3_val_curves}
\end{wrapfigure}
rather than only shifting one component of the \gls{seld} score.

Figure~\ref{fig:stage3_delta_bar} makes the component-wise effect explicit. Late cross-stitching is the only configuration that consistently improves the detection-oriented and localization-oriented quantities at the same time. Early cross-stitching produces a lower \(\mathrm{LE}_{\mathrm{CD}}\) than the no-stitch baseline, but this isolated angular improvement is not supported by the detection-dependent metrics: \(\mathrm{ER}_{20^\circ}\) worsens, \(\mathrm{F}_{20^\circ}\) slightly decreases, and the aggregate validation and test \gls{seld} scores remain worse than the regularized baseline. Enabling both early and late bridges also fails to recover the late-only behavior.

\begin{figure}[ht]
    \centering
    \includegraphics[width=0.9\linewidth]{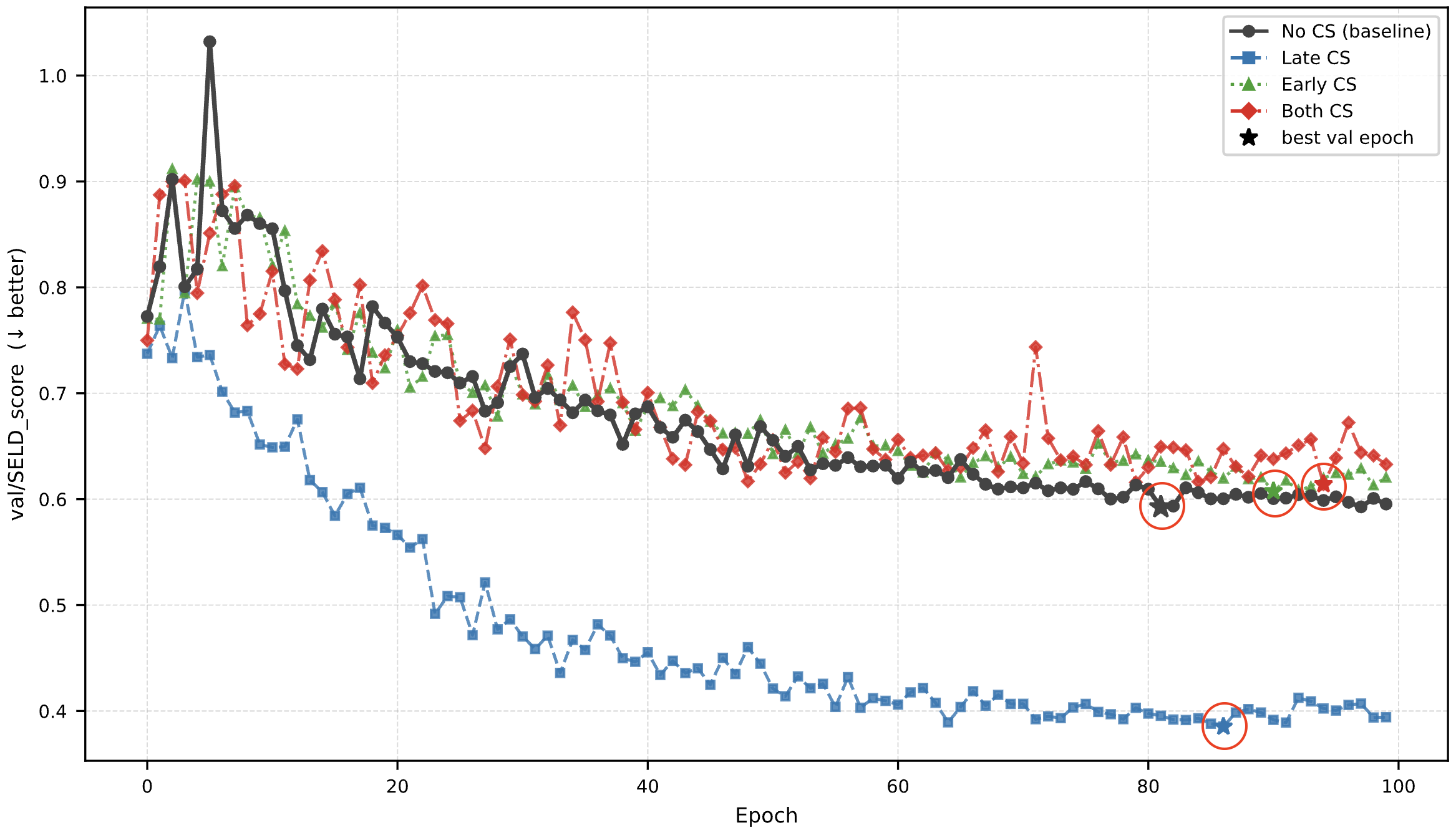}
    \caption{Stage~3 validation \gls{seld} score curves. \textcolor{red}{Red circles} denote best-score checkpoints.}
    \label{fig:stage3_val_seld_scores}
\end{figure}

The validation curves (Figure~\ref{fig:stage3_val_seld_scores}) reinforce this interpretation: the regularized baseline remains clearly separated from the late-stitch configuration, although it continues improving for a long portion of the training horizon and reaches its best checkpoint only at epoch \(81\). Late cross-stitching follows a substantially lower trajectory, reaches its best checkpoint at epoch \(86\), and remains the most competitive configuration throughout the second half of training. Early cross-stitching, either alone or combined with the late bridge, yields smoother curves, but smoother validation trajectories do not translate into better \gls{seld} performance.

The loss-component trajectories clarify why late interaction is preferable (Figure~\ref{fig:stage3_val_curves}). Relative to the Stage~2 behavior (Figure~\ref{fig:deep_losses}), Stage~3 reduces validation-loss variability, indicating that the dropout strategy is not ineffective as regularization. However, the no-stitch and early-stitch configurations fail to convert this smoother loss behavior into stronger \gls{seld} scores. The late-only bridge achieves the most favorable balance: it keeps the track-wise \gls{seld} and \gls{doa} components comparatively low while avoiding the detection degradation observed in the other configurations.

This result supports a position-dependent interpretation of semantic--spatial interaction. The late bridge acts after the spatial stream has already performed time--frequency organization and track-wise abstraction, so semantic evidence can be injected as high-level conditioning. The early bridge instead acts while the spatial stream is still processing localization-sensitive \gls{foa} feature maps. At this point, semantic features are not yet aligned with the spatial representation in a way that supports localization; they perturb low-level spatial segregation rather than regularizing it. Stage~3 therefore does not support a generic \textit{“more interaction is better”} conclusion. It supports a more specific insight: semantic--spatial interaction is beneficial only after sufficient spatial abstraction is obtained.

\subsubsection{Semantic and Per-Class Effects}
\label{subsubsec:stage3_semantic_effects}

The semantic-presence branch provides an additional view of the Stage~3 behavior. Table~\ref{tab:stage3_semantic_summary} compares the Stage~3 configurations with the best Stage~2 model. The goal is to verify whether \gls{seld} differences are accompanied by changes in semantic preservation, or whether they originate mainly from the spatial pathway.

\begin{table}[ht]
\centering
\caption{Stage~3 \gls{at2seld} semantic-presence summary.}
\label{tab:stage3_semantic_summary}
\footnotesize
\renewcommand{\arraystretch}{1.05}
\setlength{\tabcolsep}{3.8pt}
\resizebox{0.84\textwidth}{!}{%
\begin{tabular}{|p{2.85cm}|c||c|c|c|c|}
\hline
\textbf{Configuration} & \textbf{Val SELD \(\downarrow\)} & \textbf{Accuracy} & \textbf{F1} & \textbf{AUROC} & \textbf{Best epoch} \\
\hline
\texttt{e3\_l1\_s1} & 0.357 & 0.953 & 0.560 & 0.910 & 98 \\
\hline
\rowcolor{gray!15}
\texttt{e3\_t1\_cs01} & 0.385 & \cellcolor{green!16}0.955 & \cellcolor{green!8}0.562 & \cellcolor{green!12}0.912 & 86 \\
\texttt{e3\_t1\_cs00} & 0.592 & \cellcolor{gray!8}0.953 & \cellcolor{green!18}0.569 & \cellcolor{gray!8}0.910 & 81 \\
\texttt{e3\_t1\_cs10} & 0.608 & \cellcolor{red!10}0.951 & \cellcolor{red!18}0.531 & \cellcolor{red!25}0.894 & 90 \\
\texttt{e3\_t1\_cs11} & 0.614 & \cellcolor{red!18}0.948 & \cellcolor{red!28}0.520 & \cellcolor{red!8}0.907 & 94 \\
\hline
\end{tabular}%
}

\vspace{0.1cm}
\parbox{0.92\linewidth}{\centering\footnotesize
Accuracy, F1, and AUROC are computed on the auxiliary semantic-presence branch. Semantic cells are color-coded against the best Stage~2 configuration: \textcolor{green}{green} = improvement, \textcolor{red}{red} = degradation, \textcolor{gray}{gray} = reference or negligible variation. Best epochs are 0-indexed.
}
\end{table}

Semantic accuracy remains globally high across all configurations, but the interaction strategy is no longer neutral. The regularized no-stitch baseline preserves almost the same accuracy and \gls{auroc} as the Stage~2 reference and slightly improves F1. This shows that the Stage~3 degradation of \gls{seld} performance is not caused by collapse of the semantic-presence branch. The failure is mainly associated with the regularized spatial pathway. Late cross-stitching is the only interaction strategy that improves both the aggregate \gls{seld} behavior and the semantic-presence indicators, increasing accuracy from \(0.953\) to \(0.955\) and \gls{auroc} from \(0.910\) to \(0.912\) relative to the Stage~2 reference.

By contrast, early cross-stitching degrades the semantic branch as well as the spatially grounded output. The early-only configuration reduces F1 and \gls{auroc}, while the early--late configuration further degrades F1. This confirms that the interaction point is critical: injecting semantic information before the spatial stream has completed time--frequency organization perturbs not only localization-aware prediction, but also the stability of the transferred semantic representation. The late bridge provides controlled high-level coupling; the early bridge introduces premature coupling.

\begin{figure}[ht]
    \centering
    \includegraphics[width=0.95\linewidth]{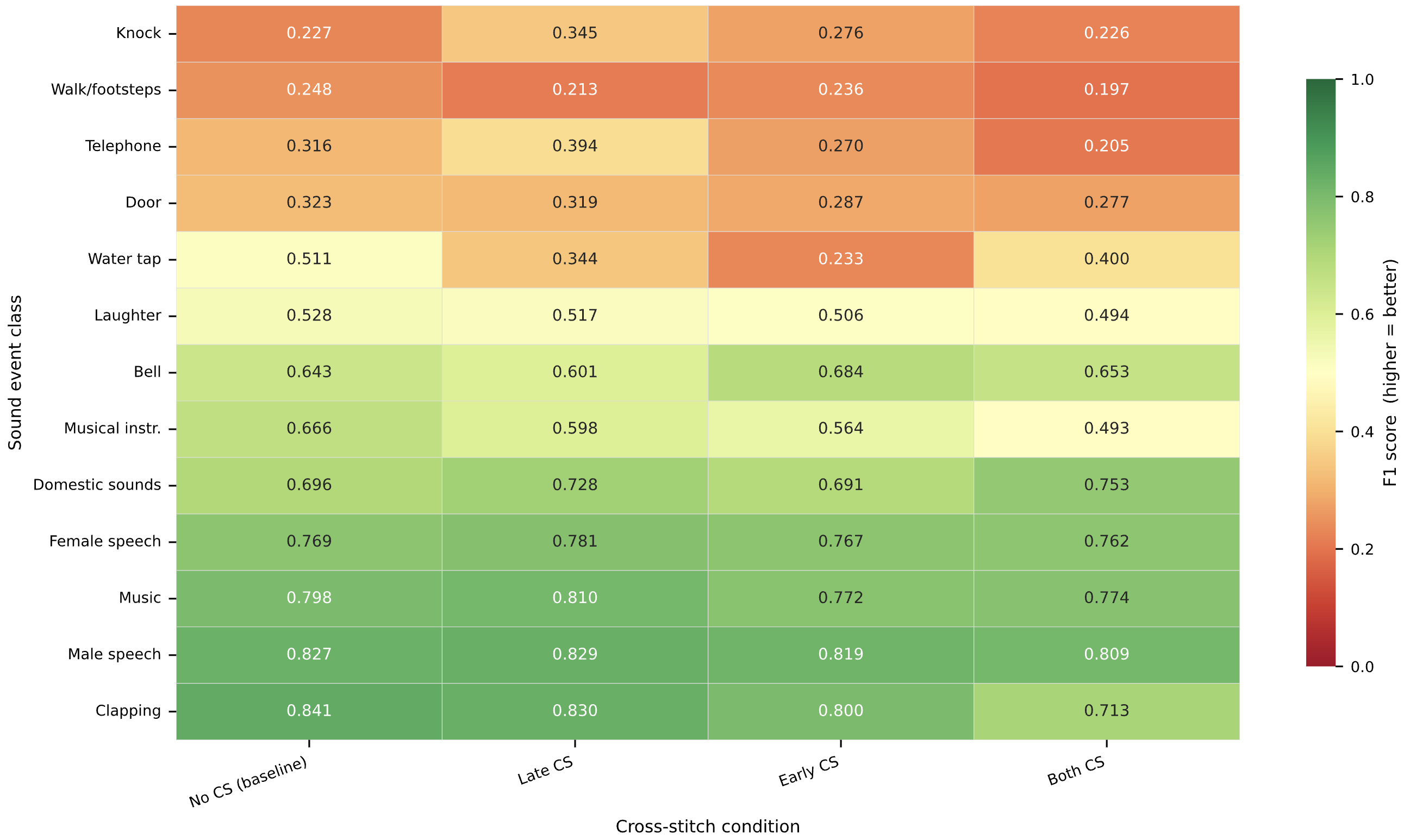}
    \caption{Stage~3 auxiliary semantic-presence per-class F1 scores.}
    \label{fig:stage3_per_class_f1}
\end{figure}

The per-class F1 heatmap, derived from \gls{sed} detections aggregation (Figure~\ref{fig:stage3_per_class_f1}), shows that the effect of semantic--spatial interaction is class-dependent. Relative to the regularized no-stitch baseline, late cross-stitching improves several sparse or difficult classes, most notably \texttt{Knock} \((0.227 \rightarrow 0.345)\) and \texttt{Telephone} \((0.316 \rightarrow 0.394)\). It also produces smaller gains for already robust categories such as \texttt{Female speech}, \texttt{Music}, and \texttt{Male speech}. However, the effect is not uniform: \texttt{Water tap}, \texttt{Walk/footsteps}, and \texttt{Musical instrument} decrease under late cross-stitching. The global benefit of late interaction therefore coexists with class-specific trade-offs.

\begin{figure}[ht]
    \centering
    \includegraphics[width=\linewidth]{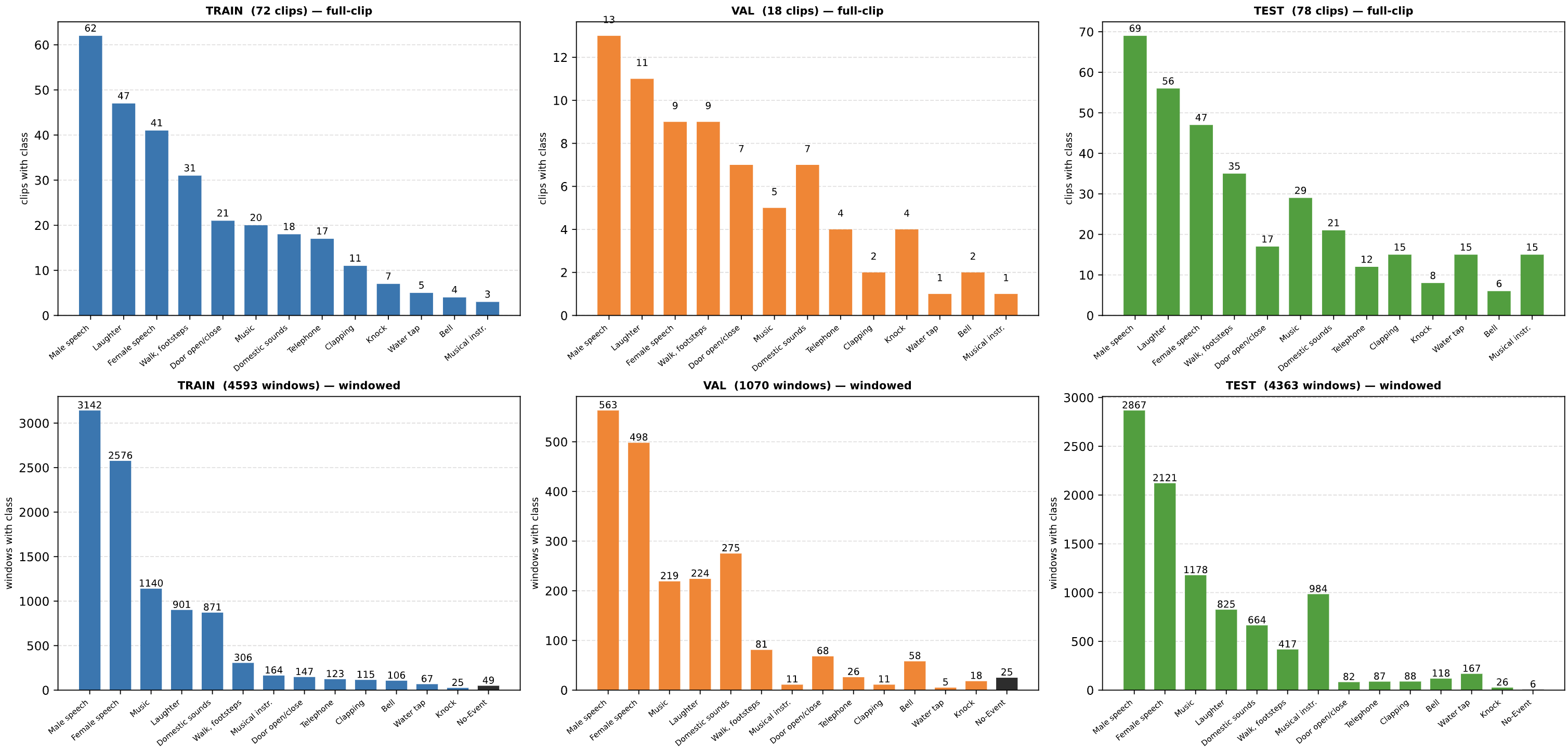}
    \caption{STARSS23 split distributions used in Stages~1--3.}
    \label{fig:starss23_distributions}
\end{figure}

These class-level effects must be interpreted together with the STARSS23 data distribution (Figure~\ref{fig:starss23_distributions}). Several classes with low or unstable F1 are poorly represented in both the native clip distribution and the \(10\)-s windowed training regime, including \texttt{Knock}, \texttt{Water tap}, \texttt{Bell}, \texttt{Clapping}, and \texttt{Musical instrument}. By contrast, dominant classes such as \texttt{Male speech}, \texttt{Female speech}, \texttt{Laughter}, \texttt{Music}, and \texttt{Domestic sounds} provide substantially more supervision. The per-class semantic behavior is therefore not only an architectural effect; it also reflects the long-tailed real-scene distribution inherited from STARSS23.

The exceptions are informative. \texttt{Musical instrument} can reach relatively high F1 despite limited coverage, likely because its spectral structure is more separable for the pretrained semantic branch. Conversely, short, impulsive, or noise-like events such as \texttt{Knock}, \texttt{Water tap}, \texttt{Clapping}, and partially \texttt{Door} remain fragile because their temporal support is short, their spectral signature is less stable, and their acoustic morphology overlaps with environmental transients. These observations anticipate the diagnostic interventions of Section~\ref{sec:diagnostic_characterization}, where class balancing, loss calibration, and threshold selection are used to separate data scarcity from architectural limitation.

Overall, Stage~3 identifies late cross-stitching as the only interaction strategy that provides a favorable trade-off between \gls{seld} performance, computational overhead, semantic preservation, and test-side behavior. Early interaction is both more expensive and less effective, while simultaneous early--late interaction inherits the drawbacks of the early bridge. The selected operating point for the subsequent diagnostic characterization is therefore the late-interaction topology, with the caveat that its class-dependent behavior remains constrained by the long-tailed STARSS23 supervision regime.


\subsection{Summary of Findings}
\label{subsec:nas_architecture_synthesis}

The informed \gls{nas} identifies a consistent progression from spatial representation selection to depth allocation and semantic--spatial interaction. Stage~1 shows that explicit spectral \gls{foa} descriptors provide the most reliable spatial input representation in the shallow regime. The combination of \gls{stft} magnitude, phase, and \glspl{iv} exposes localization-relevant structure while remaining in a time--frequency domain that is representationally compatible with the spectrogram-based processing of the pretrained \gls{gpat} branch. This coherence is central to the proposed \gls{at2seld} setting: spatial cues are not introduced as an unrelated raw-domain stream, but as structured spectral evidence that can be progressively aligned with semantic feature maps derived from \gls{gpat} pretraining. This result provides the current answer to \textbf{RQ1}. The advantage of the \gls{foa}+\gls{iv} interface is not explained by semantic-presence metrics alone, which remain comparatively stable across the shallow configurations. The decisive difference lies in how the spatial branch receives, preserves, and organizes multi-channel cues before track-wise prediction, and in how naturally these cues can later interact with high-level semantic representations through late feature-level coupling.

Stage~2 answers \textbf{RQ2} by showing that, under this feature regime, the most sensitive processing stage is the early spatial encoder. The largest performance gain is obtained by deepening the residual front-end that manipulates, segregates, and reorganizes the \gls{foa} time--frequency maps before they enter track-wise abstraction. This suggests that the early stage is where multi-channel spectral cues acquire a representation that is simultaneously spatially meaningful and compatible with the semantic structure required by the downstream \gls{seld} heads. By contrast, increasing the depth of the late \texttt{TrackTransformer} or the \texttt{BiGRU} smoother provides little or no benefit. These modules remain functionally necessary: the former organizes class-track hypotheses, whereas the latter regularizes temporal continuity. However, the Stage~2 results indicate that their role is better understood as algorithmic refinement of an already structured representation than as the main source of additional parameter capacity.

Stage~3 provides the current answer to \textbf{RQ3}. Shared \gls{seld} supervision is sufficient to train a competitive integrated model, but it does not fully exploit the availability of pretrained \gls{at} representations. Feature-level coupling becomes beneficial only when introduced after sufficient spatial abstraction. The late cross-stitch bridge is the only interaction mechanism that improves the regularized topology across validation and test behavior. By contrast, early cross-stitching is computationally more expensive and less effective for the task, because it injects semantic information while the spatial branch is still processing localization-sensitive feature maps. The \gls{nas} therefore supports a position-dependent conclusion: semantic information should condition high-level spatial representations rather than low-level feature encoding.

The answer to \textbf{RQ4} instead, remains necessarily partial at this stage. The search suggests that track-wise attentive abstraction and compact recurrent smoothing are preferable to the alternatives tested in the shallow grid. \texttt{TrackTransformer} is consistently more robust than formant-oriented refinement for organizing class-track hypotheses, and \texttt{BiGRU} smoothing provides a better performance--efficiency compromise than \texttt{MHA-MGU}. However, architecture search alone does not fully resolve the interaction between track-wise prediction, activity calibration, localization optimization, and overlapping sound event conditions. These issues depend on the supervision regime and operating-point calibration, and are therefore addressed in the diagnostic analysis.

Overall, the informed \gls{nas} selects a \gls{foa}-based \gls{at2seld} architecture composed of deep \texttt{ResNet} spatial encoders, a \texttt{TrackTransformer} abstraction module, a compact \texttt{BiGRU} smoothing layer, and late branches cross-stitch coupling. This selection is not treated as a deployment-ready endpoint, but as the most informative configuration for diagnostic characterization. The search results already show that aggregate \gls{seld} scores can hide distinct mechanisms: stable \gls{at} metrics do not guarantee spatial improvement, smoother validation losses do not necessarily imply reliable spatial detections, and lower angular errors may coexist with weak recall. The following section therefore focuses on diagnostic analysis, examining whether loss calibration, activity-conditioned \gls{doa}, dataset class balancing, threshold selection, and cross-dataset evaluation can explain and alleviate the remaining limitations of the selected \gls{at2seld} family.

\clearpage


\section{Diagnostic Characterization and Cross-Dataset Evaluation}
\label{sec:diagnostic_characterization}
The diagnostic analysis is motivated by the mechanisms exposed in the previous Section. Aggregate \gls{seld} scores do not by themselves explain whether errors arise from insufficient class exposure, weak activity calibration, inactive-target dominance in \gls{doa} regression, spatial-domain mismatch, or limitations of track-wise assignment. Moreover, the selected \gls{at2seld} design intentionally uses a compact task-facing spatial pathway compared with many dedicated \gls{seld} systems (Section~\ref{sec:related_work}). The objective is therefore not to claim leaderboard-level superiority over challenge-optimized architectures, but to determine how far a pretrained semantic prior can support a reduced spatial branch, which supervision choices make this integration effective, and where the remaining operating limits appear.

\subsection{Scopes and Methodology}
\label{subsec:diagnostic_scope}

The diagnostic stage starts from the late-interaction topology identified at the end of the \gls{nas} campaign. This configuration uses explicit spectral \gls{foa} features, a three-layer \texttt{ResNet} early spatial encoder, a single \texttt{TrackTransformer} abstraction module, a compact \texttt{BiGRU} smoother, and one late \texttt{CrossStitch} bridge between the pretrained semantic branch and the spatial pathway. It is selected because it provides the most favorable Stage~3 trade-off between validation behavior, test-side \gls{seld} score, semantic preservation, and computational overhead.

The first diagnostic axis is \textit{data-side coverage}: Stage~3 showed that several STARSS23 classes remain fragile under the native real-scene distribution, especially short, sparse, or acoustically ambiguous events. A coverage-oriented \texttt{BalancedSTARSS23Dataset} is therefore introduced to test whether additional compatible material from external \gls{foa} corpora and controlled waveform mixing can reduce class scarcity while preserving the STARSS23 label space and target interface. This intervention is not intended to replace STARSS23 with synthetic data (Section~\ref{subsec:related_conformer}), but to examine whether class exposure is a limiting factor for the selected \gls{at2seld} configuration.

The second diagnostic axis concerns the \textit{activity objective}: since the track-wise output tensor is dominated by inactive class-track entries, the activity head is evaluated under different loss formulations and calibration regimes. Standard \gls{bce}, focal re-weighting, and positive-class weighting are compared to determine whether improved event sensitivity translates into better location-aware detection (Section~\ref{subsec:related_permutation_tracking}). This distinction is necessary because a loss that improves standalone \gls{sed} metrics may still produce an unfavorable \gls{seld} operating point if false positives or poorly localized detections increase.

The third diagnostic axis concerns \textit{localization supervision}: the original regression objective evaluates \gls{doa} errors over both active and inactive output slots. This can overemphasize null-vector prediction and suppress learning on the comparatively rare active sources (Section~\ref{subsec:related_permutation_tracking}). Activity-conditioned \gls{doa} supervision is therefore evaluated as a correction of the supervision domain: the localization loss is computed only where an event is active, while optional weak regularization controls inactive outputs. This analysis is central because it tests whether the selected architecture already contains useful spatial capacity that is partially hidden by the loss formulation.

The fourth diagnostic axis concerns \textit{operating-point calibration}: activity thresholds strongly affect \gls{seld} metrics, especially when the model is under-confident or when class priors are modified by balancing. The diagnostic stage therefore separates test-side threshold sweeps, which are used only to understand sensitivity, from validation-selected threshold policies, which provide a leakage-free calibration procedure. This distinction is required for deployment-oriented interpretation: threshold tuning can recover missed activity, but it cannot compensate for inaccurate \gls{doa} estimates outside the angular tolerance.

The final diagnostic axis is coverage-aware \textit{cross-dataset evaluation}: the selected \gls{at2seld} family is evaluated across STARSS23, TAU2019, TAU-NIGENS2020, and TAU-NIGENS2021, under projected label spaces and shared evaluation conventions. These experiments do not constitute a fully controlled challenge comparison, because the corpora differ in class vocabulary, rendering procedure, source motion, acoustic realism, and spatial statistics (Section \ref{subsubsec:dataset_roles}). Instead, they quantify how the same semantic-to-spatial mechanism behaves across fixed-source synthetic scenes, dynamic synthetic scenes, intermediate reverberant moving-source conditions, and real spatial recordings.

Taken together, these analyses characterize the selected \gls{at2seld} family from complementary perspectives, evaluating not only what the selected model achieves, but also why it behaves as observed and which methodological choices most effectively improve its operating regime.


\subsection{Data Balancing and Coverage Expansion}
\label{subsec:diagnostic_balanced_starss}

The first diagnostic intervention addresses the data-side limitations exposed by the Stage~3 per-class analysis. STARSS23 provides the most coherent real-scene condition for the \gls{at2seld} framework, but its native training distribution is strongly long-tailed: several classes are represented by a small number of clips, short active intervals, or acoustically variable events (Figure~\ref{fig:starss23_distributions}). The objective of the balancing procedure is to expand class coverage while preserving the STARSS23 label space, target representation, and validation/test protocol. The resulting \texttt{BalancedSTARSS23Dataset} combines three sources of evidence: original STARSS23 clips, compatible projected clips from external \gls{foa} datasets, and controlled mixed items synthesized for classes with no reliable external counterpart.

\paragraph{Label-space projection.}

All balancing operations are performed in the \(13\)-class STARSS23 label space. STARSS23 annotations are used directly, whereas TAU-NIGENS2021 and TAU2019 events are projected through fixed class mappings before being admitted into the balanced index. The projection is \textit{many-to-one} in several cases, reflecting the fact that external datasets may distinguish event subclasses that are collapsed in STARSS23. Events without a reliable semantic correspondence are discarded from the target construction. This avoids introducing ambiguous artificial labels while still exploiting external material where the acoustic meaning is compatible with the STARSS23 taxonomy.

\begin{table}[ht]
\centering
\caption{Projection of external dataset classes into the STARSS23 target space. Empty mappings indicate classes that are not supported by the corresponding external corpus. Non-mappable source classes are discarded.}
\label{tab:stage4_projection_mapping}
\footnotesize
\renewcommand{\arraystretch}{1.05}
\setlength{\tabcolsep}{3.8pt}
\begin{tabular}{|p{3cm}|p{5.0cm}|p{5.7cm}|}
\hline
\textbf{STARSS23} & \textbf{TAU-NIGENS2021} & \textbf{TAU2019} \\
\hline
\texttt{female\_speech} & \texttt{female\_scream}, \texttt{female\_speech} & -- \\
\texttt{male\_speech} & \texttt{male\_scream}, \texttt{male\_speech} & -- \\
\texttt{clapping} & -- & -- \\
\texttt{telephone} & \texttt{phone\_ring} & \texttt{phone} \\
\texttt{laughter} & -- & \texttt{laughter} \\
\texttt{domestic\_sounds} & -- & \texttt{drawer}, \texttt{keyboard}, \texttt{keysDrop}, \texttt{pageturn} \\
\texttt{walk\_footsteps} & \texttt{footstep} & -- \\
\texttt{door\_open\_close} & -- & \texttt{doorslam} \\
\texttt{music} & \texttt{piano} & -- \\
\texttt{musical\_instrument} & \texttt{piano} & -- \\
\texttt{water\_tap} & -- & -- \\
\texttt{bell} & -- & -- \\
\texttt{knock} & \texttt{knock} & \texttt{knock} \\
\hline
-- & \texttt{alarm}, \texttt{baby\_cry}, \texttt{crash}, \texttt{dog\_bark} & \texttt{clearthroat}, \texttt{cough}, \texttt{speech} \\
\hline
\end{tabular}
\end{table}

Let \(
\mathbf{n}
=
[n_1,\ldots,n_C]
\), where \(\mathbf{n}\) is the native STARSS23 clip-presence vector, \(n_c\) is the number of STARSS23 training clips containing class \(c\), and \(C=13\) is the number of STARSS23 classes. The balancing target is defined as \(
n_{\mathrm{target}}
=
\max_{c}
n_c
\) where \(n_{\mathrm{target}}\) is the maximum native class count in the STARSS23 training split. In the considered split, this value corresponds to the dominant \texttt{male\_speech} class. All original STARSS23 training clips are retained, and external or mixed items are added only to reduce the deficit of underrepresented classes.

\paragraph{Clip-level balancing and mixed synthesis.}

The balancing procedure operates at clip level before expansion into fixed-length training windows. It is implemented in two phases: the \textit{first phase} performs greedy external selection for all STARSS23 classes that have a valid mapping from TAU-NIGENS2021 or TAU2019. Classes are processed according to their remaining deficit:
\begin{equation}
\Delta_c
=
n_{\mathrm{target}}
-
n_c
\end{equation}
where \(\Delta_c\) is the clip-level deficit of class \(c\). For each class, the algorithm first selects unused external clips containing class \(c\) after projection. This de-duplication step reduces repeated use of the same external item. If the deficit cannot be filled by unique clips, sampling with replacement is allowed from the available covering pool. Since external clips may be multi-label after projection, selecting one clip can update multiple class counts:
\begin{equation}
n_{c'}
\leftarrow
n_{c'}+1,
\qquad
\forall c'\in\mathcal{G}(x)
\end{equation}
where \(\mathcal{G}(x)\) denotes the projected ground-truth event set of clip \(x\). This mechanism can compensate several deficits simultaneously, but it can also increase already frequent classes. The procedure is therefore interpreted as coverage expansion, not as exact uniformization.

The \textit{second phase} handles STARSS23 classes with no usable external support \(
\mathcal{C}_{\mathrm{mix}}
=
\{
\texttt{clapping},
\texttt{water\_tap},
\texttt{bell}
\}
\) where \(\mathcal{C}_{\mathrm{mix}}\) is the set of classes reinforced through mixed synthesis. For each class in \(\mathcal{C}_{\mathrm{mix}}\), a rare STARSS23 clip containing the target class is paired with a mixer clip that does not contain it. The mixer pool includes STARSS23 training clips and the unique TAU-NIGENS2021/TAU2019 clips selected during the external-selection phase. This choice is useful because external clips with sparse or empty projected labels can act as low-impact acoustic material without further increasing the most frequent STARSS23 classes.

Mixer selection is controlled through a priority score:
\begin{equation}
s(x)
=
\sum_{c\in\mathcal{G}(x)}
n_c
\end{equation}
where \(s(x)\) is the score of candidate mixer clip \(x\), and \(\mathcal{G}(x)\) is its projected class-presence set. At each step, the candidate with the smallest score is selected from a \textit{min-heap}, paired with a randomly sampled rare STARSS23 clip, and reinserted with its updated score~\cite{frederickson1993optimal_selection_min_heap}. This produces a soft-reuse policy: a mixer can be used multiple times, but each reuse makes it less likely to be selected again immediately.

Given two aligned four-channel \gls{foa} waveforms \(\mathbf{x}_a\) and \(\mathbf{x}_b\), the mixed waveform is obtained by simple averaged summation:
\begin{equation}
\mathbf{x}_{\mathrm{mix}}
=
\frac{1}{2}
\left(
\mathbf{x}_a
+
\mathbf{x}_b
\right)
\end{equation}
The corresponding target is obtained by inserting the active events of both items into the same track-wise tensor:
\begin{equation}
\mathbf{Y}_{\mathrm{mix}}
=
\mathbf{Y}_{a}
\oplus
\mathbf{Y}_{b}
\end{equation}
where \(\mathbf{Y}_{a}\) and \(\mathbf{Y}_{b}\) are the track-wise Cartesian targets of the two items, and \(\oplus\) denotes source-set union under the maximum track capacity of the pipeline. The rare STARSS23 target is inserted first, followed by the mixer target. If the number of simultaneous active sources exceeds the available track capacity, excess sources are discarded according to the same maximum-polyphony rule used by the unified data pipeline (Section~\ref{subsubsec:splits_windowing_metadata}).

\begin{table}[ht]
\centering
\caption{Clip-level and window-level composition of \texttt{BalancedSTARSS23Dataset}.}
\label{tab:stage4_balanced_composition}
\footnotesize
\renewcommand{\arraystretch}{1.05}
\setlength{\tabcolsep}{4.2pt}
\begin{tabular}{|p{3.0cm}|c|c|c|c|}
\hline
\textbf{Source Dataset} & \textbf{Audio Clips} & \textbf{\% clips} & \textbf{Windowed items} & \textbf{\% windows} \\
\hline
STARSS23 & 72 & 23.2 & 4592 & 45.8 \\
TAU-NIGENS2021 & 68 & 21.9 & 1428 & 14.2 \\
TAU2019 & 44 & 14.1 & 884 & 8.8 \\
\hline
\textit{mixing} & 127 & 40.8 & 3130 & 31.2 \\
\hline
\textbf{Total} &
\textbf{311} &
\multicolumn{1}{c|}{} &
\multicolumn{1}{c|}{\textbf{10\,034}} &
\multicolumn{1}{c}{} \\
\cline{1-2}\cline{4-4}
\end{tabular}

\vspace{0.1cm}
\parbox{0.92\linewidth}{\centering\footnotesize
Windowed statistics refer to \(10~\mathrm{s}\) windows with \(2.5~\mathrm{s}\) hop.\\In windowed mode, \(120\) windows \((1.2\%)\) contain no active events.
}
\end{table}

Table~\ref{tab:stage4_balanced_composition} summarizes the resulting dataset composition: the mixed component accounts for \(40.8\%\) of the clip-level index because the rarest STARSS23-only classes cannot be supported by external corpora. After windowing, STARSS23 remains the largest contribution \((45.8\%)\), while mixed items account for \(31.2\%\), TAU-NIGENS2021 for \(14.2\%\), and TAU2019 for \(8.8\%\). The balanced dataset therefore increases external and mixed coverage while preserving STARSS23 as the dominant in-domain component.

Runtime augmentation is applied after balancing and only on the training split, following procedures described in Section~\ref{subsubsec:augmentation_batching}. For mixed items, augmentation is applied to the already combined \gls{foa} waveform and to the union target. Since the \textit{16patterns} transformations and global gain scaling are linear operations in Ambisonics B-format, applying them after waveform summation remains consistent with the physical interpretation of the mixed scene.

\paragraph{Window-level composition and spatial distribution.}

The distinction between clip-level and window-level balancing is essential. The algorithm reduces class scarcity at the clip and annotation level, but it does not explicitly optimize the final number of training windows per class. Long clips and events with longer temporal support still generate more \(10~\mathrm{s}\) windows than short events. As a result, the final window-level distribution remains non-uniform, although it is substantially less sparse for the low-resource classes that motivated the intervention.

\begin{figure}[ht]
    \centering
    \includegraphics[width=\linewidth]{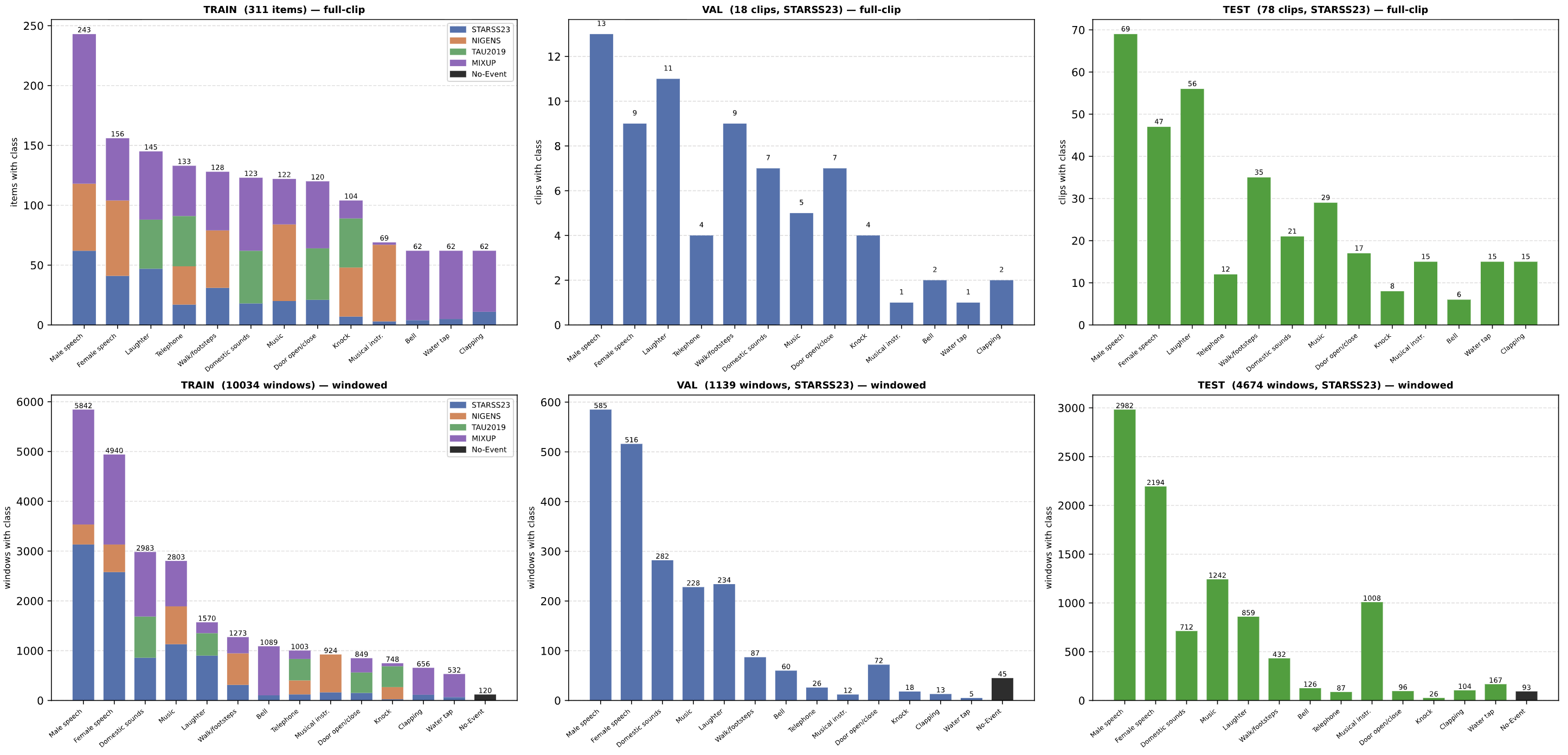}
    \caption{Class distribution of \texttt{BalancedSTARSS23Dataset} after external projection, greedy clip selection, and mixed synthesis.}
    \label{fig:balanced_starss23_class_dist}
\end{figure}

\begin{table}[ht]
\centering
\caption{Per-class window presence in \texttt{BalancedSTARSS23Dataset}.}
\label{tab:stage4_balanced_windows}
\footnotesize
\renewcommand{\arraystretch}{1.02}
\setlength{\tabcolsep}{3.6pt}
\begin{tabular}{|p{3cm}|c|c|c|c|c|}
\hline
\textbf{Class} & \textbf{Total} & \textbf{STARSS23} & \textbf{TAU-NIGENS2021} & \textbf{TAU2019} & \textbf{Mixed} \\
\hline
\texttt{male\_speech} & 5842 & 3133 & 401 & \cellcolor{black!90} & 2308 \\
\texttt{female\_speech} & 4940 & 2579 & 553 & \cellcolor{black!90} & 1808 \\
\texttt{domestic\_sounds} & 2983 & 858 & \cellcolor{black!90} & 828 & 1297 \\
\texttt{music} & 2803 & 1130 & 760 & \cellcolor{black!90} & 913 \\
\texttt{laughter} & 1570 & 901 & \cellcolor{black!90} & 451 & 218 \\
\texttt{walk\_footsteps} & 1273 & 313 & 634 & \cellcolor{black!90} & 326 \\
\texttt{bell} & 1089 & 103 & \cellcolor{black!90} & \cellcolor{black!90} & 986 \\
\texttt{telephone} & 1003 & 122 & 281 & 431 & 169 \\
\texttt{musical\_instrument} & 924 & 164 & 760 & \cellcolor{black!90} & \cellcolor{black!90} \\
\texttt{door\_open\_close} & 849 & 152 & \cellcolor{black!90} & 411 & 286 \\
\texttt{knock} & 748 & 25 & 243 & 420 & 60 \\
\texttt{clapping} & 656 & 115 & \cellcolor{black!90} & \cellcolor{black!90} & 541 \\
\texttt{water\_tap} & 532 & 67 & \cellcolor{black!90} & \cellcolor{black!90} & 465 \\
\hline
\end{tabular}

\vspace{0.1cm}
\parbox{0.92\linewidth}{\centering\footnotesize
Counts indicate windows in which each class is present; multi-label windows contribute to multiple rows. Black cells denote unavailable mappings according to Table~\ref{tab:stage4_projection_mapping}.
}
\end{table}

Figure~\ref{fig:balanced_starss23_class_dist} and Table~\ref{tab:stage4_balanced_windows} show that the intervention increases exposure for the most underrepresented classes, especially \texttt{bell}, \texttt{clapping}, \texttt{water\_tap}, \texttt{telephone}, and \texttt{knock}. However, dominant classes such as \texttt{male\_speech} and \texttt{female\_speech} remain frequent because they appear in many STARSS23 clips and can also co-occur in mixed items. The resulting dataset should therefore be interpreted as a coverage-expanded STARSS23-centered training set, not as a strict window-level uniform sampler. In the resulting index, the window-level imbalance ratio is reduced to \(10.98\times\).

In the \(10~\mathrm{s}\) windowed setting, data batches have the form:
\begin{equation}
\mathbf{X}
\in
\mathbb{R}^{B\times 4\times 10 f_{\mathrm{model}}},
\qquad
\mathbf{Y}
\in
\mathbb{R}^{B\times 1001\times 3\times 13\times 3}
\end{equation}
where \(\mathbf{X}\) is the batched \gls{foa} waveform tensor, \(B\) is the batch size, and \(f_{\mathrm{model}}\) is the model sampling rate in Hertz.

\begin{figure}[ht]
    \centering
    \includegraphics[width=0.75\linewidth]{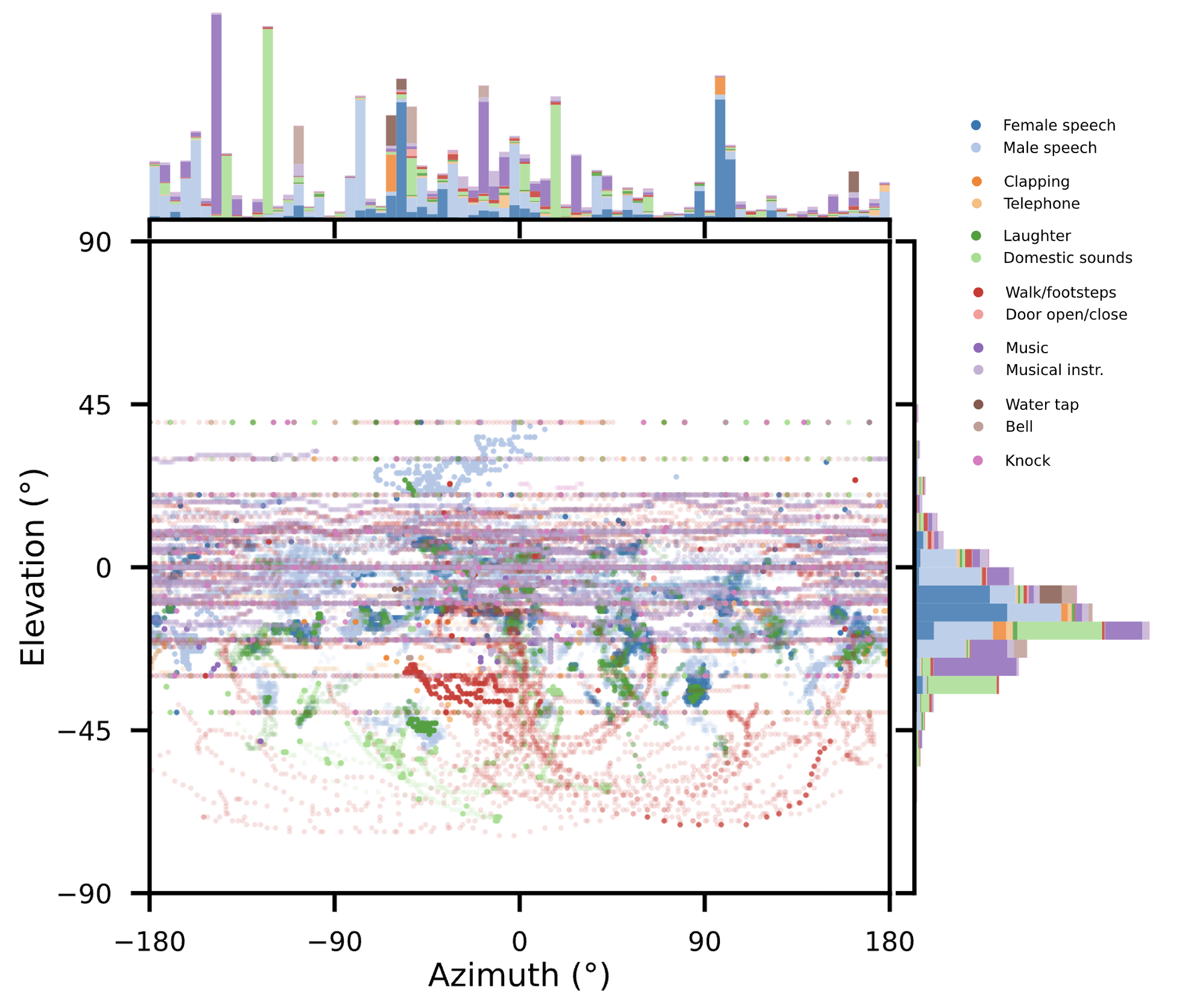}
    \caption{\Gls{doa} marginal distributions of \texttt{BalancedSTARSS23Dataset} after external projection, greedy clip selection, and mixed synthesis.}
    \label{fig:balanced_starss23_doa}
\end{figure}

External clips and mixed items increase directional diversity for several low-resource classes, while the resulting geometry remains class-dependent. Speech and music-related categories retain broad azimuth coverage, whereas classes such as \texttt{bell}, \texttt{water\_tap}, and \texttt{door\_open\_close} remain more spatially constrained (Figure~\ref{fig:balanced_starss23_doa}). This behavior is desirable for diagnostic characterization: the balancing procedure increases evidence for underrepresented classes without erasing the real-scene spatial structure that makes STARSS23 a meaningful in-domain target. The validation split remains the STARSS23 validation set, and test evaluation is performed separately on the STARSS23 test split and on projected external test loaders when required by the cross-dataset diagnostics, preserving the role of the balanced dataset as a training-side intervention. Improvements on STARSS23 quantify in-domain robustness under increased coverage, whereas projected external evaluations quantify whether the same \gls{at2seld} mechanism transfers to synthetic and fixed-source spatial-audio conditions. The balancing stage therefore establishes a clear separation between \emph{coverage} and \emph{learnability}: class exposure is substantially improved, but the following sections are required to determine whether the joint \gls{seld} objective can exploit that additional supervision.


\subsection{Activity-Loss Optimization}
\label{subsec:diagnostic_activity_loss}

The coverage expansion described in Section~\ref{subsec:diagnostic_balanced_starss} increases the exposure of underrepresented STARSS23 classes, but it does not remove the strong asymmetry between active and inactive entries in the track-wise output tensor. Even after balancing, only a small fraction of frame--track--class positions corresponds to active sound events. The activity head must therefore learn from a supervision space dominated by negative targets while remaining coupled to \gls{doa} regression through the \gls{tpit} objective. The first loss-side diagnostic consequently evaluates whether reshaping the activity loss improves the joint \gls{seld} operating point without modifying the architecture selected by the \gls{nas}.

The first comparison isolates standard \gls{bce} and focal loss~\cite{focal_loss}. Both models are trained on \texttt{BalancedSTARSS23Dataset} using the late-interaction \gls{at2seld} topology, the same augmentation policy, the same target representation, and the same original all-entry Cartesian \gls{doa} regression term. The only controlled change is the focal focusing parameter:
\begin{equation}
\gamma
=
\begin{cases}
0, & \text{standard \gls{bce}}\\
2, & \text{focal loss}
\end{cases}
\end{equation}
where \(\gamma=0\) recovers ordinary \gls{bce}, while \(\gamma=2\) reduces the contribution of confidently classified activity entries and preserves larger gradients for difficult examples. For a target-aligned probability \(p_t\), this modulation is:
\begin{equation}
\mathcal{L}_{\mathrm{focal}}
=
-
\alpha_t
\left(
1-p_t
\right)^{\gamma}
\log
\left(
p_t
\right)
\end{equation}
where \(\alpha_t\) is the optional class-balancing coefficient. In this comparison, focal loss is used to reshape the activity gradient, not to modify the \gls{tpit} assignment rule or the \gls{doa} objective.

\begin{table}[ht]
\centering
\caption{Activity-loss results for the \gls{at2seld} reference \gls{nas} model, trained on \texttt{BalancedSTARSS23Dataset} and evaluated on STARSS23. Post-hoc threshold values are diagnostic and are selected independently for \gls{seld} score and frame-level \gls{sed} F1.}
\label{tab:stage4_sed_loss_optimization}
\footnotesize
\renewcommand{\arraystretch}{1.08}
\setlength{\tabcolsep}{3.3pt}
\resizebox{\textwidth}{!}{%
\begin{tabular}{|p{2.35cm}|c|c|c|c|c|c|c|}
\hline
\textbf{Configuration} &
\textbf{Activity loss} &
\textbf{\gls{doa} loss} &
\textbf{Val SELD \(\downarrow\)} &
\textbf{Test SELD \(\downarrow\)} &
\textbf{\(F_{20^\circ}\) \(\uparrow\)} &
\textbf{SED F1 \(\uparrow\)} &
\textbf{Best thresholds} \\
\hline
\texttt{mixed\_bce} &
\gls{bce} &
All-entry MSE &
0.601 &
0.696 &
0.109 &
\textbf{0.781} &
\makecell{\(0.10\) SELD\\\(0.30\) SED} \\
\hline
\texttt{mixed\_focal} &
Focal &
All-entry MSE &
0.470 &
0.661 &
0.173 &
0.758 &
\makecell{\(0.40\) SELD\\\(0.40\) SED} \\
\hline
\hline
\rowcolor{gray!15}
\texttt{ExpA} &
Focal &
Active-only MSE &
\textbf{0.337} &
\textbf{0.607} &
\textbf{0.247} &
0.704 &
\makecell{\(0.30\) SELD\\\(0.30\) SED} \\
\hline
\texttt{ExpA-posw} &
Focal + \(w_c^+\) &
Active-only MSE &
0.677 &
0.648 &
0.168 &
0.728 &
\makecell{\(0.50\) SELD\\\(0.40\) SED} \\
\hline
\end{tabular}%
}

\vspace{0.1cm}
\parbox{0.95\linewidth}{\centering\footnotesize
\textbf{Val SELD} denotes the minimum validation \gls{seld} score observed during training. \textbf{Test SELD} and \(F_{20^\circ}\) are reported at the global threshold minimizing test \gls{seld} score. \textbf{SED F1} is computed independently at the frame level and is not an official DCASE score.\\\texttt{ExpA} and \texttt{ExpA-posw} anticipate the activity-conditioned \gls{doa} analysis in Section~\ref{subsec:diagnostic_active_doa}; they are included here to isolate the effect of positive activity weighting after the localization objective has been corrected.
}
\end{table}

The comparison shows that focal loss is more favorable for the joint \gls{seld} objective. The \gls{bce} model reaches a validation \gls{seld} score of \(0.601\), whereas the focal model reaches \(0.470\). The same tendency appears on the STARSS23 test set: the post-hoc test \gls{seld} score improves from \(0.696\) to \(0.661\), and \(F_{20^\circ}\) increases from \(0.109\) to \(0.173\). The improvement is therefore not only a matter of training speed or threshold tuning; it affects the location-aware detection component of the \gls{seld} criterion.

This result should be distinguished from isolated \gls{sed} quality. When localization is removed from the evaluation and each model is assessed at the threshold maximizing frame-level \gls{sed} F1, the \gls{bce} model obtains \(0.781\), compared with \(0.758\) for focal loss. Thus, focal loss does not maximize standalone event-activity classification. Its benefit emerges when activity estimation must cooperate with track assignment and spatial regression inside the integrated branch. This distinction is important for the proposed framework: a stronger activity classifier is not necessarily a stronger spatially grounded detector if its detections are poorly localized or poorly matched to the track-wise output space.

The localization diagnostics confirm that the activity loss alone does not explain the remaining errors. Under the original all-entry \gls{doa} objective, focal loss reduces the median angular error on target-active slots with respect to \gls{bce}, but both models still produce a substantial fraction of near-null \gls{doa} vectors for active events. This indicates that focal loss improves the joint optimization regime, but does not remove the supervision mismatch caused by inactive-target dominance in the localization activity. This issue is addressed explicitly in Section~\ref{subsec:diagnostic_active_doa}.

After introducing activity-conditioned \gls{doa} supervision, a second activity-side experiment evaluates positive-class weighting. For each class \(c\), a positive weight \(w_c^+\) is estimated from the training targets and clipped to avoid extreme rare-class factors. The weighted focal activity term is:
\begin{equation}
\mathcal{L}_{\mathrm{SED}}^{\mathrm{pw}}
=
-
\sum_{c}
\left[
w_c^+
y_c
\log
p_c
+
\left(
1-y_c
\right)
\log
\left(
1-p_c
\right)
\right]
\left(
1-p_{t,c}
\right)^{\gamma}
\end{equation}
where \(y_c\) is the binary activity target, \(p_c\) is the predicted activity probability, \(p_{t,c}\) is the target-aligned probability, and \(w_c^+\) is the class-dependent positive weight. These weights are applied only to the gradient loss and are excluded from the \gls{tpit} permutation-selection cost, so that rare-class factors do not directly dominate the track assignment.

The positive-weighted variant, \texttt{ExpA-posw}, confirms that stronger positive supervision increases activity sensitivity: frame-level \gls{sed} F1 improves from \(0.704\) to \(0.728\) relative to \texttt{ExpA}. However, this improvement does not transfer to the complete \gls{seld} criterion. The test \gls{seld} score worsens from \(0.607\) to \(0.648\), and \(F_{20^\circ}\) decreases from \(0.247\) to \(0.168\). This behavior indicates that positive weighting increases the pressure to activate rare or difficult classes, but also makes the model more sensitive to false positives and threshold calibration. In a track-wise \gls{seld} setting, additional activity recall is beneficial only when it remains associated with reliable spatial estimates.

Overall these findings motivate two subsequent analyses: activity-conditioned \gls{doa} supervision, which corrects the localization objective itself, and threshold calibration, which determines whether the activity probabilities can be converted into reliable operating points without using test-set information.


\subsection{Activity-Conditioned \glsentryshort{doa} Supervision}
\label{subsec:diagnostic_active_doa}

The previous activity-loss analysis shows that focal re-weighting improves the joint \gls{seld} operating point, but does not fully explain the remaining localization behavior. In particular, the all-entry \gls{doa} objective still evaluates Cartesian regression over both active and inactive class-track slots. This creates a supervision imbalance that is distinct from class imbalance: the target tensor is dominated by inactive entries whose correct Cartesian target is the null vector. The activity-conditioned \gls{doa} experiments therefore test whether the selected \gls{at2seld} architecture already contains useful spatial capacity that is partially suppressed by the localization loss.

The diagnostic evidence comes from a preliminary \textit{oracle-activity} evaluation. In this setting, the ground-truth activity mask is used to select the slots on which angular localization is evaluated, thereby removing errors due to predicted \gls{sed} gating. For the balanced focal reference model trained with the legacy \gls{doa} loss, only \(3.36\%\) of the evaluated frame--track--class slots are active. Under oracle activity, the median angular error is \(50.49^\circ\), and only \(18.88\%\) of active predictions fall within \(20^\circ\) of their targets. More importantly, the median norm of predicted active vectors is only \(0.224\), despite the corresponding target vectors having approximately unit norm. Approximately \(31.49\%\) of the active predictions have norm below \(0.1\). This indicates that the localization limitation is not only caused by missed detections: even when activity is supplied by the ground truth, many active \gls{doa} predictions remain close to the origin. Let:
\begin{equation}
\widehat{\mathbf{r}}_{b,t,n,c}\in[-1,1]^3,
\qquad
\mathbf{r}_{b,t,n,c}\in\mathbb{R}^{3}
\end{equation}
where \(\widehat{\mathbf{r}}_{b,t,n,c}\) is the predicted Cartesian \gls{doa} vector, \(\mathbf{r}_{b,t,n,c}\) is the target vector, and \(b\), \(t\), \(n\), and \(c\) index batch item, frame, track, and class, respectively. In the activity-coupled target representation, active entries have approximately unit-norm Cartesian vectors, whereas inactive entries are represented by the null vector. The legacy all-entry loss is:
\begin{equation}
\mathcal{L}_{\mathrm{DOA}}^{\mathrm{all}}
=
\frac{1}{BTNC}
\sum_{b=1}^{B}
\sum_{t=1}^{T}
\sum_{n=1}^{N}
\sum_{c=1}^{C}
\left\|
\widehat{\mathbf{r}}_{b,t,n,c}
-
\mathbf{r}_{b,t,n,c}
\right\|_2^2
\end{equation}
where \(B\) is the batch size, \(T\) is the number of supervision frames, \(N\) is the number of track slots, and \(C\) is the number of classes. Since inactive targets are numerically dominant, minimizing this objective can reward near-null predictions over the majority of the tensor. This is redundant with the adopted decoupled head formulation: event presence is already modeled by the activity head, whereas the localization head should estimate direction conditional on activity.

The corrected supervision domain is defined through a target-derived active mask:
\begin{equation}
A_{b,t,n,c}
=
\mathbbm{1}
\left[
\left\|
\mathbf{r}_{b,t,n,c}
\right\|_2
>
\epsilon_{\mathrm{act}}
\right],
\qquad
\epsilon_{\mathrm{act}}=0.5
\end{equation}
where \(A_{b,t,n,c}\in\{0,1\}\) denotes whether a target is active and \(\epsilon_{\mathrm{act}}\) separates inactive null vectors from approximately unit-norm active targets. For each frame--class combination \(q=(b,t,c)\), the active-only Cartesian loss is:
\begin{equation}
\ell_{\mathrm{act}}^{(q)}
=
\frac{
\sum_{n=1}^{N}
A_{q,n}
\left\|
\widehat{\mathbf{r}}_{q,n}
-
\mathbf{r}_{q,n}
\right\|_2^2
}{
\max
\left(
1,
\sum_{n=1}^{N}A_{q,n}
\right)
}
\end{equation}
where \(\ell_{\mathrm{act}}^{(q)}\) is the active localization error for frame--class combination \(q\). Combinations without active targets contribute zero to the directional part of the loss rather than pulling predictions toward the origin. The active-only loss is then averaged over all frame--class combinations:
\begin{equation}
\mathcal{L}_{\mathrm{DOA}}^{\mathrm{act}}
=
\frac{1}{BTC}
\sum_{q}
\ell_{\mathrm{act}}^{(q)}
\end{equation}
where \(BTC\) is the total number of frame--class combinations in the mini-batch.

Completely unconstrained inactive outputs could still produce large vectors associated with false-positive activity predictions. A weak inactive-norm regularizer is therefore retained:
\begin{equation}
\mathcal{L}_{\mathrm{inactive}}
=
\frac{1}{BTC}
\sum_q
\frac{
\sum_{n=1}^{N}
\left(
1-A_{q,n}
\right)
\left\|
\widehat{\mathbf{r}}_{q,n}
\right\|_2^2
}{
\max
\left(
1,
\sum_{n=1}^{N}
\left(
1-A_{q,n}
\right)
\right)
}
\end{equation}
where \(\mathcal{L}_{\mathrm{inactive}}\) penalizes large inactive \gls{doa} vectors without allowing inactive slots to dominate the localization objective. The \texttt{ExpA} localization objective is:
\begin{equation}
\mathcal{L}_{\mathrm{DOA}}^{A}
=
\mathcal{L}_{\mathrm{DOA}}^{\mathrm{act}}
+
\lambda_{\mathrm{inactive}}
\mathcal{L}_{\mathrm{inactive}},
\qquad
\lambda_{\mathrm{inactive}}=0.005
\end{equation}
where \(\lambda_{\mathrm{inactive}}\) controls the strength of the inactive-norm regularizer. The small coefficient preserves bounded inactive predictions while keeping the active targets as the dominant source of localization gradients.

The same principle is applied to track-permutation selection. For each frame--class combination, all admissible track permutations are evaluated using an activity cost and an active-conditioned \gls{doa} cost:
\begin{equation}
\pi^{\star}
=
\arg\min_{\pi}
\left[
\lambda_{\mathrm{SED}}
\mathcal{C}_{\mathrm{SED}}^{(\pi)}
+
\left(
1-\lambda_{\mathrm{SED}}
\right)
\mathcal{C}_{\mathrm{DOA}}^{(\pi)}
\right]
\end{equation}
where \(\pi^{\star}\) is the selected permutation, \(\mathcal{C}_{\mathrm{SED}}^{(\pi)}\) is the activity assignment cost, \(\mathcal{C}_{\mathrm{DOA}}^{(\pi)}\) is computed with the active-only localization criterion, and \(\lambda_{\mathrm{SED}}\) balances activity and localization during assignment. In \texttt{ExpA}, \(\lambda_{\mathrm{SED}}\) is set to \(0.50\), focal modulation remains fixed at \(\gamma=2\), and the late-interaction \gls{at2seld} architecture, balanced dataset, augmentation pipeline, and three-track output representation are preserved.

\begin{table}[ht]
\centering
\caption{Effect of activity-conditioned \gls{doa} supervision on the balanced \gls{at2seld} configuration. \Gls{seld} metrics use \(\tau=0.5\); oracle metrics use ground-truth activity.}
\label{tab:stage4_active_doa_results}
\footnotesize
\renewcommand{\arraystretch}{1.08}
\setlength{\tabcolsep}{4.0pt}
\resizebox{\textwidth}{!}{%
\begin{tabular}{|p{4.5cm}|c|c|}
\hline
\textbf{Metrics on STARSS23} &
\textbf{All-entry MSE + Focal} &
\textbf{Active-only MSE + Focal} \\
\hline
Best validation \gls{seld} score
    & \(0.470\) & \(\mathbf{0.337}\) \\
\hline
Test \gls{seld} score
    & \(0.666\) & \(\mathbf{0.635}\) \\
\(\mathrm{ER}_{20^\circ}\)
    & \(0.846\) & \(\mathbf{0.816}\) \\
\(F_{20^\circ}\)
    & \(0.175\) & \(\mathbf{0.235}\) \\
\(\mathrm{LE}_{\mathrm{CD}}\)
    & \(13.57^\circ\) & \(\mathbf{12.48^\circ}\) \\
\(\mathrm{LR}_{\mathrm{CD}}\)
    & \(0.083\) & \(\mathbf{0.108}\) \\
\hline
Oracle median angular error
    & \(50.49^\circ\) & \(\mathbf{27.24^\circ}\) \\
Oracle accuracy within \(20^\circ\)
    & \(18.88\%\) & \(\mathbf{33.67\%}\) \\
Median active-vector norm
    & \(0.224\) & \(\mathbf{0.735}\) \\
Active vectors with norm \(<0.1\)
    & \(31.49\%\) & \(\mathbf{2.49\%}\) \\
\hline
\end{tabular}%
}
\end{table}

Table~\ref{tab:stage4_active_doa_results} shows that activity-conditioned \gls{doa} supervision recovers meaningful spatial-vector prediction. The median active-vector norm increases from \(0.224\) to \(0.735\), while the proportion of near-null active vectors decreases from \(31.49\%\) to \(2.49\%\). At the same time, the oracle median angular error is almost halved, from \(50.49^\circ\) to \(27.24^\circ\), and oracle accuracy within \(20^\circ\) increases from \(18.88\%\) to \(33.67\%\). Since the oracle evaluation bypasses predicted activity decisions, this improvement directly indicates that the legacy all-entry MSE suppressed the spatial capacity of the selected \gls{at2seld} model.

The improvement also propagates to end-to-end \gls{seld} metrics at the fixed activity threshold. Relative to the balanced focal reference, \texttt{ExpA} reduces the STARSS23 test \gls{seld} score from \(0.666\) to \(0.635\), increases \(F_{20^\circ}\) from \(0.175\) to \(0.235\), and increases \(\mathrm{LR}_{\mathrm{CD}}\) from \(0.083\) to \(0.108\). The best validation \gls{seld} score decreases from \(0.470\) to \(0.337\), showing that the corrected supervision provides a substantially stronger optimization signal before any post-hoc threshold calibration. Although the comparison changes both the \gls{doa} formulation and the activity/localization mixing coefficient, the oracle angular improvement and the large increase in active-vector norms cannot be explained by activity thresholding, because oracle evaluation uses the ground-truth activity mask.

Activity-conditioned Cartesian MSE is therefore retained as the reference localization correction. It should be understood as a minimal and effective reformulation of the supervision domain, that still couples angular and magnitude errors, and the predicted trajectories may retain frame-to-frame variability. The following experiments therefore separate these factors by adding explicit direction, norm, and temporal-continuity terms without reintroducing inactive-slot dominance.

\paragraph{Directional and temporal extensions.}

\texttt{ExpB} replaces the active Cartesian error with an explicit direction--norm formulation. For active slots, predicted and target vectors are normalized as:
\begin{equation}
\widehat{\mathbf{u}}_{q,n}
=
\frac{
\widehat{\mathbf{r}}_{q,n}
}{
\max
\left(
\left\|
\widehat{\mathbf{r}}_{q,n}
\right\|_2,
\varepsilon
\right)
},
\qquad
\mathbf{u}_{q,n}
=
\frac{
\mathbf{r}_{q,n}
}{
\left\|
\mathbf{r}_{q,n}
\right\|_2
}
\end{equation}
where \(\widehat{\mathbf{u}}_{q,n}\) and \(\mathbf{u}_{q,n}\) are unit direction vectors, and \(\varepsilon>0\) avoids division by zero. The directional loss is:
\begin{equation}
\mathcal{L}_{\mathrm{dir}}
=
\frac{1}{BTC}
\sum_q
\frac{
\sum_{n=1}^{N}
A_{q,n}
\left(
1-
\widehat{\mathbf{u}}_{q,n}^{\top}
\mathbf{u}_{q,n}
\right)
}{
\max
\left(
1,
\sum_{n=1}^{N}A_{q,n}
\right)
}
\end{equation}
where \(\mathcal{L}_{\mathrm{dir}}\) penalizes angular disagreement through cosine distance. Since cosine similarity is insensitive to vector magnitude, a unit-norm regularizer is added on active slots:
\begin{equation}
\mathcal{L}_{\mathrm{norm}}
=
\frac{1}{BTC}
\sum_q
\frac{
\sum_{n=1}^{N}
A_{q,n}
\left(
\left\|
\widehat{\mathbf{r}}_{q,n}
\right\|_2
-
1
\right)^2
}{
\max
\left(
1,
\sum_{n=1}^{N}A_{q,n}
\right)
}
\end{equation}
where \(\mathcal{L}_{\mathrm{norm}}\) encourages active predictions to remain close to the unit sphere. The \texttt{ExpB} objective is:
\begin{equation}
\mathcal{L}_{\mathrm{DOA}}^{B}
=
\mathcal{L}_{\mathrm{dir}}
+
0.2\mathcal{L}_{\mathrm{norm}}
+
0.005\mathcal{L}_{\mathrm{inactive}}
\end{equation}
where all architecture, dataset, augmentation, and optimization settings are kept identical to \texttt{ExpA}, including \(\lambda_{\mathrm{SED}}=0.50\).

\texttt{ExpC} extends \texttt{ExpB} with temporal regularization. Let
\begin{equation}
P_{b,t,n,c}
=
A_{b,t,n,c}
A_{b,t-1,n,c}
\end{equation}
where \(P_{b,t,n,c}=1\) only when the same target track is active in two consecutive frames after \gls{tpit} alignment. The temporal smoothness term is:
\begin{equation}
\mathcal{L}_{\mathrm{smooth}}
=
\frac{
\sum_{b,t,n,c}
P_{b,t,n,c}
\left(
1-
\widehat{\mathbf{u}}_{b,t,n,c}^{\top}
\widehat{\mathbf{u}}_{b,t-1,n,c}
\right)
}{
\max
\left(
1,
\sum_{b,t,n,c}
P_{b,t,n,c}
\right)
}
\end{equation}
where \(\mathcal{L}_{\mathrm{smooth}}\) penalizes frame-to-frame directional changes for continuously active tracks. The complete \texttt{ExpC} objective is:
\begin{equation}
\mathcal{L}_{\mathrm{DOA}}^{C}
=
\mathcal{L}_{\mathrm{dir}}
+
0.2\mathcal{L}_{\mathrm{norm}}
+
0.005\mathcal{L}_{\mathrm{inactive}}
+
0.02\mathcal{L}_{\mathrm{smooth}}
\end{equation}
where the small temporal coefficient is intended to reduce directional jitter without forcing moving sources toward stationary trajectories.

A further \texttt{ExpB} variant increases \(\lambda_{\mathrm{SED}}\) from \(0.50\) to \(0.70\), while initially adopting an activity threshold of \(0.30\). This experiment tests whether the stabilized localization formulation permits returning more weight to activity estimation. Since both the objective balance and the configured activity threshold change, this variant is interpreted as an operating-point experiment rather than as a single-factor ablation.

\begin{table}[ht]
\centering
\caption{Validation and STARSS23 test results for the activity-conditioned \gls{doa} loss sequence. Test metrics use each experiment's configured activity threshold.}
\label{tab:stage4_doa_loss_results}
\footnotesize
\renewcommand{\arraystretch}{1.07}
\setlength{\tabcolsep}{3.1pt}
\resizebox{\textwidth}{!}{%
\begin{tabular}{|p{2.8cm}|c|c|c|c|}
\hline
\textbf{Metric}
& \textbf{ExpA}
& \textbf{ExpB}
& \textbf{ExpB \(\lambda_{\mathrm{SED}}=.70\)}
& \textbf{ExpC} \\
\hline
\gls{doa} mode
& active MSE
& cosine+norm
& cosine+norm
& cosine+norm+smooth \\
\(\lambda_{\mathrm{SED}}\)
& \(0.50\) & \(0.50\) & \(0.70\) & \(0.50\) \\
Activity threshold
& \(0.50\) & \(0.50\) & \(0.30\) & \(0.50\) \\
\hline
Validation \gls{seld}
& \(\mathbf{0.337}\) & \(0.393\) & \(0.424\) & \(0.385\) \\
\hline
Test \gls{seld}
& \(\mathbf{0.635}\) & \(0.663\) & \(0.823\) & \(0.682\) \\
\(\mathrm{ER}_{20^\circ}\)
& \(\mathbf{0.816}\) & \(0.845\) & \(1.528\) & \(0.885\) \\
\(F_{20^\circ}\)
& \(\mathbf{0.235}\) & \(0.197\) & \(0.155\) & \(0.159\) \\
\(\mathrm{LE}_{\mathrm{CD}}\)
& \(\mathbf{12.48^\circ}\) & \(13.74^\circ\)
& \(12.55^\circ\) & \(13.68^\circ\) \\
\(\mathrm{LR}_{\mathrm{CD}}\)
& \(0.108\) & \(0.075\) & \(\mathbf{0.152}\) & \(0.072\) \\
\hline
\end{tabular}%
}
\end{table}

\begin{table}[ht]
\centering
\caption{Post-hoc activity calibration and oracle-\gls{doa} diagnostics on the STARSS23 test split. The selected threshold minimizes the post-hoc \gls{seld} score independently for each checkpoint.}
\label{tab:stage4_doa_oracle_results}
\footnotesize
\renewcommand{\arraystretch}{1.07}
\setlength{\tabcolsep}{3.0pt}
\resizebox{\textwidth}{!}{%
\begin{tabular}{|p{4.25cm}|c|c|c|c|}
\hline
\textbf{Diagnostic}
& \textbf{ExpA}
& \textbf{ExpB}
& \textbf{ExpB \(\lambda_{\mathrm{SED}}=.70\)}
& \textbf{ExpC} \\
\hline
Best post-hoc threshold
& \(0.30\) & \(0.50\) & \(0.50\) & \(0.40\) \\
Tuned \gls{seld}
& \(\mathbf{0.607}\) & \(0.663\) & \(0.665\) & \(0.653\) \\
\(F_{20^\circ}\) at selected threshold
& \(\mathbf{0.247}\) & \(0.196\) & \(0.175\) & \(0.180\) \\
Prediction/target ratio
& \(1.015\) & \(0.573\) & \(0.641\) & \(0.913\) \\
\hline
Oracle median angular error
& \(\mathbf{27.24^\circ}\) & \(29.61^\circ\)
& \(34.33^\circ\) & \(32.53^\circ\) \\
Oracle accuracy within \(20^\circ\)
& \(\mathbf{33.67\%}\) & \(31.78\%\)
& \(25.34\%\) & \(29.32\%\) \\
Median active-vector norm
& \(0.735\) & \(0.761\) & \(\mathbf{0.786}\) & \(0.720\) \\
Active norm \(<0.1\)
& \(2.49\%\) & \(0.88\%\) & \(\mathbf{0.53\%}\) & \(0.61\%\) \\
Median predicted velocity
& \(\mathbf{0.174}\) & \(0.211\) & \(0.278\) & \(0.191\) \\
\hline
\end{tabular}%
}
\end{table}

Tables~\ref{tab:stage4_doa_loss_results} and~\ref{tab:stage4_doa_oracle_results} show that \texttt{ExpA} remains the strongest formulation on STARSS23. Relative to the legacy focal configuration, active-only MSE increases the median active-vector norm from \(0.224\) to \(0.735\), reduces the fraction of near-null active vectors from \(31.49\%\) to \(2.49\%\), and lowers the oracle median angular error from \(50.49^\circ\) to \(27.24^\circ\). These changes confirm that inactive-slot dominance was a primary cause of the localization suppression observed with all-entry MSE.

\texttt{ExpB} further reduces near-null predictions to \(0.88\%\) and increases the median active norm to \(0.761\). However, this does not produce more accurate directions: oracle median angular error increases to \(29.61^\circ\), accuracy within \(20^\circ\) decreases to \(31.78\%\), and the STARSS23 test \gls{seld} score rises to \(0.663\). The cosine--norm decomposition therefore improves vector-magnitude regularity without outperforming active-only Cartesian MSE in angular or end-to-end \gls{seld} terms. Better norm calibration is not, by itself, a reliable proxy for localization accuracy.

\texttt{ExpC} obtains a slightly better validation score than \texttt{ExpB}, \(0.385\) versus \(0.393\), but this advantage does not transfer to STARSS23 test behavior. Its oracle median error reaches \(32.53^\circ\), and its test \gls{seld} score reaches \(0.682\). The temporal term reduces median predicted velocity relative to \texttt{ExpB}, from \(0.211\) to \(0.191\), but it does not improve upon \texttt{ExpA}, whose median predicted velocity is \(0.174\). The selected smoothness coefficient therefore provides a measurable but insufficient regularization effect, and may also penalize legitimate source motion or incorrect track associations.

The increased-\(\lambda_{\mathrm{SED}}\) \texttt{ExpB} variant does not recover the activity--localization balance. At its configured activity threshold of \(0.30\), the prediction/target ratio becomes too high, producing \(\mathrm{ER}_{20^\circ}=1.528\) and a test \gls{seld} score of \(0.823\). Post-hoc analysis instead selects \(0.50\), improving the tuned score to \(0.665\), but the model remains weaker than \texttt{ExpA}. This demonstrates that the \(0.30\) operating point identified for \texttt{ExpA} is checkpoint-dependent and cannot be transferred directly to configurations trained with different loss balances.

Overall, the activity-conditioned \gls{doa} sequence separates three properties that should not be conflated: suppression of null-vector collapse, angular accuracy, and temporal stability. \texttt{ExpA}, based on active-only Cartesian MSE, provides the best compromise because it directly corrects the supervision imbalance while preserving a simple regression geometry. \texttt{ExpB} improves vector-norm behavior but not angular accuracy. \texttt{ExpC} partially moderates the additional temporal variability introduced by \texttt{ExpB}, but does not surpass \texttt{ExpA}. Accordingly, active-only Cartesian MSE is retained as the reference localization objective for the subsequent threshold-calibration and cross-dataset analyses.


\subsection{Operating-Point Calibration and Threshold Sensitivity}
\label{subsec:diagnostic_threshold_calibration}

The activity-conditioned \gls{doa} experiments identify \texttt{ExpA} as the strongest localization-supervision formulation, but the resulting system still depends on the event-activity threshold used to convert logits into binary decisions. This operating point is not a secondary implementation detail: in track-wise \gls{seld}, thresholding controls the number of active hypotheses exposed to Hungarian matching, and therefore affects \(\mathrm{ER}_{20^\circ}\), \(F_{20^\circ}\), \(\mathrm{LR}_{\mathrm{CD}}\), and the aggregate \gls{seld} score (Section~\ref{subsec:exp_supervision_objective}). Threshold sensitivity is also data-dependent. Different classes have different priors, temporal supports, acoustic morphologies, and spatial regularities, so a single threshold may under-activate rare but recognizable events while over-activating ambiguous transient classes. A calibration analysis therefore separates two uses of thresholding: \emph{(I)} test-side sweeps are used only as diagnostics to understand trained model sensitivity; \emph{(II)} validation-selected thresholds are used as leakage-free operating policies.

\paragraph{Diagnostic threshold sweeps.}

For a global activity threshold \(\tau\), a class-track prediction is considered active when:
\begin{equation}
\sigma
\left(
\hat{Z}_{b,t,n,c}
\right)
\geq
\tau
\end{equation}
where \(\hat{Z}_{b,t,n,c}\) is the activity logit for batch item \(b\), frame \(t\), track \(n\), and class \(c\), while \(\sigma(\cdot)\) is the sigmoid activation. A test-side threshold sweep evaluates:
\begin{equation}
\tau_{\mathrm{SELD}}^{\star}
=
\arg\min_{\tau\in\mathcal{T}}
\mathrm{SELD}_{\mathrm{test}}(\tau)
\end{equation}
where \(\mathcal{T}\) is the tested threshold grid and \(\tau_{\mathrm{SELD}}^{\star}\) is the threshold minimizing the test \gls{seld} score. For comparison, a separate event-detection operating point is computed as:
\begin{equation}
\tau_{\mathrm{SED}}^{\star}
=
\arg\max_{\tau\in\mathcal{T}}
F_{1,\mathrm{SED}}^{\mathrm{test}}(\tau)
\end{equation}
where \(F_{1,\mathrm{SED}}^{\mathrm{test}}\) is the frame-level \gls{sed} F1 score after collapsing the track dimension. These two thresholds are intentionally kept distinct because a threshold that maximizes event-activity F1 is not necessarily optimal for location-aware \gls{seld} evaluation.

\begin{table*}[ht]
\centering
\caption{Test-side diagnostic threshold sensitivity of \gls{at2seld} models on STARSS23. The two optimal thresholds are selected independently according to the test \gls{seld} score and the frame-level \gls{sed} F1 score.}
\label{tab:nas_sed_threshold_diagnostic}
\footnotesize
\setlength{\tabcolsep}{3.5pt}
\renewcommand{\arraystretch}{1.05}
\resizebox{\textwidth}{!}{%
\begin{tabular}{|p{3.25cm}|ccccc|cc|}
\hline
\textbf{Model} &
\(\boldsymbol{\tau_{\mathrm{SELD}}^\star}\) &
\textbf{SELD} &
\(\boldsymbol{F_{20^\circ}}\) &
\(\boldsymbol{LR_{\mathrm{CD}}}\) &
\textbf{Pred./Target} &
\(\boldsymbol{\tau_{\mathrm{SED}}^\star}\) &
\textbf{SED F1} \\
\hline
\rowcolor{gray!15}
\texttt{ExpA}              & 0.30 & \textbf{0.607} & \textbf{0.247} & \textbf{0.164} & 1.015 & 0.30 & 0.704 \\
\texttt{ExpB}              & 0.50 & 0.663 & 0.196 & 0.075 & 0.573 & 0.30 & 0.723 \\
\texttt{ExpB-rw}           & 0.50 & 0.665 & 0.175 & 0.096 & 0.641 & 0.30 & 0.698 \\
\texttt{ExpC}              & 0.40 & 0.653 & 0.180 & 0.110 & 0.913 & 0.30 & 0.718 \\
\texttt{ExpA-posw}         & 0.50 & 0.648 & 0.167 & 0.147 & 0.967 & 0.40 & 0.728 \\
\texttt{mixed\_focal}      & 0.40 & 0.661 & 0.173 & 0.092 & 0.895 & 0.40 & 0.758 \\
\texttt{mixed\_bce}        & 0.10 & 0.696 & 0.109 & 0.077 & 0.972 & 0.30 & \textbf{0.781} \\
\hline
\end{tabular}%
}

\vspace{0.1cm}
\parbox{0.96\linewidth}{\centering\footnotesize
Thresholds are selected directly on the test sweep and are therefore reported only as diagnostic operating points. \textbf{Pred./Target} denotes the ratio between the number of predicted active entries and the number of reference active entries. \textbf{SED F1} is computed at model-frame resolution and is not an official DCASE segment-based score.
}
\end{table*}

Table~\ref{tab:nas_sed_threshold_diagnostic} shows that \texttt{ExpA} remains the best joint operating point among the examined configurations. At the diagnostic threshold \(\tau_{\mathrm{SELD}}^\star=0.30\), it reaches a test \gls{seld} score of \(0.607\), \(F_{20^\circ}=0.247\), and \(\mathrm{LR}_{\mathrm{CD}}=0.164\). Its predicted-to-target activity ratio is \(1.015\), indicating that the selected threshold almost exactly compensates for the under-activation observed at the conventional \(\tau=0.50\).

The comparison also confirms that isolated activity quality is not sufficient for \gls{seld} selection. The legacy \texttt{mixed\_bce} and \texttt{mixed\_focal} checkpoints reach stronger frame-level \gls{sed} F1 scores than \texttt{ExpA}, with \(0.781\) and \(0.758\), respectively. However, their best location-aware F-scores remain lower, with \(F_{20^\circ}=0.109\) for \texttt{mixed\_bce} and \(0.173\) for \texttt{mixed\_focal}. Thus, a model can correctly detect event activity while still failing to produce spatially valid hypotheses.

The threshold statistics explain the global behavior of \texttt{ExpA}. Only a small fraction of the frame--track--class tensor corresponds to active targets, and this sparsity is not uniform across classes. At the fixed threshold \(0.50\), the model predicts approximately \(56.4\%\) of the reference activity count. The retained detections are relatively precise, but recall is limited, especially for classes whose posterior distributions are shifted downward by low coverage, short event duration, or ambiguous acoustic evidence. Lowering the threshold from \(0.50\) to \(0.30\) increases the predicted-to-target ratio from \(0.564\) to \(1.015\). Over the same transition, frame-level \gls{sed} recall increases from \(0.491\) to \(0.671\), while precision decreases from \(0.865\) to \(0.742\). The \gls{seld} score improves from \(0.635\) to \(0.607\), \(\mathrm{ER}_{20^\circ}\) decreases from \(0.816\) to \(0.766\), and \(F_{20^\circ}\) increases from \(0.235\) to \(0.247\).

The useful calibration region is nevertheless narrow. Thresholds below \(0.30\) lead to over-activation: at \(\tau=0.20\) predicted activity reaches approximately \(1.49\times\) the reference activity and the \gls{seld} score worsens to \(0.713\); at \(\tau=0.10\), the prediction ratio reaches \(2.88\times\). This analysis therefore identifies under-confidence as a relevant operating-point issue, while also showing that lowering the threshold is beneficial only when the additional detections remain sufficiently localized and do not introduce excessive false positives.

\paragraph{Validation-selected calibration.}

The previous sweep is diagnostic because the threshold is analyzed and selected on the test split. A deployment-compatible protocol should instead select thresholds on validation data and then freeze them before test evaluation. The global validation threshold is defined as:
\begin{equation}
\tau_g^\star
=
\arg\max_{\tau\in\mathcal{T}}
\frac{1}{C}
\sum_{c=1}^{C}
F_{1,c}^{\mathrm{val}}(\tau)
\end{equation}
where \(F_{1,c}^{\mathrm{val}}(\tau)\) is the validation F1 score for class \(c\), \(C\) is the number of classes, and \(\tau_g^\star\) is the scalar threshold maximizing validation macro-F1. Class-specific calibration instead estimates one threshold per class:
\begin{equation}
\tau_c^\star
=
\arg\max_{\tau\in\mathcal{T}}
F_{1,c}^{\mathrm{val}}(\tau)
\end{equation}
where \(\tau_c^\star\) is the validation-selected threshold for class \(c\). The selected thresholds are then applied unchanged to the STARSS23 test split.

\begin{table}[ht]
\centering
\caption{Validation-calibrated strategy comparison, evaluated with \texttt{ExpA} and on the STARSS23 test split.}
\label{tab:nas_expa_calibration}
\footnotesize
\setlength{\tabcolsep}{3.5pt}
\renewcommand{\arraystretch}{1.05}
\resizebox{\linewidth}{!}{%
\begin{tabular}{|p{3.4cm}|ccc|c|ccc|}
\hline
\textbf{Strategy} &
\textbf{M-Precision} &
\textbf{M-Recall} &
\textbf{M-F1} &
\textbf{m-F1} &
\(\boldsymbol{ER_{20^\circ}}\) &
\(\boldsymbol{F_{20^\circ}}\) &
\textbf{SELD} \\
\hline
Fixed \(\tau=0.5\)        & 0.694 & 0.293 & 0.375 & 0.626 & 0.816 & 0.235 & 0.635 \\
Global validation         & 0.570 & 0.421 & 0.435 & 0.697 & 0.768 & \textbf{0.249} & \textbf{0.610} \\
Class-specific validation & 0.609 & \textbf{0.443} & \textbf{0.441} & \textbf{0.707} & 0.768 & 0.242 & 0.611 \\
Test class oracle         & 0.533 & 0.487 & 0.480 & 0.710 & 1.035 & 0.227 & 0.680 \\
\hline
\end{tabular}%
}
\end{table}

The global validation optimum is \(\tau_g^\star=0.35\). When applied to the test split, this threshold obtains a \gls{seld} score of \(0.610\), closely reproducing the diagnostic test-side optimum of \(0.607\) without using test information for calibration. The class-specific strategy further improves macro-F1, from \(0.375\) at the fixed operating point to \(0.441\), while preserving essentially the same \gls{seld} score as the global strategy, \(0.611\). This is a positive result: under validation-selected thresholds, class-dependent calibration improves the activity-side balance without producing an appreciable degradation of location-aware matching. At the same time, the absence of a further \gls{seld} gain indicates that the remaining errors are not only threshold-dependent, but also involve spatial accuracy.

The selected threshold vector is:
\begin{equation}
\boldsymbol{\tau}^{\star}_c
=
[
0.35,
0.40,
0.25,
0.35,
0.35,
0.30,
0.35,
0.50,
0.30,
0.15,
0.55,
0.35,
0.20
]
\end{equation}
where the ordering follows the STARSS23 label space reported in Table~\ref{tab:stage4_balanced_windows}. The selected thresholds should not be interpreted only as model-tuning parameters. They also reflect class-dependent data and signal properties. Speech, domestic sounds, and music-related categories remain close to the global threshold, suggesting that their posterior distributions are sufficiently stable under the balanced training regime. By contrast, \texttt{musical\_instrument} and \texttt{knock} require lower thresholds, indicating that useful detections for these classes tend to be produced with lower confidence. This may result from limited coverage, short temporal support, or higher intra-class variability. In contrast, \texttt{water\_tap} selects \(0.55\), showing that lowering its threshold mainly introduces false positives rather than recovering reliable events. Its calibration behavior is therefore consistent with an acoustically ambiguous class whose additional low-confidence activations are not sufficiently discriminative or spatially reliable.

\begin{table}[ht]
\centering
\caption{Validation-calibrated \gls{sed} threshold comparison on the STARSS23 test split.}
\label{tab:sed_threshold_strategy_comparison}
\footnotesize
\setlength{\tabcolsep}{3.6pt}
\renewcommand{\arraystretch}{1.05}
\resizebox{\textwidth}{!}{%
\begin{tabular}{|p{3.0cm}|c|cc|cc|cc|}
\hline
\textbf{Model} &
\(\boldsymbol{\tau_g}\) &
\multicolumn{2}{c|}{\textbf{Fixed \(\tau=0.5\)}} &
\multicolumn{2}{c|}{\textbf{Global validation}} &
\multicolumn{2}{c|}{\textbf{Class-specific validation}} \\
\cline{3-8}
& &
\textbf{Macro-F1} &
\textbf{SELD} &
\textbf{Macro-F1} &
\textbf{SELD} &
\textbf{Macro-F1} &
\textbf{SELD} \\
\hline
\texttt{ExpA}         & 0.35 & 0.375 & 0.635 & 0.435 & \textbf{0.610} & \textbf{0.441} & 0.611 \\
\texttt{mixed\_focal} & 0.45 & 0.570 & 0.666 & \textbf{0.579} & \textbf{0.663} & 0.564 & 0.665 \\
\hline
\end{tabular}%
}

\vspace{0.1cm}
\parbox{0.96\linewidth}{\centering\footnotesize
\(\tau_g\) is the scalar threshold maximizing validation macro-F1. \textbf{SELD} values are computed after applying the corresponding validation-selected threshold policy to the test split.
}
\end{table}

Table~\ref{tab:sed_threshold_strategy_comparison} shows that calibration has different effects depending on the checkpoint. For \texttt{ExpA}, global validation calibration provides most of the achievable \gls{seld} gain, while class-specific calibration mainly improves macro-F1. For \texttt{mixed\_focal}, calibration produces only minor changes, because the checkpoint is already closer to its validation-selected global threshold. This confirms that threshold calibration is most useful when the model is globally under-confident, as in \texttt{ExpA} after activity-conditioned \gls{doa} supervision.

A more detailed appendix-level analysis reports the per-class thresholds obtained when the validation criterion is changed from F1 to Precision and Recall. Tables~\ref{tab:sed_thresholds_precision}--\ref{tab:sed_thresholds_f1} show that these criteria induce substantially different operating regimes. Precision-oriented thresholds are systematically conservative and suppress Recall, whereas recall-oriented thresholds activate many low-confidence events and strongly increase false-positive pressure. The F1-oriented policy provides the most balanced activity-side calibration and is therefore the only class-specific thresholding strategy retained here.

The class-level behavior remains heterogeneous. The largest test macro-F1 improvements over \(\tau=0.5\) are obtained for \texttt{musical\_instrument} \((0.255\rightarrow0.498)\), \texttt{bell} \((0.105\rightarrow0.312)\), \texttt{music} \((0.681\rightarrow0.795)\), \texttt{walk\_footsteps} \((0.242\rightarrow0.327)\), and \texttt{female\_speech} \((0.648\rightarrow0.726)\). However, class-specific calibration is not uniformly reliable. For \texttt{clapping}, the validation-selected threshold \(0.25\) increases recall to \(0.751\) but reduces precision to \(0.169\), decreasing test F1 from \(0.398\) to \(0.276\). \texttt{knock} remains unstable, with precision \(0.034\), recall \(0.077\), and F1 \(0.047\).

The residual errors of the weakest classes cannot be interpreted only as missed activity decisions. At the globally calibrated threshold, frame-level \gls{sed} F1 reaches \(0.824\) for \texttt{male\_speech}, \(0.806\) for \texttt{music}, \(0.753\) for \texttt{domestic\_sounds}, and \(0.735\) for \texttt{female\_speech}. The weakest frame-level classes are \texttt{knock} \((0.047)\), \texttt{water\_tap} \((0.141)\), \texttt{telephone} \((0.278)\), and \texttt{door\_open\_close} \((0.308)\). More importantly, several classes with non-zero activity F1 remain near zero in location-dependent F1. For \texttt{water\_tap}, \texttt{knock}, and \texttt{bell}, this indicates that detections, when produced, are often outside the \(20^\circ\) matching tolerance.

Overall, threshold calibration identifies a useful global operating region around \(0.30\)--\(0.35\) for \texttt{ExpA}. Validation-selected calibration recovers most of the test-side optimum without leakage and should therefore be preferred over fixed default thresholding. However, calibration is an operating-point correction, not a substitute for spatial learning or data coverage. It can recover under-confident activity predictions, especially for classes whose useful detections are produced below the default threshold, but it cannot convert semantically detected events into correct \gls{seld} matches when their \gls{doa} estimates remain inaccurate, when low-confidence activations are dominated by false positives, or when the class remains poorly represented in the training data.


\subsection{Cross-Dataset Evaluation}
\label{subsec:diagnostic_cross_dataset}

The previous diagnostics evaluated the selected \gls{at2seld} family under STARSS23-centered data balancing, activity-loss calibration, activity-conditioned \gls{doa} supervision, and class-specific threshold selection. This culminating analysis extends the characterization across heterogeneous spatial-audio domains. The objective is to determine whether the selected operating point learns a dataset-specific solution or retains transferable semantic and spatial behavior across real recordings, fixed-position synthetic scenes, and moving-source synthetic mixtures.

Unless otherwise specified, the STARSS23 checkpoint corresponds to \texttt{ExpA}, i.e., the late-interaction \gls{at2seld} topology trained with focal activity loss and activity-conditioned Cartesian \gls{doa} supervision (Sections~\ref{subsec:diagnostic_activity_loss} and~\ref{subsec:diagnostic_active_doa}). The comparison considers four independently trained checkpoints using the same \gls{at2seld} architecture: \texttt{BalancedSTARSS23}, TAU2019, TAU-NIGENS2020, and TAU-NIGENS2021. Each checkpoint is evaluated on all four datasets, producing a \(4\times4\) model--dataset matrix in which diagonal entries measure native-domain performance and off-diagonal entries measure label-compatible transfer.

\subsubsection{Projection Protocol and Computational Profile}
\label{subsubsec:cross_dataset_protocol_profile}

All evaluations use the \(10~\mathrm{s}\) input window and \(2.5~\mathrm{s}\) hop adopted throughout this work. Inference is executed once for each model--dataset--split pair; activity logits, Cartesian \gls{doa} predictions, and target tensors are then cached and reused for validation-selected threshold sweeps. This avoids recomputing the forward pass for each operating point and ensures that differences between policies are caused by calibration rather than stochastic inference variability.

Cross-dataset evaluation requires a coverage-aware label projection. Let \(\mathcal{M}_{s\rightarrow m}(c)\) denotes the set of model-output classes associated with source-dataset class \(c\), where \(s\) denotes the evaluation dataset and \(m\) denotes the checkpoint label space. Exact semantic correspondences are used directly, while the fixed mappings introduced for \texttt{BalancedSTARSS23} (Section~\ref{subsec:diagnostic_balanced_starss}) are retained where applicable. Unmappable source events are discarded before scoring, and output classes without source-dataset coverage are suppressed. Consequently, each off-diagonal result is coverage-aware: it measures transfer over the compatible semantic subset and must not be interpreted as a full label-space comparison.

\begin{table}[ht]
\centering
\caption{Coverage of the label-space projections used in the cross-dataset evaluation.}
\label{tab:dataset_generalization_mapping_coverage}
\footnotesize
\renewcommand{\arraystretch}{1.05}
\setlength{\tabcolsep}{3.2pt}
\resizebox{\textwidth}{!}{%
\begin{tabular}{|l|l|c|c|}
\hline
\textbf{Training dataset} & \textbf{Evaluation dataset} & \textbf{Mapped input classes} & \textbf{Covered output classes} \\
\hline
STARSS23 & STARSS23 & 13/13 & 13/13 \\
STARSS23 & TAU2019 & 8/11 & 5/13 \\
STARSS23 & TAU-NIGENS2020 & 8/14 & 7/13 \\
STARSS23 & TAU-NIGENS2021 & 8/12 & 7/13 \\
\hline
TAU2019 & STARSS23 & 6/13 & 5/11 \\
TAU2019 & TAU2019 & 11/11 & 11/11 \\
TAU2019 & TAU-NIGENS2020 & 4/14 & 3/11 \\
TAU2019 & TAU-NIGENS2021 & 4/12 & 3/11 \\
\hline
TAU-NIGENS2020 & STARSS23 & 6/13 & 6/14 \\
TAU-NIGENS2020 & TAU2019 & 2/11 & 2/14 \\
TAU-NIGENS2020 & TAU-NIGENS2020 & 14/14 & 14/14 \\
TAU-NIGENS2020 & TAU-NIGENS2021 & 12/12 & 12/14 \\
\hline
TAU-NIGENS2021 & STARSS23 & 6/13 & 6/12 \\
TAU-NIGENS2021 & TAU2019 & 2/11 & 2/12 \\
TAU-NIGENS2021 & TAU-NIGENS2020 & 12/14 & 12/12 \\
TAU-NIGENS2021 & TAU-NIGENS2021 & 12/12 & 12/12 \\
\hline
\end{tabular}%
}
\end{table}

The projection table is essential for interpreting off-diagonal results. For example, TAU2019 evaluated on STARSS23 covers only a subset of STARSS23 classes and merges male and female speech into a single \texttt{speech} output. Similarly, NIGENS checkpoints evaluated on STARSS23 cover speech, piano-related material, footsteps, telephone, and knock-related events, but not the complete STARSS23 taxonomy. Transfer results are therefore meaningful as evidence of semantic--spatial compatibility over shared label spaces.

The cross-dataset results should also be interpreted in relation to model capacity. The complete \gls{at2seld} graph contains approximately \(31\)M parameters because it includes the pretrained semantic \gls{gpat} branch (Section~\ref{sec:nas_results}). However, the task-facing \gls{seld} path is much smaller. The pretrained \gls{epanns} branch contains approximately \(27.47\)M parameters, whereas the NAS-derived track-wise \gls{seld} path contains approximately \(2.93\)M parameters. Including the late cross-stitch bridge, the effective task-facing \gls{seld} component contains approximately \(3.51\)M parameters, corresponding to about \(11.3\%\) of the processing graph.

\begin{table}[ht]
\centering
\caption{\Gls{at2seld} profiling for one \(10~\mathrm{s}\) \gls{foa} input signal. Latency is measured on \gls{mps} acceleration.}
\label{tab:dataset_generalization_profile}
\footnotesize
\renewcommand{\arraystretch}{1.05}
\setlength{\tabcolsep}{3.2pt}
\begin{tabular}{|l|c|c|c|c|c|}
\hline
\textbf{Checkpoint} &
\textbf{Total Param.} &
\textbf{SELD Param.} &
\textbf{SELD Share} &
\textbf{Total Latency} &
\textbf{SELD Latency} \\
\hline
STARSS23        & \(30.99\)M & \(3.511\)M & \(11.33\%\) & \(170.9\) ms & \(164.5\) ms \\
TAU2019         & \(30.98\)M & \(3.505\)M & \(11.31\%\) & \(177.1\) ms & \(171.0\) ms \\
TAU-NIGENS2020  & \(30.99\)M & \(3.514\)M & \(11.34\%\) & \(172.0\) ms & \(165.5\) ms \\
TAU-NIGENS2021  & \(30.98\)M & \(3.508\)M & \(11.32\%\) & \(177.7\) ms & \(171.2\) ms \\
\hline
\end{tabular}
\end{table}

Table~\ref{tab:dataset_generalization_profile} shows that the selected family is compact on the task-facing side. The spatial pathway is not a large challenge-oriented \gls{seld} backbone; it is a reduced semantic-to-spatial branch assisted by a pretrained \gls{at} representation. The measured latency corresponds to a real-time factor of approximately \(0.017\) for a \(10~\mathrm{s}\) input, or roughly \(58\)--\(61\times\) faster than real time on the profiled device. Parameter efficiency and arithmetic cost should nevertheless be distinguished: the high-resolution spectral and spatial operations dominate the estimated \gls{gflops}, so future optimization should target the spatial front-end and feature-processing stages.

\subsubsection{Native-Domain and Cross-Domain Performance}
\label{subsubsec:cross_dataset_performance}

Native-domain results are reported in Table~\ref{tab:dataset_generalization_native}. For each checkpoint, the global activity threshold is selected on validation to minimize the \gls{seld} score at the standard \(20^\circ\) angular tolerance and then applied unchanged to the corresponding test split.

\begin{table}[ht]
\centering
\caption{In-domain \gls{at2seld} performance.}
\label{tab:dataset_generalization_native}
\footnotesize
\renewcommand{\arraystretch}{1.05}
\setlength{\tabcolsep}{4.0pt}
\resizebox{\textwidth}{!}{%
\begin{tabular}{|l|c|c|c|c||c|c|c||c|}
\hline
\textbf{Dataset} &
\(\boldsymbol{\tau_{\mathrm{SELD}}}\) &
\(\mathbf{P^{SED}}\) &
\(\mathbf{R^{SED}}\) &
\(\mathbf{F_1^{SED}}\) &
\(\mathbf{F_{20}}\) &
\textbf{LE} &
\textbf{LR} &
\textbf{SELD} \\
\hline
TAU2019        & \(0.50\) & \(0.956\) & \(0.867\) & \(\mathbf{0.910}\) & \(\mathbf{0.860}\) & \(\mathbf{6.55^\circ}\) & \(\mathbf{0.792}\) & \(\mathbf{0.142}\) \\
TAU-NIGENS2021 & \(0.40\) & \(0.648\) & \(0.572\) & \(0.607\) & \(0.355\) & \(11.67^\circ\) & \(0.348\) & \(0.505\) \\
TAU-NIGENS2020 & \(0.45\) & \(0.811\) & \(0.559\) & \(0.662\) & \(0.325\) & \(11.91^\circ\) & \(0.289\) & \(0.541\) \\
STARSS23       & \(0.45\) & \(0.842\) & \(0.539\) & \(0.657\) & \(0.242\) & \(12.48^\circ\) & \(0.122\) & \(0.626\) \\
\hline
\end{tabular}%
}
\end{table}

TAU2019 is the most favorable native operating regime. Its \(\mathrm{SED}\ F_1\) of \(0.910\), \(F_{20}=0.860\), localization error of \(6.55^\circ\), localization recall of \(0.792\), and aggregate \gls{seld} score of \(0.142\) indicate that the compact \gls{at2seld} branch can solve regular spatial scenes very effectively. This result is consistent with the controlled structure of TAU2019, where source positions are mostly fixed, maximum polyphony is lower than trajectory-based corpora, and the class distribution is more regular.

NIGENS datasets occupy an intermediate regime. TAU-NIGENS2021 obtains the better native \gls{seld} score, \(0.505\), and higher localization recall, \(0.348\), whereas TAU-NIGENS2020 obtains higher \(\mathrm{SED}\ F_1\), \(0.662\) versus \(0.607\). This divergence confirms a recurring diagnostic theme: stronger event-activity detection does not necessarily imply stronger spatially grounded detection. The 2021 checkpoint preserves a larger fraction of correctly associated and localized events, despite a weaker aggregate activity F1.

STARSS23 remains the most demanding native domain. Its \(\mathrm{SED}\ F_1\) is comparable to TAU-NIGENS2020, and the localization error of successful matches remains close to the NIGENS values. The main degradation is instead localization recall, which falls to \(0.122\). The model therefore does not simply produce imprecise directions; rather, it retains fewer class-consistent spatial matches under real-scene variability. This behavior is consistent with the smaller in-domain training set, long-tailed class structure, reverberant and non-stationary scenes, and higher annotated polyphony relative to the three-tracks output capacity adopted.

\begin{figure}[ht]
    \centering
    \includegraphics[width=\linewidth]{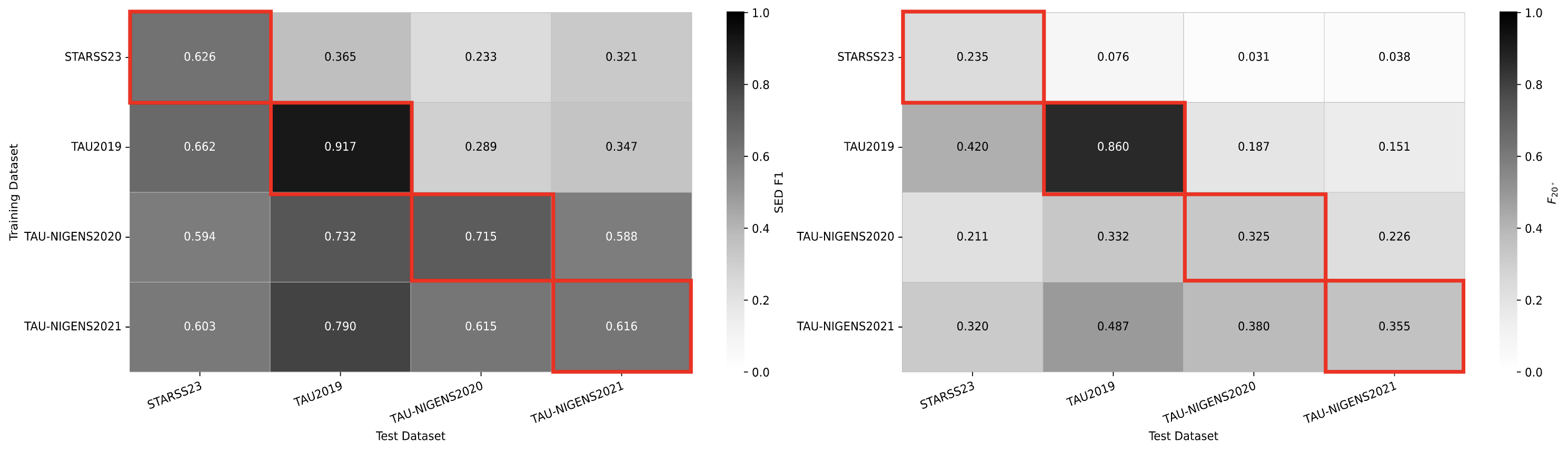}
    \caption{Coverage-aware \gls{sed} and \gls{seld} F1 scores for the selected \gls{at2seld} family. Off-diagonal cells are evaluated only over compatible classes according to Table~\ref{tab:dataset_generalization_mapping_coverage}.}
    \label{fig:dataset_generalization_seld}
\end{figure}

Figure~\ref{fig:dataset_generalization_seld} summarizes the full coverage-aware transfer matrix. The diagonal cells should be read as in-domain performance, whereas the off-diagonal cells quantify transfer only over mapped semantic subsets. The most reliable transfer occurs between TAU-NIGENS generations, which share most classes and related rendering assumptions. The TAU-NIGENS2021 checkpoint evaluated on TAU-NIGENS2020 reaches \(\mathrm{SED}\ F_1=0.615\), \(F_{20}=0.380\), and \(\mathrm{SELD}=0.485\), outperforming the native TAU-NIGENS2020 checkpoint under the same aggregate criterion. The reverse transfer is weaker, with \(\mathrm{SED}\ F_1=0.529\), \(F_{20}=0.226\), and \(\mathrm{SELD}=0.603\). This asymmetry indicates that the 2021 training distribution provides more transferable representation over the previous dataset generation.

The per-class diagnostics in Appendix~\ref{app:sed_perf_appendix}, Table~\ref{tab:dataset_generalization_class_policy}, support this interpretation: for TAU-NIGENS2021\(\rightarrow\)TAU-NIGENS2020, strong transfer is observed for \texttt{female\_speech}, \texttt{baby\_cry}, and \texttt{crash}, whereas \texttt{telephone} and \texttt{female\_scream} remain weaker. In the reverse direction, TAU-NIGENS2020 transfers well for \texttt{crash}, \texttt{alarm}, and \texttt{baby\_cry}, but remains less stable for \texttt{piano} and \texttt{female\_scream}. These differences indicate that transfer is not determined only by nominal label overlap; it also depends on rendering assumptions, event morphology, motion statistics, and class-specific acoustic variability.

TAU2019 remains a favorable evaluation domain even for non-native checkpoints. The TAU-NIGENS2021 checkpoint evaluated on TAU2019 reaches \(\mathrm{SED}\ F_1=0.773\), \(F_{20}=0.497\), and \(\mathrm{SELD}=0.427\) over the mapped classes, while the TAU-NIGENS2020 checkpoint obtains \(0.651/0.332/0.551\). These cells cover only \texttt{telephone} and \texttt{knock}; they therefore show that fixed-position spatial conditions support good transfer for events learned from moving-source synthetic regimes. Within this label space, TAU-NIGENS2021 is particularly effective on \texttt{telephone}.

The apparently strong TAU2019\(\rightarrow\)STARSS23 transfer should be interpreted with the same coverage constraint. The aggregate values, \(\mathrm{SED}\ F_1=0.679\), \(F_{20}=0.420\), and \(\mathrm{SELD}=0.505\), are dominated by the projection of STARSS23 male and female speech into the single TAU2019 \texttt{speech} output. The result is still technically meaningful: it demonstrates robust speech-direction transfer from a fixed-source synthetic domain to real spatial recordings.

The NIGENS checkpoints evaluated on STARSS23 preserve speech localization reasonably well. TAU-NIGENS2021 obtains \(F_{\mathrm{SELD}}=0.595\) for \texttt{male\_speech} and \(0.580\) for \texttt{female\_speech}; TAU-NIGENS2020 obtains \(0.535\) and \(0.381\), respectively. By contrast, \texttt{footsteps} transfers poorly despite nominal label correspondence. This shows once again that semantic compatibility does not guarantee acoustic or spatial-domain compatibility.

Overall, the cross-domain results support a positive but bounded interpretation: the selected \gls{at2seld} family is not merely fitting one dataset; it transfers meaningfully across label-compatible spatial-audio regimes, especially between related synthetic domains and for speech-related events. At the same time, real-scene transfer remains class-dependent, and the coverage-aware protocol prevents overinterpreting partial label overlap as full dataset generalization.

\subsubsection{Threshold Policies and Angular Tolerance}
\label{subsubsec:cross_dataset_thresholds}

Three scalar activity thresholds are selected independently on validation: \(\tau_{\mathrm{SED}}\), which maximizes frame-level \gls{sed} F1; \(\tau_{F_{20}}\), which maximizes the location-aware \(F_{20^\circ}\) score; and \(\tau_{\mathrm{SELD}}\), which minimizes the aggregate \(\mathrm{SELD}\) score. The selected thresholds are then applied unchanged to the test split:
\begin{equation}
\tau_{\mathrm{SED}}
=
\arg\max_{\tau\in\mathcal{T}}
F_{1}^{\mathrm{SED}}(\tau),
\qquad
\tau_{F_{20}}
=
\arg\max_{\tau\in\mathcal{T}}
F_{20^\circ}(\tau),
\qquad
\tau_{\mathrm{SELD}}
=
\arg\min_{\tau\in\mathcal{T}}
\mathrm{SELD}(\tau)
\end{equation}
where \(\mathcal{T}=\{0.05,0.10,\ldots,0.95\}\) is the validation threshold grid. These thresholds capture different operating goals and can therefore diverge, especially when event activity and localization reliability are not calibrated in the same way (Table~\ref{tab:dataset_generalization_global_policy}).

\begin{table}[ht]
\centering
\caption{Validation-selected global activity thresholds and angular-tolerance analysis.}
\label{tab:dataset_generalization_global_policy}
\footnotesize
\renewcommand{\arraystretch}{1.05}
\setlength{\tabcolsep}{3.0pt}
\resizebox{\textwidth}{!}{%
\begin{tabular}{|l|l||c|c|c|c||c|c|c|c|}
\hline
\textbf{Training model} & \textbf{Test dataset} & \(\tau_{\mathrm{SED}}\) & \(\tau_{F_{20}}\) & \(\tau_{\mathrm{SELD}}\) & \(\tau_{\mathrm{joint}}/\alpha\) & \(\mathrm{SED}\ F_1\) & \(F_{20}\) & \(\mathrm{SELD}_{20}\) & \(\mathrm{SELD}_{\mathrm{joint}}\) \\
\hline
STARSS23 & STARSS23 & 0.50 & 0.50 & 0.45 & 0.45/45$^\circ$ & 0.657 & 0.242 & 0.626 & 0.499 \\
STARSS23 & TAU2019 & 0.20 & 0.25 & 0.25 & 0.25/45$^\circ$ & 0.353 & 0.076 & 0.716 & 0.657 \\
STARSS23 & TAU-NIGENS2020 & 0.25 & 0.45 & 0.45 & 0.45/45$^\circ$ & 0.199 & 0.031 & 0.745 & 0.719 \\
STARSS23 & TAU-NIGENS2021 & 0.15 & 0.20 & 0.35 & 0.35/45$^\circ$ & 0.279 & 0.036 & 0.741 & 0.705 \\
\hline
TAU2019 & STARSS23 & 0.15 & 0.20 & 0.20 & 0.15/45$^\circ$ & 0.679 & 0.420 & 0.505 & 0.470 \\
TAU2019 & TAU2019 & 0.45 & 0.50 & 0.50 & 0.50/45$^\circ$ & 0.910 & 0.860 & 0.142 & 0.120 \\
TAU2019 & TAU-NIGENS2020 & 0.35 & 0.40 & 0.70 & 0.70/45$^\circ$ & 0.255 & 0.181 & 0.639 & 0.614 \\
TAU2019 & TAU-NIGENS2021 & 0.30 & 0.35 & 0.70 & 0.70/45$^\circ$ & 0.262 & 0.116 & 0.683 & 0.646 \\
\hline
TAU-NIGENS2020 & STARSS23 & 0.25 & 0.35 & 0.40 & 0.40/45$^\circ$ & 0.531 & 0.229 & 0.630 & 0.518 \\
TAU-NIGENS2020 & TAU2019 & 0.45 & 0.50 & 0.50 & 0.50/45$^\circ$ & 0.651 & 0.332 & 0.551 & 0.431 \\
TAU-NIGENS2020 & TAU-NIGENS2020 & 0.35 & 0.45 & 0.45 & 0.45/45$^\circ$ & 0.662 & 0.325 & 0.541 & 0.395 \\
TAU-NIGENS2020 & TAU-NIGENS2021 & 0.30 & 0.40 & 0.40 & 0.40/45$^\circ$ & 0.529 & 0.226 & 0.603 & 0.507 \\
\hline
TAU-NIGENS2021 & STARSS23 & 0.40 & 0.50 & 0.50 & 0.50/45$^\circ$ & 0.566 & 0.320 & 0.586 & 0.485 \\
TAU-NIGENS2021 & TAU2019 & 0.40 & 0.50 & 0.45 & 0.45/45$^\circ$ & 0.773 & 0.497 & 0.427 & 0.268 \\
TAU-NIGENS2021 & TAU-NIGENS2020 & 0.45 & 0.50 & 0.45 & 0.45/45$^\circ$ & 0.615 & 0.380 & 0.485 & 0.347 \\
TAU-NIGENS2021 & TAU-NIGENS2021 & 0.30 & 0.40 & 0.40 & 0.35/45$^\circ$ & 0.607 & 0.355 & 0.505 & 0.396 \\
\hline
\end{tabular}%
}
\end{table}

The global policies show that calibration is dataset-dependent. TAU2019 is already well calibrated around \(\tau=0.50\), which is consistent with its training performance and regular spatial structure. The NIGENS checkpoints tend to require lower \gls{sed}-oriented thresholds to recover activity recall, but higher \gls{seld}-oriented thresholds to limit false localized events. STARSS23 behaves similarly, although less strongly, with \(\tau_{\mathrm{SED}}=0.50\) and \(\tau_{\mathrm{SELD}}=0.45\). These differences reinforce the conclusion of Section~\ref{subsec:diagnostic_threshold_calibration}: calibration is not a universal scalar correction, but a property of the trained checkpoint, dataset prior, and class-specific detection reliability.

Angular tolerance is evaluated over:
\begin{equation}
\mathcal{A}
=
\{5^\circ,10^\circ,15^\circ,20^\circ,30^\circ,45^\circ\}
\end{equation}
where \(\mathcal{A}\) is the set of admissible angular tolerances tested during validation. The primary results retain the standard \(20^\circ\) tolerance. A second validation search jointly selects \((\tau,\alpha)\), where \(\alpha\in\mathcal{A}\), to diagnose coarse-directionality behavior. The \(45^\circ\) tolerance is used as a coarse-directionality diagnostic: by widening the angular tolerance, the analysis tests whether incorrect \(20^\circ\) matches are still concentrated around a spatially plausible directional sector, or whether they correspond to structurally wrong localization estimates. Improvements observed only at \(45^\circ\) should therefore be interpreted as evidence of coarse directional mapping.

\begin{table}[ht]
\centering
\caption{In-domain model sensitivity to angular tolerance.}
\label{tab:dataset_generalization_angular}
\footnotesize
\renewcommand{\arraystretch}{1.05}
\setlength{\tabcolsep}{4.0pt}
\begin{tabular}{|l|cc|cc|cc|}
\hline
\multirow{2}{*}{\textbf{Dataset}} &
\multicolumn{2}{c|}{\(\mathbf{20^\circ}\)} &
\multicolumn{2}{c|}{\(\mathbf{45^\circ}\)} &
\multicolumn{2}{c|}{\textbf{Difference}} \\
\cline{2-7}
& \(F_{20}\) & SELD & \(F_{45}\) & SELD &
\(\Delta F\) & \(\Delta\)SELD \\
\hline
TAU2019        & \(0.860\) & \(0.142\) & \(0.890\) & \(0.120\) & \(+0.031\) & \(-0.022\) \\
TAU-NIGENS2021 & \(0.355\) & \(0.505\) & \(0.515\) & \(0.396\) & \(+0.160\) & \(-0.109\) \\
TAU-NIGENS2020 & \(0.325\) & \(0.541\) & \(0.550\) & \(0.395\) & \(+0.225\) & \(-0.145\) \\
STARSS23       & \(0.242\) & \(0.626\) & \(0.480\) & \(0.499\) & \(+0.238\) & \(-0.128\) \\
\hline
\end{tabular}

\vspace{0.1cm}
\parbox{0.96\linewidth}{\centering\footnotesize
The \(20^\circ\) column uses the validation-selected \gls{seld} threshold.\\The \(45^\circ\) column uses the validation-selected joint threshold/tolerance policy.}
\end{table}

The small TAU2019 change confirms that most correct detections are already localized inside the \(20^\circ\) region. The larger gains for TAU-NIGENS2020, TAU-NIGENS2021, and STARSS23 indicate that many errors are not spatially random: they remain within a broader \(20^\circ\)--\(45^\circ\) neighborhood. This distinction is useful for deployment interpretation. Applications requiring accurate spatial rendering should retain the \(20^\circ\) policy, whereas coarse source-region selection may tolerate blurrier localization when the prediction remains directionally meaningful.

The class-aware angular analysis is reported in Appendix~\ref{app:sed_perf_appendix} (Table~\ref{tab:dataset_generalization_class_policy}). Almost every cross-domain class selects \(45^\circ\), reflecting broader transfer uncertainty. Native STARSS23 is more heterogeneous: speech, domestic sounds, footsteps, laughter, and clapping benefit from \(45^\circ\); music-related classes select \(30^\circ\); telephone remains at \(20^\circ\); door and bell select \(15^\circ\); and water tap and knock select \(10^\circ\). Low selected tolerances for weak classes should not be interpreted as evidence of superior localization. They often arise when increasing the tolerance does not recover additional true positives and only affects the false-positive profile. Therefore, class-aware tolerance must be interpreted jointly with support, \gls{sed} recall, \(F_{\mathrm{20^\circ}}\), and localization error.

\subsubsection{Oracle-Activity \glsentryshort{doa} Analysis}
\label{subsubsec:cross_dataset_oracle_doa}

The threshold and tolerance analyses still couple activity estimation, track assignment, and direction regression. A final oracle-activity \gls{doa} analysis isolates the spatial regression information encoded by each checkpoint. For every model--dataset pair, source annotations are projected into the checkpoint label space through the mappings used for the cross-dataset evaluation (Table~\ref{tab:stage4_projection_mapping}). Ground-truth activity and class identity are then supplied to an evaluator, so neither \gls{sed} logits nor an activity threshold can suppress a difficult event.

For each active target slot \(i\), the output tracks are assigned through the permutation that minimizes angular error over active ground-truth sources. The primary oracle accuracy at tolerance \(\alpha\) is:
\begin{equation}
\mathrm{DOA\,Acc}^{\mathrm{oracle}}_{\alpha}
=
\frac{1}{N_{\mathrm{active}}}
\sum_{i=1}^{N_{\mathrm{active}}}
\mathbbm{1}
\left[
\epsilon_i \leq \alpha
\right]
\end{equation}
where \(N_{\mathrm{active}}\) is the number of active target slots and \(\epsilon_i\) is the angular error:
\begin{equation}
\epsilon_i
=
\arccos
\left(
\frac{
\widehat{\mathbf{d}}_i^\top \mathbf{d}_i
}{
\left\|
\widehat{\mathbf{d}}_i
\right\|_2
\left\|
\mathbf{d}_i
\right\|_2
}
\right)
\end{equation}
where \(\widehat{\mathbf{d}}_i\) and \(\mathbf{d}_i\) are the predicted and target Cartesian direction vectors. This produces a \gls{doa}-only driven diagnostic: it measures whether a model track contains the correct direction when the event is known to be active. For comparison, the analysis also reports \(\mathrm{DOA\,Acc}^{\mathrm{tPIT}}_{20^\circ}\), obtained using the checkpoint's training-time \gls{tpit} permutation. Unlike the primary angular assignment, the \gls{tpit}-aligned result retains part of the assignment difficulty encountered during \gls{seld} inference.

\begin{table}[ht]
\centering
\caption{Coverage-aware \gls{doa} estimation under oracle activity.}
\label{tab:dataset_generalization_doa_oracle}
\footnotesize
\renewcommand{\arraystretch}{1.05}
\setlength{\tabcolsep}{3.0pt}
\resizebox{\textwidth}{!}{%
\begin{tabular}{|l|l||c|c||c|c|c|c|c||c|c|}
\hline
\textbf{Training dataset} & \textbf{Test dataset} & \textbf{Slots} & \textbf{Coverage} & \(\mathrm{Acc}_{10^\circ}\) & \(\mathrm{Acc}_{20^\circ}\) & \(\mathrm{Acc}_{30^\circ}\) & \(\mathrm{Acc}_{45^\circ}\) & \(\mathrm{Acc}^{\mathrm{tPIT}}_{20^\circ}\) & \textbf{Median AE} & \(\mathbf{P_{90}}\) \\
\hline
STARSS23 & STARSS23 & 6,128,934 & 13/13 & 0.142 & 0.396 & 0.632 & 0.823 & 0.337 & 23.97$^\circ$ & 60.32$^\circ$ \\
STARSS23 & TAU2019 & 1,420,618 & 8/11 & 0.047 & 0.174 & 0.334 & 0.570 & 0.141 & 40.24$^\circ$ & 78.52$^\circ$ \\
STARSS23 & TAU-NIGENS2020 & 993,127 & 8/14 & 0.048 & 0.166 & 0.330 & 0.574 & 0.127 & 40.06$^\circ$ & 82.50$^\circ$ \\
STARSS23 & TAU-NIGENS2021 & 2,190,153 & 8/12 & 0.049 & 0.169 & 0.329 & 0.565 & 0.121 & 40.57$^\circ$ & 88.22$^\circ$ \\
\hline
TAU2019 & STARSS23 & 2,868,090 & 6/13 & 0.307 & 0.669 & 0.835 & 0.949 & 0.638 & 14.47$^\circ$ & 36.20$^\circ$ \\
TAU2019 & TAU2019 & 1,921,633 & 11/11 & 0.783 & 0.955 & 0.983 & 0.995 & 0.937 & 5.96$^\circ$ & 14.22$^\circ$ \\
TAU2019 & TAU-NIGENS2020 & 344,349 & 4/14 & 0.215 & 0.504 & 0.655 & 0.782 & 0.456 & 19.80$^\circ$ & 68.03$^\circ$ \\
TAU2019 & TAU-NIGENS2021 & 407,584 & 4/12 & 0.199 & 0.519 & 0.727 & 0.884 & 0.478 & 19.28$^\circ$ & 47.15$^\circ$ \\
\hline
TAU-NIGENS2020 & STARSS23 & 3,896,419 & 6/13 & 0.141 & 0.420 & 0.669 & 0.853 & 0.412 & 22.80$^\circ$ & 54.95$^\circ$ \\
TAU-NIGENS2020 & TAU2019 & 391,392 & 2/11 & 0.183 & 0.523 & 0.759 & 0.901 & 0.515 & 19.30$^\circ$ & 44.85$^\circ$ \\
TAU-NIGENS2020 & TAU-NIGENS2020 & 2,265,292 & 14/14 & 0.166 & 0.458 & 0.697 & 0.884 & 0.448 & 21.45$^\circ$ & 47.50$^\circ$ \\
TAU-NIGENS2020 & TAU-NIGENS2021 & 3,354,658 & 12/12 & 0.148 & 0.424 & 0.635 & 0.821 & 0.413 & 23.16$^\circ$ & 56.10$^\circ$ \\
\hline
TAU-NIGENS2021 & STARSS23 & 3,896,419 & 6/13 & 0.164 & 0.497 & 0.736 & 0.880 & 0.468 & 20.10$^\circ$ & 48.67$^\circ$ \\
TAU-NIGENS2021 & TAU2019 & 391,392 & 2/11 & 0.281 & 0.667 & 0.870 & 0.977 & 0.627 & 15.27$^\circ$ & 32.40$^\circ$ \\
TAU-NIGENS2021 & TAU-NIGENS2020 & 1,636,249 & 12/14 & 0.229 & 0.594 & 0.820 & 0.940 & 0.537 & 17.26$^\circ$ & 37.30$^\circ$ \\
TAU-NIGENS2021 & TAU-NIGENS2021 & 3,354,658 & 12/12 & 0.232 & 0.596 & 0.803 & 0.924 & 0.543 & 16.99$^\circ$ & 40.34$^\circ$ \\
\hline
\end{tabular}%
}
\end{table}

\begin{figure}[ht]
    \centering
    \includegraphics[width=\linewidth]{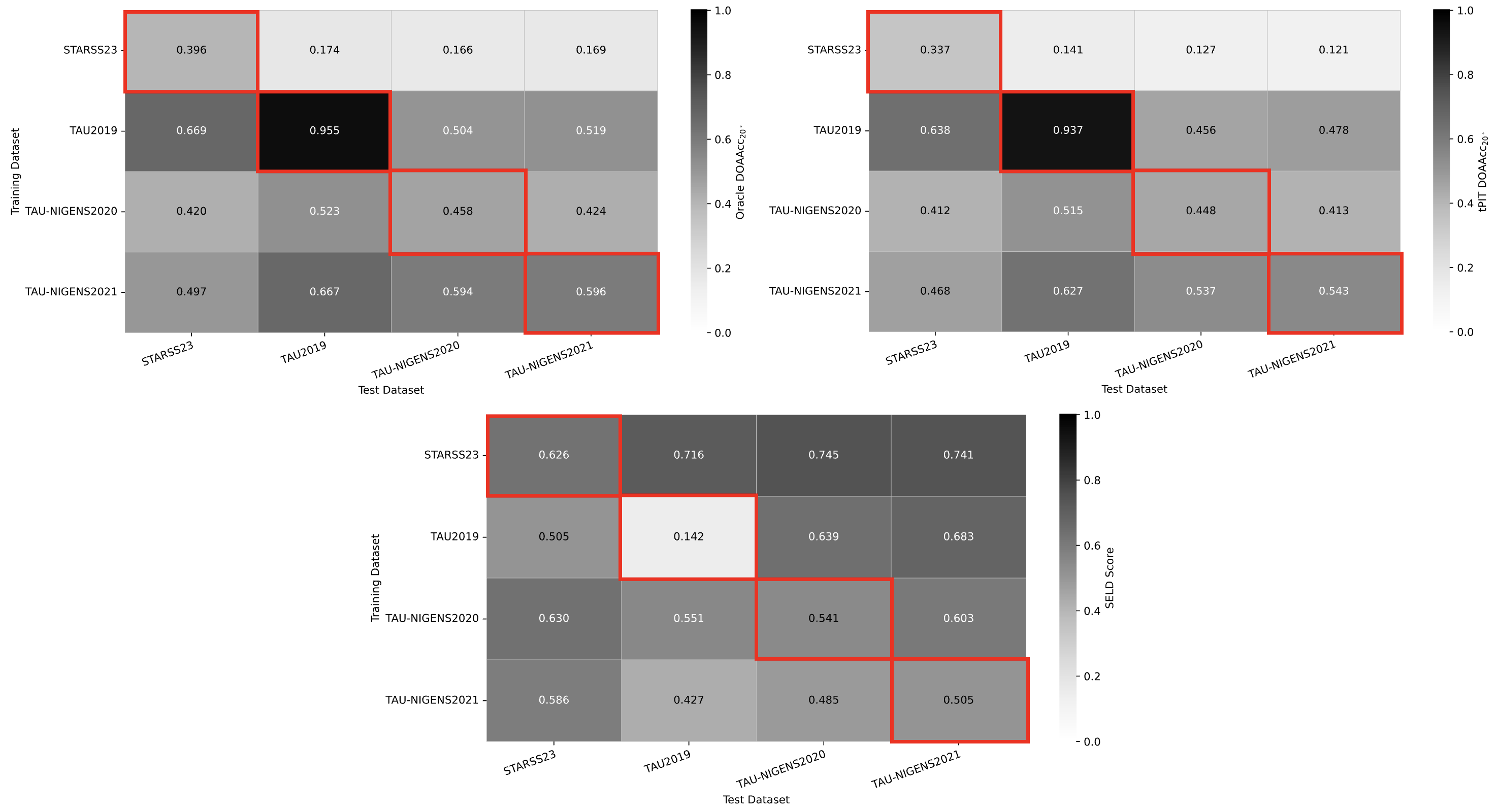}
    \caption{Coverage-aware oracle-activity \gls{doa} accuracy at \(20^\circ\) for the selected \gls{at2seld} family. Off-diagonal cells cover only the mapped semantic subsets reported in Table~\ref{tab:dataset_generalization_mapping_coverage}.}
    \label{fig:dataset_generalization_doa_oracle}
\end{figure}

Table~\ref{tab:dataset_generalization_doa_oracle} and Figure~\ref{fig:dataset_generalization_doa_oracle} confirm the favorable spatial structure of TAU2019. Its native checkpoint reaches \(\mathrm{DOA\,Acc}^{\mathrm{oracle}}_{20^\circ}=0.955\), with a median angular error of \(5.96^\circ\) and \(P_{90}=14.22^\circ\). The \gls{tpit}-aligned accuracy remains \(0.937\), indicating that track assignment introduces a negligible loss. Direction estimation is therefore already reliable before activity calibration, consistently with the small gain observed in Table~\ref{tab:dataset_generalization_angular}.

TAU-NIGENS2021 provides the second strongest in-domain spatial regression, with oracle accuracy \(0.596\), median error \(16.99^\circ\), and \(\mathrm{DOA\,Acc}^{\mathrm{oracle}}_{45^\circ}=0.924\). TAU-NIGENS2020 reaches \(0.458\) at \(20^\circ\), with a \(21.45^\circ\) median error, whereas STARSS23 reaches \(0.396\), with a \(23.97^\circ\) median and a broader \(P_{90}=60.32^\circ\). However these values should not be read negatively: STARSS23 rises to \(0.823\) at \(45^\circ\), and TAU-NIGENS2020 to \(0.884\), indicating that a major fraction of correct directions lies outside the strict \(20^\circ\) cone but within a \(\pm25^\circ\) directional neighborhood. This confirms that a consistent part of the measured \gls{doa} error is a fine-localization limitation rather than a complete failure of directional understanding.

The difference between \gls{doa}-only and \gls{tpit}-aligned accuracy further separates direction regression from track selection. The native gaps remain limited across all datasets: \(0.018\) for TAU2019, \(0.010\) for TAU-NIGENS2020, \(0.053\) for TAU-NIGENS2021, and \(0.059\) for STARSS23. These differences indicate that \gls{tpit} alignment does not introduce a substantial error. The dominant factor is instead the intrinsic angular dispersion of the predicted active directions: STARSS23 and TAU-NIGENS2021 show broader directional uncertainty than TAU2019.

The off-diagonal oracle matrix identifies TAU-NIGENS2021 as the most transferable spatial checkpoint. On TAU-NIGENS2020 it reaches \(\mathrm{DOA\,Acc}^{\mathrm{oracle}}_{20^\circ}=0.594\), essentially matching its \(0.596\) native value and exceeding the \(0.458\) accuracy of the native TAU-NIGENS2020 checkpoint. Its \(17.26^\circ\) median error and \(0.940\) accuracy at \(45^\circ\) indicate strong spatial compatibility between the two NIGENS generations. The reverse direction is weaker: TAU-NIGENS2020\(\rightarrow\)TAU-NIGENS2021 reaches \(0.424\), with a \(23.16^\circ\) median error.

TAU-NIGENS2021 also obtains \(0.497\) on the six mapped STARSS23 classes, compared with \(0.420\) for the TAU-NIGENS2020 checkpoint. Its transfer is strong for \texttt{female\_speech} and \texttt{male\_speech}, but weak for \texttt{footsteps} and \texttt{telephone}. This confirms that exact label correspondence does not ensure spatial transfer when real-scene acoustics and source trajectories differ from synthetic rendering. Similarly, the high TAU2019\(\rightarrow\)STARSS23 oracle value, \(0.669\), is supported by only six STARSS23 classes and is strongly influenced by the many-to-one speech projection. It should therefore be interpreted as robust speech-direction transfer capability.

Complete per-class oracle diagnostics are reported in Appendix~\ref{app:sed_perf_appendix} (Table~\ref{tab:dataset_generalization_doa_oracle_class}). They clarify whether weak \gls{seld} behavior is primarily semantic, spatial, or assignment-related. On native TAU-NIGENS2021, \texttt{female\_scream} has weak activity F1 but high oracle \gls{doa} accuracy, indicating that its main bottleneck is activation or class assignment rather than direction regression. The opposite pattern appears for native STARSS23 \texttt{music}: semantic detection is strong, but oracle localization is much weaker, so the bottleneck is spatial. STARSS23 \texttt{clapping} reaches usable oracle localization but low precision, indicating over-activation. Conversely, \texttt{water\_tap}, \texttt{bell}, and \texttt{musical\_instrument} remain weak even under oracle activity, showing that their poor location-dependent scores could arise from underrepresentation.

\clearpage


\section{Conclusions and Future Works}
\label{sec:conclusions}

This work investigated the transition from pretrained \gls{gpat} representations to spatially grounded \gls{seld}. The central objective was not to build a challenge-optimized architecture by scaling a dedicated \gls{seld} backbone, but to characterize whether semantic priors learned by a \gls{gpat} model can support a compact task-facing auxiliary branch. The resulting \gls{at2seld} framework integrates spectro-temporal semantic representations, explicit \gls{foa} spatial descriptors, track-wise \gls{sed}/\gls{doa} prediction, permutation-aware supervision, and diagnostic calibration into a unified experimental pipeline. The main contribution is therefore not only the resulting architecture, but the characterization of the conditions under which semantic audio priors become useful for spatially grounded event analysis.

The stage-informed \gls{nas} provided a controlled answer to the first two research questions. Regarding \textbf{RQ1}, the experiments showed that explicit spectral \gls{foa} descriptors, combining \gls{stft} magnitude, phase, and \glspl{iv}, constitute the most effective spatial input interface among the evaluated alternatives. This result is coherent with the representational nature of the pretrained \gls{gpat} branch: both semantic and spatial streams operate in spectro-temporal domains that can be aligned progressively, rather than forcing the spatial pathway to learn signal structure from arbitrary domains. Regarding \textbf{RQ2}, the strongest gains were obtained by allocating capacity to the early residual spatial encoder. The early stage is where multi-channel time--frequency maps are manipulated, segregated, and reorganized into a representation that is both localization-sensitive and semantically usable. Late track-wise abstraction and compact recurrent smoothing remained functional, but they behave primarily as algorithmic refinement stages rather than as sources of additional content understanding.

The third and fourth research questions concerned semantic--spatial interaction and track-wise temporal organization. For \textbf{RQ3}, the results showed that shared \gls{seld} supervision is sufficient to train a competitive integrated model, but explicit feature-level coupling improves the selected neural network topology only when introduced after sufficient spatial abstraction. Late cross-stitch interaction provides useful high-level semantic conditioning, whereas early coupling perturbs localization-sensitive feature maps and increases computational cost without improving the operating point. For \textbf{RQ4}, the architecture search and the oracle diagnostics indicate that track-wise attentive abstraction combined with compact \texttt{BiGRU} smoothing provides a favorable representation of sequential consistency and overlapping events. The remaining limits are not primarily caused by the \gls{tpit} mechanism itself: the difference between \gls{doa}-only and \gls{tpit}-aligned oracle accuracy are small, showing that permutation alignment does not introduce a substantial additional error source. The residual factors are instead influenced by activity calibration, spatial uncertainty, acoustic class ambiguity, and dataset-dependent coverage.

The diagnostic stage refined the selected architecture into a more interpretable operating system. \texttt{BalancedSTARSS23Dataset} increased class coverage while preserving the STARSS23 label space and real-scene validation protocol, but also confirmed that clip-level balancing does not remove window-level long-tail behavior. Activity-loss diagnostics showed that loss selection must be evaluated within the complete detection--localization system: focal loss was preferable to standard \gls{bce} for joint \gls{seld}, even though \gls{bce} remained stronger under isolated frame-level \gls{sed}. Positive weighting increased activity sensitivity but destabilized the location-aware trade-off. The decisive localization correction was activity-conditioned Cartesian \gls{doa} supervision, which reduced inactive-target dominance, recovered active-vector norms, and substantially improved oracle angular behavior. Threshold calibration then showed that validation-selected global and class-dependent thresholds can recover useful operating points without test leakage, although calibration cannot substitute for spatial learning when detected events remain outside the angular matching tolerance.

The cross-dataset evaluation completed the profile of the model. TAU2019 confirmed that the compact task-facing \gls{seld} branch can solve regular fixed-source spatial scenes with high accuracy. TAU-NIGENS2021 provided the most transferable synthetic representation, especially toward TAU-NIGENS2020, while TAU-NIGENS2020 occupied an intermediate regime between regular fixed-source scenes and more dynamic synthetic conditions. STARSS23 remained the most demanding domain because it combines real-scene variability, long-tailed coverage, higher acoustic ambiguity, track-assignment difficulty, and broader directional errors. However, the oracle-activity analysis showed that the dataset ordering observed in joint metrics is not a generic failure of the approach. It explains where each regime is limited: some classes are activity-limited, others are spatially limited, and others suffer mainly from poor coverage or ambiguous transient structure. The \(45^\circ\) directionality diagnostic further showed that several errors outside the strict \(20^\circ\) criterion still fall within a spatially plausible directional sector, suggesting positive directional understanding and no random guessing.

These findings support a positive but bounded interpretation of \gls{at2seld}. The selected family is not a full-scale challenge system. Instead, it demonstrates that pretrained semantic audio representations can be connected to a compact spatial branch without sacrificing the capacity to produce meaningful \gls{seld} behavior across heterogeneous spatial-audio regimes. The task-facing \gls{seld} component remains small relative to the complete graph, while the semantic branch supplies transferable class-level structure. The resulting model is therefore best understood as a semantically informed spatial analysis system whose reliability depends on the interaction between feature representation, architectural designing, spatial activity supervision, and validation-aware calibration.

Several extensions follow directly from these evidences: \emph{\textbf{(I)}} synthetic spatial pretraining should be revisited under a multi-domain or \textit{curriculum} strategy. The complementary behavior of TAU-NIGENS2020 and TAU-NIGENS2021 suggests that their aggregation could support stronger out-domain transfer, especially if training progresses from regular or moderately complex spatial conditions toward moving-source mixtures and real-scene fine-tuning; \emph{\textbf{(II)}} the training horizon should be studied systematically, because these experiments intentionally constrained training cost, whereas many challenge-oriented \gls{seld} systems use substantially longer schedules; extending the number of epochs, refining learning-rate policies, and monitoring convergence could clarify whether the compact \gls{at2seld} branch is optimization-limited or capacity-limited; \emph{\textbf{(III)}} augmentation should become more semantically aware. Global gain scaling for example, preserves \gls{foa} directional geometry when applied uniformly, so the \gls{doa} target remains valid, but it may still alter absolute loudness, event salience, and source--context plausibility. Future augmentation policies should therefore consider not only spatial validity, but also semantic and ecological consistency.

Further work should also address class-specific supervision and spatial uncertainty: \emph{\textbf{(IV)}} weak or ambiguous classes such as short transients, water-related events, bells, knocks, and some musical sources would likely benefit from targeted data expansion, uncertainty-aware thresholding, and localization losses that distinguish fine angular accuracy from coarse directional agreement at the training level. \emph{\textbf{(V)}} Activity-conditioned \gls{doa} supervision could be extended with confidence-aware angular objectives, trajectory-consistency terms designed for moving sources, and geometry-aware front-ends capable of operating across different microphone layouts. Finally, deployment-oriented optimization \emph{\textbf{(VI)}} should target the high-resolution spatial front-end and early spatial encoder, since these stages dominate computation despite the compact parameterization of the \gls{seld} path.

Overall, the results indicate that \gls{gpat} priors can provide a promising foundation for compact yet reliable \gls{seld} systems, especially when semantic transfer is not treated as a standalone initialization strategy, but as part of an integrated spatial-aware design and optimization process.
\clearpage


\section{Appendix}
\subsection{Reference SELD Baselines}
\label{app:reference_baselines}

{
\small
\renewcommand{\arraystretch}{1.08}
\setlength{\tabcolsep}{3.5pt}

\begin{longtable}{|
>{\RaggedRight\arraybackslash}p{0.6\textwidth}|
>{\centering\arraybackslash}p{0.10\textwidth}|
>{\centering\arraybackslash}p{0.10\textwidth}|
>{\centering\arraybackslash}p{0.10\textwidth}|
>{\centering\arraybackslash}p{0.10\textwidth}|}

\caption{Reference results for reviewed \gls{seld} systems. Rate-like metrics are reported in the \([0,1]\) range, while localization error is reported in degrees.}
\label{tab:app_reference_baselines}\\

\hline
\textbf{Dataset / \texttt{split} -- annotations} &
\textbf{ER\(\leq20^\circ\)} &
\textbf{F\(\leq20^\circ\)} &
\textbf{LE\(_{\mathrm{CD}}\)} &
\textbf{LR\(_{\mathrm{CD}}\)} \\
\hline
\endfirsthead

\hline
\textbf{Dataset / \texttt{split} -- annotations} &
\textbf{ER\(\leq20^\circ\)} &
\textbf{F\(\leq20^\circ\)} &
\textbf{LE\(_{\mathrm{CD}}\)} &
\textbf{LR\(_{\mathrm{CD}}\)} \\
\hline
\endhead

\hline
\multicolumn{5}{|r|}{\textit{Continued on next page}} \\
\hline
\endfoot

\hline
\endlastfoot

\hline
ANSYN - Max events polyphony $=1$ & 0.04 & 0.977 & 3.4 & 0.994 \\
ANSYN - Max events polyphony $=2$ & 0.16 & 0.890 & 13.8 & 0.856 \\
ANSYN - Max events polyphony $=3$ & 0.19 & 0.856 & 17.3 & 0.702 \\
\hline
RESYN / \texttt{Room 1} - Max events polyphony $=1$ & 0.10 & 0.925 & 9.2 & 0.958 \\
RESYN / \texttt{Room 1} - Max events polyphony $=2$ & 0.29 & 0.796 & 20.2 & 0.749 \\
RESYN / \texttt{Room 1} - Max events polyphony $=3$ & 0.32 & 0.765 & 26.0 & 0.564 \\
\hline
RESYN / \texttt{Room 2} - Max events polyphony $=1$ & 0.11 & 0.916 & 11.5 & 0.962 \\
RESYN / \texttt{Room 2} - Max events polyphony $=2$ & 0.33 & 0.795 & 26.0 & 0.789 \\
RESYN / \texttt{Room 2} - Max events polyphony $=3$ & 0.35 & 0.758 & 33.1 & 0.612 \\
\hline
RESYN / \texttt{Room 3} - Max events polyphony $=1$ & 0.13 & 0.898 & 12.1 & 0.959 \\
RESYN / \texttt{Room 3} - Max events polyphony $=2$ & 0.32 & 0.791 & 25.4 & 0.782 \\
RESYN / \texttt{Room 3} - Max events polyphony $=3$ & 0.34 & 0.755 & 31.9 & 0.607 \\
\hline
REAL - Max events polyphony $=1$ & 0.40 & 0.603 & 26.6 & 0.649 \\
REAL - Max events polyphony $=2$ & 0.49 & 0.531 & 33.7 & 0.415 \\
REAL - Max events polyphony $=3$ & 0.53 & 0.511 & 36.1 & 0.246 \\
\hline
REALBIG - Max events polyphony $=1$ & 0.37 & 0.654 & 23.1 & 0.680 \\
REALBIG - Max events polyphony $=2$ & 0.42 & 0.615 & 31.3 & 0.452 \\
REALBIG - Max events polyphony $=3$ & 0.50 & 0.565 & 34.9 & 0.283 \\
\hline
REALBIGAMB / \texttt{20dB} - Max events polyphony $=1$ & 0.34 & 0.656 & 25.4 & 0.691 \\
REALBIGAMB / \texttt{20dB} - Max events polyphony $=2$ & 0.46 & 0.585 & 32.5 & 0.428 \\
REALBIGAMB / \texttt{20dB} - Max events polyphony $=3$ & 0.52 & 0.550 & 36.1 & 0.258 \\
\hline
REALBIGAMB / \texttt{10dB} - Max events polyphony $=1$ & 0.37 & 0.663 & 27.2 & 0.669 \\
REALBIGAMB / \texttt{10dB} - Max events polyphony $=2$ & 0.49 & 0.554 & 32.5 & 0.400 \\
REALBIGAMB / \texttt{10dB} - Max events polyphony $=3$ & 0.52 & 0.533 & 36.1 & 0.273 \\
\hline
REALBIGAMB / \texttt{0dB} - Max events polyphony $=1$ & 0.46 & 0.579 & 30.7 & 0.625 \\
REALBIGAMB / \texttt{0dB} - Max events polyphony $=2$ & 0.58 & 0.486 & 33.7 & 0.352 \\
REALBIGAMB / \texttt{0dB} - Max events polyphony $=3$ & 0.59 & 0.490 & 36.7 & 0.234 \\
\hline
CANSYN - Max events polyphony $=1$ & 0.11 & 0.930 & 29.5 & 0.979 \\
CANSYN - Max events polyphony $=2$ & 0.18 & 0.866 & 31.3 & 0.788 \\
CANSYN - Max events polyphony $=3$ & 0.19 & 0.853 & 34.3 & 0.670 \\
\hline
CRESYN - Max events polyphony $=1$ & 0.13 & 0.904 & 28.4 & 0.964 \\
CRESYN - Max events polyphony $=2$ & 0.22 & 0.822 & 33.7 & 0.757 \\
CRESYN - Max events polyphony $=3$ & 0.30 & 0.780 & 41.0 & 0.607 \\
\hline
\hline
\multicolumn{5}{|c|}{\textbf{SELDnet (Moving Sources)}} \\
\hline
\hline
MANSYN - Max events polyphony $=1$ & 0.07 & 0.953 & 6.0 & 0.985 \\
MANSYN - Max events polyphony $=2$ & 0.10 & 0.932 & 12.3 & 0.946 \\
MANSYN - Max events polyphony $=3$ & 0.20 & 0.874 & 18.6 & 0.807 \\
\hline
MREAL - Max events polyphony $=1$ & 0.37 & 0.644 & 36.5 & 0.696 \\
MREAL - Max events polyphony $=2$ & 0.45 & 0.564 & 39.6 & 0.428 \\
MREAL - Max events polyphony $=3$ & 0.49 & 0.523 & 38.5 & 0.289 \\
\hline
\hline
\multicolumn{5}{|c|}{\textbf{URNN}} \\
\hline
\hline
TAU Spatial Sound Events 2019 Ambisonics / \texttt{Fold 1} & 0.118 & 0.927 & 5.149 & 0.915 \\
TAU Spatial Sound Events 2019 Ambisonics / \texttt{Fold 2} & 0.147 & 0.918 & 6.296 & 0.900 \\
TAU Spatial Sound Events 2019 Ambisonics / \texttt{Fold 3} & 0.130 & 0.923 & 5.149 & 0.908 \\
TAU Spatial Sound Events 2019 Ambisonics / \texttt{Fold 4} & 0.203 & 0.874 & 5.501 & 0.886 \\
TAU Spatial Sound Events 2019 Ambisonics / \texttt{All} & 0.150 & 0.910 & 5.527 & 0.902 \\
\hline
\hline
\multicolumn{5}{|c|}{\textbf{EIN-V1}} \\
\hline
\hline
TAU-NIGENS Spatial Sound Events 2020 - EAD: \texttt{False}, tPIT: \texttt{False} & 0.654 & 0.461 & 19.947 & 0.616 \\
TAU-NIGENS Spatial Sound Events 2020 - EAD: \texttt{True}, tPIT: \texttt{False}, SED mask & 0.604 & 0.488 & 24.376 & 0.671 \\
TAU-NIGENS Spatial Sound Events 2020 - EAD: \texttt{True}, tPIT: \texttt{False}, SED+EAD mask & 0.599 & 0.490 & 24.416 & 0.674 \\
TAU-NIGENS Spatial Sound Events 2020 - EAD: \texttt{True}, tPIT: \texttt{True}, SED+EAD mask & 0.47 & 0.615 & 16.7 & 0.754 \\
\hline
\hline
\multicolumn{5}{|c|}{\textbf{EIN-V2}} \\
\hline
\hline
TAU-NIGENS Spatial Sound Events 2020 - Sharing: \texttt{False}, Aug.: \texttt{False}, tPIT: \texttt{False} & 0.340 & 0.737 & 11.9 & 0.838 \\
TAU-NIGENS Spatial Sound Events 2020 - Sharing: \texttt{Hard}, Aug.: \texttt{False}, tPIT: \texttt{False} & 0.339 & 0.739 & 10.4 & 0.813 \\
TAU-NIGENS Spatial Sound Events 2020 - Sharing: \texttt{Soft}, Aug.: \texttt{False}, tPIT: \texttt{True} & 0.299 & 0.770 & 8.9 & 0.838 \\
TAU-NIGENS Spatial Sound Events 2020 - Sharing: \texttt{False}, Aug.: \texttt{True}, tPIT: \texttt{False} & 0.323 & 0.758 & 10.2 & 0.831 \\
TAU-NIGENS Spatial Sound Events 2020 - Sharing: \texttt{Hard}, Aug.: \texttt{True}, tPIT: \texttt{False} & 0.264 & 0.805 & 7.9 & 0.843 \\
TAU-NIGENS Spatial Sound Events 2020 - Sharing: \texttt{Soft}, Aug.: \texttt{True}, tPIT: \texttt{True} & 0.233 & 0.832 & 6.8 & 0.861 \\
\hline
\hline
\multicolumn{5}{|c|}{\textbf{NERC-SLIP (DCASE 2022 Leading System)}} \\
\hline
\hline
STARSS22 / \texttt{Development} - ACCDOA & 0.40 & 0.650 & 15.0 & 0.770 \\
STARSS22 / \texttt{Development} - Multi-ACCDOA & 0.41 & 0.610 & 15.3 & 0.740 \\
\hline
STARSS22 / \texttt{Evaluation} - Challenge & 0.35 & 0.583 & 14.6 & 0.737 \\
\hline
\hline
\multicolumn{5}{|c|}{\textbf{MSCA-RCnet (DCASE 2023 Leading System)}} \\
\hline
\hline
STARSS23 / \texttt{Evaluation} - RCnet & 0.50 & 0.461 & 16.9 & 0.613 \\
STARSS23 / \texttt{Evaluation} - + MSCA & 0.48 & 0.484 & 17.1 & 0.645 \\
STARSS23 / \texttt{Evaluation} - + Attentive Statistics Pooling & 0.45 & 0.520 & 15.4 & 0.652 \\
STARSS23 / \texttt{Evaluation} - + Augmentation & 0.45 & 0.511 & 15.7 & 0.672 \\
\hline
STARSS23 / \texttt{Test} - Challenge & 0.43 & 0.548 & 14.7 & 0.680 \\
\hline
\hline
\multicolumn{5}{|c|}{\textbf{SoundDet / SoundDoA}} \\
\hline
\hline
TAU-NIGENS Spatial Sound Events 2020 - SoundDet  & 0.25 & 0.810 & 8.3 & 0.820 \\
TAU-NIGENS Spatial Sound Events 2020 - SoundDoA  & 0.23 & 0.850 & 7.9 & 0.870 \\
\hline

\end{longtable}
}
\clearpage


\begin{landscape}
\subsection{Neural Spatial Modules Summary}
\label{app:spatial_modules}

\footnotesize
\setlength{\LTleft}{0pt}
\setlength{\LTright}{0pt}
\renewcommand{\arraystretch}{1.2}

\begin{longtable}{|>{\RaggedRight\arraybackslash}p{2.5cm}
                  |>{\RaggedRight\arraybackslash}p{5.7cm}
                  |>{\RaggedRight\arraybackslash}p{3.8cm}
                  |>{\RaggedRight\arraybackslash}p{4.2cm}
                  |>{\RaggedRight\arraybackslash}p{5.7cm}|}
\caption{Parametrized \textit{spatial modules} ablated for NAS experiments.}
\label{tab:spatial_modules_inventory}\\
\hline
\textbf{Module name} & \textbf{Parameters} & \textbf{Input shape} & \textbf{Output shape} & \textbf{Notes} \\
\hline
\endfirsthead

\hline
\textbf{Module name} & \textbf{Parameters} & \textbf{Input shape} & \textbf{Output shape} & \textbf{Notes} \\
\hline
\endhead

\hline
\multicolumn{5}{|r|}{\emph{Continued on next page}}\\
\hline
\endfoot

\hline
\endlastfoot

\multicolumn{5}{|c|}{\textbf{SELDnet / CRNN}}\\
\hline

SpectralProcessor &
\makecell[tl]{\texttt{n\_fft}: STFT size/window length\\
\texttt{hop\_length}: hop samples}
&
\makecell[tl]{\((B,C,S)\)\\
\(B\): batch size\\
\(C\): audio channels\\
\(S\): time samples}
&
\makecell[tl]{\((B,2C,T,F)\)\\
\(B\): batch size\\
\(T\): time frames\\
\(F\): positive-frequency bins\\
first \(C\) chs: magnitude spectra\\
last \(C\) chs: phase spectra}
&
Differentiable STFT front-end. The implementation explicitly removes the DC bin and interleaves magnitude and phase channels. It differs from the original Keras pre-processing that kept the full one-sided spectrum. \\
\hline

ConvBlock &
\makecell[tl]{\texttt{in\_channels}: input channels\\
\texttt{out\_channels}: number of filters\\
\texttt{kernel\_size}: tuple, 2D kernel\\
\texttt{pool\_size}: frequency pooling bins\\
\texttt{dropout}: dropout factor}
&
\makecell[tl]{\((B,C_{\text{in}},T,F)\)\\
\(B\): batch size\\
\(C_{\text{in}}\): input feature channels\\
\(T\): time frames\\
\(F\): frequency bins}
&
\makecell[tl]{\((B,C_{\text{out}},T,F')\)\\
\(B\): batch size\\
\(C_{\text{out}}\): output feature channels\\
\(T\): input time frames\\
\(F' = F/\texttt{pool\_size}\) if pooled}
&
Convolutional block: 2DConv \(\rightarrow\) BatchNorm \(\rightarrow\) ReLU \(\rightarrow\) MaxPool (on frequency axis) \(\rightarrow\) Dropout. Temporal resolution remains available to the recurrent stages. \\
\hline

BiGRU &
\makecell[tl]{\texttt{in\_size}: feature dimension (per frame)\\
\texttt{hidden\_size}: units per direction\\
\texttt{num\_layers}: stacked layers\\
\texttt{dropout}: inter-layer dropout factor}
&
\makecell[tl]{\((B,T,D_{\text{in}})\)\\
\(B\): batch size\\
\(T\): sequence length\\
\(D_{\text{in}}\): \texttt{in\_size}}
&
\makecell[tl]{\((B,T,H)\)\\
\(B\): batch size\\
\(T\): sequence length\\
\(H=\texttt{hidden\_size}\)}
&
Implements bi-directionality through two explicit GRUs and merges forward/backward outputs by element-wise multiplication, reproducing Keras \texttt{merge\_mode='mul'}. Recurrent dropout is unavailable in fused PyTorch semantics. \\
\hline

\multicolumn{5}{|c|}{\textbf{EIN-V2}}\\
\hline

DoubleConv &
\makecell[tl]{\texttt{in\_channels}: input channels\\
\texttt{out\_channels}: output channels\\
\texttt{kernel\_size}: tuple, 2D kernel\\
\texttt{bias}: bias flag}
&
\makecell[tl]{\((B,C_{\text{in}},T,F)\)\\
\(B\): batch size\\
\(C_{\text{in}}\): input feature channels\\
\(T\): time frames\\
\(F\): frequency bins}
&
\makecell[tl]{\((B,C_{\text{out}},T,F)\)\\
\(B\): batch size\\
\(C_{\text{out}}\): output feature channels\\
\(T\): time frames\\
\(F\): frequency bins}
&
Two consecutive 2DConv \(\rightarrow\) BatchNorm \(\rightarrow\) ReLU blocks with same-padding. This is a pure local feature extractor. \\
\hline

CrossStitching &
\makecell[tl]{\texttt{n\_channels}: input channels (per branch)\\
\texttt{init\_range}: tuple, weights init interval}
&
\makecell[tl]{\(2\times(B,C_{\text{in}},\ldots)\)\\
\(B\): batch size\\
\(C_{\text{in}}\): input feature channels\\
\textit{same trailing dimension}}
&
\makecell[tl]{\(2\times(B,C_{\text{out}},\ldots)\)\\
\(B\): batch size\\
\(C_{\text{out}}\)= \(C_{\text{in}}\)\\
\textit{same trailing dimension}}
&
Channel-wise \(2\times2\) learnable mixing matrix for semantic and localization features fusion. It preserves a detail often omitted: branch A is updated first, and branch B is computed from the updated A, so the coupling is cascade-like rather than symmetric parallel cross-stitching. \\
\hline

PositionalEncoder &
\makecell[tl]{\texttt{d\_model}: int, feature dimension\\
\texttt{max\_len}: maximum length\\
\texttt{pe\_type}: \texttt{[t]}ime or \texttt{[f]}requency axis\\
\texttt{dropout}: dropout factor}
&
\makecell[tl]{\((B,C_{\text{in}},T)\) or \((B,C_{\text{in}},T,F)\)\\
\(B\): batch size\\
\(C_{\text{in}}\): input feature channels\\
\(T\): time frames\\
\(F\): frequency bins}
&
\makecell[tl]{\((B,C_{\text{out}},T)\) or \((B,C_{\text{out}},T,F)\)\\
\(B\): batch size\\
\(C_{\text{out}}\)= \(C_{\text{in}}\)\\
\(T\): time frames\\
\(F\): frequency bins}
&
Sinusoidal positional encoding scaled by \(0.1\), supporting either time- or frequency-axis injection. \\
\hline

TrackTransformer &
\makecell[tl]{\texttt{d\_model}: transformer hidden size\\
\texttt{n\_heads}: attention heads\\
\texttt{n\_tracks}: number of track encoders\\
\texttt{num\_layers}: encoder depth\\
\texttt{dim\_feedforward}: FFN network size\\
\texttt{dropout}: dropout factor\\
\texttt{d\_input}: (optional) input projection dim\\
\texttt{pe\_enable}: positional encoding flag\\
\texttt{pe\_type}: positional encoding axis\\
\texttt{pe\_dropout}: PE dropout factor}
&
\makecell[tl]{\((B,T,D_{\text{in}})\)\\
\(B\): batch size\\
\(T\): time frames\\
\(D_{\text{in}}\): input embedding size}
&
\makecell[tl]{\((B,T,M,D_{\text{model}})\)\\
\(B\): batch size\\
\(T\): time frames\\
\(M\): number of tracks\\
\(D_{\text{model}}\)= \texttt{d\_model}}
&
Implements \(M\) independent Transformer encoders, all receiving the same sequence and producing track-specific latent representations. This module generalizes the EINV2 design with optional input projection and arbitrary track count. \\
\hline

\multicolumn{5}{|c|}{\textbf{PI-RNN}}\\
\hline

MHA\_MGU &
\makecell[tl]{\texttt{d\_model}: track embedding size\\
\texttt{n\_heads}: attention heads\\
\texttt{n\_tracks}: maintained track states\\
\texttt{d\_input}: detection embedding size\\
\texttt{dropout}: dropout factor}
&
\makecell[tl]{\texttt{x}: \((B,T,M_X,D_{\text{in}})\)\\
\(B\): batch size\\
\(T\): time frames\\
\(M_X\): detections per frame\\
\(D_{\text{in}}\)= \texttt{d\_input}\\
\\
\texttt{h0}: \((B,M_H,D_{\text{model}})\)\\
\(M_H\): tracks per frame\\
\(D_{\text{model}}\)= \texttt{d\_model}}
&
\makecell[tl]{\texttt{H}: \((B,T,M_H,D_{\text{model}})\)\\
\(B\): batch size\\
\(T\): time frames\\
\(M_H\): tracks per frame\\
\(D_{\text{model}}\)= \texttt{d\_model}\\
\\
\texttt{h\_final}: \((B,M_H,D_{\text{model}})\)}
&
Implements the PI-RNN MHA-based soft assignment with \(Q=H(t-1)\), \(K=V=X(t)\cup H(t-1)\), followed by Minimal Gated Unit update. Two practical extensions are added: optional input projection from arbitrary detection dimension, and learnable per-track initial states. \\
\hline

\multicolumn{5}{|c|}{\textbf{Conformer-based}}\\
\hline

MSCA &
\makecell[tl]{\texttt{n\_channels}: channel count\\
\texttt{reduction\_ratio}: bottleneck ratio}
&
\makecell[tl]{\((B,C,T,F)\)\\
\(B\): batch size\\
\(C\): input feature channels\\
\(T\): time frames\\
\(F\): frequency bins}
&
\makecell[tl]{\((B,C,T,F)\)\\
\(B\): batch size\\
\(C\): input feature channels\\
\(T\): time frames\\
\(F\): frequency bins}
&
Implements the dual-path MSCA: a global path on GAP-pooled context and a local path on the full feature map, both using \(1\times1\) bottlenecks and combined through sigmoid gating. \\
\hline

ResNetBlock &
\makecell[tl]{\texttt{in\_channels}: input channels\\
\texttt{out\_channels}: output channels\\
\texttt{stride}: temporal/frequency stride\\
\texttt{use\_msca}: MSCA flag\\
\texttt{msca\_reduction}: MSCA bottleneck}
&
\makecell[tl]{\((B,C_{\text{in}},T,F)\)\\
\(B\): batch size\\
\(C_{\text{in}}\): input feature channels\\
\(T\): time frames\\
\(F\): frequency bins}
&
\makecell[tl]{\((B,C_{\text{out}},T',F')\)\\
\(B\): batch size\\
\(C_{\text{out}}\): output feature channels\\
\(T'\): time frames (strided)\\
\(F'\): frequency bins (strided)}
&
Standard two-layer residual Conv \(\rightarrow\) BatchNorm \(\rightarrow\) ReLU block with optional MSCA before residual addition. \\
\hline

ConformerBlock &
\makecell[tl]{\texttt{d\_model}: hidden size\\
\texttt{n\_heads}: attention heads\\
\texttt{ffn\_expansion}: FFN expansion factor\\
\texttt{conv\_kernel}: depthwise conv kernel\\
\texttt{dropout}: dropout factor\\
\texttt{activation}: FFN nonlinearity}
&
\makecell[tl]{\((B,T,D_{\text{model}})\)\\
\(B\): batch size\\
\(T\): time frames\\
\(D_{\text{model}}\)= \texttt{d\_model}}
&
\makecell[tl]{\((B,T,D_{\text{model}})\)\\
\(B\): batch size\\
\(T\): time frames\\
\(D_{\text{model}}\)= \texttt{d\_model}}
&
Sandwich Conformer block: \( \frac12\)FFN \(\rightarrow\) MHSA \(\rightarrow\) convolution module \(\rightarrow \frac12\)FFN \(\rightarrow\) LN. \\
\hline

ASP &
\makecell[tl]{\texttt{d\_input}: input feature size\\
\texttt{hidden\_dim}: attention MLP hidden size}
&
\makecell[tl]{\((B,T,D)\)\\
\(B\): batch size\\
\(T\): time frames\\
\(D\)= \texttt{d\_input}}
&
\makecell[tl]{\((B,2D)\)\\
\(B\): batch size\\
\(2D\): weighted mean + std dev}
&
Implements Attentive Statistics Pooling as weighted first- and second-order temporal statistics. \\
\hline

\multicolumn{5}{|c|}{\textbf{Raw-waveform front ends}}\\
\hline

GaborFilterBank &
\makecell[tl]{\texttt{n\_filters}: number of Gabor filters\\
\texttt{kernel\_length}: kernel size (samples)\\
\texttt{stride}: hop size\\
\texttt{sr}: sample rate\\
\texttt{fmin}, \texttt{fmax}: initialization range}
&
\makecell[tl]{\((B,S)\)\\
\(B\): batch size\\
\(S\): waveform samples}
&
\makecell[tl]{\((B,2N,T_{\text{out}})\)\\
\(B\): batch size\\
first \(N\): real responses\\
last \(N\): imaginary responses\\
\(T_{\text{out}}\): \(\left\lfloor \frac{S-1}{\texttt{stride}} + 1 \right\rfloor\)}
&
Implements the SoundDoA learnable complex Gabor filter bank. Kernels are rebuilt every forward pass from learnable center frequency \(\eta\) and bandwidth \(\sigma\), with mel-scale initialization. The implementation realizes the complex convolution through paired real/imaginary Conv1D filters rather than complex-valued kernels. \\
\hline

GaborFrontEnd &
\makecell[tl]{\texttt{n\_filters}: number of Gabor filters\\
\texttt{kernel\_length}: kernel size (samples)\\
\texttt{stride}: hop size\\
\texttt{n\_mics}: number of mics/channels\\
\texttt{format}: str, \texttt{foa} or \texttt{mic}\\
\texttt{sr}: sample rate \\
\texttt{fmin}, \texttt{fmax}: initialization range}
&
\makecell[tl]{\((B,C_{\text{mic}},S)\)\\
\(B\): batch size\\
\(C_{\text{mic}}\): microphones\\
\(S\): waveform samples}
&
\makecell[tl]{\((B,C_{\text{spatial}},T_{\text{out}},N)\)\\
\(B\): batch size\\
\(C_{\text{spatial}} = 1 + 2(C_{\text{mic}}-1)\)\\
\(T_{\text{out}}\): \(\left\lfloor \frac{S-1}{\texttt{stride}} + 1 \right\rfloor\)\\
\(N\): \texttt{n\_filters}}
&
Applies the same Gabor filterbank independently to each channel, then forms cross-spectral spatial features relative to channel 0. For FOA, this yields an omni-magnitude plus real/imaginary cross-spectra with directional channels; for MIC, a channel-0-reference logic is used. \\
\hline

FormantEnhancer &
\makecell[tl]{\texttt{n\_channels}: input/output channels\\
\texttt{kernel\_size}: branch conv kernel}
&
\makecell[tl]{\((B,C,T,F)\)\\
\(B\): batch size\\
\(C\): input feature channels\\
\(T\): time frames\\
\(F\): frequency bins}
&
\makecell[tl]{\((B,C,T,F)\)\\
\(B\): batch size\\
\(C\): input feature channels\\
\(T\): time frames\\
\(F\): frequency bins}
&
Shape-preserving SoundDoA-inspired enhancement block. A learnable IIR smoother sweeps along the frequency axis to extract a formant trace, the detail signal is computed by subtraction, and two 2D Conv branches recombine them through sigmoid/tanh nonlinearities. \\
\hline
\hline

\multicolumn{5}{|c|}{\textbf{Neural-SRP}}\\
\hline

Window &
\makecell[tl]{\texttt{win\_size}: frame length in samples\\
\texttt{hop\_size}: frame hop in samples}
&
\makecell[tl]{\((B,C_{\text{mic}},S)\)\\
\(B\): batch size\\
\(C_{\text{mic}}\): microphones\\
\(S\): waveform samples}
&
\makecell[tl]{\((B,T,C_{\text{mic}},W)\)\\
\(B\): batch size\\
\(T\): extracted frames\\
\(C_{\text{mic}}\): microphones\\
\(W=\texttt{win\_size}\)}
&
Non-trainable Hann-windowed framing module. It unfolds the time axis into overlapping frames and applies a registered Hann window before reordering the output as frame-major. This defines the temporal granularity later processed by GCC-PHAT and by the pairwise CRNN encoder.\\
\hline

GCC-PHAT &
\makecell[tl]{\texttt{win\_size}: input frame length\\
\texttt{tau\_max}: maximum retained lag\\
\texttt{n\_dft}: next power-of-two FFT size\\
(internal, derived from \texttt{win\_size})}
&
\makecell[tl]{\((B,T,C_{\text{mic}},W)\)\\
\(B\): batch size\\
\(T\): time frames\\
\(C_{\text{mic}}\): microphones\\
\(W\): windowed samples}
&
\makecell[tl]{\((B,T,N_{\text{pairs}},2\tau_{\max})\)\\
\(B\): batch size\\
\(T\): time frames\\
\(N_{\text{pairs}}=C_{\text{mic}}(C_{\text{mic}}-1)/2\)\\
\(2\tau_{\max}\): centered GCC lags}
&
Generalized Cross-Correlation with PHAT normalization for all unique microphone pairs. The implementation applies \texttt{rfft}, magnitude normalization, pairwise cross-spectra, \texttt{irfft}, lag centering, and absolute value. Unlike global SRP-map approaches, this produces pairwise local spatial cues that remain independent of the total array geometry.\\
\hline

PairwiseConvBlock &
\makecell[tl]{\texttt{in\_channels}: input channels\\
\texttt{out\_channels}: convolution filters\\
\texttt{n\_metadata}: metadata dimension\\
\texttt{pool\_size}: \((t_{\text{pool}},f_{\text{pool}})\)\\
\texttt{dropout\_rate}: Dropout2d factor}
&
\makecell[tl]{\((B_{\text{pair}},C_{\text{in}},T,F)\)\\
\(B_{\text{pair}}=B\cdot N_{\text{pairs}}\)\\
\(C_{\text{in}}\): feature channels\\
\(T\): time frames\\
\(F\): GCC-lag bins}
&
\makecell[tl]{\((B_{\text{pair}},C_{\text{out}},T',F')\)\\
\(B_{\text{pair}}\): batch$\times$pairs\\
\(C_{\text{out}}\): output channels\\
\(T'\): pooled temporal length\\
\(F'\): pooled lag resolution}
&
Conv2d \(\rightarrow\) BatchNorm \(\rightarrow\) PReLU \(\rightarrow\) optional pooling/dropout block for pairwise GCC features. It also supports metadata-conditioned bias injection before normalization through a learned linear projection of microphone-position metadata. \\
\hline

PairwiseBiGRU &
\makecell[tl]{\texttt{rnn\_hidden}: hidden size per direction\\
\texttt{n\_rnn\_layers}: stacked recurrent layers\\
\texttt{bidirectional}: dual-stream flag}
&
\makecell[tl]{\((B_{\text{pair}},T,C_{\text{conv}})\)\\
\(B_{\text{pair}}=B\cdot N_{\text{pairs}}\)\\
\(T\): time frames\\
\(C_{\text{conv}}\): conv feature size}
&
\makecell[tl]{\((B_{\text{pair}},T,H)\)\\
\(B_{\text{pair}}\): batch$\times$pairs\\
\(T\): time frames\\
\(H=\texttt{rnn\_hidden}\)}
&
Recurrent stage applied after collapsing the lag axis by mean pooling. In the default configuration it uses explicit forward and backward GRU stacks and merges them by element-wise multiplication after \(\tanh\). If \texttt{bidirectional=False}, the implementation falls back to a single unidirectional GRU with \(\tanh\)-bounded output. \\
\hline

PairwiseMLP &
\makecell[tl]{\texttt{in\_features}: input size\\
\texttt{out\_features}: output size\\
\texttt{hidden\_features}: hidden size\\
\texttt{num\_layers}: number of linear layers\\
\texttt{batch\_norm}: BatchNorm1d flag\\
\texttt{output\_activation}: final activation}
&
\makecell[tl]{\((B_{\text{pair}},T,D_{\text{in}})\)\\
\(B_{\text{pair}}=B\cdot N_{\text{pairs}}\)\\
\(T\): frames\\
\(D_{\text{in}}\): pairwise latent size}
&
\makecell[tl]{\((B_{\text{pair}},T,D_{\text{out}})\)\\
\(B_{\text{pair}}\): batch$\times$pairs\\
\(T\): time frames\\
\(D_{\text{out}}=\texttt{out\_features}\)}
&
Per-pair multilayer perceptron with PReLU activations and optional BatchNorm1d on hidden layers. In the implemented encoder it is used after recurrent processing, optionally concatenating microphone-position metadata, to produce pairwise spatial embeddings before aggregation. \\
\hline

\multicolumn{5}{|c|}{\textbf{Hungarian Network}}\\
\hline

AttentionLayer &
\makecell[tl]{\texttt{in\_channels}: input channels\\
\texttt{out\_channels}: value/output channels\\
\texttt{key\_channels}: query/key channels}
&
\makecell[tl]{\((B,C_{\text{in}},L)\)\\
\(B\): batch size\\
\(C_{\text{in}}=\texttt{in\_channels}\)\\
\(L\): sequence length}
&
\makecell[tl]{\((B,C_{\text{out}},L)\)\\
\(B\): batch size\\
\(C_{\text{out}}=\texttt{out\_channels}\)\\
\(L\): sequence length}
&
Single-head key--value attention along the sequence axis. The module applies three \(1\times1\) convolutions to obtain \(Q\), \(K\), and \(V\), computes a row-wise softmax attention matrix \(A=Q^\top K\), and returns the attended value stream. In the HNet implementation it is used after the GRU to refine assignment features before the final linear projection. \\
\hline

HNetGRU &
\makecell[tl]{\texttt{max\_len}: n° of tracks / sequence length\\
(internal: hidden size fixed to 128\\
and one GRU layer)}
&
\makecell[tl]{\((B_{\text{frm}},N,N)\)\\
\(B_{\text{frm}}\): batch$\times$frames\\
\(N\): number of tracks
}
&
\makecell[tl]{\texttt{out1}: \((B_{\text{frm}},N^2)\)\\
\texttt{out2}: \((B_{\text{frm}},N)\)\\
\texttt{out3}: \((B_{\text{frm}},N)\)}
&
Hungarian Network based on GRU \(\rightarrow\) attention \(\rightarrow\) Linear. It learns a soft assignment structure from pairwise distance matrices between predicted and target tracks. In downstream use, \texttt{out1} is passed through a sigmoid and reshaped to \((B_{\text{frm}},N,N)\) to obtain the soft assignment matrix employed by the HNet-guided training losses. \\
\hline

\end{longtable}
\end{landscape}
\clearpage


\subsection{SELD Dataset Statistics}
\label{app:dataset_statistics}

\vspace*{\fill}

\begin{center}
\captionsetup{type=table}
\captionof{table}{STARSS23 descriptive statistics (percentages refer to the corresponding split).}
\vspace{0.2cm}
\label{tab:starss23_native_window_stats}
\footnotesize
\renewcommand{\arraystretch}{1.05}
\setlength{\tabcolsep}{2.6pt}
\resizebox{\textwidth}{!}{%
\begin{tabular}{|p{2.5cm}|c|ccccccc|ccccccc|}
\hline
\multirow{2}{*}{\textbf{Class}} & \multirow{2}{*}{\textbf{Split}} &
\multicolumn{7}{c|}{\textbf{Native framewise statistics}} &
\multicolumn{7}{c|}{\textbf{Window-based statistics (\(10\) s)}} \\
\cline{3-16}
& &
\textbf{Cov.} & \textbf{P$_{max}$} & \textbf{P$_{mean}$} & \textbf{P${_1}$} & \textbf{P${_2}$} & \textbf{P${_3}$} & \textbf{P${_{4+}}$} &
\textbf{Cov.} & \textbf{P$_{max}$} & \textbf{P$_{mean}$} & \textbf{P${_1}$} & \textbf{P${_2}$} & \textbf{P${_3}$} & \textbf{P${_{4+}}$} \\
\hline

\multirow{2}{*}{Global}
& Dev & 83.77 & 5 & 1.410 & 65.02 & 29.66 & 4.70 & 0.62 & \textbf{81.55} & 5 & 1.339 & \textbf{70.13} & \textbf{26.01} & \textbf{3.65} & 0.21 \\
& Eval  & 89.62 & 5 & 1.504 & 61.90 & 28.11 & 7.80 & 2.18 & \textbf{88.01} & 5 & 1.412 & \textbf{66.50} & \textbf{26.98} & \textbf{5.45} & \textbf{1.07} \\
\hline
\hline

\multirow{2}{*}{Female speech}
& Dev  & 28.86 & 2 & 1.045 & 95.51 & 4.49 & 0.00 & 0.00 & 29.69 & 2 & 1.053 & 94.67 & 5.33 & 0.00 & 0.00 \\
& Eval & 23.16 & 2 & 1.027 & 97.28 & 2.72 & 0.00 & 0.00 & 22.75 & 2 & 1.028 & 97.24 & 2.76 & 0.00 & 0.00 \\
\hline

\multirow{2}{*}{Male speech}
& Dev  & 31.92 & 3 & 1.052 & 94.94 & 4.92 & 0.14 & 0.00 & \textbf{28.96} & 3 & 1.063 & \textbf{93.91} & 5.91 & 0.18 & 0.00 \\
& Eval & 33.03 & 3 & 1.058 & 94.30 & 5.57 & 0.13 & 0.00 & 32.43 & 3 & 1.059 & 94.22 & 5.65 & 0.13 & 0.00 \\
\hline

\multirow{2}{*}{Clapping}
& Dev  & 0.54 & 2 & 1.063 & 93.74 & 6.26 & 0.00 & 0.00 & 0.38 & 2 & 1.114 & \textbf{88.62} & \textbf{11.38} & 0.00 & 0.00 \\
& Eval & 0.49 & 2 & 1.270 & 72.97 & 27.03 & 0.00 & 0.00 & 0.48 & 2 & 1.271 & 72.92 & 27.08 & 0.00 & 0.00 \\
\hline

\multirow{2}{*}{Telephone}
& Dev  & 1.04 & 1 & 1.000 & 100.0 & 0.00 & 0.00 & 0.00 & 0.98 & 1 & 1.000 & 100.0 & 0.00 & 0.00 & 0.00 \\
& Eval & 0.79 & 1 & 1.000 & 100.0 & 0.00 & 0.00 & 0.00 & 0.77 & 1 & 1.000 & 100.0 & 0.00 & 0.00 & 0.00 \\
\hline

\multirow{2}{*}{Laughter}
& Dev  & 2.93 & 4 & 1.247 & 78.66 & 18.48 & 2.39 & 0.46 & 3.07 & 3 & 1.260 & \textbf{77.12} & \textbf{19.76} & \textbf{3.12} & 0.00 \\
& Eval & 3.10 & 3 & 1.141 & 86.92 & 12.06 & 1.02 & 0.00 & 3.05 & 3 & 1.141 & 86.91 & 12.06 & 1.03 & 0.00 \\
\hline

\multirow{2}{*}{Domestic sounds}
& Dev  & 18.89 & 1 & 1.000 & 100.0 & 0.00 & 0.00 & 0.00 & \textbf{15.10} & 2 & 1.043 & \textbf{95.70} & \textbf{4.30} & 0.00 & 0.00 \\
& Eval & 14.20 & 1 & 1.000 & 100.0 & 0.00 & 0.00 & 0.00 & 13.94 & 2 & 1.011 & \textbf{98.87} & \textbf{1.13} & 0.00 & 0.00 \\
\hline

\multirow{2}{*}{Walk/footsteps}
& Dev  & 2.21 & 3 & 1.023 & 97.87 & 1.94 & 0.18 & 0.00 & 2.31 & 2 & 1.014 & 98.56 & 1.44 & 0.00 & 0.00 \\
& Eval & 3.98 & 2 & 1.022 & 97.77 & 2.23 & 0.00 & 0.00 & 3.90 & 2 & 1.023 & 97.75 & 2.25 & 0.00 & 0.00 \\
\hline

\multirow{2}{*}{Door}
& Dev  & 0.71 & 1 & 1.000 & 100.0 & 0.00 & 0.00 & 0.00 & 0.82 & 1 & 1.000 & 100.0 & 0.00 & 0.00 & 0.00 \\
& Eval & 0.50 & 1 & 1.000 & 100.0 & 0.00 & 0.00 & 0.00 & 0.48 & 1 & 1.000 & 100.0 & 0.00 & 0.00 & 0.00 \\
\hline

\multirow{2}{*}{Music}
& Dev  & 22.59 & 1 & 1.000 & 100.0 & 0.00 & 0.00 & 0.00 & 23.19 & 2 & 1.245 & \textbf{75.49} & \textbf{24.51} & 0.00 & 0.00 \\
& Eval & 25.81 & 1 & 1.000 & 100.0 & 0.00 & 0.00 & 0.00 & 25.35 & 2 & 1.071 & \textbf{92.85} & \textbf{7.15} & 0.00 & 0.00 \\
\hline

\multirow{2}{*}{Musical instrument}
& Dev  & 2.35 & 2 & 1.283 & 71.66 & 28.34 & 0.00 & 0.00 & 2.80 & 2 & 1.300 & \textbf{70.03} & \textbf{29.97} & 0.00 & 0.00 \\
& Eval & 18.06 & 4 & 1.281 & 85.41 & 3.46 & 8.76 & 2.36 & 17.73 & 3 & 1.257 & 85.41 & 3.46 & \textbf{11.12} & 0.00 \\
\hline

\multirow{2}{*}{Water tap}
& Dev  & 0.59 & 1 & 1.000 & 100.0 & 0.00 & 0.00 & 0.00 & 0.61 & 1 & 1.000 & 100.0 & 0.00 & 0.00 & 0.00 \\
& Eval & 2.35 & 1 & 1.000 & 100.0 & 0.00 & 0.00 & 0.00 & 2.31 & 1 & 1.000 & 100.0 & 0.00 & 0.00 & 0.00 \\
\hline

\multirow{2}{*}{Bell}
& Dev  & 1.00 & 1 & 1.000 & 100.0 & 0.00 & 0.00 & 0.00 & 1.26 & 1 & 1.000 & 100.0 & 0.00 & 0.00 & 0.00 \\
& Eval & 1.03 & 1 & 1.000 & 100.0 & 0.00 & 0.00 & 0.00 & 1.01 & 1 & 1.000 & 100.0 & 0.00 & 0.00 & 0.00 \\
\hline

\multirow{2}{*}{Knock}
& Dev  & 0.06 & 1 & 1.000 & 100.0 & 0.00 & 0.00 & 0.00 & 0.06 & 1 & 1.000 & 100.0 & 0.00 & 0.00 & 0.00 \\
& Eval & 0.06 & 1 & 1.000 & 100.0 & 0.00 & 0.00 & 0.00 & 0.06 & 1 & 1.000 & 100.0 & 0.00 & 0.00 & 0.00 \\
\hline
\end{tabular}
}

\vspace{0.35em}
\parbox{0.92\textwidth}{\centering\footnotesize
Cov. = frame coverage (\%), \textbf{P$_{max}$} = maximum polyphony, \textbf{P$_{mean}$} = mean polyphony over active regions, \textbf{P$_{1-4+}$} = proportion (\%) of active regions with polyphony of 1, 2, 3, or \(\geq 4\) events.\\
Values in \textbf{bold} denote percentage statistics that differ from the corresponding native value by \(\geq1\)pp.
}
\end{center}

\vspace*{\fill}
\clearpage


\vspace*{\fill}

\begin{center}
\captionsetup{type=table}
\captionof{table}{TAU-NIGENS2021 descriptive statistics (percentages refer to the corresponding split).}
\vspace{0.2cm}
\label{tab:tau_nigens_native_window_stats}
\footnotesize
\renewcommand{\arraystretch}{0.98}
\setlength{\tabcolsep}{2.6pt}
\resizebox{\textwidth}{!}{%
\begin{tabular}{|p{2.35cm}|c|ccccccc|ccccccc|}
\hline
\multirow{2}{*}{\textbf{Class}} & \multirow{2}{*}{\textbf{Split}} &
\multicolumn{7}{c|}{\textbf{Native framewise statistics}} &
\multicolumn{7}{c|}{\textbf{Window-based statistics (\(10\) s)}} \\
\cline{3-16}
& &
\textbf{Cov.} & \textbf{P$_{max}$} & \textbf{P$_{mean}$} & \textbf{P${_1}$} & \textbf{P${_2}$} & \textbf{P${_3}$} & \textbf{P${_{4+}}$} &
\textbf{Cov.} & \textbf{P$_{max}$} & \textbf{P$_{mean}$} & \textbf{P${_1}$} & \textbf{P${_2}$} & \textbf{P${_3}$} & \textbf{P${_{4+}}$} \\
\hline

\multirow{3}{*}{Global}
& Train & 84.52 & 3 & 1.753 & 41.69 & 41.35 & 16.96 & 0.00 & 84.39 & 3 & 1.616 & \textbf{48.90} & 40.61 & \textbf{10.49} & 0.00 \\
& Val   & 84.90 & 3 & 1.793 & 40.51 & 39.71 & 19.78 & 0.00 & 84.83 & 3 & 1.716 & \textbf{43.13} & \textbf{42.17} & \textbf{14.70} & 0.00 \\
& Test  & 81.45 & 3 & 1.760 & 41.61 & 40.75 & 17.64 & 0.00 & 81.04 & 3 & 1.634 & \textbf{46.25} & \textbf{44.08} & \textbf{9.67} & 0.00 \\
\hline
\hline

\multirow{3}{*}{Alarm}
& Train & 24.28 & 3 & 1.107 & 89.36 & 10.56 & 0.09 & 0.00 & 24.24 & 3 & 1.107 & 89.36 & 10.56 & 0.09 & 0.00 \\
& Val   & 23.89 & 2 & 1.000 & 99.97 & 0.03 & 0.00 & 0.00 & 23.86 & 2 & 1.000 & 99.97 & 0.03 & 0.00 & 0.00 \\
& Test  & 24.13 & 2 & 1.107 & 89.30 & 10.70 & 0.00 & 0.00 & 24.01 & 2 & 1.107 & 89.30 & 10.70 & 0.00 & 0.00 \\
\hline

\multirow{3}{*}{Crying baby}
& Train & 13.68 & 2 & 1.063 & 93.65 & 6.35 & 0.00 & 0.00 & 13.66 & 2 & 1.064 & 93.63 & 6.37 & 0.00 & 0.00 \\
& Val   & 20.38 & 2 & 1.018 & 98.22 & 1.78 & 0.00 & 0.00 & 20.37 & 2 & 1.018 & 98.22 & 1.78 & 0.00 & 0.00 \\
& Test  & 15.20 & 2 & 1.071 & 92.88 & 7.12 & 0.00 & 0.00 & 15.13 & 2 & 1.071 & 92.85 & 7.15 & 0.00 & 0.00 \\
\hline

\multirow{3}{*}{Crash}
& Train & 12.43 & 2 & 1.043 & 95.68 & 4.32 & 0.00 & 0.00 & 12.42 & 2 & 1.043 & 95.68 & 4.32 & 0.00 & 0.00 \\
& Val   & 22.83 & 2 & 1.026 & 97.36 & 2.64 & 0.00 & 0.00 & 22.81 & 2 & 1.026 & 97.37 & 2.63 & 0.00 & 0.00 \\
& Test  & 12.06 & 2 & 1.120 & 88.02 & 11.98 & 0.00 & 0.00 & 12.00 & 2 & 1.120 & 88.02 & 11.98 & 0.00 & 0.00 \\
\hline

\multirow{3}{*}{Barking dog}
& Train & 8.96 & 2 & 1.021 & 97.94 & 2.06 & 0.00 & 0.00 & 8.95 & 2 & 1.021 & 97.94 & 2.06 & 0.00 & 0.00 \\
& Val   & 5.60 & 2 & 1.023 & 97.70 & 2.30 & 0.00 & 0.00 & 5.59 & 2 & 1.023 & 97.70 & 2.30 & 0.00 & 0.00 \\
& Test  & 10.84 & 2 & 1.017 & 98.32 & 1.68 & 0.00 & 0.00 & 10.78 & 2 & 1.017 & 98.33 & 1.67 & 0.00 & 0.00 \\
\hline

\multirow{3}{*}{Female scream}
& Train & 6.18 & 2 & 1.017 & 98.32 & 1.68 & 0.00 & 0.00 & 6.18 & 2 & 1.017 & 98.32 & 1.68 & 0.00 & 0.00 \\
& Val   & 4.54 & 2 & 1.007 & 99.27 & 0.73 & 0.00 & 0.00 & 4.54 & 2 & 1.007 & 99.27 & 0.73 & 0.00 & 0.00 \\
& Test  & 1.68 & 2 & 1.049 & 95.11 & 4.89 & 0.00 & 0.00 & 1.67 & 2 & 1.049 & 95.11 & 4.89 & 0.00 & 0.00 \\
\hline

\multirow{3}{*}{Female speech}
& Train & 3.83 & 2 & 1.001 & 99.87 & 0.13 & 0.00 & 0.00 & 3.82 & 2 & 1.001 & 99.87 & 0.13 & 0.00 & 0.00 \\
& Val   & 1.94 & 1 & 1.000 & 100.0 & 0.00 & 0.00 & 0.00 & 1.94 & 1 & 1.000 & 100.0 & 0.00 & 0.00 & 0.00 \\
& Test  & 1.66 & 1 & 1.000 & 100.0 & 0.00 & 0.00 & 0.00 & 1.65 & 1 & 1.000 & 100.0 & 0.00 & 0.00 & 0.00 \\
\hline

\multirow{3}{*}{Footsteps}
& Train & 27.76 & 2 & 1.137 & 86.32 & 13.68 & 0.00 & 0.00 & 27.72 & 2 & 1.137 & 86.33 & 13.67 & 0.00 & 0.00 \\
& Val   & 32.24 & 2 & 1.141 & 85.91 & 14.09 & 0.00 & 0.00 & 32.21 & 2 & 1.141 & 85.90 & 14.10 & 0.00 & 0.00 \\
& Test  & 26.35 & 2 & 1.181 & 81.95 & 18.05 & 0.00 & 0.00 & 26.22 & 2 & 1.181 & 81.94 & 18.06 & 0.00 & 0.00 \\
\hline

\multirow{3}{*}{Knocking on door}
& Train & 2.81 & 2 & 1.020 & 98.00 & 2.00 & 0.00 & 0.00 & 2.81 & 2 & 1.020 & 98.00 & 2.00 & 0.00 & 0.00 \\
& Val   & 3.17 & 1 & 1.000 & 100.0 & 0.00 & 0.00 & 0.00 & 3.17 & 1 & 1.000 & 100.0 & 0.00 & 0.00 & 0.00 \\
& Test  & 4.10 & 2 & 1.020 & 98.04 & 1.96 & 0.00 & 0.00 & 4.08 & 2 & 1.020 & 98.02 & 1.98 & 0.00 & 0.00 \\
\hline

\multirow{3}{*}{Male scream}
& Train & 3.90 & 2 & 1.000 & 99.97 & 0.03 & 0.00 & 0.00 & 3.89 & 2 & 1.000 & 99.97 & 0.03 & 0.00 & 0.00 \\
& Val   & 3.48 & 1 & 1.000 & 100.0 & 0.00 & 0.00 & 0.00 & 3.48 & 1 & 1.000 & 100.0 & 0.00 & 0.00 & 0.00 \\
& Test  & 7.64 & 2 & 1.004 & 99.56 & 0.44 & 0.00 & 0.00 & 7.60 & 2 & 1.004 & 99.56 & 0.44 & 0.00 & 0.00 \\
\hline

\multirow{3}{*}{Male speech}
& Train & 2.71 & 2 & 1.044 & 95.60 & 4.40 & 0.00 & 0.00 & 2.71 & 2 & 1.044 & 95.60 & 4.40 & 0.00 & 0.00 \\
& Val   & 3.45 & 2 & 1.051 & 94.87 & 5.13 & 0.00 & 0.00 & 3.44 & 2 & 1.051 & 94.88 & 5.12 & 0.00 & 0.00 \\
& Test  & 3.57 & 2 & 1.016 & 98.36 & 1.64 & 0.00 & 0.00 & 3.55 & 2 & 1.016 & 98.37 & 1.63 & 0.00 & 0.00 \\
\hline

\multirow{3}{*}{Ringing phone}
& Train & 9.27 & 2 & 1.027 & 97.33 & 2.67 & 0.00 & 0.00 & 9.25 & 2 & 1.027 & 97.33 & 2.67 & 0.00 & 0.00 \\
& Val   & 9.95 & 1 & 1.000 & 100.0 & 0.00 & 0.00 & 0.00 & 9.94 & 1 & 1.000 & 100.0 & 0.00 & 0.00 & 0.00 \\
& Test  & 8.10 & 1 & 1.000 & 100.0 & 0.00 & 0.00 & 0.00 & 8.08 & 1 & 1.000 & 100.0 & 0.00 & 0.00 & 0.00 \\
\hline

\multirow{3}{*}{Piano}
& Train & 20.76 & 2 & 1.146 & 85.35 & 14.65 & 0.00 & 0.00 & 20.73 & 2 & 1.146 & 85.36 & 14.64 & 0.00 & 0.00 \\
& Val   & 14.19 & 2 & 1.049 & 95.15 & 4.85 & 0.00 & 0.00 & 14.18 & 2 & 1.049 & 95.15 & 4.85 & 0.00 & 0.00 \\
& Test  & 17.76 & 1 & 1.000 & 100.0 & 0.00 & 0.00 & 0.00 & 17.67 & 1 & 1.000 & 100.0 & 0.00 & 0.00 & 0.00 \\
\hline

\end{tabular}
}

\vspace{0.35em}
\parbox{0.92\textwidth}{\centering\footnotesize
Cov. = frame coverage (\%), \textbf{P$_{max}$} = maximum polyphony, \textbf{P$_{mean}$} = mean polyphony over active regions, \textbf{P$_{1-4+}$} = proportion (\%) of active regions with polyphony of 1, 2, 3, or \(\geq 4\) events.\\
Values in \textbf{bold} denote percentage statistics that differ from the corresponding native value by \(\geq1\)pp.
}
\end{center}

\vspace*{\fill}
\clearpage


\vspace*{\fill}

\begin{center}
\captionsetup{type=table}
\captionof{table}{TAU-NIGENS2020 descriptive statistics (percentages refer to the corresponding split).}
\vspace{0.2cm}
\label{tab:tau_nigens2020_native_window_stats}
\footnotesize
\renewcommand{\arraystretch}{0.98}
\setlength{\tabcolsep}{2.6pt}
\resizebox{\textwidth}{!}{%
\begin{tabular}{|p{2.35cm}|c|ccccccc|ccccccc|}
\hline
\multirow{2}{*}{\textbf{Class}} & \multirow{2}{*}{\textbf{Split}} &
\multicolumn{7}{c|}{\textbf{Native framewise statistics}} &
\multicolumn{7}{c|}{\textbf{Window-based statistics (\(10\) s)}} \\
\cline{3-16}
& &
\textbf{Cov.} & \textbf{P$_{max}$} & \textbf{P$_{mean}$} & \textbf{P${_1}$} & \textbf{P${_2}$} & \textbf{P${_3}$} & \textbf{P${_{4+}}$} &
\textbf{Cov.} & \textbf{P$_{max}$} & \textbf{P$_{mean}$} & \textbf{P${_1}$} & \textbf{P${_2}$} & \textbf{P${_3}$} & \textbf{P${_{4+}}$} \\
\hline

\multirow{3}{*}{Global}
& Train & 82.51 & 2 & 1.325 & 67.47 & 32.53 & 0.00 & 0.00 & \textbf{74.75} & 2 & 1.296 & \textbf{70.43} & \textbf{29.57} & 0.00 & 0.00 \\
& Val   & 81.05 & 2 & 1.306 & 69.39 & 30.61 & 0.00 & 0.00 & \textbf{72.62} & 2 & 1.271 & \textbf{72.87} & \textbf{27.13} & 0.00 & 0.00 \\
& Test & 83.44 & 2 & 1.329 & 67.14 & 32.86 & 0.00 & 0.00 & \textbf{74.72} & 2 & 1.290 & \textbf{71.00} & \textbf{29.00} & 0.00 & 0.00 \\
\hline
\hline

\multirow{3}{*}{Alarm}
& Train & 12.96 & 2 & 1.017 & 98.33 & 1.67 & 0.00 & 0.00 & \textbf{11.74} & 2 & 1.017 & 98.33 & 1.67 & 0.00 & 0.00 \\
& Val   & 8.51 & 2 & 1.007 & 99.30 & 0.70 & 0.00 & 0.00 & 7.63 & 2 & 1.007 & 99.30 & 0.70 & 0.00 & 0.00 \\
& Test & 11.89 & 2 & 1.003 & 99.66 & 0.34 & 0.00 & 0.00 & \textbf{10.64} & 2 & 1.003 & 99.66 & 0.34 & 0.00 & 0.00 \\
\hline

\multirow{3}{*}{Crying baby}
& Train & 8.71 & 2 & 1.010 & 99.00 & 1.00 & 0.00 & 0.00 & 7.89 & 2 & 1.010 & 99.00 & 1.00 & 0.00 & 0.00 \\
& Val   & 8.54 & 1 & 1.000 & 100.0 & 0.00 & 0.00 & 0.00 & 7.65 & 1 & 1.000 & 100.0 & 0.00 & 0.00 & 0.00 \\
& Test & 9.42 & 1 & 1.000 & 100.0 & 0.00 & 0.00 & 0.00 & 8.43 & 1 & 1.000 & 100.0 & 0.00 & 0.00 & 0.00 \\
\hline

\multirow{3}{*}{Crash}
& Train & 7.96 & 2 & 1.037 & 96.25 & 3.75 & 0.00 & 0.00 & 7.21 & 2 & 1.037 & 96.25 & 3.75 & 0.00 & 0.00 \\
& Val   & 10.13 & 2 & 1.001 & 99.94 & 0.06 & 0.00 & 0.00 & \textbf{9.07} & 2 & 1.001 & 99.94 & 0.06 & 0.00 & 0.00 \\
& Test & 8.43 & 2 & 1.006 & 99.43 & 0.57 & 0.00 & 0.00 & 7.55 & 2 & 1.006 & 99.43 & 0.57 & 0.00 & 0.00 \\
\hline

\multirow{3}{*}{Barking dog}
& Train & 7.46 & 2 & 1.002 & 99.84 & 0.16 & 0.00 & 0.00 & 6.76 & 2 & 1.002 & 99.84 & 0.16 & 0.00 & 0.00 \\
& Val   & 3.18 & 2 & 1.016 & 98.36 & 1.64 & 0.00 & 0.00 & 2.85 & 2 & 1.016 & 98.38 & 1.62 & 0.00 & 0.00 \\
& Test & 7.23 & 2 & 1.110 & 89.03 & 10.97 & 0.00 & 0.00 & 6.47 & 2 & 1.110 & 89.03 & 10.97 & 0.00 & 0.00 \\
\hline

\multirow{3}{*}{Running engine}
& Train & 12.68 & 2 & 1.046 & 95.41 & 4.59 & 0.00 & 0.00 & \textbf{11.49} & 2 & 1.046 & 95.41 & 4.59 & 0.00 & 0.00 \\
& Val   & 12.06 & 2 & 1.048 & 95.24 & 4.76 & 0.00 & 0.00 & \textbf{10.81} & 2 & 1.048 & 95.24 & 4.76 & 0.00 & 0.00 \\
& Test & 13.24 & 2 & 1.033 & 96.68 & 3.32 & 0.00 & 0.00 & \textbf{11.85} & 2 & 1.033 & 96.68 & 3.32 & 0.00 & 0.00 \\
\hline

\multirow{3}{*}{Female scream}
& Train & 3.56 & 2 & 1.012 & 98.84 & 1.16 & 0.00 & 0.00 & 3.23 & 2 & 1.012 & 98.84 & 1.16 & 0.00 & 0.00 \\
& Val   & 2.67 & 2 & 1.010 & 98.96 & 1.04 & 0.00 & 0.00 & 2.39 & 2 & 1.010 & 98.97 & 1.03 & 0.00 & 0.00 \\
& Test & 1.02 & 1 & 1.000 & 100.0 & 0.00 & 0.00 & 0.00 & 0.92 & 1 & 1.000 & 100.0 & 0.00 & 0.00 & 0.00 \\
\hline

\multirow{3}{*}{Female speech}
& Train & 2.19 & 2 & 1.003 & 99.71 & 0.29 & 0.00 & 0.00 & 1.98 & 2 & 1.003 & 99.70 & 0.30 & 0.00 & 0.00 \\
& Val   & 1.05 & 1 & 1.000 & 100.0 & 0.00 & 0.00 & 0.00 & 0.94 & 1 & 1.000 & 100.0 & 0.00 & 0.00 & 0.00 \\
& Test & 1.02 & 1 & 1.000 & 100.0 & 0.00 & 0.00 & 0.00 & 0.92 & 1 & 1.000 & 100.0 & 0.00 & 0.00 & 0.00 \\
\hline

\multirow{3}{*}{Burning fire}
& Train & 14.99 & 2 & 1.046 & 95.38 & 4.62 & 0.00 & 0.00 & \textbf{13.59} & 2 & 1.046 & 95.38 & 4.62 & 0.00 & 0.00 \\
& Val   & 14.27 & 1 & 1.000 & 100.0 & 0.00 & 0.00 & 0.00 & \textbf{12.79} & 1 & 1.000 & 100.0 & 0.00 & 0.00 & 0.00 \\
& Test & 15.32 & 2 & 1.119 & 88.11 & 11.89 & 0.00 & 0.00 & \textbf{13.72} & 2 & 1.119 & 88.11 & 11.89 & 0.00 & 0.00 \\
\hline

\multirow{3}{*}{Footsteps}
& Train & 14.95 & 2 & 1.026 & 97.40 & 2.60 & 0.00 & 0.00 & \textbf{13.55} & 2 & 1.026 & 97.40 & 2.60 & 0.00 & 0.00 \\
& Val   & 15.65 & 2 & 1.059 & 94.10 & 5.90 & 0.00 & 0.00 & \textbf{14.02} & 2 & 1.059 & 94.10 & 5.90 & 0.00 & 0.00 \\
& Test & 12.07 & 2 & 1.006 & 99.35 & 0.65 & 0.00 & 0.00 & \textbf{10.81} & 2 & 1.006 & 99.35 & 0.65 & 0.00 & 0.00 \\
\hline

\multirow{3}{*}{Knocking on door}
& Train & 1.58 & 2 & 1.003 & 99.74 & 0.26 & 0.00 & 0.00 & 1.43 & 2 & 1.003 & 99.74 & 0.26 & 0.00 & 0.00 \\
& Val   & 1.74 & 1 & 1.000 & 100.0 & 0.00 & 0.00 & 0.00 & 1.56 & 1 & 1.000 & 100.0 & 0.00 & 0.00 & 0.00 \\
& Test & 2.38 & 1 & 1.000 & 100.0 & 0.00 & 0.00 & 0.00 & 2.13 & 1 & 1.000 & 100.0 & 0.00 & 0.00 & 0.00 \\
\hline

\multirow{3}{*}{Male scream}
& Train & 2.05 & 1 & 1.000 & 100.0 & 0.00 & 0.00 & 0.00 & 1.86 & 1 & 1.000 & 100.0 & 0.00 & 0.00 & 0.00 \\
& Val   & 4.08 & 1 & 1.000 & 100.0 & 0.00 & 0.00 & 0.00 & 3.66 & 1 & 1.000 & 100.0 & 0.00 & 0.00 & 0.00 \\
& Test & 4.75 & 1 & 1.000 & 100.0 & 0.00 & 0.00 & 0.00 & 4.25 & 1 & 1.000 & 100.0 & 0.00 & 0.00 & 0.00 \\
\hline

\multirow{3}{*}{Male speech}
& Train & 1.54 & 2 & 1.009 & 99.13 & 0.87 & 0.00 & 0.00 & 1.39 & 2 & 1.009 & 99.13 & 0.87 & 0.00 & 0.00 \\
& Val   & 2.04 & 1 & 1.000 & 100.0 & 0.00 & 0.00 & 0.00 & 1.83 & 1 & 1.000 & 100.0 & 0.00 & 0.00 & 0.00 \\
& Test & 2.08 & 1 & 1.000 & 100.0 & 0.00 & 0.00 & 0.00 & 1.86 & 1 & 1.000 & 100.0 & 0.00 & 0.00 & 0.00 \\
\hline

\multirow{3}{*}{Ringing phone}
& Train & 8.39 & 2 & 1.012 & 98.82 & 1.18 & 0.00 & 0.00 & 7.60 & 2 & 1.012 & 98.82 & 1.18 & 0.00 & 0.00 \\
& Val   & 12.54 & 2 & 1.094 & 90.57 & 9.43 & 0.00 & 0.00 & \textbf{11.24} & 2 & 1.094 & 90.57 & 9.43 & 0.00 & 0.00 \\
& Test & 11.44 & 1 & 1.000 & 100.0 & 0.00 & 0.00 & 0.00 & \textbf{10.25} & 1 & 1.000 & 100.0 & 0.00 & 0.00 & 0.00 \\
\hline

\multirow{3}{*}{Piano}
& Train & 7.88 & 1 & 1.000 & 100.0 & 0.00 & 0.00 & 0.00 & 7.14 & 1 & 1.000 & 100.0 & 0.00 & 0.00 & 0.00 \\
& Val   & 6.55 & 1 & 1.000 & 100.0 & 0.00 & 0.00 & 0.00 & 5.87 & 1 & 1.000 & 100.0 & 0.00 & 0.00 & 0.00 \\
& Test & 7.35 & 1 & 1.000 & 100.0 & 0.00 & 0.00 & 0.00 & 6.58 & 1 & 1.000 & 100.0 & 0.00 & 0.00 & 0.00 \\
\hline

\end{tabular}
}

\vspace{0.35em}
\parbox{0.92\textwidth}{\centering\footnotesize
Cov. = frame coverage (\%), \textbf{P$_{max}$} = maximum polyphony, \textbf{P$_{mean}$} = mean polyphony over active regions, \textbf{P$_{1-4+}$} = proportion (\%) of active regions with polyphony of 1, 2, 3, or \(\geq 4\) events.\\
Values in \textbf{bold} denote percentage statistics that differ from the corresponding native value by \(\geq1\)pp.
}
\end{center}

\vspace*{\fill}
\clearpage


\vspace*{\fill}

\begin{center}
\captionsetup{type=table}
\captionof{table}{TAU2019 descriptive statistics (percentages refer to the corresponding split).}
\vspace{0.2cm}
\label{tab:tau2019_native_window_stats}
\footnotesize
\renewcommand{\arraystretch}{0.98}
\setlength{\tabcolsep}{2.6pt}
\resizebox{\textwidth}{!}{%
\begin{tabular}{|p{2.35cm}|c|ccccccc|ccccccc|}
\hline
\multirow{2}{*}{\textbf{Class}} & \multirow{2}{*}{\textbf{Split}} &
\multicolumn{7}{c|}{\textbf{Native framewise statistics}} &
\multicolumn{7}{c|}{\textbf{Window-based statistics (\(10\) s)}} \\
\cline{3-16}
& &
\textbf{Cov.} & \textbf{P$_{max}$} & \textbf{P$_{mean}$} & \textbf{P${_1}$} & \textbf{P${_2}$} & \textbf{P${_3}$} & \textbf{P${_{4+}}$} &
\textbf{Cov.} & \textbf{P$_{max}$} & \textbf{P$_{mean}$} & \textbf{P${_1}$} & \textbf{P${_2}$} & \textbf{P${_3}$} & \textbf{P${_{4+}}$} \\
\hline

\multirow{3}{*}{Global}
& Train & 70.09 & 2 & 1.237 & 76.30 & 23.70 & 0.00 & 0.00 & 70.02 & 2 & 1.215 & \textbf{78.55} & \textbf{21.45} & 0.00 & 0.00 \\
& Val   & 71.20 & 2 & 1.239 & 76.12 & 23.88 & 0.00 & 0.00 & 71.16 & 2 & 1.214 & \textbf{78.60} & \textbf{21.40} & 0.00 & 0.00 \\
& Test  & 70.79 & 2 & 1.228 & 77.17 & 22.83 & 0.00 & 0.00 & 70.72 & 2 & 1.202 & \textbf{79.81} & \textbf{20.19} & 0.00 & 0.00 \\
\hline
\hline

\multirow{3}{*}{Clearthroat}
& Train & 5.65 & 2 & 1.009 & 99.14 & 0.86 & 0.00 & 0.00 & 5.64 & 2 & 1.009 & 99.14 & 0.86 & 0.00 & 0.00 \\
& Val   & 5.52 & 2 & 1.010 & 99.02 & 0.98 & 0.00 & 0.00 & 5.51 & 2 & 1.010 & 99.01 & 0.99 & 0.00 & 0.00 \\
& Test  & 4.52 & 2 & 1.011 & 98.87 & 1.13 & 0.00 & 0.00 & 4.51 & 2 & 1.011 & 98.86 & 1.14 & 0.00 & 0.00 \\
\hline

\multirow{3}{*}{Cough}
& Train & 7.52 & 2 & 1.014 & 98.64 & 1.36 & 0.00 & 0.00 & 7.52 & 2 & 1.014 & 98.64 & 1.36 & 0.00 & 0.00 \\
& Val   & 6.50 & 2 & 1.020 & 98.02 & 1.98 & 0.00 & 0.00 & 6.49 & 2 & 1.020 & 98.02 & 1.98 & 0.00 & 0.00 \\
& Test  & 5.94 & 2 & 1.015 & 98.46 & 1.54 & 0.00 & 0.00 & 5.94 & 2 & 1.015 & 98.45 & 1.55 & 0.00 & 0.00 \\
\hline

\multirow{3}{*}{Doorslam}
& Train & 3.39 & 2 & 1.011 & 98.92 & 1.08 & 0.00 & 0.00 & 3.39 & 2 & 1.011 & 98.92 & 1.08 & 0.00 & 0.00 \\
& Val   & 3.67 & 2 & 1.006 & 99.40 & 0.60 & 0.00 & 0.00 & 3.67 & 2 & 1.006 & 99.40 & 0.60 & 0.00 & 0.00 \\
& Test  & 4.43 & 2 & 1.008 & 99.16 & 0.84 & 0.00 & 0.00 & 4.43 & 2 & 1.008 & 99.16 & 0.84 & 0.00 & 0.00 \\
\hline

\multirow{3}{*}{Drawer}
& Train & 6.71 & 2 & 1.013 & 98.69 & 1.31 & 0.00 & 0.00 & 6.70 & 2 & 1.013 & 98.68 & 1.32 & 0.00 & 0.00 \\
& Val   & 6.61 & 2 & 1.007 & 99.33 & 0.67 & 0.00 & 0.00 & 6.62 & 2 & 1.007 & 99.33 & 0.67 & 0.00 & 0.00 \\
& Test  & 7.83 & 2 & 1.023 & 97.70 & 2.30 & 0.00 & 0.00 & 7.82 & 2 & 1.023 & 97.71 & 2.29 & 0.00 & 0.00 \\
\hline

\multirow{3}{*}{Keyboard}
& Train & 11.89 & 2 & 1.025 & 97.54 & 2.46 & 0.00 & 0.00 & 11.88 & 2 & 1.025 & 97.54 & 2.46 & 0.00 & 0.00 \\
& Val   & 11.71 & 2 & 1.036 & 96.40 & 3.60 & 0.00 & 0.00 & 11.71 & 2 & 1.036 & 96.40 & 3.60 & 0.00 & 0.00 \\
& Test  & 12.71 & 2 & 1.028 & 97.21 & 2.79 & 0.00 & 0.00 & 12.70 & 2 & 1.028 & 97.21 & 2.79 & 0.00 & 0.00 \\
\hline

\multirow{3}{*}{Keys drop}
& Train & 2.74 & 2 & 1.005 & 99.50 & 0.50 & 0.00 & 0.00 & 2.73 & 2 & 1.005 & 99.51 & 0.49 & 0.00 & 0.00 \\
& Val   & 2.25 & 2 & 1.005 & 99.55 & 0.45 & 0.00 & 0.00 & 2.25 & 2 & 1.004 & 99.56 & 0.44 & 0.00 & 0.00 \\
& Test  & 2.47 & 2 & 1.002 & 99.79 & 0.21 & 0.00 & 0.00 & 2.47 & 2 & 1.002 & 99.79 & 0.21 & 0.00 & 0.00 \\
\hline

\multirow{3}{*}{Knock}
& Train & 5.80 & 2 & 1.010 & 99.00 & 1.00 & 0.00 & 0.00 & 5.79 & 2 & 1.010 & 98.99 & 1.01 & 0.00 & 0.00 \\
& Val   & 4.65 & 2 & 1.013 & 98.69 & 1.31 & 0.00 & 0.00 & 4.65 & 2 & 1.013 & 98.68 & 1.32 & 0.00 & 0.00 \\
& Test  & 5.64 & 2 & 1.010 & 99.01 & 0.99 & 0.00 & 0.00 & 5.63 & 2 & 1.010 & 99.01 & 0.99 & 0.00 & 0.00 \\
\hline

\multirow{3}{*}{Laughter}
& Train & 10.33 & 2 & 1.023 & 97.69 & 2.31 & 0.00 & 0.00 & 10.33 & 2 & 1.023 & 97.69 & 2.31 & 0.00 & 0.00 \\
& Val   & 11.48 & 2 & 1.020 & 98.01 & 1.99 & 0.00 & 0.00 & 11.48 & 2 & 1.020 & 98.01 & 1.99 & 0.00 & 0.00 \\
& Test  & 11.26 & 2 & 1.030 & 96.98 & 3.02 & 0.00 & 0.00 & 11.25 & 2 & 1.030 & 96.98 & 3.02 & 0.00 & 0.00 \\
\hline

\multirow{3}{*}{Page turn}
& Train & 7.91 & 2 & 1.019 & 98.06 & 1.94 & 0.00 & 0.00 & 7.90 & 2 & 1.019 & 98.07 & 1.93 & 0.00 & 0.00 \\
& Val   & 4.81 & 2 & 1.003 & 99.68 & 0.32 & 0.00 & 0.00 & 4.81 & 2 & 1.003 & 99.69 & 0.31 & 0.00 & 0.00 \\
& Test  & 6.70 & 2 & 1.014 & 98.63 & 1.37 & 0.00 & 0.00 & 6.69 & 2 & 1.014 & 98.63 & 1.37 & 0.00 & 0.00 \\
\hline

\multirow{3}{*}{Phone}
& Train & 11.19 & 2 & 1.020 & 98.03 & 1.97 & 0.00 & 0.00 & 11.19 & 2 & 1.020 & 98.03 & 1.97 & 0.00 & 0.00 \\
& Val   & 16.50 & 2 & 1.039 & 96.09 & 3.91 & 0.00 & 0.00 & 16.49 & 2 & 1.039 & 96.09 & 3.91 & 0.00 & 0.00 \\
& Test  & 11.74 & 2 & 1.029 & 97.09 & 2.91 & 0.00 & 0.00 & 11.73 & 2 & 1.029 & 97.08 & 2.92 & 0.00 & 0.00 \\
\hline

\multirow{3}{*}{Speech}
& Train & 11.97 & 2 & 1.028 & 97.20 & 2.80 & 0.00 & 0.00 & 11.97 & 2 & 1.028 & 97.20 & 2.80 & 0.00 & 0.00 \\
& Val   & 12.72 & 2 & 1.011 & 98.87 & 1.13 & 0.00 & 0.00 & 12.72 & 2 & 1.011 & 98.86 & 1.14 & 0.00 & 0.00 \\
& Test  & 11.83 & 2 & 1.028 & 97.23 & 2.77 & 0.00 & 0.00 & 11.82 & 2 & 1.028 & 97.23 & 2.77 & 0.00 & 0.00 \\
\hline
\end{tabular}
}
\vspace{0.35em}
\parbox{0.92\textwidth}{\centering\footnotesize
Cov. = frame coverage (\%), \textbf{P$_{max}$} = maximum polyphony, \textbf{P$_{mean}$} = mean polyphony over active regions, \textbf{P$_{1-4+}$} = proportion (\%) of active regions with polyphony of 1, 2, 3, or \(\geq 4\) events.\\
Values in \textbf{bold} denote percentage statistics that differ from the corresponding native value by \(\geq1\)pp.
}
\end{center}

\vspace*{\fill}
\clearpage


\subsection{AT2SELD Performance Diagnosis}
\label{app:sed_perf_appendix}

\vspace{1cm}

\begingroup
\scriptsize
\setlength{\tabcolsep}{2.1pt}
\renewcommand{\arraystretch}{1.08}
\captionsetup{type=table,font=scriptsize,skip=2pt}

\begin{center}

\captionof{table}{Per-class thresholds selected on validation to maximize Precision.}
\label{tab:sed_thresholds_precision}
\resizebox{\textwidth}{!}{%
\begin{tabular}{|l|c|ccc|ccc|c|ccc|ccc|}
\hline
\textbf{Class} & \multicolumn{7}{c|}{\textbf{ExpA}} & \multicolumn{7}{c|}{\textbf{mixed focal}} \\
\cline{2-15}
 & \(\boldsymbol{\tau^\star}\) & \multicolumn{3}{c|}{\textbf{Validation}} & \multicolumn{3}{c|}{\textbf{Test}} & \(\boldsymbol{\tau^\star}\) & \multicolumn{3}{c|}{\textbf{Validation}} & \multicolumn{3}{c|}{\textbf{Test}} \\
 &  & \textbf{P} & \textbf{R} & \textbf{F1} & \textbf{P} & \textbf{R} & \textbf{F1} &  & \textbf{P} & \textbf{R} & \textbf{F1} & \textbf{P} & \textbf{R} & \textbf{F1} \\
\hline
\texttt{female\_speech} & 0.95 & 1.000 & 0.005 & 0.010 & 0.988 & 0.003 & 0.007 & 0.95 & 1.000 & 0.005 & 0.010 & 0.902 & 0.009 & 0.017 \\
\texttt{male\_speech} & 0.95 & 0.998 & 0.069 & 0.129 & 0.997 & 0.020 & 0.040 & 0.95 & 1.000 & 0.006 & 0.012 & 1.000 & 0.003 & 0.007 \\
\texttt{clapping} & 0.95 & 1.000 & 0.001 & 0.002 & 0.000 & 0.000 & 0.000 & 0.95 & 1.000 & 0.258 & 0.411 & 0.992 & 0.129 & 0.228 \\
\texttt{telephone} & 0.75 & 1.000 & 0.099 & 0.180 & 0.510 & 0.017 & 0.033 & 0.75 & 1.000 & 0.387 & 0.558 & 0.589 & 0.356 & 0.444 \\
\texttt{laughter} & 0.90 & 1.000 & 0.005 & 0.011 & 0.000 & 0.000 & 0.000 & 0.90 & 1.000 & 0.016 & 0.032 & 0.994 & 0.005 & 0.010 \\
\texttt{domestic\_sounds} & 0.95 & 1.000 & 0.154 & 0.266 & 1.000 & 0.044 & 0.084 & 0.95 & 0.983 & 0.702 & 0.819 & 1.000 & 0.129 & 0.228 \\
\texttt{walk\_footsteps} & 0.85 & 1.000 & 0.107 & 0.193 & 0.930 & 0.011 & 0.021 & 0.85 & 1.000 & 0.034 & 0.066 & 0.382 & 0.000 & 0.000 \\
\texttt{door\_open\_close} & 0.80 & 1.000 & 0.144 & 0.251 & 1.000 & 0.010 & 0.020 & 0.90 & 1.000 & 0.003 & 0.007 & 0.978 & 0.002 & 0.004 \\
\texttt{music} & 0.95 & 1.000 & 0.083 & 0.153 & 0.999 & 0.010 & 0.020 & 0.90 & 1.000 & 0.652 & 0.789 & 0.990 & 0.363 & 0.531 \\
\texttt{musical\_instrument} & 0.60 & 0.933 & 0.004 & 0.007 & 0.981 & 0.100 & 0.182 & 0.85 & 1.000 & 0.032 & 0.062 & 0.990 & 0.111 & 0.200 \\
\texttt{water\_tap} & 0.75 & 1.000 & 0.820 & 0.901 & 1.000 & 0.063 & 0.118 & 0.70 & 1.000 & 0.774 & 0.872 & 0.983 & 0.121 & 0.215 \\
\texttt{bell} & 0.60\textsuperscript{*} & 0.000 & 0.000 & 0.000 & 1.000 & 0.013 & 0.025 & 0.70\textsuperscript{*} & 0.000 & 0.000 & 0.000 & 1.000 & 0.058 & 0.110 \\
\texttt{knock} & 0.25 & 0.158 & 0.063 & 0.090 & 0.040 & 0.031 & 0.035 & 0.50 & 1.000 & 0.615 & 0.762 & 0.482 & 0.475 & 0.478 \\
\hline
\textbf{Best model by test macro-Precision: mixed focal} & -- & -- & -- & -- & -- & -- & -- & -- & 0.922 & 0.268 & 0.339 & 0.868 & 0.135 & 0.190 \\
\hline
\end{tabular}%
}

\vspace{1cm}

\captionof{table}{Per-class thresholds selected on validation to maximize Recall.}
\label{tab:sed_thresholds_recall}
\resizebox{\textwidth}{!}{%
\begin{tabular}{|l|c|ccc|ccc|c|ccc|ccc|}
\hline
\textbf{Class} & \multicolumn{7}{c|}{\textbf{ExpA}} & \multicolumn{7}{c|}{\textbf{mixed focal}} \\
\cline{2-15}
 & \(\boldsymbol{\tau^\star}\) & \multicolumn{3}{c|}{\textbf{Validation}} & \multicolumn{3}{c|}{\textbf{Test}} & \(\boldsymbol{\tau^\star}\) & \multicolumn{3}{c|}{\textbf{Validation}} & \multicolumn{3}{c|}{\textbf{Test}} \\
 &  & \textbf{P} & \textbf{R} & \textbf{F1} & \textbf{P} & \textbf{R} & \textbf{F1} &  & \textbf{P} & \textbf{R} & \textbf{F1} & \textbf{P} & \textbf{R} & \textbf{F1} \\
\hline
\texttt{female\_speech} & 0.05 & 0.314 & 0.998 & 0.478 & 0.321 & 0.989 & 0.485 & 0.05 & 0.313 & 0.996 & 0.476 & 0.284 & 0.997 & 0.442 \\
\texttt{male\_speech} & 0.05 & 0.466 & 0.999 & 0.636 & 0.415 & 0.995 & 0.585 & 0.05 & 0.451 & 1.000 & 0.622 & 0.435 & 0.993 & 0.605 \\
\texttt{clapping} & 0.05 & 0.136 & 0.995 & 0.239 & 0.024 & 0.983 & 0.048 & 0.05 & 0.130 & 0.994 & 0.230 & 0.041 & 0.995 & 0.079 \\
\texttt{telephone} & 0.05 & 0.065 & 0.964 & 0.121 & 0.019 & 0.831 & 0.037 & 0.05 & 0.097 & 0.919 & 0.175 & 0.032 & 0.869 & 0.061 \\
\texttt{laughter} & 0.05 & 0.054 & 0.990 & 0.102 & 0.073 & 0.937 & 0.135 & 0.05 & 0.047 & 0.969 & 0.090 & 0.075 & 0.928 & 0.139 \\
\texttt{domestic\_sounds} & 0.05 & 0.616 & 0.999 & 0.762 & 0.312 & 0.917 & 0.465 & 0.05 & 0.804 & 0.996 & 0.890 & 0.527 & 0.870 & 0.656 \\
\texttt{walk\_footsteps} & 0.05 & 0.052 & 0.906 & 0.098 & 0.080 & 0.809 & 0.145 & 0.05 & 0.035 & 0.989 & 0.067 & 0.055 & 0.828 & 0.103 \\
\texttt{door\_open\_close} & 0.05 & 0.011 & 1.000 & 0.022 & 0.013 & 0.909 & 0.026 & 0.05 & 0.023 & 1.000 & 0.045 & 0.048 & 0.963 & 0.091 \\
\texttt{music} & 0.05 & 0.482 & 0.999 & 0.651 & 0.501 & 0.969 & 0.660 & 0.05 & 0.389 & 0.995 & 0.560 & 0.413 & 0.918 & 0.569 \\
\texttt{musical\_instrument} & 0.05 & 0.046 & 0.623 & 0.085 & 0.434 & 0.677 & 0.529 & 0.05 & 0.038 & 0.853 & 0.072 & 0.432 & 0.890 & 0.582 \\
\texttt{water\_tap} & 0.10 & 0.105 & 1.000 & 0.190 & 0.047 & 0.094 & 0.062 & 0.10 & 0.424 & 1.000 & 0.596 & 0.401 & 0.380 & 0.390 \\
\texttt{bell} & 0.05\textsuperscript{*} & 0.000 & 0.000 & 0.000 & 0.078 & 0.861 & 0.143 & 0.05\textsuperscript{*} & 0.000 & 0.000 & 0.000 & 0.159 & 0.753 & 0.262 \\
\texttt{knock} & 0.05 & 0.012 & 0.895 & 0.024 & 0.003 & 0.486 & 0.007 & 0.05 & 0.009 & 1.000 & 0.017 & 0.004 & 0.836 & 0.008 \\
\hline
\textbf{Best model by test macro-Recall: mixed focal} & -- & -- & -- & -- & -- & -- & -- & -- & 0.212 & 0.901 & 0.295 & 0.223 & 0.863 & 0.307 \\
\hline
\end{tabular}%
}

\vspace{1cm}

\captionof{table}{Per-class thresholds selected on validation to maximize F1-score.}
\label{tab:sed_thresholds_f1}
\resizebox{\textwidth}{!}{%
\begin{tabular}{|l|c|ccc|ccc|c|ccc|ccc|}
\hline
\textbf{Class} & \multicolumn{7}{c|}{\textbf{ExpA}} & \multicolumn{7}{c|}{\textbf{mixed focal}} \\
\cline{2-15}
 & \(\boldsymbol{\tau^\star}\) & \multicolumn{3}{c|}{\textbf{Validation}} & \multicolumn{3}{c|}{\textbf{Test}} & \(\boldsymbol{\tau^\star}\) & \multicolumn{3}{c|}{\textbf{Validation}} & \multicolumn{3}{c|}{\textbf{Test}} \\
 &  & \textbf{P} & \textbf{R} & \textbf{F1} & \textbf{P} & \textbf{R} & \textbf{F1} &  & \textbf{P} & \textbf{R} & \textbf{F1} & \textbf{P} & \textbf{R} & \textbf{F1} \\
\hline
\texttt{female\_speech} & 0.35 & 0.836 & 0.790 & 0.812 & 0.749 & 0.703 & 0.726 & 0.45 & 0.879 & 0.859 & 0.869 & 0.738 & 0.860 & 0.794 \\
\texttt{male\_speech} & 0.40 & 0.850 & 0.912 & 0.880 & 0.864 & 0.770 & 0.814 & 0.50 & 0.903 & 0.905 & 0.904 & 0.918 & 0.753 & 0.827 \\
\texttt{clapping} & 0.25 & 0.720 & 0.830 & 0.771 & 0.169 & 0.751 & 0.276 & 0.45 & 0.856 & 0.919 & 0.886 & 0.734 & 0.893 & 0.806 \\
\texttt{telephone} & 0.35 & 0.796 & 0.553 & 0.653 & 0.190 & 0.451 & 0.268 & 0.35 & 0.935 & 0.606 & 0.736 & 0.283 & 0.650 & 0.395 \\
\texttt{laughter} & 0.35 & 0.638 & 0.545 & 0.588 & 0.612 & 0.443 & 0.514 & 0.45 & 0.654 & 0.546 & 0.595 & 0.657 & 0.434 & 0.523 \\
\texttt{domestic\_sounds} & 0.30 & 0.927 & 0.955 & 0.941 & 0.762 & 0.744 & 0.753 & 0.45 & 0.955 & 0.976 & 0.965 & 0.958 & 0.652 & 0.776 \\
\texttt{walk\_footsteps} & 0.35 & 0.581 & 0.496 & 0.536 & 0.590 & 0.226 & 0.327 & 0.50 & 0.791 & 0.423 & 0.551 & 0.353 & 0.193 & 0.250 \\
\texttt{door\_open\_close} & 0.50 & 0.761 & 0.593 & 0.666 & 0.566 & 0.172 & 0.263 & 0.50 & 0.744 & 0.703 & 0.723 & 0.656 & 0.356 & 0.462 \\
\texttt{music} & 0.30 & 0.887 & 0.946 & 0.916 & 0.803 & 0.787 & 0.795 & 0.50 & 0.975 & 0.945 & 0.959 & 0.883 & 0.673 & 0.764 \\
\texttt{musical\_instrument} & 0.15 & 0.242 & 0.277 & 0.259 & 0.776 & 0.367 & 0.498 & 0.50 & 0.692 & 0.323 & 0.441 & 0.919 & 0.578 & 0.710 \\
\texttt{water\_tap} & 0.55 & 0.984 & 0.928 & 0.955 & 0.879 & 0.076 & 0.140 & 0.35 & 0.981 & 0.962 & 0.971 & 0.946 & 0.254 & 0.400 \\
\texttt{bell} & 0.35\textsuperscript{*} & 0.000 & 0.000 & 0.000 & 0.928 & 0.188 & 0.312 & 0.45\textsuperscript{*} & 0.000 & 0.000 & 0.000 & 0.985 & 0.200 & 0.333 \\
\texttt{knock} & 0.20 & 0.088 & 0.123 & 0.103 & 0.034 & 0.077 & 0.047 & 0.35 & 0.919 & 0.798 & 0.855 & 0.195 & 0.583 & 0.293 \\
\hline
\textbf{Best model by test macro-F1-score: mixed focal} & -- & -- & -- & -- & -- & -- & -- & -- & 0.791 & 0.690 & 0.727 & 0.710 & 0.545 & 0.564 \\
\hline
\end{tabular}%
}

\vspace{0.5cm}

\parbox{0.96\linewidth}{\centering\scriptsize
Each threshold is selected independently on the STARSS23 validation split.\\Precision (P), Recall (R), and F1 report the metrics obtained after applying that threshold on validation and test data.\\The final row reports test macro-averages for the model with the highest test macro-score after validation-only threshold selection.\\\textbf{\textsuperscript{(*)}} denotes a class with negligible positive validation support, for which the model-level validation optimum is used as fallback.
}

\end{center}
\endgroup

\clearpage


\begin{landscape}
\begin{longtable}{|l|l|l|c|c|r|c|c|c|c|c|}
\caption{Class-aware deployment diagnostics. Activity thresholds and angular tolerances are selected on validation and applied unchanged to test. Only classes supported by each projection are included.}
\label{tab:dataset_generalization_class_policy}\\
\hline
\textbf{Training dataset} & \textbf{Test dataset} & \textbf{Class} & \(\tau_c\) & \(\alpha_c\) & \textbf{Support} & \(\mathrm{SED}\ P\) & \(\mathrm{SED}\ R\) & \(\mathrm{SED}\ F_1\) & \(F_{\mathrm{SELD}}\) & \(\mathrm{LE}\) \\
\hline
\endfirsthead
\hline
\textbf{Training dataset} & \textbf{Test dataset} & \textbf{Class} & \(\tau_c\) & \(\alpha_c\) & \textbf{Support} & \(\mathrm{SED}\ P\) & \(\mathrm{SED}\ R\) & \(\mathrm{SED}\ F_1\) & \(F_{\mathrm{SELD}}\) & \(\mathrm{LE}\) \\
\hline
\endhead
STARSS23 & STARSS23 & \texttt{male\_speech} & 0.50 & 45$^\circ$ & 1486350 & 0.910 & 0.681 & 0.779 & 0.702 & 18.46$^\circ$ \\
STARSS23 & STARSS23 & \texttt{music} & 0.40 & 30$^\circ$ & 1155060 & 0.832 & 0.684 & 0.751 & 0.241 & 19.26$^\circ$ \\
STARSS23 & STARSS23 & \texttt{female\_speech} & 0.50 & 45$^\circ$ & 1044489 & 0.822 & 0.534 & 0.648 & 0.491 & 15.54$^\circ$ \\
STARSS23 & STARSS23 & \texttt{musical\_instrument} & 0.15 & 30$^\circ$ & 818413 & 0.776 & 0.367 & 0.498 & 0.204 & 21.32$^\circ$ \\
STARSS23 & STARSS23 & \texttt{domestic\_sounds} & 0.50 & 45$^\circ$ & 627966 & 0.901 & 0.542 & 0.677 & 0.563 & 27.73$^\circ$ \\
STARSS23 & STARSS23 & \texttt{footsteps} & 0.55 & 45$^\circ$ & 176889 & 0.793 & 0.120 & 0.208 & 0.198 & 17.91$^\circ$ \\
STARSS23 & STARSS23 & \texttt{laughter} & 0.45 & 45$^\circ$ & 141315 & 0.727 & 0.348 & 0.471 & 0.331 & 18.49$^\circ$ \\
STARSS23 & STARSS23 & \texttt{water\_tap} & 0.50 & 10$^\circ$ & 101696 & 0.762 & 0.078 & 0.141 & 0.000 & 0.00$^\circ$ \\
STARSS23 & STARSS23 & \texttt{bell} & 0.55 & 15$^\circ$ & 45799 & 1.000 & 0.030 & 0.059 & 0.000 & 0.00$^\circ$ \\
STARSS23 & STARSS23 & \texttt{telephone} & 0.70 & 20$^\circ$ & 35953 & 0.444 & 0.035 & 0.064 & 0.013 & 18.36$^\circ$ \\
STARSS23 & STARSS23 & \texttt{clapping} & 0.20 & 45$^\circ$ & 22123 & 0.129 & 0.814 & 0.222 & 0.197 & 20.58$^\circ$ \\
STARSS23 & STARSS23 & \texttt{door} & 0.45 & 15$^\circ$ & 21429 & 0.461 & 0.212 & 0.291 & 0.083 & 7.01$^\circ$ \\
STARSS23 & STARSS23 & \texttt{knock} & 0.40 & 10$^\circ$ & 1717 & 0.000 & 0.000 & 0.000 & 0.000 & 0.00$^\circ$ \\
\hline
STARSS23 & TAU2019 & \texttt{domestic\_sounds} & 0.05 & 45$^\circ$ & 629458 & 0.306 & 0.581 & 0.400 & 0.090 & 30.02$^\circ$ \\
STARSS23 & TAU2019 & \texttt{telephone} & 0.20 & 45$^\circ$ & 260418 & 0.607 & 0.676 & 0.639 & 0.255 & 25.21$^\circ$ \\
STARSS23 & TAU2019 & \texttt{laughter} & 0.35 & 45$^\circ$ & 253864 & 0.638 & 0.560 & 0.596 & 0.413 & 22.62$^\circ$ \\
STARSS23 & TAU2019 & \texttt{knock} & 0.05 & 45$^\circ$ & 122065 & 0.126 & 0.498 & 0.201 & 0.061 & 25.99$^\circ$ \\
STARSS23 & TAU2019 & \texttt{door} & 0.15 & 45$^\circ$ & 96588 & 0.089 & 0.276 & 0.135 & 0.056 & 26.50$^\circ$ \\
\hline
STARSS23 & TAU-NIGENS2020 & \texttt{footsteps} & 0.20 & 45$^\circ$ & 255673 & 0.287 & 0.384 & 0.328 & 0.042 & 33.48$^\circ$ \\
STARSS23 & TAU-NIGENS2020 & \texttt{telephone} & 0.25 & 45$^\circ$ & 232735 & 0.204 & 0.201 & 0.202 & 0.068 & 25.52$^\circ$ \\
STARSS23 & TAU-NIGENS2020 & \texttt{music} & 0.25 & 45$^\circ$ & 138222 & 0.085 & 0.269 & 0.130 & 0.063 & 29.44$^\circ$ \\
STARSS23 & TAU-NIGENS2020 & \texttt{musical\_instrument} & 0.45 & 45$^\circ$ & 138222 & 0.523 & 0.336 & 0.409 & 0.223 & 23.36$^\circ$ \\
STARSS23 & TAU-NIGENS2020 & \texttt{male\_speech} & 0.60 & 45$^\circ$ & 132502 & 0.268 & 0.199 & 0.229 & 0.164 & 25.26$^\circ$ \\
STARSS23 & TAU-NIGENS2020 & \texttt{knock} & 0.05 & 45$^\circ$ & 49894 & 0.060 & 0.447 & 0.106 & 0.031 & 26.65$^\circ$ \\
STARSS23 & TAU-NIGENS2020 & \texttt{female\_speech} & 0.45 & 45$^\circ$ & 40936 & 0.096 & 0.172 & 0.123 & 0.104 & 20.73$^\circ$ \\
\hline
STARSS23 & TAU-NIGENS2021 & \texttt{footsteps} & 0.05 & 45$^\circ$ & 619638 & 0.303 & 0.800 & 0.439 & 0.084 & 32.02$^\circ$ \\
STARSS23 & TAU-NIGENS2021 & \texttt{music} & 0.75 & 45$^\circ$ & 419592 & 0.432 & 0.094 & 0.154 & 0.040 & 27.62$^\circ$ \\
STARSS23 & TAU-NIGENS2021 & \texttt{musical\_instrument} & 0.20 & 45$^\circ$ & 419592 & 0.352 & 0.575 & 0.437 & 0.217 & 21.99$^\circ$ \\
STARSS23 & TAU-NIGENS2021 & \texttt{male\_speech} & 0.40 & 45$^\circ$ & 252567 & 0.330 & 0.278 & 0.302 & 0.181 & 26.56$^\circ$ \\
STARSS23 & TAU-NIGENS2021 & \texttt{telephone} & 0.25 & 45$^\circ$ & 191188 & 0.162 & 0.298 & 0.210 & 0.071 & 22.13$^\circ$ \\
STARSS23 & TAU-NIGENS2021 & \texttt{knock} & 0.05 & 45$^\circ$ & 93513 & 0.081 & 0.479 & 0.139 & 0.042 & 23.89$^\circ$ \\
STARSS23 & TAU-NIGENS2021 & \texttt{female\_speech} & 0.30 & 45$^\circ$ & 69001 & 0.101 & 0.237 & 0.142 & 0.101 & 22.81$^\circ$ \\
\hline
\hline
TAU2019 & STARSS23 & \texttt{speech} & 0.15 & 45$^\circ$ & 2388153 & 0.926 & 0.731 & 0.817 & 0.652 & 14.59$^\circ$ \\
TAU2019 & STARSS23 & \texttt{laughter} & 0.20 & 45$^\circ$ & 141315 & 0.169 & 0.272 & 0.209 & 0.162 & 15.01$^\circ$ \\
TAU2019 & STARSS23 & \texttt{telephone} & 0.15 & 45$^\circ$ & 35953 & 0.048 & 0.668 & 0.090 & 0.051 & 26.15$^\circ$ \\
TAU2019 & STARSS23 & \texttt{door} & 0.15 & 45$^\circ$ & 21429 & 0.093 & 0.210 & 0.129 & 0.108 & 15.84$^\circ$ \\
TAU2019 & STARSS23 & \texttt{knock} & 0.50 & 45$^\circ$ & 1717 & 0.163 & 0.422 & 0.235 & 0.235 & 10.97$^\circ$ \\
\hline
TAU2019 & TAU2019 & \texttt{keyboard} & 0.50 & 45$^\circ$ & 279200 & 0.991 & 0.851 & 0.916 & 0.902 & 6.82$^\circ$ \\
TAU2019 & TAU2019 & \texttt{speech} & 0.50 & 45$^\circ$ & 261542 & 0.974 & 0.930 & 0.952 & 0.923 & 6.89$^\circ$ \\
TAU2019 & TAU2019 & \texttt{telephone} & 0.45 & 45$^\circ$ & 260418 & 0.982 & 0.970 & 0.976 & 0.964 & 6.55$^\circ$ \\
TAU2019 & TAU2019 & \texttt{laughter} & 0.55 & 45$^\circ$ & 253864 & 0.963 & 0.913 & 0.937 & 0.912 & 7.07$^\circ$ \\
TAU2019 & TAU2019 & \texttt{drawer} & 0.50 & 45$^\circ$ & 172812 & 0.930 & 0.739 & 0.824 & 0.805 & 8.45$^\circ$ \\
TAU2019 & TAU2019 & \texttt{page\_turn} & 0.50 & 45$^\circ$ & 149463 & 0.901 & 0.892 & 0.897 & 0.875 & 8.07$^\circ$ \\
TAU2019 & TAU2019 & \texttt{cough} & 0.50 & 45$^\circ$ & 129894 & 0.952 & 0.853 & 0.900 & 0.885 & 6.86$^\circ$ \\
TAU2019 & TAU2019 & \texttt{knock} & 0.50 & 45$^\circ$ & 122065 & 0.946 & 0.816 & 0.876 & 0.865 & 7.46$^\circ$ \\
TAU2019 & TAU2019 & \texttt{clearthroat} & 0.50 & 45$^\circ$ & 98877 & 0.927 & 0.863 & 0.894 & 0.869 & 7.73$^\circ$ \\
TAU2019 & TAU2019 & \texttt{door} & 0.50 & 45$^\circ$ & 96588 & 0.920 & 0.665 & 0.772 & 0.754 & 7.49$^\circ$ \\
TAU2019 & TAU2019 & \texttt{keys\_drop} & 0.50 & 45$^\circ$ & 55008 & 0.929 & 0.760 & 0.836 & 0.818 & 7.92$^\circ$ \\
\hline
TAU2019 & TAU-NIGENS2020 & \texttt{telephone} & 0.30 & 45$^\circ$ & 232735 & 0.152 & 0.412 & 0.222 & 0.169 & 16.93$^\circ$ \\
TAU2019 & TAU-NIGENS2020 & \texttt{speech} & 0.50 & 45$^\circ$ & 61720 & 0.866 & 0.713 & 0.782 & 0.751 & 12.41$^\circ$ \\
TAU2019 & TAU-NIGENS2020 & \texttt{knock} & 0.50 & 45$^\circ$ & 49894 & 0.311 & 0.069 & 0.113 & 0.106 & 17.02$^\circ$ \\
\hline
TAU2019 & TAU-NIGENS2021 & \texttt{telephone} & 0.25 & 45$^\circ$ & 191188 & 0.189 & 0.855 & 0.310 & 0.176 & 19.64$^\circ$ \\
TAU2019 & TAU-NIGENS2021 & \texttt{speech} & 0.45 & 45$^\circ$ & 118403 & 0.668 & 0.485 & 0.562 & 0.480 & 15.46$^\circ$ \\
TAU2019 & TAU-NIGENS2021 & \texttt{knock} & 0.35 & 45$^\circ$ & 93513 & 0.197 & 0.156 & 0.174 & 0.142 & 16.41$^\circ$ \\
\hline
\hline
TAU-NIGENS2020 & STARSS23 & \texttt{male\_speech} & 0.35 & 45$^\circ$ & 1486350 & 0.855 & 0.593 & 0.700 & 0.535 & 20.33$^\circ$ \\
TAU-NIGENS2020 & STARSS23 & \texttt{female\_speech} & 0.25 & 45$^\circ$ & 1044489 & 0.681 & 0.704 & 0.692 & 0.381 & 15.73$^\circ$ \\
TAU-NIGENS2020 & STARSS23 & \texttt{piano} & 0.45 & 30$^\circ$ & 818413 & 0.499 & 0.277 & 0.357 & 0.232 & 17.60$^\circ$ \\
TAU-NIGENS2020 & STARSS23 & \texttt{footsteps} & 0.35 & 45$^\circ$ & 176889 & 0.112 & 0.087 & 0.098 & 0.031 & 35.06$^\circ$ \\
TAU-NIGENS2020 & STARSS23 & \texttt{telephone} & 0.45 & 45$^\circ$ & 35953 & 0.125 & 0.434 & 0.195 & 0.130 & 28.02$^\circ$ \\
TAU-NIGENS2020 & STARSS23 & \texttt{knock} & 0.40 & 45$^\circ$ & 1717 & 0.019 & 0.211 & 0.035 & 0.018 & 24.60$^\circ$ \\
\hline
TAU-NIGENS2020 & TAU2019 & \texttt{telephone} & 0.50 & 45$^\circ$ & 260418 & 0.867 & 0.624 & 0.725 & 0.617 & 17.53$^\circ$ \\
TAU-NIGENS2020 & TAU2019 & \texttt{knock} & 0.45 & 45$^\circ$ & 122065 & 0.621 & 0.547 & 0.582 & 0.365 & 19.88$^\circ$ \\
\hline
TAU-NIGENS2020 & TAU-NIGENS2020 & \texttt{burning\_fire} & 0.45 & 45$^\circ$ & 311809 & 0.803 & 0.852 & 0.827 & 0.693 & 18.61$^\circ$ \\
TAU-NIGENS2020 & TAU-NIGENS2020 & \texttt{running\_engine} & 0.45 & 45$^\circ$ & 271335 & 0.812 & 0.645 & 0.719 & 0.612 & 18.94$^\circ$ \\
TAU-NIGENS2020 & TAU-NIGENS2020 & \texttt{footsteps} & 0.45 & 45$^\circ$ & 255673 & 0.750 & 0.448 & 0.561 & 0.455 & 17.37$^\circ$ \\
TAU-NIGENS2020 & TAU-NIGENS2020 & \texttt{alarm} & 0.40 & 45$^\circ$ & 239733 & 0.977 & 0.515 & 0.674 & 0.575 & 19.57$^\circ$ \\
TAU-NIGENS2020 & TAU-NIGENS2020 & \texttt{telephone} & 0.45 & 45$^\circ$ & 232735 & 0.631 & 0.321 & 0.425 & 0.328 & 19.39$^\circ$ \\
TAU-NIGENS2020 & TAU-NIGENS2020 & \texttt{baby\_cry} & 0.45 & 45$^\circ$ & 198898 & 0.948 & 0.629 & 0.756 & 0.658 & 20.25$^\circ$ \\
TAU-NIGENS2020 & TAU-NIGENS2020 & \texttt{crash} & 0.45 & 45$^\circ$ & 171730 & 0.916 & 0.571 & 0.704 & 0.599 & 17.47$^\circ$ \\
TAU-NIGENS2020 & TAU-NIGENS2020 & \texttt{dog\_bark} & 0.50 & 45$^\circ$ & 152404 & 0.890 & 0.567 & 0.693 & 0.543 & 18.30$^\circ$ \\
TAU-NIGENS2020 & TAU-NIGENS2020 & \texttt{piano} & 0.45 & 45$^\circ$ & 138222 & 0.683 & 0.264 & 0.381 & 0.318 & 18.82$^\circ$ \\
TAU-NIGENS2020 & TAU-NIGENS2020 & \texttt{male\_scream} & 0.40 & 45$^\circ$ & 93841 & 0.848 & 0.651 & 0.736 & 0.592 & 19.88$^\circ$ \\
TAU-NIGENS2020 & TAU-NIGENS2020 & \texttt{knock} & 0.45 & 45$^\circ$ & 49894 & 0.510 & 0.598 & 0.550 & 0.412 & 18.19$^\circ$ \\
TAU-NIGENS2020 & TAU-NIGENS2020 & \texttt{male\_speech} & 0.40 & 45$^\circ$ & 41519 & 0.904 & 0.691 & 0.783 & 0.527 & 20.68$^\circ$ \\
TAU-NIGENS2020 & TAU-NIGENS2020 & \texttt{female\_scream} & 0.40 & 45$^\circ$ & 21135 & 0.446 & 0.507 & 0.474 & 0.350 & 18.86$^\circ$ \\
TAU-NIGENS2020 & TAU-NIGENS2020 & \texttt{female\_speech} & 0.45 & 45$^\circ$ & 20201 & 0.938 & 0.689 & 0.795 & 0.674 & 16.44$^\circ$ \\
\hline
TAU-NIGENS2020 & TAU-NIGENS2021 & \texttt{footsteps} & 0.20 & 45$^\circ$ & 619638 & 0.694 & 0.563 & 0.621 & 0.348 & 19.24$^\circ$ \\
TAU-NIGENS2020 & TAU-NIGENS2021 & \texttt{alarm} & 0.30 & 45$^\circ$ & 568124 & 0.849 & 0.641 & 0.730 & 0.455 & 19.58$^\circ$ \\
TAU-NIGENS2020 & TAU-NIGENS2021 & \texttt{piano} & 0.40 & 45$^\circ$ & 419592 & 0.518 & 0.209 & 0.298 & 0.195 & 17.54$^\circ$ \\
TAU-NIGENS2020 & TAU-NIGENS2021 & \texttt{baby\_cry} & 0.35 & 45$^\circ$ & 361962 & 0.732 & 0.702 & 0.717 & 0.448 & 19.10$^\circ$ \\
TAU-NIGENS2020 & TAU-NIGENS2021 & \texttt{crash} & 0.40 & 45$^\circ$ & 276813 & 0.734 & 0.629 & 0.677 & 0.531 & 16.39$^\circ$ \\
TAU-NIGENS2020 & TAU-NIGENS2021 & \texttt{dog\_bark} & 0.45 & 45$^\circ$ & 250633 & 0.609 & 0.501 & 0.550 & 0.392 & 17.61$^\circ$ \\
TAU-NIGENS2020 & TAU-NIGENS2021 & \texttt{telephone} & 0.45 & 45$^\circ$ & 191188 & 0.427 & 0.308 & 0.358 & 0.228 & 20.50$^\circ$ \\
TAU-NIGENS2020 & TAU-NIGENS2021 & \texttt{male\_scream} & 0.40 & 45$^\circ$ & 179070 & 0.855 & 0.469 & 0.606 & 0.447 & 19.00$^\circ$ \\
TAU-NIGENS2020 & TAU-NIGENS2021 & \texttt{knock} & 0.40 & 45$^\circ$ & 93513 & 0.332 & 0.569 & 0.419 & 0.241 & 18.67$^\circ$ \\
TAU-NIGENS2020 & TAU-NIGENS2021 & \texttt{male\_speech} & 0.30 & 45$^\circ$ & 79850 & 0.675 & 0.635 & 0.655 & 0.356 & 20.61$^\circ$ \\
TAU-NIGENS2020 & TAU-NIGENS2021 & \texttt{female\_speech} & 0.45 & 45$^\circ$ & 39707 & 0.877 & 0.422 & 0.570 & 0.419 & 23.31$^\circ$ \\
TAU-NIGENS2020 & TAU-NIGENS2021 & \texttt{female\_scream} & 0.35 & 45$^\circ$ & 29294 & 0.154 & 0.514 & 0.236 & 0.153 & 18.73$^\circ$ \\
\hline
\hline
TAU-NIGENS2021 & STARSS23 & \texttt{male\_speech} & 0.45 & 45$^\circ$ & 1486350 & 0.917 & 0.576 & 0.707 & 0.595 & 16.97$^\circ$ \\
TAU-NIGENS2021 & STARSS23 & \texttt{female\_speech} & 0.35 & 45$^\circ$ & 1044489 & 0.653 & 0.761 & 0.703 & 0.580 & 16.52$^\circ$ \\
TAU-NIGENS2021 & STARSS23 & \texttt{piano} & 0.15 & 45$^\circ$ & 818413 & 0.353 & 0.630 & 0.453 & 0.322 & 21.67$^\circ$ \\
TAU-NIGENS2021 & STARSS23 & \texttt{footsteps} & 0.60 & 45$^\circ$ & 176889 & 0.152 & 0.085 & 0.109 & 0.050 & 36.43$^\circ$ \\
TAU-NIGENS2021 & STARSS23 & \texttt{telephone} & 0.45 & 45$^\circ$ & 35953 & 0.255 & 0.209 & 0.230 & 0.208 & 27.57$^\circ$ \\
TAU-NIGENS2021 & STARSS23 & \texttt{knock} & 0.65 & 30$^\circ$ & 1717 & 0.084 & 0.227 & 0.123 & 0.104 & 21.75$^\circ$ \\
\hline
TAU-NIGENS2021 & TAU2019 & \texttt{telephone} & 0.40 & 45$^\circ$ & 260418 & 0.991 & 0.750 & 0.854 & 0.817 & 15.40$^\circ$ \\
TAU-NIGENS2021 & TAU2019 & \texttt{knock} & 0.55 & 45$^\circ$ & 122065 & 0.807 & 0.511 & 0.625 & 0.585 & 18.69$^\circ$ \\
\hline
TAU-NIGENS2021 & TAU-NIGENS2020 & \texttt{footsteps} & 0.50 & 45$^\circ$ & 255673 & 0.585 & 0.533 & 0.558 & 0.515 & 16.86$^\circ$ \\
TAU-NIGENS2021 & TAU-NIGENS2020 & \texttt{alarm} & 0.45 & 45$^\circ$ & 239733 & 0.831 & 0.412 & 0.550 & 0.531 & 18.43$^\circ$ \\
TAU-NIGENS2021 & TAU-NIGENS2020 & \texttt{telephone} & 0.40 & 45$^\circ$ & 232735 & 0.626 & 0.353 & 0.451 & 0.429 & 18.84$^\circ$ \\
TAU-NIGENS2021 & TAU-NIGENS2020 & \texttt{baby\_cry} & 0.40 & 45$^\circ$ & 198898 & 0.696 & 0.779 & 0.735 & 0.698 & 15.73$^\circ$ \\
TAU-NIGENS2021 & TAU-NIGENS2020 & \texttt{crash} & 0.60 & 45$^\circ$ & 171730 & 0.856 & 0.586 & 0.695 & 0.690 & 16.26$^\circ$ \\
TAU-NIGENS2021 & TAU-NIGENS2020 & \texttt{dog\_bark} & 0.50 & 45$^\circ$ & 152404 & 0.895 & 0.556 & 0.686 & 0.642 & 15.06$^\circ$ \\
TAU-NIGENS2021 & TAU-NIGENS2020 & \texttt{piano} & 0.55 & 45$^\circ$ & 138222 & 0.851 & 0.432 & 0.573 & 0.510 & 20.16$^\circ$ \\
TAU-NIGENS2021 & TAU-NIGENS2020 & \texttt{male\_scream} & 0.40 & 45$^\circ$ & 93841 & 0.790 & 0.691 & 0.737 & 0.690 & 17.63$^\circ$ \\
TAU-NIGENS2021 & TAU-NIGENS2020 & \texttt{knock} & 0.55 & 45$^\circ$ & 49894 & 0.570 & 0.533 & 0.551 & 0.535 & 16.21$^\circ$ \\
TAU-NIGENS2021 & TAU-NIGENS2020 & \texttt{male\_speech} & 0.50 & 45$^\circ$ & 41519 & 0.735 & 0.650 & 0.690 & 0.608 & 18.39$^\circ$ \\
TAU-NIGENS2021 & TAU-NIGENS2020 & \texttt{female\_scream} & 0.35 & 45$^\circ$ & 21135 & 0.242 & 0.679 & 0.357 & 0.337 & 16.51$^\circ$ \\
TAU-NIGENS2021 & TAU-NIGENS2020 & \texttt{female\_speech} & 0.60 & 45$^\circ$ & 20201 & 0.925 & 0.854 & 0.888 & 0.869 & 15.70$^\circ$ \\
\hline
TAU-NIGENS2021 & TAU-NIGENS2021 & \texttt{footsteps} & 0.35 & 45$^\circ$ & 619638 & 0.684 & 0.590 & 0.634 & 0.503 & 17.01$^\circ$ \\
TAU-NIGENS2021 & TAU-NIGENS2021 & \texttt{alarm} & 0.30 & 45$^\circ$ & 568124 & 0.712 & 0.664 & 0.687 & 0.608 & 17.31$^\circ$ \\
TAU-NIGENS2021 & TAU-NIGENS2021 & \texttt{piano} & 0.50 & 45$^\circ$ & 419592 & 0.616 & 0.301 & 0.404 & 0.364 & 18.59$^\circ$ \\
TAU-NIGENS2021 & TAU-NIGENS2021 & \texttt{baby\_cry} & 0.35 & 45$^\circ$ & 361962 & 0.601 & 0.884 & 0.716 & 0.575 & 16.09$^\circ$ \\
TAU-NIGENS2021 & TAU-NIGENS2021 & \texttt{crash} & 0.40 & 45$^\circ$ & 276813 & 0.802 & 0.656 & 0.722 & 0.650 & 13.63$^\circ$ \\
TAU-NIGENS2021 & TAU-NIGENS2021 & \texttt{dog\_bark} & 0.45 & 45$^\circ$ & 250633 & 0.758 & 0.714 & 0.735 & 0.671 & 16.20$^\circ$ \\
TAU-NIGENS2021 & TAU-NIGENS2021 & \texttt{telephone} & 0.40 & 45$^\circ$ & 191188 & 0.499 & 0.447 & 0.472 & 0.371 & 19.46$^\circ$ \\
TAU-NIGENS2021 & TAU-NIGENS2021 & \texttt{male\_scream} & 0.50 & 45$^\circ$ & 179070 & 0.871 & 0.409 & 0.557 & 0.536 & 17.20$^\circ$ \\
TAU-NIGENS2021 & TAU-NIGENS2021 & \texttt{knock} & 0.35 & 45$^\circ$ & 93513 & 0.314 & 0.566 & 0.404 & 0.336 & 17.78$^\circ$ \\
TAU-NIGENS2021 & TAU-NIGENS2021 & \texttt{male\_speech} & 0.35 & 45$^\circ$ & 79850 & 0.637 & 0.642 & 0.639 & 0.473 & 18.52$^\circ$ \\
TAU-NIGENS2021 & TAU-NIGENS2021 & \texttt{female\_speech} & 0.55 & 45$^\circ$ & 39707 & 0.886 & 0.600 & 0.715 & 0.634 & 20.17$^\circ$ \\
TAU-NIGENS2021 & TAU-NIGENS2021 & \texttt{female\_scream} & 0.40 & 45$^\circ$ & 29294 & 0.226 & 0.576 & 0.325 & 0.306 & 16.21$^\circ$ \\
\hline
\end{longtable}
\end{landscape}
\clearpage


\begin{landscape}
\begin{longtable}{|l|l|l|r|c|c|c|c|c|c|}
\caption{Per-class coverage-aware DOA estimation under oracle activity. Primary accuracies use DOA-only angular track matching; the tPIT-aligned result is reported separately.}\label{tab:dataset_generalization_doa_oracle_class}\\
\hline
\textbf{Training dataset} & \textbf{Test dataset} & \textbf{Output class} & \textbf{Slots} & \(\mathrm{Acc}_{10^\circ}\) & \(\mathrm{Acc}_{20^\circ}\) & \(\mathrm{Acc}_{30^\circ}\) & \(\mathrm{Acc}^{\mathrm{tPIT}}_{20^\circ}\) & \textbf{Median AE} & \(\mathbf{P_{90}}\) \\
\hline
\endfirsthead
\hline
\textbf{Training dataset} & \textbf{Test dataset} & \textbf{Output class} & \textbf{Slots} & \(\mathrm{Acc}_{10^\circ}\) & \(\mathrm{Acc}_{20^\circ}\) & \(\mathrm{Acc}_{30^\circ}\) & \(\mathrm{Acc}^{\mathrm{tPIT}}_{20^\circ}\) & \textbf{Median AE} & \(\mathbf{P_{90}}\) \\
\hline
\endhead
STARSS23 & STARSS23 & \texttt{male\_speech} & 1,574,286 & 0.227 & 0.578 & 0.860 & 0.503 & 17.80$^\circ$ & 32.44$^\circ$ \\
STARSS23 & STARSS23 & \texttt{music} & 1,238,924 & 0.061 & 0.290 & 0.529 & 0.234 & 28.32$^\circ$ & 60.81$^\circ$ \\
STARSS23 & STARSS23 & \texttt{female\_speech} & 1,073,433 & 0.281 & 0.620 & 0.785 & 0.547 & 16.07$^\circ$ & 57.87$^\circ$ \\
STARSS23 & STARSS23 & \texttt{musical\_instrument} & 1,030,057 & 0.022 & 0.107 & 0.320 & 0.073 & 38.76$^\circ$ & 91.01$^\circ$ \\
STARSS23 & STARSS23 & \texttt{domestic\_sounds} & 635,253 & 0.042 & 0.245 & 0.534 & 0.209 & 28.86$^\circ$ & 43.75$^\circ$ \\
STARSS23 & STARSS23 & \texttt{footsteps} & 180,973 & 0.179 & 0.477 & 0.707 & 0.391 & 20.88$^\circ$ & 45.61$^\circ$ \\
STARSS23 & STARSS23 & \texttt{laughter} & 161,272 & 0.201 & 0.455 & 0.670 & 0.408 & 21.86$^\circ$ & 47.70$^\circ$ \\
STARSS23 & STARSS23 & \texttt{water\_tap} & 101,696 & 0.096 & 0.262 & 0.441 & 0.154 & 37.65$^\circ$ & 101.22$^\circ$ \\
STARSS23 & STARSS23 & \texttt{bell} & 45,799 & 0.056 & 0.158 & 0.335 & 0.081 & 38.31$^\circ$ & 65.75$^\circ$ \\
STARSS23 & STARSS23 & \texttt{telephone} & 35,953 & 0.049 & 0.291 & 0.612 & 0.285 & 24.81$^\circ$ & 57.53$^\circ$ \\
STARSS23 & STARSS23 & \texttt{clapping} & 28,142 & 0.173 & 0.421 & 0.686 & 0.413 & 23.36$^\circ$ & 50.92$^\circ$ \\
STARSS23 & STARSS23 & \texttt{door} & 21,429 & 0.144 & 0.402 & 0.710 & 0.387 & 21.89$^\circ$ & 54.61$^\circ$ \\
STARSS23 & STARSS23 & \texttt{knock} & 1,717 & 0.020 & 0.218 & 0.274 & 0.185 & 51.94$^\circ$ & 87.41$^\circ$ \\
\hline
STARSS23 & TAU2019 & \texttt{domestic\_sounds} & 670,620 & 0.028 & 0.100 & 0.215 & 0.063 & 50.08$^\circ$ & 84.00$^\circ$ \\
STARSS23 & TAU2019 & \texttt{telephone} & 268,153 & 0.056 & 0.199 & 0.382 & 0.166 & 36.24$^\circ$ & 68.91$^\circ$ \\
STARSS23 & TAU2019 & \texttt{laughter} & 261,141 & 0.091 & 0.334 & 0.570 & 0.323 & 27.00$^\circ$ & 52.58$^\circ$ \\
STARSS23 & TAU2019 & \texttt{knock} & 123,239 & 0.036 & 0.161 & 0.320 & 0.115 & 39.73$^\circ$ & 97.89$^\circ$ \\
STARSS23 & TAU2019 & \texttt{door} & 97,465 & 0.051 & 0.199 & 0.398 & 0.156 & 35.50$^\circ$ & 76.32$^\circ$ \\
\hline
STARSS23 & TAU-NIGENS2020 & \texttt{footsteps} & 257,358 & 0.017 & 0.067 & 0.170 & 0.033 & 53.52$^\circ$ & 89.66$^\circ$ \\
STARSS23 & TAU-NIGENS2020 & \texttt{telephone} & 232,735 & 0.033 & 0.139 & 0.277 & 0.107 & 45.94$^\circ$ & 87.35$^\circ$ \\
STARSS23 & TAU-NIGENS2020 & \texttt{music} & 138,222 & 0.040 & 0.163 & 0.363 & 0.116 & 36.70$^\circ$ & 66.59$^\circ$ \\
STARSS23 & TAU-NIGENS2020 & \texttt{musical\_instrument} & 138,222 & 0.073 & 0.216 & 0.422 & 0.170 & 33.49$^\circ$ & 84.66$^\circ$ \\
STARSS23 & TAU-NIGENS2020 & \texttt{male\_speech} & 135,360 & 0.107 & 0.281 & 0.495 & 0.251 & 30.22$^\circ$ & 55.04$^\circ$ \\
STARSS23 & TAU-NIGENS2020 & \texttt{knock} & 49,894 & 0.043 & 0.191 & 0.375 & 0.126 & 36.52$^\circ$ & 92.23$^\circ$ \\
STARSS23 & TAU-NIGENS2020 & \texttt{female\_speech} & 41,336 & 0.089 & 0.362 & 0.607 & 0.310 & 24.94$^\circ$ & 52.69$^\circ$ \\
\hline
STARSS23 & TAU-NIGENS2021 & \texttt{footsteps} & 733,399 & 0.016 & 0.072 & 0.170 & 0.039 & 53.78$^\circ$ & 98.05$^\circ$ \\
STARSS23 & TAU-NIGENS2021 & \texttt{music} & 419,592 & 0.037 & 0.148 & 0.331 & 0.119 & 39.69$^\circ$ & 80.50$^\circ$ \\
STARSS23 & TAU-NIGENS2021 & \texttt{musical\_instrument} & 419,592 & 0.096 & 0.320 & 0.517 & 0.209 & 29.02$^\circ$ & 86.42$^\circ$ \\
STARSS23 & TAU-NIGENS2021 & \texttt{male\_speech} & 261,118 & 0.075 & 0.229 & 0.432 & 0.188 & 32.88$^\circ$ & 61.61$^\circ$ \\
STARSS23 & TAU-NIGENS2021 & \texttt{telephone} & 191,188 & 0.041 & 0.131 & 0.280 & 0.110 & 41.48$^\circ$ & 72.19$^\circ$ \\
STARSS23 & TAU-NIGENS2021 & \texttt{knock} & 95,450 & 0.068 & 0.183 & 0.423 & 0.126 & 34.08$^\circ$ & 91.30$^\circ$ \\
STARSS23 & TAU-NIGENS2021 & \texttt{female\_speech} & 69,814 & 0.082 & 0.268 & 0.464 & 0.231 & 31.58$^\circ$ & 60.54$^\circ$ \\
\hline
\hline
TAU2019 & STARSS23 & \texttt{speech} & 2,647,719 & 0.317 & 0.682 & 0.843 & 0.652 & 14.17$^\circ$ & 35.54$^\circ$ \\
TAU2019 & STARSS23 & \texttt{laughter} & 161,272 & 0.231 & 0.595 & 0.783 & 0.539 & 16.61$^\circ$ & 41.84$^\circ$ \\
TAU2019 & STARSS23 & \texttt{telephone} & 35,953 & 0.030 & 0.219 & 0.552 & 0.191 & 27.95$^\circ$ & 50.78$^\circ$ \\
TAU2019 & STARSS23 & \texttt{door} & 21,429 & 0.159 & 0.470 & 0.703 & 0.422 & 21.00$^\circ$ & 45.50$^\circ$ \\
TAU2019 & STARSS23 & \texttt{knock} & 1,717 & 0.373 & 0.726 & 0.886 & 0.713 & 13.93$^\circ$ & 31.84$^\circ$ \\
\hline
TAU2019 & TAU2019 & \texttt{keyboard} & 287,223 & 0.800 & 0.967 & 0.989 & 0.954 & 5.81$^\circ$ & 13.24$^\circ$ \\
TAU2019 & TAU2019 & \texttt{speech} & 269,172 & 0.806 & 0.967 & 0.985 & 0.950 & 5.75$^\circ$ & 12.83$^\circ$ \\
TAU2019 & TAU2019 & \texttt{telephone} & 268,153 & 0.849 & 0.984 & 0.993 & 0.975 & 5.31$^\circ$ & 11.69$^\circ$ \\
TAU2019 & TAU2019 & \texttt{laughter} & 261,141 & 0.807 & 0.962 & 0.985 & 0.938 & 5.71$^\circ$ & 12.83$^\circ$ \\
TAU2019 & TAU2019 & \texttt{drawer} & 176,643 & 0.686 & 0.916 & 0.974 & 0.896 & 7.07$^\circ$ & 18.68$^\circ$ \\
TAU2019 & TAU2019 & \texttt{page\_turn} & 151,626 & 0.755 & 0.945 & 0.987 & 0.920 & 6.29$^\circ$ & 15.87$^\circ$ \\
TAU2019 & TAU2019 & \texttt{cough} & 131,939 & 0.818 & 0.959 & 0.987 & 0.940 & 5.60$^\circ$ & 13.48$^\circ$ \\
TAU2019 & TAU2019 & \texttt{knock} & 123,239 & 0.734 & 0.936 & 0.973 & 0.921 & 6.28$^\circ$ & 16.36$^\circ$ \\
TAU2019 & TAU2019 & \texttt{clearthroat} & 99,904 & 0.753 & 0.941 & 0.972 & 0.929 & 6.50$^\circ$ & 15.41$^\circ$ \\
TAU2019 & TAU2019 & \texttt{door} & 97,465 & 0.719 & 0.913 & 0.963 & 0.885 & 6.66$^\circ$ & 18.42$^\circ$ \\
TAU2019 & TAU2019 & \texttt{keys\_drop} & 55,128 & 0.712 & 0.931 & 0.978 & 0.910 & 6.73$^\circ$ & 17.09$^\circ$ \\
\hline
TAU2019 & TAU-NIGENS2020 & \texttt{telephone} & 232,735 & 0.161 & 0.399 & 0.549 & 0.347 & 26.12$^\circ$ & 78.51$^\circ$ \\
TAU2019 & TAU-NIGENS2020 & \texttt{speech} & 61,720 & 0.395 & 0.812 & 0.908 & 0.793 & 11.78$^\circ$ & 28.57$^\circ$ \\
TAU2019 & TAU-NIGENS2020 & \texttt{knock} & 49,894 & 0.248 & 0.614 & 0.836 & 0.547 & 16.77$^\circ$ & 36.83$^\circ$ \\
TAU2019 & TAU-NIGENS2021 & \texttt{telephone} & 191,188 & 0.160 & 0.463 & 0.667 & 0.427 & 21.56$^\circ$ & 52.63$^\circ$ \\
TAU2019 & TAU-NIGENS2021 & \texttt{speech} & 120,946 & 0.252 & 0.602 & 0.799 & 0.553 & 16.58$^\circ$ & 41.83$^\circ$ \\
TAU2019 & TAU-NIGENS2021 & \texttt{knock} & 95,450 & 0.208 & 0.529 & 0.757 & 0.486 & 18.99$^\circ$ & 41.38$^\circ$ \\
\hline
\hline
TAU-NIGENS2020 & STARSS23 & \texttt{male\_speech} & 1,574,286 & 0.176 & 0.477 & 0.712 & 0.465 & 20.87$^\circ$ & 42.37$^\circ$ \\
TAU-NIGENS2020 & STARSS23 & \texttt{female\_speech} & 1,073,433 & 0.213 & 0.524 & 0.713 & 0.519 & 19.06$^\circ$ & 61.76$^\circ$ \\
TAU-NIGENS2020 & STARSS23 & \texttt{piano} & 1,030,057 & 0.044 & 0.305 & 0.664 & 0.299 & 24.62$^\circ$ & 46.66$^\circ$ \\
TAU-NIGENS2020 & STARSS23 & \texttt{footsteps} & 180,973 & 0.003 & 0.025 & 0.111 & 0.021 & 51.38$^\circ$ & 76.77$^\circ$ \\
TAU-NIGENS2020 & STARSS23 & \texttt{telephone} & 35,953 & 0.002 & 0.101 & 0.475 & 0.100 & 30.70$^\circ$ & 53.63$^\circ$ \\
TAU-NIGENS2020 & STARSS23 & \texttt{knock} & 1,717 & 0.118 & 0.277 & 0.651 & 0.268 & 25.04$^\circ$ & 40.92$^\circ$ \\
\hline
TAU-NIGENS2020 & TAU2019 & \texttt{telephone} & 268,153 & 0.215 & 0.579 & 0.808 & 0.570 & 17.62$^\circ$ & 40.13$^\circ$ \\
TAU-NIGENS2020 & TAU2019 & \texttt{knock} & 123,239 & 0.115 & 0.402 & 0.651 & 0.394 & 23.43$^\circ$ & 51.09$^\circ$ \\
\hline
TAU-NIGENS2020 & TAU-NIGENS2020 & \texttt{burning\_fire} & 348,259 & 0.200 & 0.503 & 0.737 & 0.497 & 19.89$^\circ$ & 46.82$^\circ$ \\
TAU-NIGENS2020 & TAU-NIGENS2020 & \texttt{running\_engine} & 280,784 & 0.193 & 0.527 & 0.752 & 0.512 & 19.15$^\circ$ & 40.50$^\circ$ \\
TAU-NIGENS2020 & TAU-NIGENS2020 & \texttt{footsteps} & 257,358 & 0.153 & 0.408 & 0.633 & 0.399 & 23.21$^\circ$ & 56.17$^\circ$ \\
TAU-NIGENS2020 & TAU-NIGENS2020 & \texttt{alarm} & 240,614 & 0.168 & 0.448 & 0.714 & 0.436 & 21.57$^\circ$ & 44.85$^\circ$ \\
TAU-NIGENS2020 & TAU-NIGENS2020 & \texttt{telephone} & 232,735 & 0.111 & 0.346 & 0.570 & 0.333 & 26.61$^\circ$ & 58.40$^\circ$ \\
TAU-NIGENS2020 & TAU-NIGENS2020 & \texttt{baby\_cry} & 198,898 & 0.214 & 0.469 & 0.728 & 0.466 & 21.13$^\circ$ & 42.09$^\circ$ \\
TAU-NIGENS2020 & TAU-NIGENS2020 & \texttt{crash} & 172,370 & 0.194 & 0.553 & 0.804 & 0.541 & 18.38$^\circ$ & 35.82$^\circ$ \\
TAU-NIGENS2020 & TAU-NIGENS2020 & \texttt{dog\_bark} & 169,462 & 0.145 & 0.474 & 0.718 & 0.466 & 20.94$^\circ$ & 47.89$^\circ$ \\
TAU-NIGENS2020 & TAU-NIGENS2020 & \texttt{piano} & 138,222 & 0.093 & 0.316 & 0.552 & 0.301 & 27.63$^\circ$ & 56.17$^\circ$ \\
TAU-NIGENS2020 & TAU-NIGENS2020 & \texttt{male\_scream} & 93,841 & 0.143 & 0.443 & 0.690 & 0.442 & 21.96$^\circ$ & 43.59$^\circ$ \\
TAU-NIGENS2020 & TAU-NIGENS2020 & \texttt{knock} & 49,894 & 0.170 & 0.540 & 0.792 & 0.536 & 18.72$^\circ$ & 40.16$^\circ$ \\
TAU-NIGENS2020 & TAU-NIGENS2020 & \texttt{male\_speech} & 41,519 & 0.103 & 0.387 & 0.651 & 0.373 & 23.65$^\circ$ & 50.74$^\circ$ \\
TAU-NIGENS2020 & TAU-NIGENS2020 & \texttt{female\_scream} & 21,135 & 0.172 & 0.548 & 0.769 & 0.524 & 18.33$^\circ$ & 42.89$^\circ$ \\
TAU-NIGENS2020 & TAU-NIGENS2020 & \texttt{female\_speech} & 20,201 & 0.198 & 0.590 & 0.801 & 0.589 & 16.84$^\circ$ & 36.67$^\circ$ \\
\hline
TAU-NIGENS2020 & TAU-NIGENS2021 & \texttt{footsteps} & 733,399 & 0.126 & 0.359 & 0.547 & 0.348 & 27.06$^\circ$ & 65.94$^\circ$ \\
TAU-NIGENS2020 & TAU-NIGENS2021 & \texttt{alarm} & 629,828 & 0.141 & 0.431 & 0.663 & 0.419 & 22.60$^\circ$ & 52.64$^\circ$ \\
TAU-NIGENS2020 & TAU-NIGENS2021 & \texttt{piano} & 419,592 & 0.109 & 0.332 & 0.533 & 0.318 & 28.19$^\circ$ & 62.45$^\circ$ \\
TAU-NIGENS2020 & TAU-NIGENS2021 & \texttt{baby\_cry} & 387,944 & 0.180 & 0.483 & 0.691 & 0.479 & 20.71$^\circ$ & 49.37$^\circ$ \\
TAU-NIGENS2020 & TAU-NIGENS2021 & \texttt{crash} & 311,355 & 0.215 & 0.563 & 0.757 & 0.549 & 17.86$^\circ$ & 48.60$^\circ$ \\
TAU-NIGENS2020 & TAU-NIGENS2021 & \texttt{dog\_bark} & 254,970 & 0.171 & 0.460 & 0.677 & 0.454 & 21.63$^\circ$ & 49.29$^\circ$ \\
TAU-NIGENS2020 & TAU-NIGENS2021 & \texttt{telephone} & 191,188 & 0.107 & 0.363 & 0.595 & 0.351 & 25.60$^\circ$ & 58.21$^\circ$ \\
TAU-NIGENS2020 & TAU-NIGENS2021 & \texttt{male\_scream} & 179,879 & 0.173 & 0.510 & 0.735 & 0.501 & 19.65$^\circ$ & 41.57$^\circ$ \\
TAU-NIGENS2020 & TAU-NIGENS2021 & \texttt{knock} & 95,450 & 0.166 & 0.470 & 0.704 & 0.455 & 21.13$^\circ$ & 46.12$^\circ$ \\
TAU-NIGENS2020 & TAU-NIGENS2021 & \texttt{male\_speech} & 81,239 & 0.122 & 0.398 & 0.614 & 0.392 & 24.20$^\circ$ & 50.22$^\circ$ \\
TAU-NIGENS2020 & TAU-NIGENS2021 & \texttt{female\_speech} & 39,707 & 0.121 & 0.381 & 0.669 & 0.370 & 23.97$^\circ$ & 46.41$^\circ$ \\
TAU-NIGENS2020 & TAU-NIGENS2021 & \texttt{female\_scream} & 30,107 & 0.206 & 0.481 & 0.693 & 0.460 & 20.64$^\circ$ & 48.87$^\circ$ \\
\hline
\hline
TAU-NIGENS2021 & STARSS23 & \texttt{male\_speech} & 1,574,286 & 0.193 & 0.584 & 0.841 & 0.556 & 17.67$^\circ$ & 35.05$^\circ$ \\
TAU-NIGENS2021 & STARSS23 & \texttt{female\_speech} & 1,073,433 & 0.249 & 0.593 & 0.794 & 0.566 & 16.64$^\circ$ & 43.58$^\circ$ \\
TAU-NIGENS2021 & STARSS23 & \texttt{piano} & 1,030,057 & 0.063 & 0.359 & 0.630 & 0.323 & 24.66$^\circ$ & 60.91$^\circ$ \\
TAU-NIGENS2021 & STARSS23 & \texttt{footsteps} & 180,973 & 0.004 & 0.030 & 0.147 & 0.009 & 46.41$^\circ$ & 68.32$^\circ$ \\
TAU-NIGENS2021 & STARSS23 & \texttt{telephone} & 35,953 & 0.019 & 0.141 & 0.418 & 0.131 & 34.41$^\circ$ & 54.31$^\circ$ \\
TAU-NIGENS2021 & STARSS23 & \texttt{knock} & 1,717 & 0.296 & 0.630 & 0.860 & 0.540 & 16.84$^\circ$ & 33.78$^\circ$ \\
\hline
TAU-NIGENS2021 & TAU2019 & \texttt{telephone} & 268,153 & 0.328 & 0.734 & 0.917 & 0.689 & 13.63$^\circ$ & 28.55$^\circ$ \\
TAU-NIGENS2021 & TAU2019 & \texttt{knock} & 123,239 & 0.178 & 0.519 & 0.766 & 0.492 & 19.42$^\circ$ & 37.95$^\circ$ \\
\hline
TAU-NIGENS2021 & TAU-NIGENS2020 & \texttt{footsteps} & 257,358 & 0.233 & 0.553 & 0.747 & 0.518 & 18.10$^\circ$ & 46.56$^\circ$ \\
TAU-NIGENS2021 & TAU-NIGENS2020 & \texttt{alarm} & 240,614 & 0.220 & 0.626 & 0.878 & 0.523 & 17.08$^\circ$ & 31.02$^\circ$ \\
TAU-NIGENS2021 & TAU-NIGENS2020 & \texttt{telephone} & 232,735 & 0.142 & 0.418 & 0.666 & 0.360 & 23.02$^\circ$ & 51.31$^\circ$ \\
TAU-NIGENS2021 & TAU-NIGENS2020 & \texttt{baby\_cry} & 198,898 & 0.304 & 0.735 & 0.926 & 0.673 & 14.03$^\circ$ & 27.57$^\circ$ \\
TAU-NIGENS2021 & TAU-NIGENS2020 & \texttt{crash} & 172,370 & 0.236 & 0.669 & 0.904 & 0.627 & 15.89$^\circ$ & 29.65$^\circ$ \\
TAU-NIGENS2021 & TAU-NIGENS2020 & \texttt{dog\_bark} & 169,462 & 0.320 & 0.692 & 0.879 & 0.635 & 14.61$^\circ$ & 31.81$^\circ$ \\
TAU-NIGENS2021 & TAU-NIGENS2020 & \texttt{piano} & 138,222 & 0.164 & 0.468 & 0.721 & 0.419 & 20.90$^\circ$ & 44.58$^\circ$ \\
TAU-NIGENS2021 & TAU-NIGENS2020 & \texttt{male\_scream} & 93,841 & 0.246 & 0.576 & 0.843 & 0.520 & 17.53$^\circ$ & 33.70$^\circ$ \\
TAU-NIGENS2021 & TAU-NIGENS2020 & \texttt{knock} & 49,894 & 0.225 & 0.667 & 0.881 & 0.630 & 15.45$^\circ$ & 31.55$^\circ$ \\
TAU-NIGENS2021 & TAU-NIGENS2020 & \texttt{male\_speech} & 41,519 & 0.145 & 0.492 & 0.791 & 0.477 & 20.26$^\circ$ & 40.46$^\circ$ \\
TAU-NIGENS2021 & TAU-NIGENS2020 & \texttt{female\_scream} & 21,135 & 0.272 & 0.749 & 0.929 & 0.670 & 14.09$^\circ$ & 26.89$^\circ$ \\
TAU-NIGENS2021 & TAU-NIGENS2020 & \texttt{female\_speech} & 20,201 & 0.269 & 0.747 & 0.957 & 0.711 & 13.92$^\circ$ & 27.12$^\circ$ \\
\hline
TAU-NIGENS2021 & TAU-NIGENS2021 & \texttt{footsteps} & 733,399 & 0.196 & 0.510 & 0.708 & 0.458 & 19.61$^\circ$ & 53.49$^\circ$ \\
TAU-NIGENS2021 & TAU-NIGENS2021 & \texttt{alarm} & 629,828 & 0.257 & 0.669 & 0.880 & 0.588 & 15.60$^\circ$ & 31.72$^\circ$ \\
TAU-NIGENS2021 & TAU-NIGENS2021 & \texttt{piano} & 419,592 & 0.178 & 0.530 & 0.764 & 0.463 & 19.07$^\circ$ & 44.13$^\circ$ \\
TAU-NIGENS2021 & TAU-NIGENS2021 & \texttt{baby\_cry} & 387,944 & 0.289 & 0.658 & 0.855 & 0.622 & 15.08$^\circ$ & 34.17$^\circ$ \\
TAU-NIGENS2021 & TAU-NIGENS2021 & \texttt{crash} & 311,355 & 0.299 & 0.670 & 0.827 & 0.637 & 14.50$^\circ$ & 38.83$^\circ$ \\
TAU-NIGENS2021 & TAU-NIGENS2021 & \texttt{dog\_bark} & 254,970 & 0.283 & 0.682 & 0.869 & 0.646 & 14.75$^\circ$ & 32.99$^\circ$ \\
TAU-NIGENS2021 & TAU-NIGENS2021 & \texttt{telephone} & 191,188 & 0.193 & 0.532 & 0.757 & 0.472 & 18.93$^\circ$ & 45.59$^\circ$ \\
TAU-NIGENS2021 & TAU-NIGENS2021 & \texttt{male\_scream} & 179,879 & 0.188 & 0.584 & 0.861 & 0.532 & 17.93$^\circ$ & 32.93$^\circ$ \\
TAU-NIGENS2021 & TAU-NIGENS2021 & \texttt{knock} & 95,450 & 0.200 & 0.558 & 0.781 & 0.538 & 18.04$^\circ$ & 39.43$^\circ$ \\
TAU-NIGENS2021 & TAU-NIGENS2021 & \texttt{male\_speech} & 81,239 & 0.181 & 0.505 & 0.727 & 0.487 & 19.84$^\circ$ & 44.07$^\circ$ \\
TAU-NIGENS2021 & TAU-NIGENS2021 & \texttt{female\_speech} & 39,707 & 0.142 & 0.495 & 0.740 & 0.485 & 20.15$^\circ$ & 41.30$^\circ$ \\
TAU-NIGENS2021 & TAU-NIGENS2021 & \texttt{female\_scream} & 30,107 & 0.299 & 0.737 & 0.912 & 0.683 & 14.00$^\circ$ & 28.77$^\circ$ \\
\hline
\end{longtable}
\end{landscape}


\clearpage

\addcontentsline{toc}{section}{References}
\bibliographystyle{IEEEtran}
\bibliography{references}

\end{document}